UNIVERSITY OF CAMBRIDGE

# A type-theoretic approach to semistrict higher categories

Alexander Rice

Darwin College

18th April 2024

This thesis is submitted for the degree of Doctor of Philosophy

# Declaration

This thesis is the result of my own work and includes nothing which is the outcome of work done in collaboration except as declared in the preface and specified in the text. It is not substantially the same as any work that has already been submitted, or, is being concurrently submitted, for any degree, diploma or other qualification at the University of Cambridge or any other University or similar institution except as declared in the preface and specified in the text. It does not exceed the prescribed word limit for the relevant Degree Committee.

Alexander Rice
18th April 2024

# Abstract

## A type-theoretic approach to semistrict higher categories

*Alexander Rice*

Weak $\infty$-categories are known to be more expressive than their strict counterparts, but are more difficult to work with, as constructions in such a category involve the manipulation of explicit coherence data. This motivates the search for definitions of semistrict $\infty$-categories, where some, but not all, of the operations have been strictified.

We introduce a general framework for adding definitional equality to the type theory Catt, a type theory whose models correspond to globular weak $\infty$-categories, which was introduced by Finster and Mimram. Adding equality to this theory causes the models to exhibit *semistrict* behaviour, trivialising some operations while leaving others weak. The framework consists of a generalisation of Catt extended with an equality relation generated by an arbitrary set of equality rules $\mathcal{R}$, which we name $\text{Catt}_{\mathcal{R}}$. We study this framework in detail, formalising much of its metatheory in the proof assistant Agda, and studying how certain operations of Catt behave in the presence of definitional equality.

The main contribution of this thesis is to introduce two type theories, $\text{Catt}_{\text{su}}$ and $\text{Catt}_{\text{sua}}$, which are instances of this general framework. $\text{Catt}_{\text{su}}$, short for Catt with strict units, is a variant of Catt where the unitor isomorphisms trivialise to identities. It is primarily generated by a reduction we call *pruning*, which removes identities from composites, simplifying their structure. $\text{Catt}_{\text{sua}}$, which stands for Catt with strict units and associators, trivialises both the associativity and unitality operations of Catt, and is generated by a generalisation of pruning called *insertion*. Insertion merges multiple composites into a single operation, flattening the structure of terms in the theory.

Further, we provide reduction systems that generate the equality of both $\text{Catt}_{\text{su}}$ and $\text{Catt}_{\text{sua}}$ respectively, and prove that these reductions systems are strongly terminating and confluent. We therefore prove that the equality, and hence typechecking, of both theories is decidable. This is used to give an implementation of these type theories, which uses an approach inspired by normalisation by evaluation to efficiently find normal forms for terms. We further introduce a bidirectional typechecking algorithm used by the implementation which allows for terms to be defined in a convenient syntax where many arguments can be left implicit.

# Acknowledgements

I would firstly like to thank everyone that I have collaborated with over the course of my PhD, both for their contributions to the work that appears in this thesis, but also for their contributions to my development as a researcher. I would especially like to thank my supervisor, Jamie Vicary, whose guidance throughout was invaluable, for keeping my research on track despite the disruptions caused by the pandemic during the first years of my PhD.

I would also like to thank all the friends who have been with me at any point in this journey. I particularly want to show my appreciation (and apologise) to everyone who was bombarded with technical questions throughout the writing up of this text; I thoroughly enjoyed our discussions on correct typesetting and use of the English language.

Lastly, I would like to thank my family for supporting me throughout my entire education. I would not have made it to this point without them.

# Contents

# Introduction

The study of higher-dimensional structures is becoming more prevalent in both mathematics and computer science. *Higher categories* [Lei04; RV22], a broad term for many different generalisations categories which capture these higher-dimensional ideas, are a central tool for studying these structures. The "higher" nature of these categories typically corresponds to the existence of morphisms whose source and target may be other morphisms, instead of just objects. A common method of organising this data is by giving a set of $n$-cells for each $n \in \mathbb{N}$. A $0$-cell then corresponds to the objects of an ordinary category, and the source and target of an $(n+1)$-cell are given by $n$-cells.

These higher categories present in many forms, and have been characterised into a periodic table of categories [CG07a; CG07b]. Of particular interest are the $(n, k)$-categories for $n, k \in \mathbb{N} \cup \{\infty\}$, higher categories which contain $m$-cells for $m \leq n$, and whose $m$-cells are invertible for $m \leq k$. In mathematics, the study of $(\infty, 0)$-categories, known as $\infty$-groupoids, is motivated by the study of the homotopy structure of topological spaces [Bou16], where $n$-cells are given by paths in the topological space, with higher cells taking the form of homotopies between lower cells. In computer science, many applications have been found for $(n, n)$-categories for smaller $n$, more commonly referred to as $n$-categories, including quantum computing [HV19], logic [Bar91; Mel09], physics [BD95], and game theory [GHWZ18], among others [Str12].

The composition of 1-cells in an $n$-category functions identically to the composition of morphisms in a 1 category; two morphisms $f : x \to y$ and $g : y \to z$ can be composed to form a 1-cell $f * g : x \to z$. However, there are two distinct ways of composing 2-cells, depicted by the diagrams below:

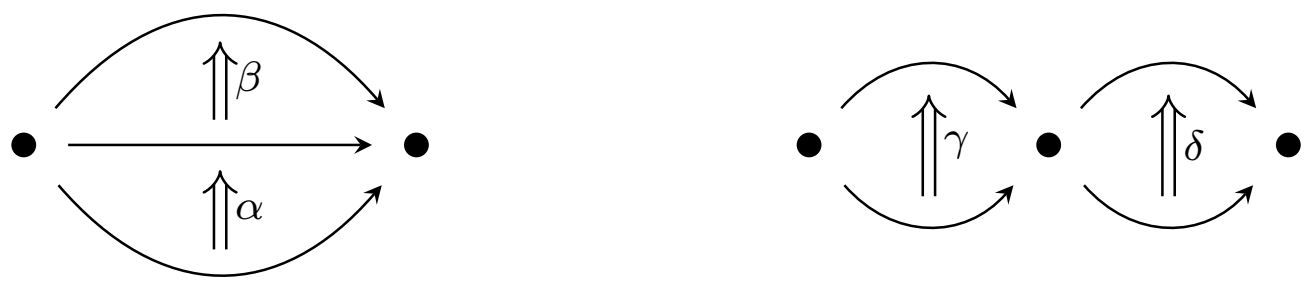

These diagrams mirror the concept of commutative diagrams for 1-categories, where spaces in the commutative diagram representing an equality have been replaced by 2-cell arrows. The first of these composites composes two 2-cells $\alpha$ and $\beta$ along a shared 1-cell boundary creating the vertical composite $\alpha \star_1 \beta$. The second composes the 2-cells $\gamma$ and $\delta$ along a 0-cell boundary and creates the horizontal composite $\gamma \star_0 \delta$. In higher dimensions, the pattern continues of having $n$ distinct ways of composing two $n$-cells. For each $n$-cell, there is also an identity $(n+1)$-cell.

Similarly to 1-categories, $n$-categories must satisfy various laws concerning their operations. These can be roughly organised into 3 groups:

- Associativity laws: Each of the composition operations in an $n$-category is associative.
- Unitality laws: The identity morphisms are a left and right unit for the appropriate composition operations.
- Interchange laws: These laws govern the relation between different compositions on the same cells. For any four 2-cells that form the following diagram:

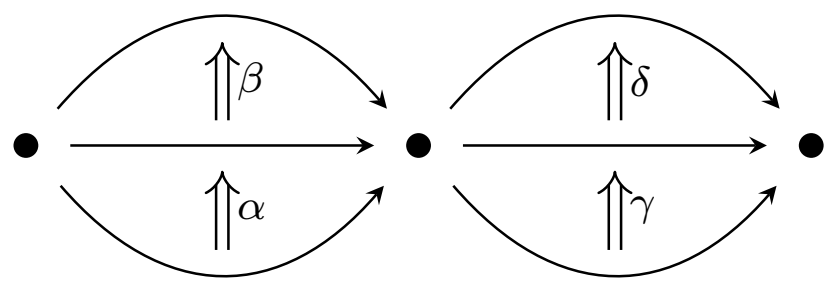

the first of the interchange laws states that two composites below are related:

$$(\alpha \star_1 \beta) \star_0 (\gamma \star_1 \delta) \simeq (\alpha \star_0 \gamma) \star_1 (\beta \star_0 \delta)$$

These laws can be combined to create non-trivial emergent behaviour in a form not seen in the theory of 1-categories. One critical example of this is known as the *Eckmann-Hilton* argument [EH62], which states that the composition of two scalars, morphisms from the identity to the identity, commute. The argument proceeds by moving the two scalars around each other, as depicted in Figure 1. This crucially uses both the interchange and unitality laws.

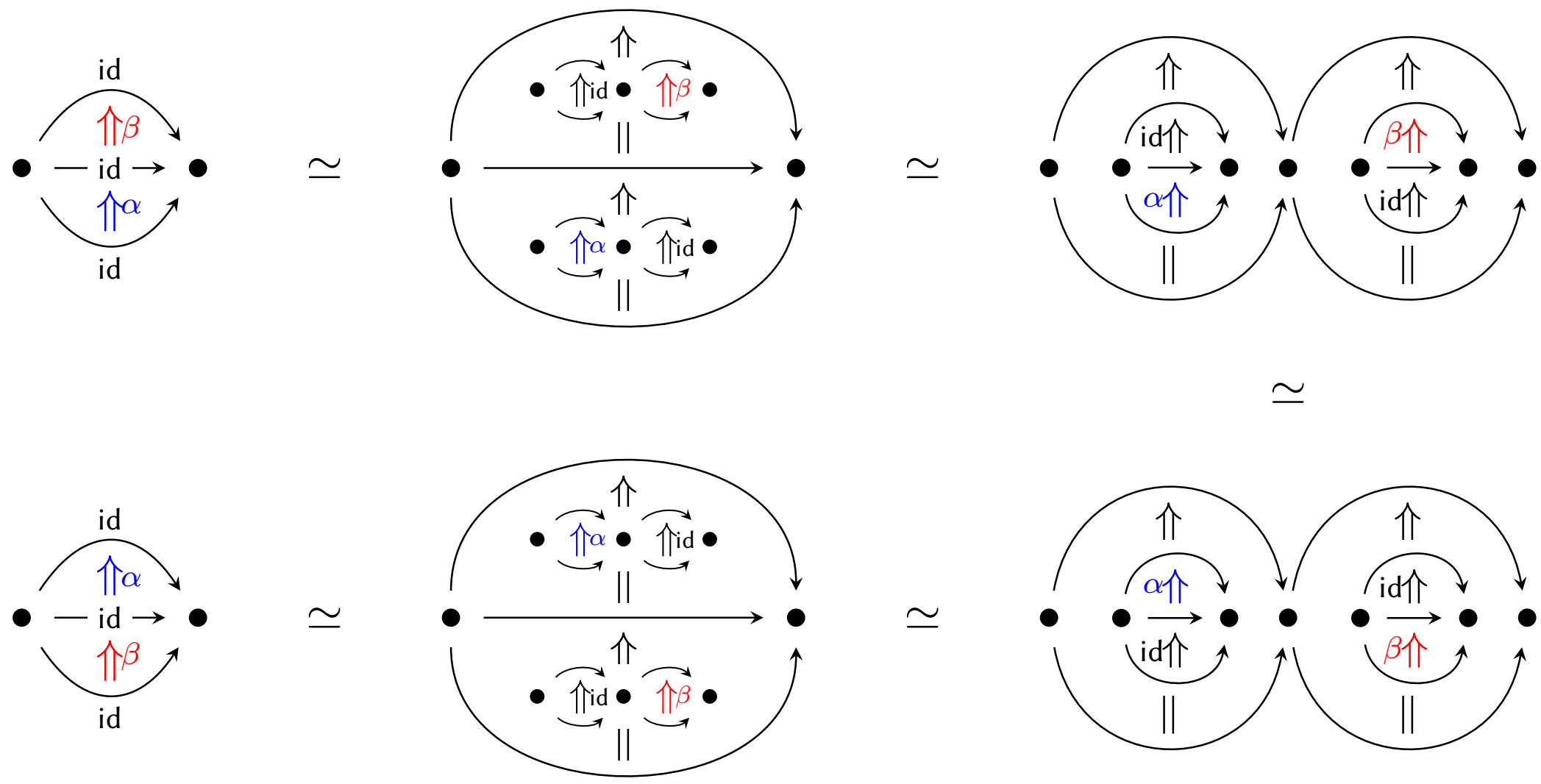

Figure 1: The Eckmann-Hilton argument.

**Semistrict higher categories** While we have given the types of laws that must hold in $n$-categories, we have not yet stated the full nature of these laws. By taking each of these laws to hold up to equality, one obtains the notion of a *strict* $n$-category. It is often the case in category theory that equality is the incorrect notion by which to compare objects, with the coarser relation of isomorphism being preferable. In the presence of higher-dimensional cells, arrows themselves can be compared up to isomorphism. This allows the laws for an $n$-category to be stated with isomorphism replacing equality, giving rise to the notion of *weak* $n$-category.

In such a weak $n$-category, each law is given by a set of isomorphisms, which are given as part of the data of the category. For the associativity law of three 1-cells $f$, $g$, and $h$, an invertible 2-cell known as the *associator* must be given, which takes the following form:

$$\alpha_{f,g,h} : (f * g) * h \to f * (g * h)$$

Similarly, the unit laws for a 1-cell $f$ are given by the *left unitor* $\lambda_f$ and the *right unitor* $\rho_f$ which take the following form:

$$\lambda_f : \mathsf{id} * f \to f \qquad \rho_f : f * \mathsf{id} \to f$$

Whereas two morphisms being equal is a property of those morphisms, an isomorphism between the same morphisms is a form of data, and the choice of isomorphism may not be unique. Weak higher categories therefore contain higher *coherence laws* which govern the interaction of these isomorphisms. These coherence laws can also be given as isomorphisms instead of equalities, and must satisfy their own coherence laws, leading to a tower of coherence laws. The amount of data needed to define an $n$-category therefore increases exponentially as $n$ increases.

In addition to the difficulty in defining a weak $n$-category, it is also more difficult to give proofs in a weak environment, due to the bureaucracy of working around the various coherence isomorphisms. Consider the proof of Eckmann-Hilton given in Figure 1. In a weak environment, we would hope to be able to simply replace each equality by the appropriate isomorphism, however doing so for the first equality in the proof would require us to give an isomorphism:

$$\alpha \cong \alpha * \mathsf{id}$$

Each side of this isomorphism has a different source and target, and hence no such isomorphism can be given in the globular setting used in this thesis. A full proof of Eckmann-Hilton is still possible but far more involved.

Weak categories are a more general notion than their strict counterparts, with every strict $n$-category generating a corresponding weak category by letting every coherence isomorphism be given by the identity morphism. For 2-categories, the converse is in fact possible; every weak 2-category is equivalent to a strict 2-category, allowing proofs for weak 2-categories to be given by instead proving the same property for strict 2-categories.

This is no longer possible in $n$-categories where $n \geq 3$. It was shown by Simpson [Sim98] that strict $n$-categories do not model the homotopy structure of all topological spaces, with the topological space $S^2$ having no interpretation. More concretely, we consider the morphism $\mathsf{EH}_{\alpha,\beta} : \alpha \star_1 \beta \to \beta \star_1 \alpha$ generated by the Eckmann-Hilton argument for scalars $\alpha$ and $\beta$. In a strict 3-category, this morphism is given by the identity and so:

$$\mathsf{EH}_{\alpha,\beta} \star_2 \mathsf{EH}_{\beta,\alpha} = \mathsf{id}$$

This equality does not hold in a general weak 3-category (even up to isomorphism), contradicting that each weak 3-category is equivalent to a strict 3-category.

This motivates the search for semistrict definitions of $n$-category: definitions where some operations are strict, yet do not lose the expressivity of weak $n$-categories. For 3-categories, two such definitions have been proposed:

- Joyal and Kock [JK07; JK13] define a monoidal 2-category (which can be viewed as a 3-category with a single 0-cell) which only has weak units and unitors, and is otherwise strict. They prove that all braided monoidal categories (weak 3-categories with a unique 0-cell and unique 1-cell) can be interpreted in this setting as the category of endomorphisms on the weak unit morphism.
- Gray-categories are a form of semistrict 3-categories for which all structure is strict except the interchanger, the isomorphism witnessing the interchange law. Gordon, Power, and Street [GPS95] prove that every weak 3-category is equivalent to a Gray-category.

It is non-trivial to even define such a notion of semistrict $n$-category for $n > 3$, let alone prove that it loses no expressivity over its weak counterpart. Simpson conjectures [Sim98] that having only the unit laws weak is sufficient to model all homotopy groupoids, $\infty$-groupoids arising from the homotopy of topological spaces, though it is unclear if such a definition has been given. Hadzihasanovic [Had19] defines weak higher categories based on *diagrammatic sets*. It could be argued that such a definition can model strict interchange, though the classes of diagrams that can be composed in this theory are restricted to those that are *spherical*, which disallows horizontal composites in the form stated above and makes comparison difficult. Batanin, Cisinski, and Weber [BCW13] define a notion of $\infty$-category with strict units based on the language of operads.

Definitions of semistrict $n$-categories which are strictly unital and associative have also been defined, primarily inspired by the graphical language of *string diagrams*. Bar and Vicary [BV17] define *quasi-strict* $4$*-categories*, where the associativity and unitality laws hold strictly up to equality. Dorn [Dor18] defines *associative* $n$*-categories*: a definition of strictly associative and unital $n$-category similarly based on geometric principles. Associative $n$-categories are further studied by Heidemann, Reutter, Tataru, and Vicary [RV19; HRV22; TV24], which has recently led to the construction of the graphical proof assistant homotopy.io [CHH+24] for manipulating higher-dimensional string diagrams. Similarly to the case for diagrammatic sets, the composition operations in these theories have a different form to those of strict $n$-categories, making comparison difficult. The connection between these definitions and geometry is studied by Dorn and Douglas [DD21] and Heidemann [Hei23].

**Type theory and higher categories** Deep links exist between higher category theory and type theory. The identity type in Martin-Löf type theory (MLTT) [Mar75] naturally leads to higher-dimensional structure; the identity type $s =_A t$ can be formed for any two terms $s$ and $t$ of type $A$, but this construction can be iterated since the identity type is a type itself, leading to higher identity types $p =_{s =_A t} q$ for $p, q : s =_A t$. Operations on this type are generated by the J-rule, an induction principle for the identity type. Independent proofs by Lumsdaine [Lum10] and Garner and van den Berg [GvdB10] show that the J-rule is sufficient to equip identity types with the appropriate operations to form a weak $\infty$-groupoid.

Terms of the identity type $s =_A t$ correspond to witnesses of the fact that $s$ and $t$ are equal, or can even be viewed as proofs of the equality. The study of these proofs as objects of study in their own right is known as *proof relevance*. Although the axiom of uniqueness of identity proofs (UIP), which states that any two terms of the identity type are themselves equal, is consistent with Mltt, it was shown that it is not provable by Hofmann and Streicher, who constructed a model of Mltt where types are interpreted as 1-groupoids, and identity types are non-trivial.

The $\infty$-groupoidal nature of Mltt is embraced in Homotopy type theory (Hott) [Uni13], where types are interpreted as topological spaces. The key component of Hott, the *univalence axiom*, which is incompatible with UIP, states that the identities between types are given by equivalences between these types, which need not be unique.

The models of Hott are equipped with more structure than is present in an $\infty$-groupoid, and are given by $\infty$-toposes [Shu19]. In the appendices of his thesis [Bru16], Brunerie defines a type theory for $\infty$-groupoids by removing all structure from Mltt which does not concern the identity type. This theory constructs the identity type similarly to Mltt, but replaces the J-rule with a rule stating that all terms over *contractible contexts* are equal. Finster and Mimram further refine this idea to produce the type theory Catt [FM17], a type theory for weak $\infty$-categories, using techniques from a definition of weak $\infty$-categories due to Maltsiniotis [Mal10] which itself is based on an earlier definition of $\infty$-groupoids which was given by Grothendieck [Gro83]. It was later shown [BFM24] that type-theoretic models of Catt coincide with $\infty$-categories defined by Maltsiniotis.

The type theory Catt is unusual, due to having no computation or equality rules. In the current work we leverage this to define new notions of semistrict $\infty$-category, by adding definitional equality to Catt. This equality unifies certain terms, which correspond to operations in a weak $\infty$-category, causing the semistrict behaviour of the resulting theories. This thesis develops a framework for working with equality relations in Catt, and uses this to define two new type theories, $\text{Catt}_{\text{su}}$ and $\text{Catt}_{\text{sua}}$:

- $\text{Catt}_{\text{su}}$ is a version of Catt which is strictly unital. It is primarily generated by the *pruning* reduction, a computation rule which removes unnecessary identities from more complex terms.
- $\text{Catt}_{\text{sua}}$ is Catt with strict unitors and associators. In this theory, pruning is replaced by a more general reduction which we call *insertion*, which merges multiple composites into a single composite, flattening the structure of terms in the theory. We claim to give the first algebraic definition of an $\infty$-category where the unitality and associativity laws hold strictly as models of $\text{Catt}_{\text{sua}}$.

The majority of the technical content of this thesis is concerned with proving standard metatheoretic properties of these type theories. This includes defining a notion of computation for each theory, given by demonstrating the existence of a confluent and terminating reduction system, which allows these theories to be implemented. This is used to produce interpreters for both theories, allowing complex constructions to be checked mechanically. We demonstrate the utility of this by formalising a proof of the *syllepsis*, a 5-dimensional term witnessing a commutativity property of the Eckmann-Hilton argument.

**Overview** We now give an overview of the content contained in each of the following chapters of the thesis.

- Chapter 1 gives an introduction to $\infty$-category theory. It defines strict $\infty$-categories and continues to define the definition of weak $\infty$-categories due to Maltsiniotis. The chapter ends by giving a definition of the type theory Catt, as defined by Finster and Mimram, and describing some preliminary well-known constructions in Catt.

- Chapter 2 introduces a general framework for studying variants of Catt with definitional equality relations generated from a set of rules $\mathcal{R}$, which we name $\text{Catt}_{\mathcal{R}}$. The chapter also states various properties concerning the metatheory of $\text{Catt}_{\mathcal{R}}$, including specifying conditions on the set of equality rules $\mathcal{R}$, under which the theory is well-behaved. The description of Catt in this chapter is comprehensive and self-contained, although lacks some exposition of the previous chapter. The type theory $\text{Catt}_{\mathcal{R}}$ is accompanied by an Agda formalisation, which is introduced in this chapter.

- Chapter 3 takes an arbitrary well-behaved variant of $\text{Catt}_{\mathcal{R}}$, and explores various constructions that can be formed in this setting. The primary purpose of this chapter is to introduce the *pruning operation*, which is done in Section 3.1, and the *insertion operation*, which is introduced in Section 3.4. Sections 3.2 and 3.3 build up theory about a certain class of contexts represented by trees, and terms that appear in these contexts. This theory is vital for a complete understanding of insertion.

- In Chapter 4, the type theories $\text{Catt}_{\text{su}}$ and $\text{Catt}_{\text{sua}}$ are finally defined in Sections 4.2 and 4.3 respectively, as variants of the framework $\text{Catt}_{\mathcal{R}}$. Preliminary results about both theories are proved, primarily by compiling results that have been stated in the previous two chapters. The main technical contribution of this section involves giving reduction systems for both theories, and giving proofs that these reductions systems are strongly terminating and globally confluent, hence making equality in these theories decidable.

  In Section 4.4, the decidability of equality is used to implement a typechecker for both theories $\text{Catt}_{\text{su}}$ and $\text{Catt}_{\text{sua}}$. The typechecker uses *normalisation by evaluation* (NbE) to reduce terms to a canonical form where they can be checked for equality. The section discusses the interaction of NbE with Catt, as well as discussing limitations of this approach in this setting.

  Section 4.5 discusses some properties of the models of these type theories, introducing a technique which we call *rehydration*, which "pads out" terms of the semistrict theory with the necessary coherences to produce a term of Catt which is equivalent to the original term. Rehydration can be seen as a conservativity result for the semistrict theories introduced at the start of the chapter. A proof of rehydration is given for the restricted case of terms over a certain class of context known as ps-contexts. This partial rehydration result is sufficient to determine that the semistrictness defined by $\text{Catt}_{\text{su}}$ and $\text{Catt}_{\text{sua}}$ is a property, a model of Catt can be a model of $\text{Catt}_{\text{su}}$ or $\text{Catt}_{\text{sua}}$ in at most one way. We further explore some obstructions to rehydration in a generic context.

  The thesis ends with a discussion of further variants of Catt and other options for future work.

Although results of later chapters depend on definitions and results of the preceding chapters,

a linear reading of this thesis is not essential. A reader who is already familiar with the type theory Catt can safely skip Chapter 1, and a reader who is only interested in the type theory $\textsc{Catt}_{\mathsf{su}}$ could read Chapter 2 followed by Sections 3.1 and 4.2. Similarly, a reader only interested in $\textsc{Catt}_{\mathsf{sua}}$ can ignore any content on the pruning construction. Section 4.4 may be of interest to a reader who is purely interested in the type-theoretic techniques used, and not the type theory Catt itself.

**Statement of authorship** The type theory $\textsc{Catt}_{\mathsf{su}}$ was originally developed in collaboration with Eric Finster, David Reutter, and Jamie Vicary, and was presented by the author at the Logic in Computer Science conference in 2022 [FRVR22]. $\textsc{Catt}_{\mathsf{sua}}$ will be presented at Logic in Computer Science 2024 [FRV24] and was developed in collaboration with Eric Finster and Jamie Vicary.

The author claims the development of the framework $\textsc{Catt}_{\mathcal{R}}$ and its accompanying Agda formalisation as individual contribution, as well as the implementation of $\textsc{Catt}_{\mathsf{su}}$ and $\textsc{Catt}_{\mathsf{sua}}$ which appears in Section 4.4.

# Chapter 1

# Background

We begin with an overview of the important concepts required for the rest of the thesis. Throughout, we will assume knowledge of various basic concepts from computer science, as well as a basic knowledge of category theory (including functor categories, presheaves, and (co)limits) and type theory. The primary purpose of the following sections is to introduce weak $\infty$-categories. While there are many differing definitions of $\infty$-categories (see [Lei01]), we focus here on models of the type theory Catt [FM17], which are known to be equivalent to a definition of Maltsiniotis [Mal10] based off an earlier definition by Grothendieck [Gro83], which we introduce in Section 1.1.2. In Section 1.2, we define the type theory Catt, similarly to how it was originally defined.

This section additionally serves as a place to introduce various syntax and notations which will be used throughout the rest of the thesis.

## 1.1 Higher categories

A higher category is a generalisation of the ordinary notion of a category to allow higher-dimensional structure. This manifests in the form of allowing arrows or morphisms to have their source or target be another morphism instead of an object. In this thesis, we are primarily concerned with models of $\infty$-categories, which are equipped with the notion of an $n$-cell for each $n \in \mathbb{N}$, where each $(n+1)$-cell has a source and target $n$-cell, and 0-cells play the role of objects in an ordinary category.

The role of objects is played by 0-cells, with 1-cells as the morphisms between these objects. For 0-cells $x$ and $y$, a 1-cell $f$ with source $x$ and target $y$ will be drawn as:

$$x \xrightarrow{f} y$$

or may be written as $f : x \to y$. Two cells are *parallel* if they have the same source and target. Between any two parallel $n$-cells $f$ and $g$, we have a set of $(n+1)$-cells between them. A 2-cell $\alpha : f \to g$ may be drawn as:

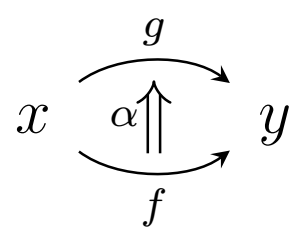

A 3-cell $\gamma$ between parallel 2-cells $\alpha$ and $\beta$ could be drawn as:

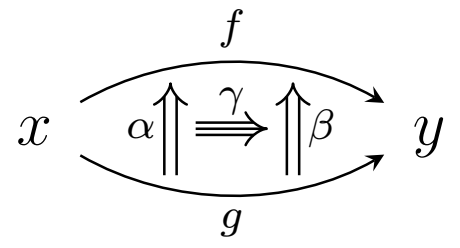

Just as in ordinary 1-category theory, we expect to be able to compose morphisms whose boundaries are compatible. For 1-cells, nothing has changed, given 1-cells $f : x \to y$ and $g : y \to z$ we form the composition $f * g$:

$$x \xrightarrow{f} y \xrightarrow{g} z$$

which has source $x$ and target $z$. We pause here to note that composition will be given in "diagrammatic order" throughout the whole thesis, which is the opposite of the order of function composition yet the same as the order of the arrows as drawn above. This is chosen as it will be common for us to draw higher-dimensional arrows in a diagram, and rare for us to consider categories where the higher arrows are given by functions. In an attempt to avoid confusion, we use an asterisk ($*$) to represent composition of arrows or cells in a higher category, and will use a circle ($\circ$) only for function composition.

In two dimensions, there is no longer a unique composition operation. For 2-cells $\alpha : f \to g$ and $\beta : g \to h$, the composite $\alpha *_1 \beta$ can be formed as before:

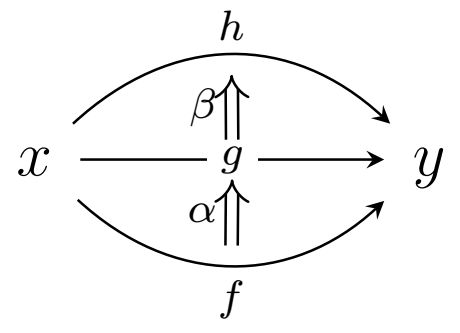

We refer to this composition as *vertical composition*. The cells $\gamma : i \to j$ and $\delta : k \to l$ can also be composed in the following way:

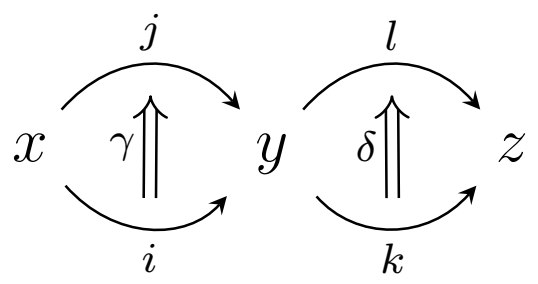

This composition is called the *horizontal composition*, and is written $\gamma *_0 \delta$. The subscript refers to the dimension of the shared boundary in the composition, with the 1-cell $g$ being the shared boundary in the vertical composition example and the 0-cell $y$ being the shared boundary in the horizontal composition example. The difference between the dimension of the cells being composed and the dimension of this shared boundary is known as the *codimension* of the composition, due to the similarity to the definition of the codimension of a vector subspace.

This pattern continues with 3-cells, which can be composed at codimension 0, 1, or 2, as depicted below:

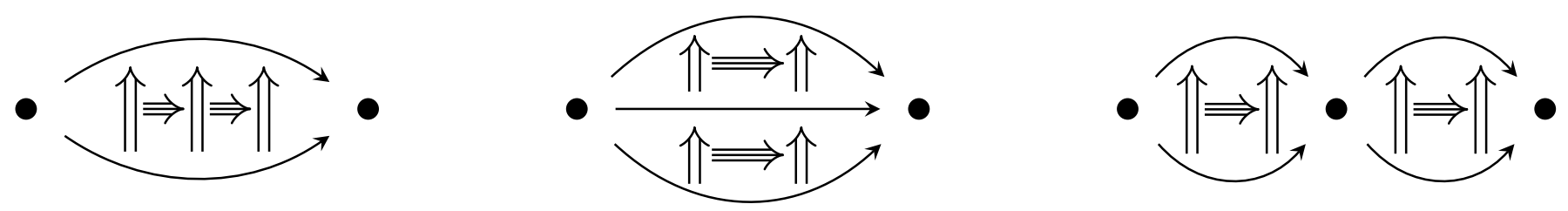

where the unlabelled arrows and objects (which are written •) are assumed to represent arbitrary potentially-distinct cells.

For every $n$-cell $x$, there is an $(n+1)$-cell $\mathsf{id}(x) : x \to x$, called the *identity morphism*.

Similarly to 1-categories, $\infty$-categories need to satisfy certain laws, which fall into 3 groups: associativity, unitality, and interchange. These laws can hold strictly, meaning that they hold up to equality, or weakly, meaning that they hold up to a higher-dimensional isomorphism. We delay the discussion of weak $\infty$-categories to Section 1.1.2, and begin with the discussion of strict $\infty$-categories.

In these strict categories, associativity laws are the same as for 1-categories, only now a law is needed for each composition (in every dimension and codimension). Unitality is again similar to the case for 1-categories, except we again need unitality laws for each composition. We note that for higher-codimensional compositions, an iterated identity is needed. For example, given a 2-cell $\alpha : f \to g$, the appropriate equation for left unitality of horizontal composition is:

$$\mathsf{id}(\mathsf{id}(x)) *_0 \alpha = \alpha$$

In general for a unit to be cancelled, it must be iterated a number of times equal to the codimension of the composition.

Interchange laws do not appear in 1-categories, and specify how compositions of different dimensions interact. The first interchange law states that for suitable 2-cells $\alpha$, $\beta$, $\gamma$, and $\delta$, that:

$$(\alpha *_0 \gamma) *_1 (\beta *_0 \delta) = (\alpha *_1 \beta) *_0 (\gamma *_1 \delta)$$

This can be diagrammatically depicted as:

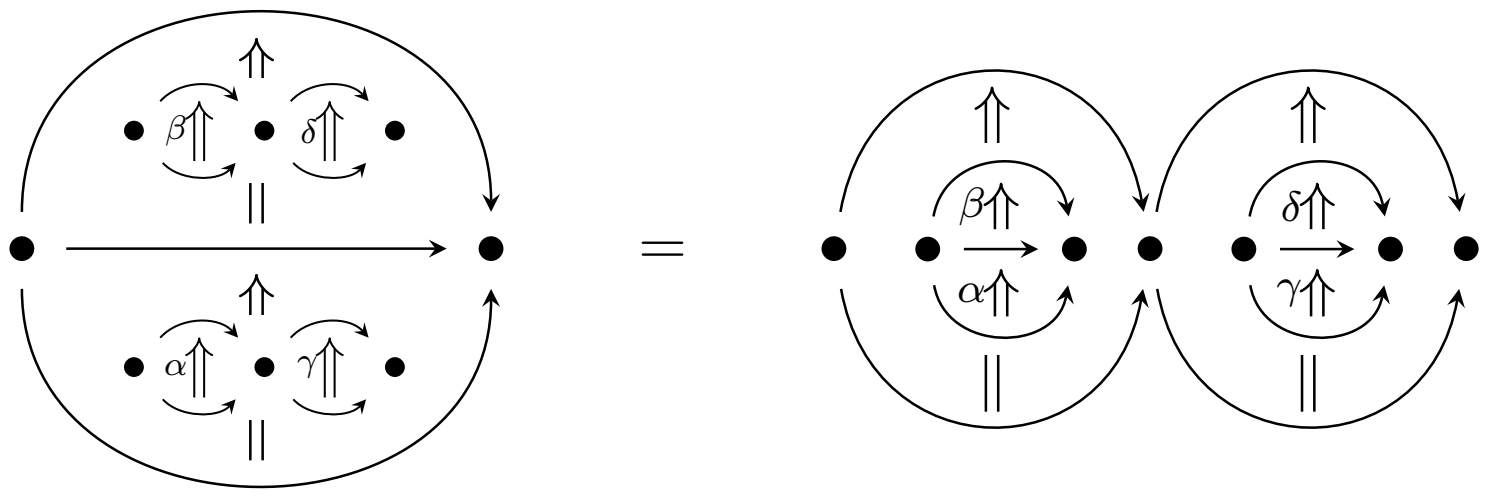

There are also interchange laws for the interaction of composition and identities; A composition of two identities is the same as an identity on the composition of the underlying cells.

The $\infty$-categories that we study in this thesis will be globular, meaning that their cells form a globular set. A globular set can be seen as natural extension of the data of a category, whose data can be arranged into the following diagram:

$$M \underset{t}{\overset{s}{\rightrightarrows}} O$$

where $O$ is a set of objects, $M$ is a set of all morphisms, and $s$ and $t$ are functions assigning each morphism to its source and target object respectively. 2-cells can be added to this diagram in a natural way:

$$C_2 \overset{s_1}{\underset{t_1}{\rightrightarrows}} C_1 \overset{s_0}{\underset{t_0}{\rightrightarrows}} C_0$$

In a globular set, the source and target of any cell must be parallel, meaning they share the same source and target. This condition is imposed by *globularity conditions.* Adding these and iterating the process leads to the following definition.

**Definition 1.1.1.** The category of globes $\mathbf{G}$ has objects given by the natural numbers and morphisms generated from $\mathbf{s}_n, \mathbf{t}_n : n \to n+1$ quotiented by the *globularity conditions*:

$$\mathbf{s}_{n+1} \circ \mathbf{s}_n = \mathbf{t}_{n+1} \circ \mathbf{s}_n$$
$$\mathbf{s}_{n+1} \circ \mathbf{t}_n = \mathbf{t}_{n+1} \circ \mathbf{t}_n$$

The category of globular sets **Glob**, is the presheaf category $[\mathbf{G}^{\mathrm{op}}, \mathbf{Set}]$.

Unwrapping this definition, a globular set $G$ consists of sets $G(n)$ for each $n \in \mathbb{N}$, with source and target maps $s_n, t_n : G(n+1) \to G(n)$, forming the following diagram:

$$\cdots \rightrightarrows G(3) \overset{s_2}{\underset{t_2}{\rightrightarrows}} G(2) \overset{s_1}{\underset{t_1}{\rightrightarrows}} G(1) \overset{s_0}{\underset{t_0}{\rightrightarrows}} G(0)$$

and satisfying the globularity conditions. A morphism of globular sets $F : G \to H$ is a collection of functions $G(n) \to H(n)$ which commute with the source and target maps.

Given a globular set $G$, we will call the elements of $G(n)$ the $n$-cells and write $f : x \to y$ for an $(n+1)$-cell $f$ where $s_n(f) = x$ and $t_n(f) = y$. We further define the $n$-boundary operators $\delta_n^-$ and $\delta_n^+$ which take the source or target respectively of a $(n+k)$-cell $k$ times, returning an $n$-cell.

*Example* 1.1.2. The $n$-disc $D^n$ is a finite globular set given by $Y(n)$, where $Y$ is the Yoneda embedding $\mathbf{G} \to \mathbf{Glob}$. $D^n$ has no $k$-cells for $k > n$, a single $n$-cell $d_n$, and two $m$-cells $d_m^-$ and $d_m^+$ for $m < n$. Every $(m+1)$-cell of $D^n$ has source $d_m^-$ and target $d_m^+$. The first few discs are depicted in Figure 1.1. The Yoneda lemma tells us that a map of globular sets $D^n \to G$ is the same as an $n$-cell of $G$. For an $n$-cell $x$ of $G$, we let $\{x\}$ be the unique map $D^n \to G$ which sends $d_n$ to $x$.

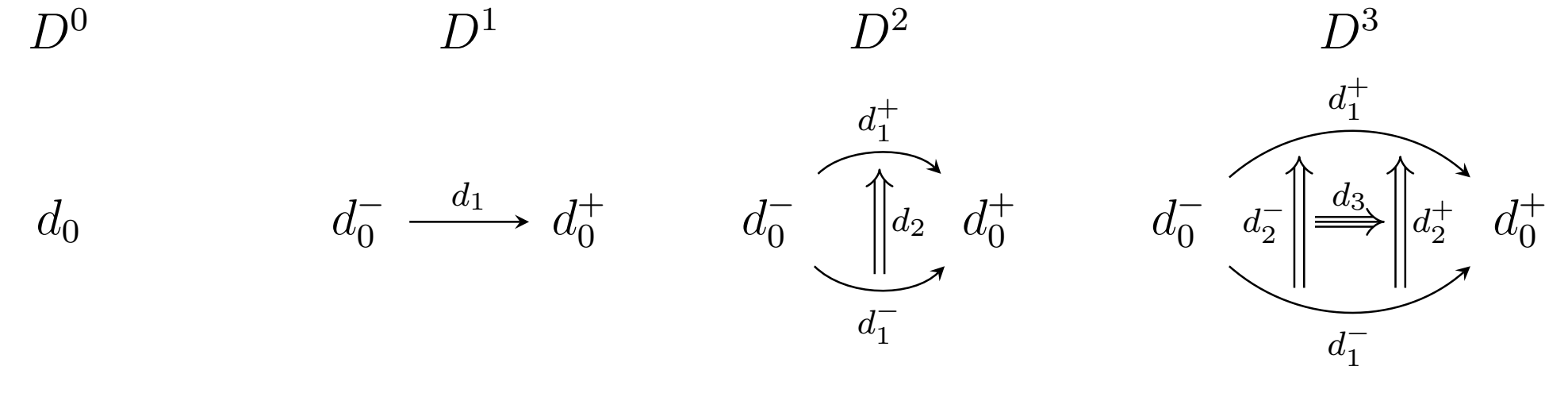

Figure 1.1: The first disc globular sets.

*Remark* 1.1.3. Globular sets are not the only natural extension of the data of a 1-category. The form of this data in a definition of a higher category is referred to as the *shape* of the cells. Notable alternatives to globular sets include simplicial sets, opetopic sets, and cubical sets.

We can now give the definition of a strict $\infty$-category.

**Definition 1.1.4.** A *strict $\infty$-category* is a globular set $G$ with the following operations:

- For $m < n$, a composition $*_m$ taking $n$-cells $f$ and $g$ with $\delta_m^+(f) = \delta_m^-(g)$ and producing an $n$-cell $f *_m g$ with:

$$s(f *_m g) = \begin{cases} s(f) & \text{if } m = n - 1 \\ s(f) *_m s(g) & \text{otherwise} \end{cases}$$
$$t(f *_m g) = \begin{cases} t(g) & \text{if } m = n - 1 \\ t(f) *_m t(g) & \text{otherwise} \end{cases}$$

- For any $n$-cell $x$, an identity $(n+1)$-cell $\mathsf{id}(x) : x \to x$.

and satisfying equalities:

- Associativity: Given $m < n$ and $n$-cells $f$, $g$, and $h$ with $\delta_m^+(f) = \delta_m^-(g)$ and $\delta_m^+(g) = \delta_m^-(h)$:
$$(f *_m g) *_m h = f *_m (g *_m h)$$

- Unitality: Given $m < n$ and $n$-cell $f$:

$$\mathsf{id}^{n-m}(\delta_m^-(f)) *_m f = f$$
$$f *_m \mathsf{id}^{n-m}(\delta_m^+(f)) = f$$

- Composition interchange: If $o < m < n$ and $\alpha$, $\beta$, $\gamma$, and $\delta$ be $n$-cells with

$$\delta_m^+(\alpha) = \delta_m^-(\beta) \qquad \delta_m^+(\gamma) = \delta_m^-(\delta) \qquad \delta_o^+(\alpha) = \delta_o^-(\gamma)$$

  then:
$$(\alpha *_o \gamma) *_m (\beta *_o \delta) = (\alpha *_m \beta) *_o (\gamma *_m \delta)$$

- Identity interchange: Let $m < n$ and $f$ and $g$ be $n$-cells with $\delta_m^+(f) = \delta_m^-(g)$. Then:

$$\mathsf{id}(f) *_m \mathsf{id}(g) = \mathsf{id}(f *_m g)$$

A morphism of $\infty$ categories is a morphism of the underlying globular sets which preserves composition and identities.

There is a clear forgetful functor from the category of strict $\infty$-categories to the category of globular sets, which has a left adjoint given by taking the free strict $\infty$-category over a globular set.

We end this section with an example of a non-trivial application of the axioms of an $\infty$-category, known as the Eckmann-Hilton argument. The argument shows that any two scalars (morphisms from the identity to the identity) commute.

**Proposition 1.1.5** (Eckmann-Hilton). *Let $x$ be an $n$-cell in an $\infty$-category and let $\alpha$ and $\beta$ be $(n+2)$-cells with source and target* $\mathsf{id}(x)$. *Then $\alpha *_{n+1} \beta = \beta *_{n+1} \alpha$.*

*Proof.* The cells $\alpha$ and $\beta$ can be manoeuvred around each other as follows:

$$\begin{aligned} &\alpha *_{n+1} \beta \\ &= (\alpha *_n i) *_{n+1} (i *_n \beta) && \text{Unitality} \\ &= (\alpha *_{n+1} i) *_n (i *_{n+1} \beta) && \text{Interchange} \\ &= \alpha *_n \beta && \text{Unitality} \\ &= (i *_{n+1} \alpha) *_n (\beta *_{n+1} i) && \text{Unitality} \\ &= (i *_n \beta) *_{n+1} (\alpha *_n i) && \text{Interchange} \\ &= \beta *_{n+1} \alpha && \text{Unitality} \end{aligned}$$

where $i = \mathsf{id}(\mathsf{id}(x))$. □

We give a more graphical representation of the proof in Figure 1, which appeared in the introduction. In this proof the $\alpha$ is moved to the left of $\beta$, though we equally could have moved it round the right, and the choice made was arbitrary.

### 1.1.1 Pasting diagrams

The definition of $\infty$-categories given in the previous section is close in spirit to the ordinary definitions of 1-categories and clearly demonstrates the different families of axioms present. However, we will see in Section 1.1.2 that these sorts of definitions do not scale well to our eventual setting of weak higher categories.

There is a special class of (finite) globular sets known as *pasting diagrams*, sometimes known as *pasting schemes*. The elements of the free strict $\infty$-category on a globular set $G$ can instead be represented by a pasting diagram equipped with a map into $G$. To do this, it must be possible to obtain a canonical composite from each pasting diagram.

Informally, we can define an $n$-dimensional pasting diagram to be a finite globular set which admits a unique full composite of dimension $n$, where a full composite of a globular set $G$ is an element of the free $\infty$-category over $G$ which uses all the maximal elements. This functions as the primary intuition on the role of pasting diagrams.

Pasting diagrams were used directly by Batanin [Bat98b] to give a definition of weak $\infty$-categories, and will be pivotal in Section 1.1.2 to define the variety of $\infty$-categories that Catt is based on. A more in-depth discussion of pasting diagrams, representations of free strict $\infty$-categories using them, and their use in the definition of weak $\infty$-categories can be found in *Higher operads, higher categories* [Lei04].

Before giving a more formal definition of pasting diagrams, we explore some examples and non-examples. In contrast to Leinster, we consider pasting diagrams as a full subcategory of globular sets, rather than a separate category with a function sending each pasting diagram to a globular set.

The disc globular sets introduced in Example 1.1.2 are all examples of pasting diagrams. The unique "composite" of these globular sets is just given by their maximal element, noting that we allow a singular cell in our informal definition of composite. The uniqueness of this is trivial as the only possible operations we could apply are compositions with units, which gives the same cell under the laws of an $\infty$-category.

The diagrams used to graphically represent our composition operations (of which we recall three below) are also pasting diagrams.

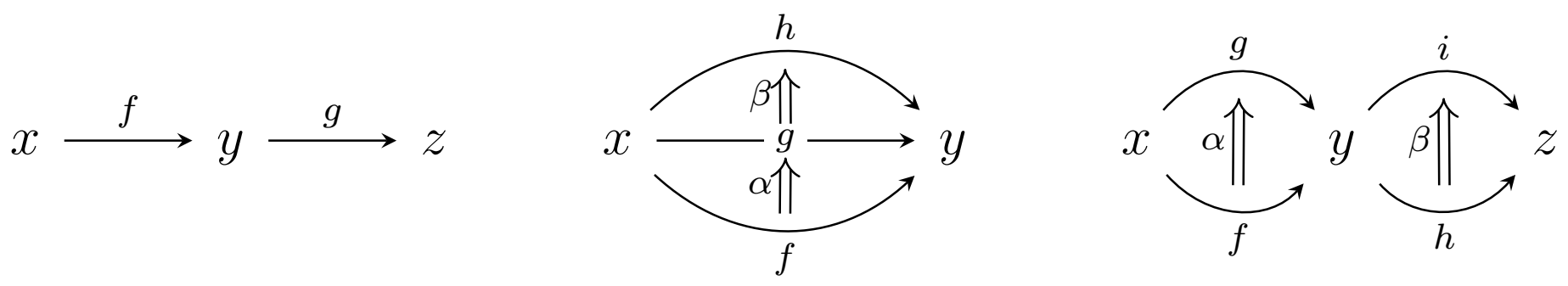

The composite of these diagrams is just the composite of the two maximal cells with the appropriate codimension.

We can also consider composites which are not binary composites of two cells of equal dimension. For example the following globular set is a pasting diagram:

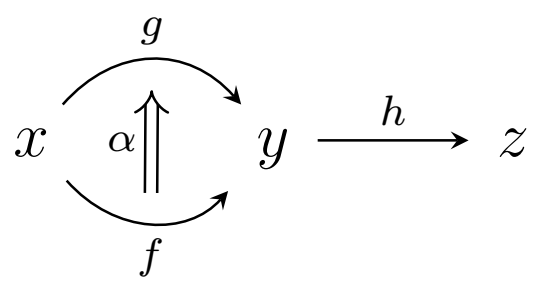

with a composite given by $\alpha *_0 \mathsf{id}(h)$. This operation is fairly common (in fact we have already seen it in Proposition 1.1.5) and is known as *whiskering*. In this case we would say that the composite is given by the right whiskering of $\alpha$ with $h$.

The 1-dimensional pasting diagrams are all given by chains of 1-cells of the form:

$$x_0 \xrightarrow{f_0} x_1 \xrightarrow{f_1} x_2 \xrightarrow{f_2} \cdots \xrightarrow{f_n} x_{n+1}$$

There are multiple ways to form a composite over these diagrams by repeated binary composition, however these all have the same result due to associativity.

Lastly we look at the following diagram, where all the 0-cells and 1-cells are assumed to be distinct:

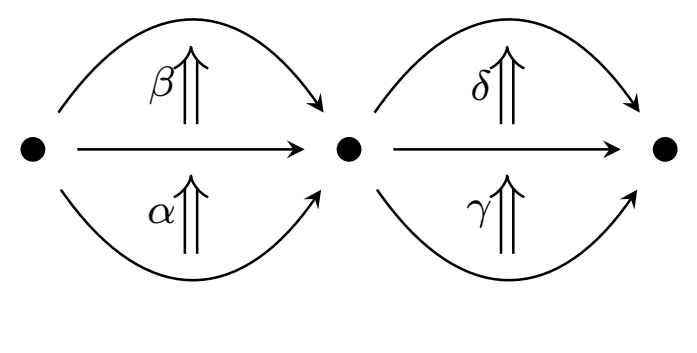

We get a composite given by $(\alpha *_1 \beta) *_0 (\gamma *_1 \delta)$. The uniqueness of this composite is due to the interchange law.

Non-examples of pasting diagrams roughly fall into two groups: those that do not admit a composite, and those that admit many distinct composites. The following three globular sets fail to admit a composite (the last is drawn in a box to emphasise that $z$ is part of the same globular set as $x$, $y$, $f$, $g$, and $\alpha$):

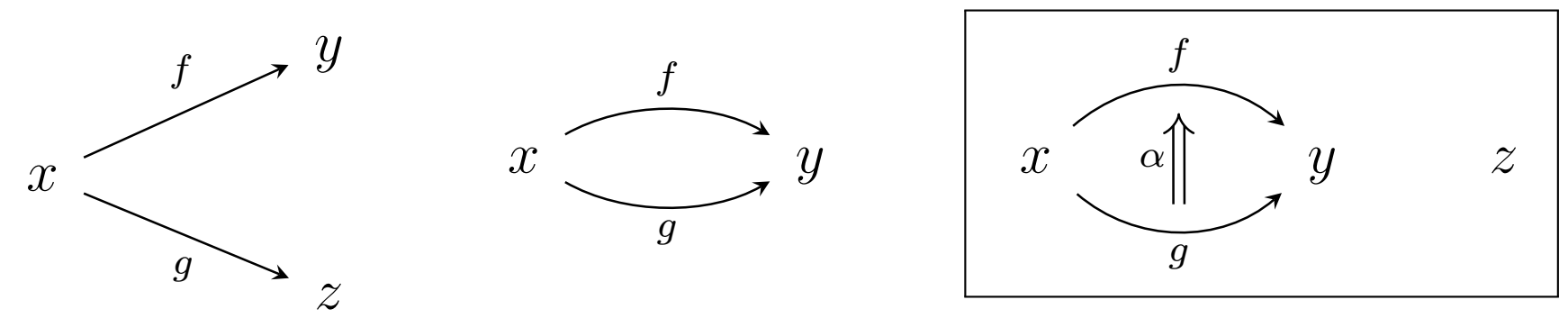

The globular set with a single 0-cell $x$, and a single 1-cell $f : x \to x$ has too many composites: $f$ and $f *_0 f$ need not be equal in an $\infty$-category.

To describe the free $\infty$-category in terms of pasting diagrams we need to be able to extract a composite from a pasting diagram, and construct a pasting diagram from an arbitrary composite. Each pasting diagram having a unique composite solves the former issue.

To be able to construct a pasting diagram from a composite, we wish to equip our set of pasting diagrams itself with the structure of an $\infty$-category. We therefore need our pasting diagrams to have a notion of boundary and a notion of composition. A natural candidate for composition is given by colimits, as **Glob** has all colimits due to being a presheaf category, and so it is sufficient for our class of pasting diagrams to be closed under these specific colimits. In fact, it is sufficient to contain a class of colimits known as *globular sums*.

**Definition 1.1.6.** A *globular structure* on a category $\mathcal{C}$ is a functor $D : \mathbf{G} \to \mathcal{C}$, specifying certain objects as discs in the category. A *globular sum* is a colimit of a diagram of the form:

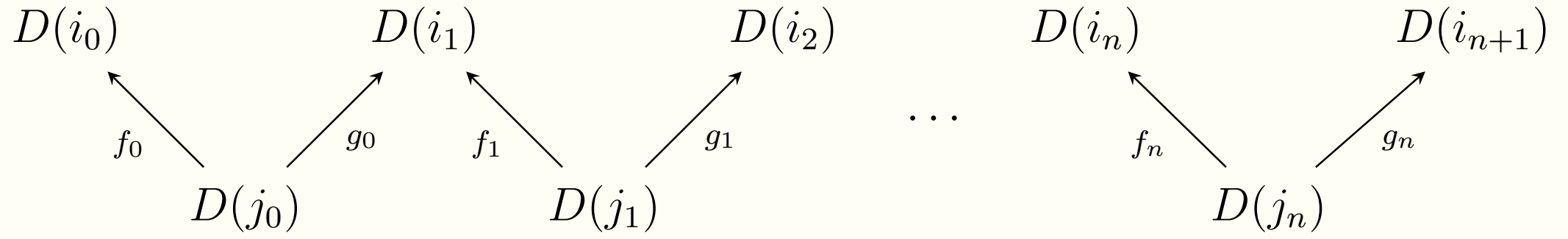

Where all morphisms $g_i$ are a composite of source maps ($D(\mathbf{s}_n)$ for some $n$) and the morphisms $f_i$ are a composite of target maps ($D(\mathbf{t}_n)$ for some $n$). Given that the maps $f_i$ and $g_i$ are uniquely determined, we may write such a globular sum as:

$$D(i_0) \amalg_{D(j_0)} D(i_1) \amalg_{D(j_1)} D(i_2) \cdots D(i_n) \amalg_{D(j_n)} D(i_{n+1})$$

A *globular extension* is a category equipped with a globular structure where all globular sums exist, and a morphism of globular extensions is a functor of the underlying categories commuting with the disc functors and preserving globular sums.

We can now give our first definition of a pasting diagram.

**Definition 1.1.7.** The category **Glob** has a globular structure $\mathbf{G} \to \mathbf{Glob}$ given by the Yoneda embedding. The category of *pasting diagrams*, **Pd**, is the full subcategory containing the globular sets which are globular sums. The boundary of an $(n+1)$-dimensional pasting diagram is given by replacing each instance of $D^{n+1}$ by $D^n$ in its globular sum representation. There are two canonical maps including the boundary into the original pasting diagram, whose images give the source and target of the pasting diagram.

The category of pasting diagrams has a globular structure $\mathbf{G} \to \mathbf{Pd}$ sending $n$ to $D^n$. It is a globular extension and is in fact the universal globular extension; it is initial in the category of globular extensions [Ara10].

We finish this section with one larger example.

*Example* 1.1.8. The following depicts a 2-dimensional pasting diagram.

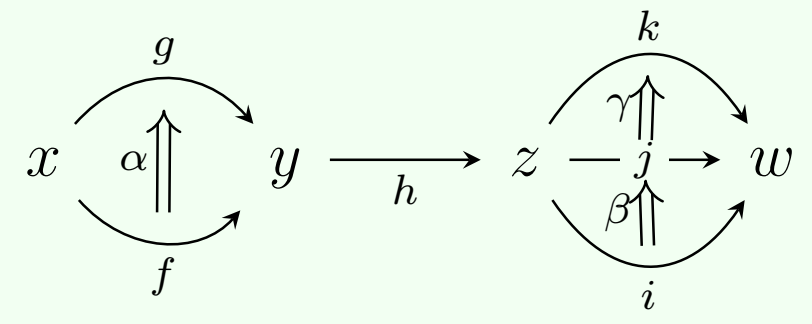

This has the following globular sum decomposition:

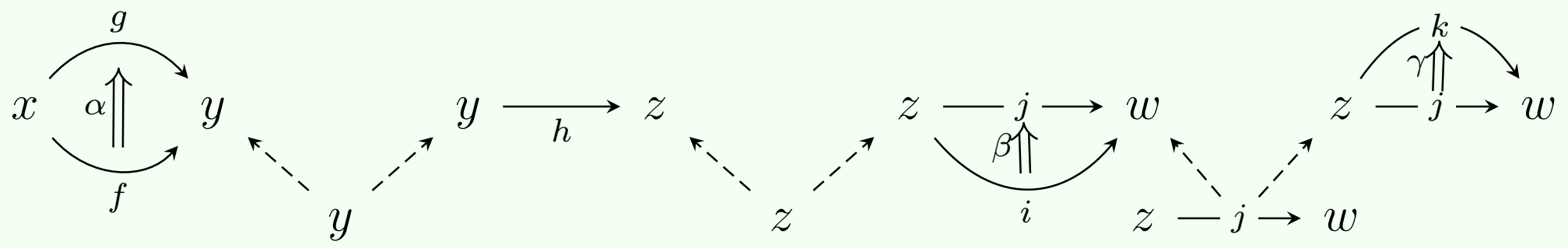

The source and target of the diagram are given by the isomorphic pasting diagrams:

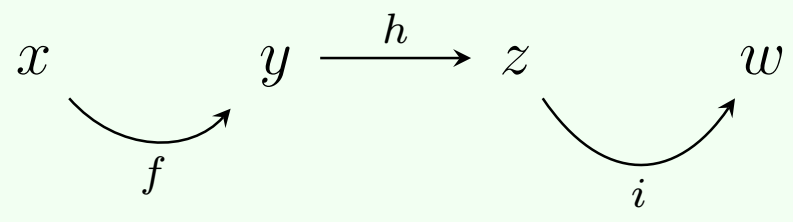

and

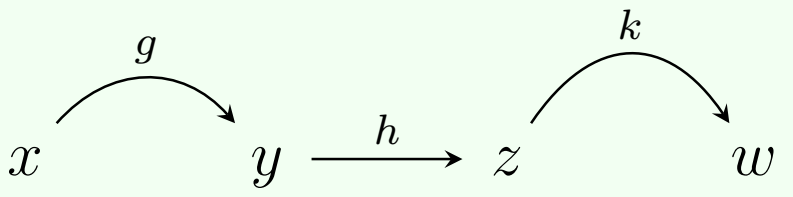

### 1.1.2 Weak higher categories

The $\infty$-categories we have defined so far have all been strict $\infty$-categories, meaning that the laws are required to hold up to equality. In ordinary 1-category theory, isomorphism is usually preferred over equality for comparing objects. Similarly, when we have access to higher-dimensional arrows, it follows that we can also consider isomorphisms between morphisms, and therefore consider laws such as associativity up to isomorphism instead of equality.

Topological spaces provide one of the primary examples for where it is useful to consider weak laws. Given a topological space $X$, we can define a globular set of paths and homotopies. Let the 0-cells be given by points $x$ of the topological space, let morphisms from $x$ to $y$ be given as paths $I \to X$ (where $I$ is the topological interval $[0,1]$) which send 0 to $x$ and 1 to $y$, and let higher cells be given by homotopies. The natural composition of two paths $p$ and $q$ is the following path:

$$(p * q)(i) = \begin{cases} p(2i) & \text{when } i < 0.5 \\ q(2i-1) & \text{when } i \geq 0.5 \end{cases}$$

which effectively lines up the paths end to end. Given 3 paths $p$, $q$, and $r$, the compositions $(p * q) * r$ and $p * (q * r)$ are not identical but are equal up to homotopy, meaning the two compositions are isomorphic. Therefore, in this case the composition $p * q$ does not form a strict $\infty$-category structure, but rather a weak structure.

**Weak 2-categories** We start our exploration of weak higher categories by considering the lower dimension case of bicategories (weak 2-categories). Here, interchange must still be given by a strict equality, as there are no non-trivial 3-cells in a 2-category. However, associativity and unitality can be given by isomorphisms known as associators and unitors:

$$\begin{aligned} \alpha_{f,g,h} &: (f *_0 g) *_0 h \to f *_0 (g *_0 h) \\ \lambda_f &: \mathsf{id}(x) *_0 f \to f \\ \rho_f &: f *_0 \mathsf{id}(y) \to f \end{aligned}$$

for $f : x \to y$, $g : y \to z$, and $h : z \to w$.

*Example* 1.1.9. All strict 2-categories are also bicategories. The bicategory of spans is an example of a bicategory which is not strict. Starting with a category $\mathcal{C}$ equipped with chosen pullbacks, we define the bicategory of spans over $\mathcal{C}$ to be:

- Objects are the same as $\mathcal{C}$
- Morphisms $A$ to $B$ are spans $A \leftarrow C \to B$.
- A 2-morphism from $A \leftarrow C \to B$ to $A \leftarrow C' \to B$ is a morphism $C \to C'$ such that the following diagram commutes:

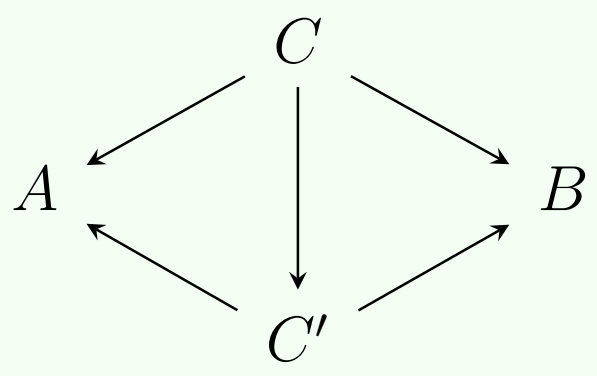

- Compositions and identities of 2-morphisms is given by composition and identities of the underlying morphisms in $\mathcal{C}$.
- The identity on an object $A$ is the span $A \leftarrow A \to A$.
- Given spans $A \leftarrow D \to B$ and $B \leftarrow E \to C$, their composite is given by the pullback:

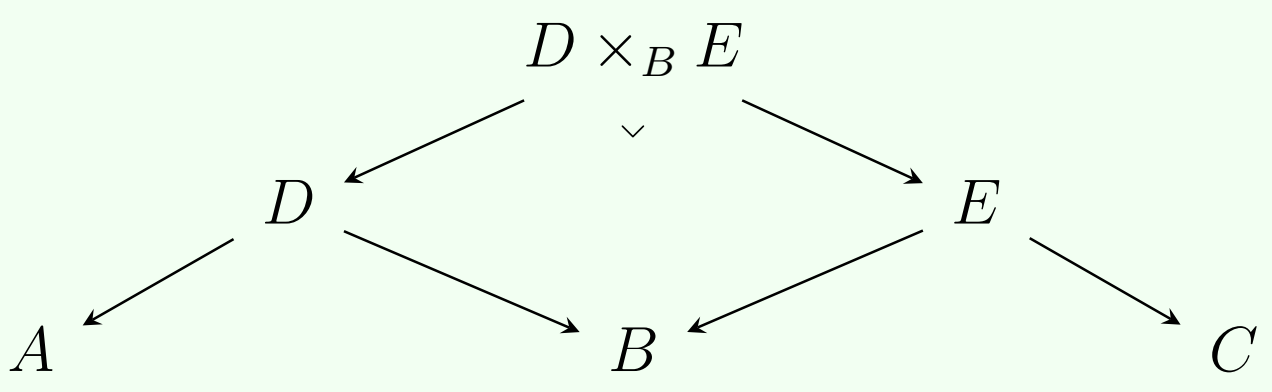

- Associators and unitors are given by the universal property of the pullback.

In general, there could be many possible isomorphisms between $(f * g) * h$ and $f * (g * h)$, and we require that the chosen morphisms satisfy certain compatibility properties. The first is that each of the associator, left unitor, and right unitor should be a natural isomorphism. The second is a property known as *coherence*, saying that any two parallel morphisms built purely from naturality moves, associators, and unitors must be equal.

For bicategories it is sufficient to give two coherence laws: the triangle equality and pentagon equality. The triangle equality identifies two ways of cancelling the identity in the composite $f * \mathsf{id} * g$, giving a compatibility between the left and right unitors. It is given by the following commutative diagram:

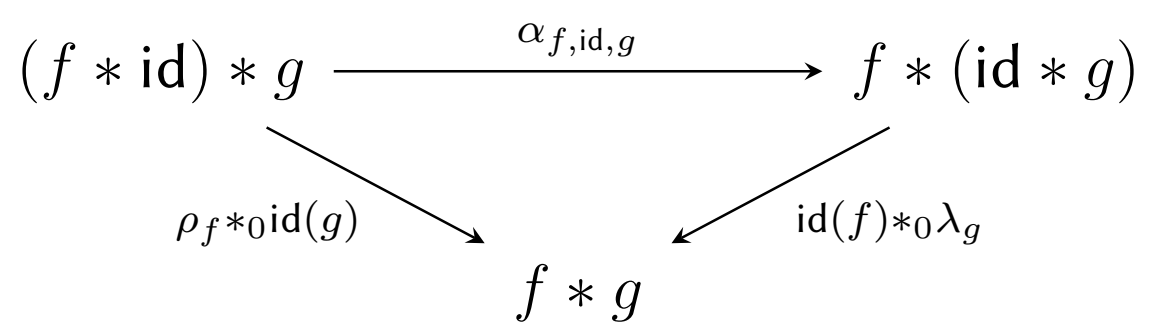

The pentagon equation identifies two ways of associating $((f * g) * h) * k$ to $f * (g * (h * k))$. It is given by the diagram below:

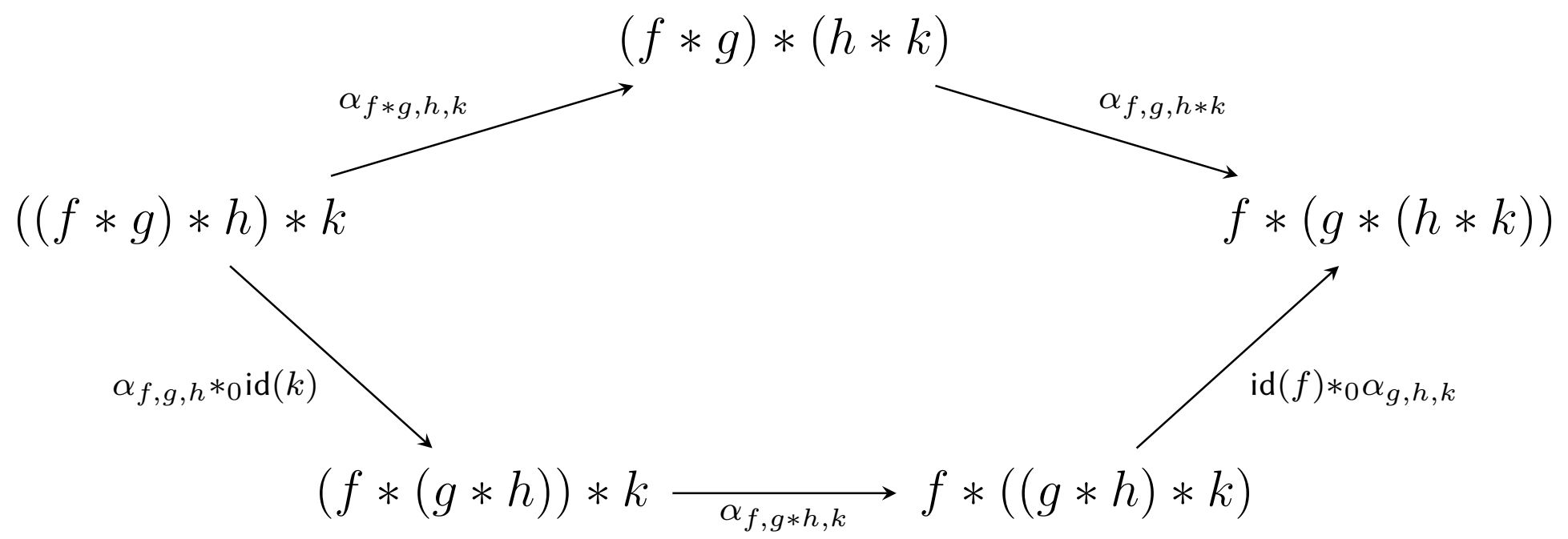

Surprisingly, these two equations are enough to give full coherence. For the example of spans from Example 1.1.9, these two equations follow from the uniqueness of the universal morphism.

**Weak $\infty$-categories** To move from weak 2-categories to weak 3-categories, new coherence cells for interchangers are added to replace the interchanger equalities, and new equalities must be added to specify the interaction between the interchangers and other coherence morphisms. Furthermore, the triangle and pentagon equations from 2-categories will become isomorphisms in a weak 3-category, causing more coherence equations to be added.

As we move up in dimension, the number of coherence morphisms and equalities required increases rapidly. A bicategory has 11 operations (1-identity, 2-identity, 1-composition, vertical composition, horizontal composition, left unitor (and inverse), right unitor (and inverse), and associator (and inverse)), whereas a fully weak tricategory already has around 51 operations [Gur06]. These numbers are obtained by unwrapping various subdefinitions and should be treated as approximate. Comparisons between the size of partially weak definitions can be found in [BV17].

Because of this complexity, we look for more uniform ways to represent the operations and axioms of an $\infty$-category. In this thesis, we will work with the type theory Catt, which is based on a definition of $\infty$-categories due to Maltsiniotis [Mal10], which is itself based on a definition of $\infty$-groupoid by Grothendieck [Gro83]. We will sketch the ideas behind these definitions here, and give a definition of Catt in Section 1.2.

The key insight behind Grothendieck's definition is that pasting diagrams should be contractible, instead of containing a unique composite. Whereas in a strict $\infty$-category, each pasting diagram effectively has 1 composite, in a weak $\infty$-category there can be many operations over a pasting diagram.

These operations are assembled into a globular extension called a *coherator*. A weak $\infty$-groupoid is then a presheaf on this coherator for which the opposite functor preserves globular sums (alternatively, the dual notion of globular product could be defined, and such a presheaf could be asked to preserve globular products). The objects of a coherator are given by pasting diagrams, with $D^n$ being sent to the $n$-cells of the category and other pasting diagrams being sent to composable sets of cells (as determined by the preservation of globular sums).

Operations over a pasting diagram $P$ in the coherator are given by morphisms $D^n \to P$. When we take a presheaf over this, we obtain a function that takes an $P$-shaped collection of cells to a single $n$-cell. Operations can be precomposed with source and target maps $D^{n-1} \to D^n$ to get the source and target of an operation. To build the coherator, we start by taking the category of pasting diagrams. The "operations" of this category consist solely of the inclusions of discs into pasting diagrams, which correspond to picking a single element from the pasting diagram. Other operations are then built using the following guiding principle.

**Guiding principle for groupoids.** Let $f$ and $g$ be two parallel operations over a pasting diagram $P$. Then there is an operation $h$ over $P$ with source $f$ and target $g$.

We define a pair of operations $f, g : D^n \to X$ to be *parallel* if $n = 0$ or both $n > 0$ and $f \circ \mathbf{s}_{n-1} = g \circ \mathbf{s}_{n-1}$ and $f \circ \mathbf{t}_{n-1} = g \circ \mathbf{t}_{n-1}$. A *lift* for such a pair of parallel operations is an operation $h : D^{n+1} \to X$ such that $h \circ \mathbf{s}_n = f$ and $h \circ \mathbf{t}_n = g$. Closing under this principle then amounts to inductively adding lifts for all parallel operations, while ensuring that the category remains a globular extension.

We start with some basic operations: Consider the pasting diagram $A = D^1 \amalg D^1$ given by:

$$x \xrightarrow{a} y \xrightarrow{b} z$$

Our rule now tells us that since $x$ and $z$ are elements of $A$, that there should be an operation returning a cell with source $x$ and target $z$, namely the composition of $a$ and $b$. In the language of coherators, there are operations $f, g : D^0 \to A$, where $f$ includes into the source of the first disc of $A$, and $g$ includes into the target of the second disc of $A$. These are trivially parallel, and so there exists a lift $h : D^1 \to A$, giving 1-composition. Similarly, if we take the pasting diagram with a single 0-cell $x$ and no other cells, then applying our rule with $f, g$ both being the operation returning the element $x$ produces an operation with source and target $x$, the identity on $x$.

We can generate more complicated operations with this principle, consider pasting diagram $B$:

$$x \xrightarrow{f} y \xrightarrow{g} z \xrightarrow{h} w$$

We already know the coherator contains 1-composition, and using composition and the universal property of globular sums, we can generate operations realising the compound composites $(f * g) * h$ and $f * (g * h)$. The principle then gives us an operation returning the 2-cell $(f * g) * h \to f * (g * h)$, which is of course the associator. This one principle allows us to generate all the structure we need, as well as structure that is arguably unnecessary, such as ternary compositions that did not appear in the definition of bicategory.

Unfortunately, as we have already mentioned, Grothendieck's definition is for $\infty$-groupoids, where everything is invertible, instead of $\infty$-categories in full generality, as we want to study in this thesis. This can be seen by taking the pasting diagram $C$:

$$x \xrightarrow{f} y$$

and applying the rule with $f$ returning $y$ and $g$ returning $x$, giving an operation that returns a 1-cell $f^{-1} : y \to x$, the inverse of $f$. The rule as we have stated it is too powerful.

Maltsiniotis' definition provides a solution to this problem by giving a more refined version of the principle. Whereas Grothendieck's definition treats all operations as coherences, Maltsiniotis' definition splits operations into two classes: compositions and equivalences. Both classes are obtained by restricting the classes of parallel operations that admit lifts.

We begin by defining what it means for an operation to be algebraic:

**Definition 1.1.10.** Let $\mathcal{C}$ be a globular extension for which the canonical functor $P : \mathbf{Pd} \to \mathcal{C}$ is faithful and the identity on objects. Then an operation $f : D^n \to X$ in $\mathcal{C}$ is *algebraic* if whenever $f = P(g) \circ f'$, $g = \mathsf{id}$.

Intuitively, an operation is algebraic when it does not factor through any proper inclusion. Algebraicity is equivalent to requiring that an operation makes use of all the locally maximal elements of the pasting diagram, elements which do not appear in the source or target of a higher-dimensional element of the diagram.

Equivalences contain the various invertible laws of our $\infty$-categories such as associators, unitors, identities, and interchangers. For two operations $f, g : D^n \to X$ to admit a lift under the rule for equivalences, they must both be algebraic. This gives the following rule:

**Guiding principle for categories** (Equivalences)**.** Let $f$ and $g$ be two parallel operations over a pasting diagram $P$. If both $f$ and $g$ use all locally maximal variables of $P$, then there is an operation over $P$ with source $f$ and target $g$.

Clearly any operations generated by this principle are invertible, as the extra condition imposed is symmetric. For compositions, we introduce the following asymmetric principle, recalling that pasting diagrams are equipped with source and target inclusions, and letting $\partial^-(P)$ and $\partial^+(P)$ be the images of these inclusions:

**Guiding principle for categories** (Composites)**.** Let $f$ and $g$ are parallel operations over a (non-singleton) pasting diagram $P$ such that $f$ uses all locally maximal cells of $\partial^-(P)$ and no cells outside of $\partial^-(P)$ and $g$ uses all locally maximal cells of $\partial^+(P)$ and no cells outside of $\partial^+(P)$. Then there is an operation over $P$ with source $f$ and target $g$.

The condition required to form a composite can be expressed by the operation $f : D^n \to P$ factoring into an algebraic map composed with the source inclusion into $P$, and similar for $g$ with the target inclusion. It can be easily checked that the inverse operation given above does not satisfy the criteria for being an equivalence or composite.

As with Grothendieck's definition, a coherator can be made by closing the globular extension of pasting diagrams under these restricted principles, and then weak $\infty$-categories can be defined to be presheaves on this coherator such that the opposite functor preserves globular sums.

*Remark* 1.1.11. We have claimed that a coherator can be formed by closing under adding lifts to parallel operations, though this is not precise and there are actually multiple ways of performing this closure that lead to different coherators. For example, one could add the lift for 1-composition twice, to get two distinct 1-composition operations, as long as one also added a lift between these now parallel operations. Grothendieck gives a general schema for producing (invertible) coherators, and conjectures that any two of these give rise to equivalent models of $\infty$-groupoids. Despite the similarities in their constructions, the author is not aware of a more general definition of coherator which includes both the Grothendieck and Maltsiniotis variants and has more structure than a globular extension.

We now turn our attention back to the proof of Eckmann-Hilton from Figure 1. Given a 0-cell $x$ and two scalars $\alpha, \beta : \mathsf{id}(x) \to \mathsf{id}(x)$, we expect the Eckmann-Hilton argument to give us an isomorphism in a weak higher category, rather than the equality obtained in the strict case. In fact, we immediately see that equalities 2, 3, and 4 in the proof can be immediately replaced by isomorphisms (interchangers and unitors).

The first and last equalities however are more problematic, although at first we may believe that there should exist some horizontal unitor isomorphism, upon closer inspection the two compositions do not even have the same boundary and so are not parallel. The composition $\alpha *_1 \beta$ has source and target $\mathsf{id}(x)$, whereas the source of $\alpha *_0 \mathsf{id}(\mathsf{id}(x))$ is $\mathsf{id}(x) *_0 \mathsf{id}(x)$.

To recover the proof in a weak setting, the intermediate composites must be composed with unitors so that they all have source and target $\mathsf{id}(x)$. To give equivalences for the first and last step, these unitors must be moved around with naturality moves, and at a critical point the isomorphism $\lambda_{\mathsf{id}(x)} \simeq \rho_{\mathsf{id}(x)}$ is required. Multiple full proofs of Eckmann-Hilton will be given in Section 4.4.4. The proof of Eckmann-Hilton is vastly simpler in the strict case, mainly due to the presence of the equation $\mathsf{id}(x) *_0 \mathsf{id}(x) = \mathsf{id}(x)$.

### 1.1.3 Computads

A free group is generated by a set, and a free category is generated by a directed graph, and so it is a natural question what the generating data for a free $\infty$-category is. We have already seen that a free $\infty$-category can be generated by a globular set, but free $\infty$-categories can also be generated by data that does not form a globular set.

Consider the minimum data needed to state the Eckmann-Hilton principle (see Figure 1 or Proposition 1.1.5). We require a single 0-cell $x$, and two 2-cells $\alpha, \beta : \mathsf{id}(x) \to \mathsf{id}(x)$. This data does not form a globular set as, for example, the source of the 2-cell $\alpha$ is not in the generating data, but is rather an operation applied to the data. We could try to remedy this by adding a new 1-cell $f$ to the data to represent $\mathsf{id}(x)$, but then the connection between $\mathsf{id}(x)$ and $f$ would be lost and $f$ and $\mathsf{id}(x)$ would be distinct in any free $\infty$-category generated on this data.

The correct generating data for an $\infty$-category is a *computad*. A version for 2-categories was introduced by Street [Str76], which allows a generating 2-cell to have a composite or identity as its source or target. These were extended to strict $\infty$-categories by Burroni [Bur93] and weak $\infty$-categories by Batanin [Bat98a], which allow the source and target of an $n$-cell to be any $(n-1)$-cell of the free $\infty$-category generated by the lower-dimensional data.

A modern approach to computads for weak $\infty$-categories is given by Dean, Finster, Markakis, Reutter, and Vicary [DFM+24], which avoids much of the complexity of globular operads, relying only on (mutual) structural induction. This definition of a computad is much closer in style (and is inspired by) the type theory Catt which we review in Section 1.2.

## 1.2 The type theory Catt

In this section we give an overview of the dependent type theory Catt [FM17]. Catt serves as a definition of weak $\infty$-categories, by defining a weak $\infty$-category to be a model of the type theory (e.g. using categories with families [Dyb96]). In Chapter 2, we give a more general and comprehensive presentation of Catt, allowing the addition of equality relations to the type theory, pre-empting Chapter 4. In contrast, this section presents the version of Catt closer to the one found in the literature, and compares its various constructions to the ideas introduced in Section 1.1.2.

### 1.2.1 Syntax of Catt

Catt has 4 classes of syntax: contexts, terms, types, and substitutions.

- Contexts contain a list of variables with an associated type. We can consider contexts as finite computads, the generating data for a weak $\infty$-category (see Section 1.1.3). It is alternatively valid to consider contexts in Catt as finitely generated $\infty$-categories. The set of contexts contains all finite globular sets (and hence all pasting diagrams).
- Terms over a context $\Gamma$ correspond to the operations from Section 1.1.2. Terms can either be a variable, which corresponds to the operations which pick a single cell out of a globular set, or those generated by the unique constructor Coh, which correspond to the operations generated by lifting. A term over a context $\Gamma$ can also be seen as an element of the free $\infty$-category generated from $\Gamma$.
- Types over a context $\Gamma$ consist of a collection of terms over the same context, and contain the boundary information for a term. Types either take the form of the constructor $*$, the type of 0-cells (which have no boundary data), or an arrow type $s \to_A t$, where $s$ and $t$ are terms giving the source and target of the boundary and the type $A$ gives lower-dimensional boundary information. This can be viewed as a directed version of the equality type $s =_A t$ from Martin-Löf type theory.
- Substitutions from a context $\Gamma$ to a context $\Delta$ are a mapping from variables of $\Gamma$ to terms of $\Delta$. These play the role of functors between the $\infty$-categories generated by $\Gamma$ and $\Delta$ and are also syntactically crucial for forming compound composites in the theory.

$$\frac{}{\emptyset : \mathsf{Ctx}} \qquad \frac{\Gamma : \mathsf{Ctx} \qquad A : \mathsf{Type}_\Gamma}{\Gamma, (x : A) : \mathsf{Ctx}}$$

$$\frac{}{\langle\rangle : \emptyset \to \Gamma} \qquad \frac{\sigma : \Delta \to \Gamma \qquad t : \mathsf{Term}_\Gamma \qquad A : \mathsf{Type}_\Delta}{\langle \sigma, t\rangle : \Delta, (x : A) \to \Gamma}$$

$$\frac{}{\star : \mathsf{Type}_\Gamma} \qquad \frac{A : \mathsf{Type}_\Gamma \qquad s : \mathsf{Term}_\Gamma \qquad t : \mathsf{Term}_\Gamma}{s \to_A t : \mathsf{Type}_\Gamma}$$

$$\frac{x \in \mathsf{Var}(\Gamma)}{x : \mathsf{Term}_\Gamma} \qquad \frac{\Delta : \mathsf{Ctx} \qquad A : \mathsf{Type}_\Delta \qquad \sigma : \Delta \to \Gamma}{\mathsf{Coh}_{(\Delta\,;\,A)}[\sigma] : \mathsf{Term}_\Gamma}$$

Figure 1.2: Syntax constructions in Catt.

The rules for constructing each piece of syntax are given in Figure 1.2. To simplify the notation, we may avoid writing substitutions in a fully nested fashion, writing $\langle \sigma, s, t \rangle$ instead of $\langle \langle \sigma, s \rangle, t \rangle$, or $\langle s \rangle$ instead of $\langle \langle \rangle, s \rangle$. We may also omit the subscript in the arrow type. As opposed to the original paper on Catt, we index terms, types, and substitutions over contexts, allowing us to avoid any problems with substitution only extending to a partial operation on terms. We write Ctx for the set of contexts, $\mathsf{Term}_\Gamma$ for the set of terms in a context $\Gamma$, $\mathsf{Type}_\Gamma$ for the set of types in a context $\Gamma$, and write $\sigma : \Delta \to \Gamma$ when $\sigma$ is a substitution taking variables of $\Delta$ to terms of $\Gamma$.

*Remark* 1.2.1. In the literature, substitutions are often written as going in the opposite direction. We emphasise here that the direction of our substitution morphisms agrees with the direction of the function from variables to terms, the direction of the induced functor between the $\infty$-categories freely generated from the domain and codomain contexts, and the direction of arrows in a Grothendieck coherator.

We write $\equiv$ for *syntactic equality*, up to renaming of variables and $\alpha$-equivalence. The various pieces of syntax will be considered as equal up to this relation, which can be achieved by using a de Bruijn index representation of the syntax as we present in Chapter 2 for the formalisation. However, we continue to use named variables in the prose of the thesis to aid readability, assuming that all variables in a context are always distinct. We contrast this with the equality symbol, $=$, which will represent the equality derived from extra equality rules we have placed on Catt in Section 2.2, and will be referred to as *definitional equality*.

The action of a substitution $\sigma : \Delta \to \Gamma$ can be extended from variables to all terms $t \in \mathsf{Term}_\Delta$, types $A \in \mathsf{Type}_\Delta$, and substitutions $\tau : \Theta \to \Delta$ by mutual recursion:

$$\begin{aligned}
x[\![\sigma]\!] &= t && \text{if } (x \mapsto t) \in \sigma \\
\mathsf{Coh}_{(\Theta\,;\,A)}[\tau][\![\sigma]\!] &= \mathsf{Coh}_{(\Theta\,;\,A)}[\tau \bullet \sigma] \\
\star[\![\sigma]\!] &= \star \\
s \to_A t[\![\sigma]\!] &= s[\![\sigma]\!] \to_{A[\![\sigma]\!]} t[\![\sigma]\!] \\
\langle\rangle \bullet \sigma &= \langle\rangle \\
\langle \tau, t \rangle \bullet \sigma &= \langle \tau \bullet \sigma, t[\![\sigma]\!] \rangle
\end{aligned}$$

For every context $\Gamma$, there is an identity substitution $\mathsf{id}_\Gamma$, which sends every variable to itself, which along with composition of substitutions above gives a category of contexts and substitutions.

The coherence constructor $\mathsf{Coh}_{(\Delta\,;\,A)}[\sigma]$ allows us to construct lifts between parallel operations over pasting diagrams. The context $\Delta$ plays the role of the pasting diagram. The type $A$ will always be of the form $s \to_B t$, and the terms $s$ and $t$ play the role of the parallel operation (with the type $s \to_B t$ being well-formed ensuring that $s$ and $t$ are parallel). The substitution $\sigma : \Delta \to \Gamma$ holds the data of a set of arguments to the coherence, allowing compound composites/operations to be formed and taking the role of composition of morphisms in the coherator.

We next define the free variables of each piece of syntax. These will be used to encode the condition of an operation being algebraic from the theory of non-invertible coherators. Let $\mathrm{Var}(\Gamma)$ denote the variables of $\Gamma$. For a term $t \in \mathsf{Term}_\Gamma$, a type $A \in \mathsf{Type}_\Gamma$ and a substitution $\sigma : \Delta \to \Gamma$ we define their free variables $\mathrm{FV}(t), \mathrm{FV}(A), \mathrm{FV}(\sigma) \subseteq \mathrm{Var}(\Gamma)$ by mutual recursion.

$$\begin{aligned}
\mathrm{FV}(x) &= \{x\} && \text{if } x \text{ is a variable}\\
\mathrm{FV}(\mathsf{Coh}_{(\Delta\,;\,A)}[\sigma]) &= \mathrm{FV}(\sigma)\\
\mathrm{FV}(\star) &= \{\}\\
\mathrm{FV}(s \to_A t) &= \mathrm{FV}(s) \cup \mathrm{FV}(A) \cup \mathrm{FV}(t)\\
\mathrm{FV}(\langle\rangle) &= \{\}\\
\mathrm{FV}(\langle \sigma, t\rangle) &= \mathrm{FV}(\sigma) \cup \mathrm{FV}(t)
\end{aligned}$$

The free variables of a term are often the wrong notion to use for testing algebraicity. For example in the context $D^1$, the term $d_1$ has free variables $\{d_1\}$, whereas the unary composite of $d_1$, $\mathsf{Coh}_{(D_1\,;\,d_0^- \to_\star d_0^+)}[\mathsf{id}_{D^1}]$, has free variables $\{d_0^-, d_0^+, d_1\}$. To remedy this, the original paper considers $\mathrm{FV}(t) \cup \mathrm{FV}(A)$, for a term $t$ of type $A$. In this thesis we instead define the support of each piece of syntax as a purely syntactic construction.

**Definition 1.2.2.** Fix a context $\Gamma$. The subset $V \subseteq \mathrm{Var}(\Gamma)$ is *downwards closed* if for all $(x : A) \in \Gamma$ we have:

$$x \in V \implies \mathrm{FV}(A) \subseteq V$$

The downwards closure of a set $V$ in a context $\Gamma$, $\mathrm{DC}_\Gamma(V)$ can be defined by induction on the context:

$$\mathrm{DC}_\emptyset(\emptyset) = \emptyset$$

$$\mathrm{DC}_{\Gamma,x:A}(V) = \begin{cases} \mathrm{DC}_\Gamma(V) & \text{if } x \notin V\\ \{x\} \cup \mathrm{DC}_\Gamma(V \cup \mathrm{FV}(A)) & \text{if } x \in V \end{cases}$$

The support of a term, type, or substitution is then defined as the downwards closure of its free variables:

$$\mathrm{Supp}(t) = \mathrm{DC}_\Gamma(\mathrm{FV}(t)) \qquad \mathrm{Supp}(A) = \mathrm{DC}_\Gamma(\mathrm{FV}(A)) \qquad \mathrm{Supp}(\sigma) = \mathrm{DC}_\Gamma(\mathrm{FV}(\sigma))$$

for terms $t \in \mathsf{Term}_\Gamma$, types $A \in \mathsf{Type}_\Gamma$, and substitutions $\sigma : \Delta \to \Gamma$.

We will see later (Lemma 2.4.25(iv)) that for well-formed terms $t$ of typed $A$ that the support of $t$ is equal to $\mathrm{FV}(t) \cup \mathrm{FV}(A)$ and that $\mathrm{Supp}(A) = \mathrm{FV}(A)$ for well-formed types. Modifying Catt to use the support operation therefore does not change the theory.

We lastly define the *dimension* of types, contexts, and terms. For types this is defined recursively:

$$\dim(\star) = 0 \qquad \dim(s \to_A t) = 1 + \dim(A)$$

For contexts, we define $\dim(\Gamma)$ to be the maximum of the dimension of each type in $\Gamma$. For coherences $\mathsf{Coh}_{(\Gamma\,;\,A)}[\sigma]$, the dimension is given by $\dim(A)$, and for variables the dimension is given by the dimension of the associated type in the context.

### 1.2.2 Ps-contexts

We need to be able to describe pasting diagrams within the theory Catt. As contexts model globular sets it is natural to treat pasting diagrams as a subset of contexts. We will build pasting diagrams by iteratively attaching discs to a context, which is done by introducing the judgements:

$$\Delta \vdash_{\mathsf{ps}} x : A \qquad \text{and} \qquad \Delta \vdash_{\mathsf{ps}}$$

If the first judgement holds, then $\Delta$ is a pasting diagram for which a disc can be attached to the variable $x$, called a *dangling variable*, which has type $A$. The contexts $\Delta$ for which the second judgement holds are fully formed pasting diagrams, which we call *ps-contexts* (short for pasting scheme contexts). The rules for these judgements are given in Figure 1.3.

We note that these rules do not just specify which globular sets are pasting diagrams, but they also specify an ordering on the elements of the pasting diagram, ensuring that there is a unique ps-context for each pasting diagram. For example, the following judgement holds:

$$(x : \star), (y : \star), (f : x \to_\star y), (z : \star), (g : y \to_\star z) \vdash_{\mathsf{ps}} \tag{1.2.3}$$

However, the context:

$$(y : \star), (z : \star), (g : y \to_\star z), (x : \star), (f : x \to_\star y)$$

represents the same globular set but is not a ps-context.

$$\frac{}{(x : \star) \vdash_{\mathsf{ps}} x : \star}\,(\textsc{pss}) \qquad \frac{\Gamma \vdash_{\mathsf{ps}} x : A}{\Gamma, (y : A), (f : x \to_A y) \vdash_{\mathsf{ps}} f : x \to_A y}\,(\textsc{pse})$$

$$\frac{\Gamma \vdash_{\mathsf{ps}} x : s \to_A t}{\Gamma \vdash_{\mathsf{ps}} t : A}\,(\textsc{psd}) \qquad \frac{\Gamma \vdash_{\mathsf{ps}} x : \star}{\Gamma \vdash_{\mathsf{ps}}}\,(\textsc{ps})$$

Figure 1.3: Rules for ps-contexts.

*Example* 1.2.4. Judgement 1.2.3 is given by the following derivation:

$$\dfrac{\dfrac{\dfrac{\dfrac{\dfrac{\dfrac{}{(x : \star) \vdash_{\mathsf{ps}} x : \star}\,(\textsc{pss})}{(x : \star), (y : \star), (f : x \to_\star y) \vdash_{\mathsf{ps}} f : x \to_\star y}\,(\textsc{pse})}{(x : \star), (y : \star), (f : x \to_\star y) \vdash_{\mathsf{ps}} y : \star}\,(\textsc{psd})}{(x : \star), (y : \star), (f : x \to_\star y), (z : \star), (g : y \to_\star z) \vdash_{\mathsf{ps}} g : y \to_\star z}\,(\textsc{pse})}{(x : \star), (y : \star), (f : x \to_\star y), (z : \star), (g : y \to_\star z) \vdash_{\mathsf{ps}} z : \star}\,(\textsc{psd})}{(x : \star), (y : \star), (f : x \to_\star y), (z : \star), (g : y \to_\star z) \vdash_{\mathsf{ps}}}\,(\textsc{ps})$$

The applications of (pse) allow new variables to be added to the context, by adding a fresh variable, and attaching a variable from the dangling variable to the new fresh variable. The rule (psd) encodes that if we can attach a variable to $f : x \to y$, then we can also attach a variable to $y$. The rule (ps) forces as many (psd) rules to be applied as possible before completing the derivation, ensuring that derivations of ps-contexts are unique.

We now state the following theorem, which follows immediately from [BFM24, Theorem 53].

**Theorem 1.2.5.** *The set of ps-contexts is in bijection with the set of pasting diagrams.*

In order to use ps-contexts as our notion of pasting diagram, we need to be able to identify the source and target variables of each ps-context. This will be done by specifying the dimension $i$ source and target of each pasting context.

More precisely, for each ps-context $\Gamma$ and $i \in \mathbb{N}$, we define a ps-context $\partial_i(\Gamma)$ and subcontext inclusions:

$$\delta_i^-(\Gamma) : \partial_i(\Gamma) \to \Gamma \qquad \text{and} \qquad \delta_i^+(\Gamma) : \partial_i(\Gamma) \to \Gamma$$

Intuitively, the context $\partial_i(\Gamma)$ can be constructed by removing any variables of dimension greater than $i$ from $\Gamma$, and quotienting the dimension $i$ variables by the (symmetric transitive closure of the) relation $x \sim y$ if there exists an $f : x \to y$. The inclusions then send this quotiented variable to the variable appearing first in the equivalence class for the source inclusion, and the variable appearing last in the class for the target inclusion.

These contexts and substitutions can be defined by recursion on the context $\Gamma$:

$$\begin{aligned}
\partial_i((x : \star)) &= (x : \star) \\
\partial_i(\Gamma, (y : A), (f : x \to_A y)) &= \begin{cases} \partial_i(\Gamma) & \text{if } i \leq \dim(A) \\ \partial_i(\Gamma), (y : A), (f : x \to_A y) & \text{otherwise} \end{cases} \\
\delta_i^\epsilon((x : \star)) &= \langle x \rangle \\
\delta_i^\epsilon(\Gamma, (y : A), (f : x \to_A y)) &= \begin{cases} \delta_i^\epsilon(\Gamma) & \text{if } i < \dim(A) \\ \delta_i^-(\Gamma) & \text{if } i = \dim(A) \text{ and } \epsilon = - \\ \text{replace}(\delta_i^+(\Gamma), y) & \text{if } i = \dim(A) \text{ and } \epsilon = + \\ \langle \delta_i^\epsilon(\Gamma), y, f \rangle & \text{otherwise} \end{cases}
\end{aligned}$$

where $\epsilon \in \{-, +\}$ and $\text{replace}(\langle \sigma, s \rangle, t) = \langle \sigma, t \rangle$. As it will be common to take the boundary of $\Gamma$ at the dimension below the dimension of $\Gamma$ itself, we write

$$\delta^\epsilon(\Gamma) = \delta^\epsilon_{\dim(\Gamma)-1}(\Gamma)$$

when $\dim(\Gamma)$ is not zero.

In the original Catt paper, these inclusion substitutions are not given and instead the source and target variables are given directly as subcontexts. It can be easily checked that the free variables of the inclusions are equal to the subcontexts, and that the free variable sets of these inclusions are downwards closed. It is known, e.g. from [BFM24, Lemma 55], that these constructions agree with the constructions of the source and target pasting diagrams in Section 1.1.1.

We state the following well-known result (see [FM17]) about isomorphisms between pasting contexts.

**Proposition 1.2.6.** *Let $\Gamma$ and $\Delta$ be ps-contexts and suppose $\sigma : \Gamma \to \Delta$ is an isomorphism. Then $\Gamma \equiv \Delta$ and $\sigma$ is the identity substitution.*

### 1.2.3 Typing for Catt

We now have all the prerequisites in place to state the typing rules for Catt. These take the form of 4 judgements (not including the judgements for ps-contexts introduced in Section 1.2.2):

| | |
|---|---|
| $\Gamma \vdash$ | $\Gamma \in \mathsf{Ctx}$ is a well-formed context. |
| $\Gamma \vdash A$ | $A \in \mathsf{Type}_\Gamma$ is a well-formed type in context $\Gamma$. |
| $\Gamma \vdash t : A$ | $t \in \mathsf{Term}_\Gamma$ is a well-formed term of type $A \in \mathsf{Type}_\Gamma$. |
| $\Gamma \vdash \sigma : \Delta$ | $\sigma : \Delta \to \Gamma$ is a well-formed substitution. |

The typing rules for these judgements are then given in Figure 1.4. As most of these are standard we draw attention to a couple of the key rules. The rule for arrow types ensures that both the source and target of the arrow themselves have the same type, namely the one given in the subscript of the arrow. This effectively ensures the globular nature of the type theory, as given a term $f : s \to_{x \to_A y} t$, both the source of the source and source of the target are $x$, and both the target of the source and target of the target are $y$.

$$\frac{}{\emptyset \vdash} \qquad \frac{\Gamma \vdash \qquad \Gamma \vdash A}{\Gamma, (x : A) \vdash} \qquad \frac{}{\Gamma \vdash \star} \qquad \frac{\Gamma \vdash s : A \qquad \Gamma \vdash A \qquad \Gamma \vdash t : A}{\Gamma \vdash s \to_A t}$$

$$\frac{}{\Gamma \vdash \langle\rangle : \emptyset} \qquad \frac{\Gamma \vdash \sigma : \Delta \qquad \Gamma \vdash t : A[\![\sigma]\!]}{\Gamma \vdash \langle \sigma, t\rangle : \Delta, (x : A)} \qquad \frac{(x : A) \in \Gamma}{\Gamma \vdash x : A}$$

$$\frac{\Gamma \vdash \sigma : \Delta \qquad \dim(\Delta) \neq 0 \qquad \begin{array}{c}\Delta \vdash_{\mathsf{ps}} \qquad \Delta \vdash s \to_A t \\ \mathrm{Supp}(s) = \mathrm{Supp}(\delta^-(\Delta))\end{array} \qquad \mathrm{Supp}(t) = \mathrm{Supp}(\delta^+(\Delta))}{\Gamma \vdash \mathsf{Coh}_{(\Delta\,;\, s \to_A t)}[\sigma] : s[\![\sigma]\!] \to_{A[\![\sigma]\!]} t[\![\sigma]\!]}$$

$$\frac{\Delta \vdash_{\mathsf{ps}} \qquad \Delta \vdash s \to_A t \qquad \Gamma \vdash \sigma : \Delta \qquad \mathrm{Supp}(s) = \mathrm{Supp}(t) = \mathrm{Var}(\Delta)}{\Gamma \vdash \mathsf{Coh}_{(\Delta\,;\, s \to_A t)}[\sigma] : s[\![\sigma]\!] \to_{A[\![\sigma]\!]} t[\![\sigma]\!]}$$

Figure 1.4: Typing rules for Catt.

There are two rules given for typing coherence, corresponding to the two guiding principles for categories from Section 1.1.2. The first rule allows composites to be typed and the second allows equivalences to be typed. In both, the ps-context $\Delta$ corresponds to the pasting diagram $P$, the terms $s$ and $t$ correspond to the operations $f$ and $g$ over $P$ (with the judgement $\Delta \vdash s \to_A t$ enforcing that they are parallel), and the conditions involving support give the remaining side conditions.

By a straightforward mutual induction we can prove that application of substitution to terms, types, and other substitutions preserves typing. Therefore, the *syntactic category* of Catt can be formed, which contains well-formed contexts as objects and well-formed substitutions between these contexts as morphisms, which by an abuse of notation we call Catt. There is a full subcategory $\mathsf{Catt}^{\mathsf{ps}}$, which only contains the contexts which are ps-contexts.

**Theorem 1.2.7.** *The category* $\mathsf{Catt}^{\mathsf{ps}}$ *is a coherator for* $\infty$*-categories.*

*Proof.* Follows from [BFM24, Theorem 73], noting that the opposite convention for substitution is used in that paper. □

Thus, we immediately get that a presheaf over $\mathsf{Catt}^{\mathsf{ps}}$ which preserves globular products is an $\infty$-category (using the Maltsiniotis definition). Further, presheaves of this form are equivalent to type-theoretic models of Catt by [BFM24, Theorem 88], meaning type-theoretic models of Catt are $\infty$-categories.

### 1.2.4 Basic constructions

We now introduce some examples of basic categorical operations in order to give some early examples. Suppose we have terms $a : s \to_\star t$ and $b : t \to_\star u$ in some context $\Gamma$. Then the ps-context

$$\Delta = (x : \star), (y : \star), (f : x \to_\star y), (z : \star), (g : y \to_\star z)$$

from Judgement 1.2.3 can be used to form the 1-composite:

$$a *_0 b = \mathsf{Coh}_{(\Delta\,;\,x \to_\star z)}[\langle s, t, a, u, b \rangle]$$

It is often not necessary to give all the terms in a substitution, especially when the substitution is from a pasting diagram (or more generally a globular set). In these cases it is sufficient to give terms for the *locally maximal* variables of the context, those that do not appear as the source or target of another variable. For $\Delta$, the locally maximal variables are $f$ and $g$, and so it suffices to give the substitution above as $\langle a, b \rangle$, with the rest of the terms being inferable.

The disc contexts $D^n$ can be formed in Catt as the analogue of the disc globular sets given in Example 1.1.2 and satisfy the property that a substitution from a disc context $D^n$ contains the same data as a term and $n$-dimensional type. Given a term $t$ of type $A$ in context $\Gamma$, we write this substitution $\{A, t\} : D^{\dim(A)} \to \Gamma$. All disc contexts are ps-contexts.

Using these, the identity can be formed on a term $t$ of type $A$ in $\Gamma$:

$$\mathsf{id}(A, t) = \mathsf{Coh}_{(D^n\,;\,d_n \to d_n)}[\{A, t\}]$$

where $\dim(A) = n$, which is typed using the rule for equivalences. The structure of this term changes for different values of $n$, and we will relate these different terms in Section 1.2.5. As before, the non-locally maximal elements of a substitution can be inferred, and so we may write $\mathsf{id}(t)$ or $\{t\}$ when the type $A$ is inferable. In Catt, all types are inferable, though later when we consider semistrict variations of Catt it may be necessary to specify the exact type we are using up to syntactic equality.

**Standard coherences** The composite and identity above form part of a more general collection of coherences, which we call *standard coherences.*

**Definition 1.2.8.** Given a pasting diagram $\Delta$, we mutually define for all $n$ the *standard*

*coherence* $\mathcal{C}^n_\Delta$, the *standard term* $\mathcal{T}^n_\Delta$, and the *standard type* $\mathcal{U}^n_\Delta$:

$$\begin{aligned}
\mathcal{C}^n_\Delta &= \mathsf{Coh}_{(\Delta\,;\,\mathcal{U}^n_\Delta)}[\mathsf{id}_\Delta] \\
\mathcal{T}^n_\Delta &= \begin{cases} d^n & \text{when } \Delta \text{ is the disc } D^n \\ \mathcal{C}^n_\Delta & \text{otherwise} \end{cases} \\
\mathcal{U}^0_\Delta &= \star \\
\mathcal{U}^{n+1}_\Delta &= \mathcal{T}^n_{\partial_n(\Delta)}[\![\delta^-_n(\Delta)]\!] \to_{\mathcal{U}^n_\Delta} \mathcal{T}^n_{\partial_n(\Delta)}[\![\delta^+_n(\Delta)]\!]
\end{aligned}$$

The standard type takes the standard term over each boundary of $\Delta$, includes these all back into $\Delta$ and assembles them into a type. When $n = \dim(\Delta)$ we will refer to the standard coherence as the *standard composite.*

Intuitively, the standard coherence $\mathcal{C}^n_\Delta$ is the canonical composite in dimension $n$ of the pasting diagram $\Delta$. To give this a type is needed to form the coherence, for which the standard type $\mathcal{U}^n_\Delta$ is used. The standard term $\mathcal{T}^n_\Delta$ is used as a variant of the standard coherence which special cases disc contexts. This avoids the standard type containing unary composites and allows standard composites (of non-disc contexts) to be normal forms of the reduction systems that will be described in Chapter 4.

It is immediate that the composite of 1-cells $a *_0 b$ is given by $\mathcal{C}^1_\Delta[\![\langle a, b\rangle]\!]$ and the identity on a term $t$ of dimension $n$ is given by $\mathcal{C}^{n+1}_{D^n}[\![\{t\}]\!]$. This construction can be used to generate all the composites in the definition of a strict $\infty$-category. For example the vertical composite of 2-cells is the standard composite over the context given by the diagram:

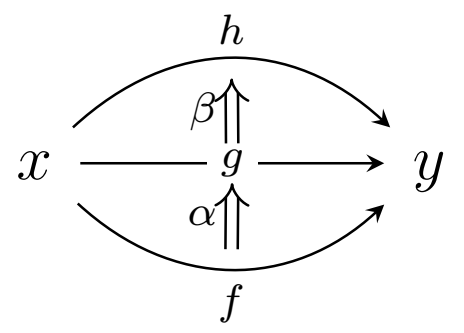

and the horizontal composite of 2-cells is the standard composite over:

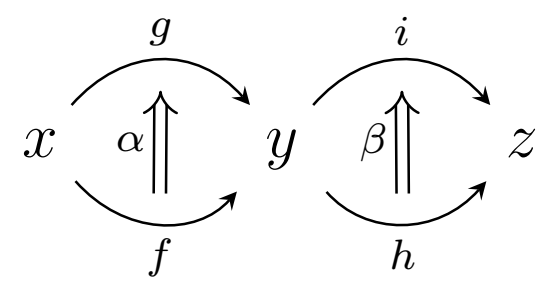

Noting that the standard type over the above diagram has source $f * h$ and target $g * i$, themselves being standard compositions demonstrating the mutual recursive behaviour of these constructions.

*Remark* 1.2.9. Above we gave two ps-contexts by drawing a diagram of the globular set that they represent. Ps-contexts fix the order that variables occur in and as such the mapping from ps-contexts to globular sets is injective. The use of diagrams to define ps-contexts is therefore unambiguous.

**Further examples** The substitution component of a coherence allows operations to be combined into compound operations. Consider the (Ps-)context given by the following diagram:

$$\Gamma = \quad s \xrightarrow{a} t \xrightarrow{b} u \xrightarrow{c} v$$

There are (at least) 3 ways to compose together the elements of this context. We could take the unbiased ternary composite $a * b * c = \mathcal{C}^1_\Gamma[\![\langle a, b, c\rangle]\!]$, but could also construct either biased composite:

$$(a * b) * c = \mathcal{C}^1_\Delta[\![\langle \mathcal{C}^1_\Delta[\![\langle a, b\rangle]\!], c\rangle]\!]$$
$$a * (b * c) = \mathcal{C}^1_\Delta[\![\langle a, \mathcal{C}^1_\Delta[\![\langle b, c\rangle]\!]\rangle]\!]$$

Using the equivalence typing rule, we can relate these biased composite with the following term:

$$\alpha_{a,b,c} = \mathsf{Coh}_{(\Gamma\,;\,(a*b)*c \to a*(b*c))}[\mathsf{id}_\Gamma]$$

which is the associator. Similarly, for a term $f : x \to_\star y$, unitors can be formed over the disc context $D^1$ using the equivalence rule:

$$\lambda_f = \mathsf{Coh}_{(D^1\,;\,\mathsf{id}(d_0^-)*d_1 \to d_1)}[\{f\}]$$
$$\rho_f = \mathsf{Coh}_{(D^1\,;\,d_1*\mathsf{id}(d_0^-) \to d_1)}[\{f\}]$$

The remainder of the operations for a 2-category can be defined similarly, as each displays the equivalence of two terms built over a pasting diagram. We observe that both the unitors and associator (as well as any coherence typed with the equivalence rule) are trivially invertible.

### 1.2.5 Suspension

To end this section, we introduce the meta-operation of *suspension*, as described for Catt by Benjamin [Ben20]. Suspension takes any piece of syntax as input and produces one with a dimension one higher. It can be used as an aid to defining operations in Catt, but will also form a key part of the formal development of the constructions described in Chapter 3.

Suspension is inspired by the identically named operation on topological spaces. Given a topological space $X$, its suspension $\Sigma X$ is formed by quotienting the space $X \times [0, 1]$ by the relation that identifies all points of the form $(x, 0)$ for $x \in X$ and identifies points $(x, 1)$ for $x \in X$.

The suspension on a space $X$ can be alternatively viewed as the space containing two distinguished points $N$ and $S$, and a path from $N$ to $S$ for each point $x \in X$. The names $N$ and $S$ stand for north and south, as the suspension of a circle can be visualised as a globe, with $N$ and $S$ being the north and south pole and each of the paths between them being a meridian.

A similar operation can be applied to globular sets. Given a globular set $G$, its suspension $\Sigma G$ is obtained by shifting the dimension of every $n$-cell up by one (making it into an $(n+1)$-cell), adding two new 0-cells $N$ and $S$, and letting the source of every 1-cell be $N$ and the target be $S$. The globularity conditions for this construction can be quickly verified.

This construction extends to all computads [BM24], and can be defined in Catt by mutually defining the operation on contexts, types, terms, and substitutions.

**Definition 1.2.10.** For contexts $\Gamma \in \mathsf{Ctx}$, types $A \in \mathsf{Type}_\Gamma$, terms $t \in \mathsf{Term}_\Gamma$, and substitutions $\sigma : \Delta \to \Gamma$, we define their *suspensions* $\Sigma(\Gamma) \in \mathsf{Ctx}$, $\Sigma(A) \in \mathsf{Type}_{\Sigma(\Gamma)}$,

$\Sigma(t) \in \mathsf{Term}_{\Sigma(\Gamma)}$, and $\Sigma(\sigma) : \Sigma(\Delta) \to \Sigma(\Gamma)$ by mutual recursion.

$$\begin{aligned} \Sigma(\emptyset) &= (N : \star), (S : \star) & \Sigma(\Gamma, (x : A)) &= \Sigma\Gamma, (x : \Sigma A) \\ \Sigma(\star) &= N \to_\star S & \Sigma(s \to_A t) &= \Sigma s \to_{\Sigma A} \Sigma t \\ \Sigma(\langle\rangle) &= \langle N, S\rangle & \Sigma(\langle \sigma, x\rangle) &= \langle \Sigma(\sigma), \Sigma(t)\rangle \\ \Sigma(x) &= x & \Sigma(\mathsf{Coh}_{(\Delta\,;\,A)}[\sigma]) &= \mathsf{Coh}_{(\Sigma(\Delta)\,;\,\Sigma(A))}[\Sigma(\sigma)] \end{aligned}$$

where $x$ is a variable of $\Gamma$.

The dimension shift of suspension is driven by the cases for types, especially the case for the base type $\star$, which returns a type of dimension 1, namely $N \to_\star S$, using the two new variables $N$ and $S$. We note that the suspension of any ps-context is also a ps-context, and in general the suspension of any piece of well-formed Catt syntax can be well-formed. These results are given in [Ben20, Section 3.2], but will be proved in Section 2.4 in more generality.

We can now investigate the action of suspension on the operations we have already defined. Take the context:

$$(x : \star), (y : \star), (f : x \to_\star y), (z : \star), (g : y \to_\star z)$$

used in Section 1.2.4 to generate 1-composition. Applying suspension to this context gives:

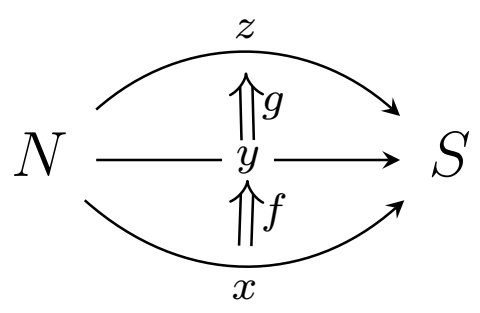

which is the context used to generate vertical 2-composition. Furthermore, applying suspension directly to 1-composition operation forms the vertical 2-composition operation.

The suspension of each disc context $D^n$ is (up to $\alpha$-renaming) $D^{n+1}$. It can be checked that applying suspension to the identity operation for $n$-dimensional terms returns the identity operation for $(n+1)$-dimensional terms. Repeating this logic, all identity operations can be obtained as iterated suspensions of the identity for 0-cells. The following more general result about standard coherences holds:

**Proposition 1.2.11.** *The following syntactic equalities hold:*

$$\Sigma(\mathcal{C}^n_\Delta) = \mathcal{C}^{n+1}_{\Sigma(\Delta)} \qquad \Sigma(\mathcal{T}^n_\Delta) = \mathcal{T}^{n+1}_{\Sigma(\Delta)} \qquad \Sigma(\mathcal{U}^n_\Delta) = \mathcal{U}^{n+1}_{\Sigma(\Delta)}$$

*for all ps-contexts $\Delta$ and $n \in \mathbb{N}$.*

The proof of these results is delayed to Chapter 3, where we will have more tools for dealing with these constructions.

# Chapter 2

# A formalised presentation of Catt with equality

The main purpose of this chapter will be to define the family of type theories $\textsc{Catt}_{\mathcal{R}}$, which extend the base type theory Catt with a specified set $\mathcal{R}$ of equality rules. These equality rules equate various terms of the theory, which unifies the corresponding operations their models, allowing us in Chapter 4 to generate type theories that model semistrict categories, categories where some but not all structure is strictified.

This chapter will also introduce the Agda formalisation [Ric24a] which accompanies this thesis, which compiles with Agda v2.6.4 and standard library v2.0. The formalisation implements the syntax and typing judgements of $\textsc{Catt}_{\mathcal{R}}$, and contains proofs of most results in this chapter and Chapter 3. By formalising $\textsc{Catt}_{\mathcal{R}}$, instead of the more specific type theories $\textsc{Catt}_{\mathsf{su}}$ and $\textsc{Catt}_{\mathsf{sua}}$ introduced in Sections 4.2 and 4.3, the formalisation of many results can be applied to both type theories. This also allows these results to be applied to any future type theories of this form.

A dependency graph of the formalisation is given in Figure 2.2, and an online version of this graph can be found at `https://alexarice.github.io/catt-agda/dep-graph.svg` for which each node is a clickable link to an HTML version of the code. This graph was generated by processing the dependency graph output of Agda with the tool sd-visualiser [HRT24].

## 2.1 Extended substitution

$\textsc{Catt}_{\mathcal{R}}$ uses the same syntax as Catt with one exception. In $\textsc{Catt}_{\mathcal{R}}$ we make a natural generalisation to substitutions, which will allow more operations to be defined for working with the suspension operation introduced in Section 1.2.5. Unfortunately, the full utility of this generalisation will not be realised until Section 3.3, but we choose to introduce it here as it forms a core part of the syntax, and requires little modification to the rules of the type theory.

We recall that the suspension operation $\Sigma$ acts on contexts, substitutions, types, and terms. Given a substitution $\sigma : \Delta \to \Gamma$, its suspension $\Sigma(\sigma)$ has domain $\Sigma(\Delta)$ and codomain $\Sigma(\Gamma)$. When we define trees and tree labellings in Chapter 3, which will be used to define the insertion operation in Section 3.4, we will need to be able to define substitutions from suspended contexts to arbitrary contexts. More generally, we would like to be able to describe substitu-

tions of the form:

$$\Sigma^n(\Delta) \to \Gamma$$

where $\Sigma^n(\Delta)$ is the operation that applies suspension $n$ times to $\Delta$.

Consider the data contained in a substitution $\tau : \Sigma(\Delta) \to \Gamma$. There are two terms $N[\![\tau]\!]$ and $S[\![\tau]\!]$ of type $\star$, and then a term for each variable of $\Delta$. Temporarily ignoring the typing conditions for substitutions, we see that the data is equivalent to a substitution from $\Delta$ to $\Gamma$ and two additional terms.

If we now consider a substitution $\tau : \Sigma(\Sigma(\Delta)) \to \Gamma$, we notice that there is a term in $\Gamma$ for each variable of $\Delta$, as well as two terms $s = N[\![\tau]\!]$ and $t = S[\![\tau]\!]$ for the outer suspension and terms $u = N'[\![\tau]\!]$ and $v = S'[\![\tau]\!]$ for the inner suspension. As before, the terms $s$ and $t$ should have type $\star$, but the terms $u$ and $v$ should have type $s \to_\star t$. We note that this is the exact condition needed for $u \to_{s \to_\star t} v$ to be a well-formed type. This motivates the notion of an *extended substitution*, which is obtained by equipping a substitution with a type.

We have not yet determined the typing conditions required on the substitution part of these extended substitutions. We return to the example of a substitution $\tau : \Sigma^2(\Delta) \to \Gamma$, and suppose that $\Delta$ has a variable $x$ of type $\star$. In $\Sigma^2(\Delta)$, $x$ has the type $N' \to_{N \to_\star S} S'$, and so $x$ should be sent to a term of type $u \to_{s \to_\star t} v$, the type portion of the extended substitution. In a substitution $\sigma : \Delta \to \Gamma$, $x$ would be sent to a term of type $\star[\![\sigma]\!]$, which suggests that $\star[\![\sigma]\!]$ should be redefined to send $\star$ to the type part of the extended substitution.

This one change to the application of substitution to types is sufficient to generalise from substitutions to extended substitutions. An extended substitution $\sigma : \Delta \to \Gamma$ then has the following intuition: The substitution part specifies where each variable in $\Delta$ should be sent, and the type part specifies where the base type $\star$ should be sent. The other cases for the application of substitution extend this to all terms, types, and (extended) substitutions as before. The extended substitution $\sigma$ then represents a standard substitution $\Sigma^n(\Delta)$ to $\Gamma$, where $n$ is the dimension of the type part of $\sigma$. Hence, a regular substitution can be recovered as an extended substitution with type part $\star$.

We modify the syntax of Catt as follows, and will refer to these extended substitutions simply as substitutions, as extended substitutions are a direct generalisation of substitutions, and the notion of substitution is still recoverable by setting the type part to $\star$:

- Substitutions will now be indexed by a type of their codomain context, which we will write $\sigma : \Delta \to_A \Gamma$ where $A \in \mathsf{Type}_\Gamma$. We note that this allows us to specify that $\sigma$ is a regular substitution by writing $\sigma : \Delta \to_\star \Gamma$.
- The constructor $\langle\rangle$ is removed, and is replaced by the constructor $\langle A \rangle : \emptyset \to_A \Gamma$, where $A \in \mathsf{Type}_\Gamma$. Adding a term to a substitution preserves the type of the substitution. As before we may write a substitution $\langle\langle\langle A \rangle, s\rangle, t\rangle$ as $\langle A, s, t\rangle$. We let $\mathrm{FV}(\langle A \rangle) = \mathrm{FV}(A)$.
- An operation $\mathrm{Ty}(\sigma)$ is introduced that returns the type portion of a substitution. For $\sigma : \Delta \to_A \Gamma$, we have $\mathrm{Ty}(\sigma) = A$.
- Coherences $\mathsf{Coh}_{(\Delta\,;\,A)}[\sigma] \in \mathsf{Term}_\Gamma$ are restricted so that $\sigma$ is a regular substitution. In other words $\mathrm{Ty}(\sigma)$ must be $\star$ for $\sigma$ to appear in a substitution. While this condition could be dropped, it is convenient to keep the same operations as Catt.

To witness the equivalence of extended substitutions $\Delta \to \Gamma$ and regular substitutions $\Sigma^n(\Delta) \to \Gamma$, we introduce new operations.

**Definition 2.1.1.** For a substitution $\sigma : \Delta \to_{s \to_A t} \Gamma$, we define its *unrestriction*:

$$\downarrow \sigma : \Sigma(\Delta) \to_A \Gamma$$

by induction on the length of $\Delta$:

$$\begin{aligned} \downarrow\langle s \to_A t \rangle &= \langle A, s, t \rangle \\ \downarrow\langle \sigma', u \rangle &= \langle \downarrow \sigma', u \rangle \end{aligned}$$

The unrestrict operation simply moves two terms from the type part of the substitution into the main body of the substitution.

To define the second operation, we need to first specify the changes to application of substitution:

- The composition of substitutions takes substitutions $\sigma : \Theta \to_A \Delta$ and $\tau : \Delta \to_B \Gamma$ to a substitution $\sigma \bullet \tau : \Theta \to_{A[\![\tau]\!]} \Gamma$.
- For a substitution $\sigma : \Delta \to_A \Gamma$, we define $\star[\![\sigma]\!] = A$.
- As the substitution in a coherence must have type $\star$, we define the application of an extended substitution $\tau : \Delta \to_{s \to_A t} \Gamma$ to a coherence as:

  $$\mathsf{Coh}_{(\Theta\,;\,A)}[\sigma][\![\tau]\!] = \mathsf{Coh}_{(\Sigma(\Theta)\,;\,\Sigma(A))}[\Sigma(\sigma)][\![\downarrow \tau]\!]$$

  The case for applying a regular substitution to a coherence remains unchanged.

We can now define an inverse to the unrestriction operation.

**Definition 2.1.2.** For a substitution $\sigma : \Sigma(\Delta) \to_A \Gamma$, its *restriction*

$$\uparrow \sigma : \Delta \to_{N[\![\sigma]\!] \to_A S[\![\sigma]\!]} \Gamma$$

is defined by induction on the length of $\Delta$:

$$\begin{aligned} \uparrow\langle A, s, t \rangle &= \langle s \to_A t \rangle \\ \uparrow\langle \sigma', u \rangle &= \langle \uparrow \sigma', u \rangle \end{aligned}$$

Inversely to the unrestrict operation, the restrict operation moves two terms into the type part of the substitution.

As restriction and unrestriction cancel each other, the suspension of the substitution $\sigma : \Delta \to_\star \Gamma$ can be factored into $(\downarrow \circ (\uparrow \circ \Sigma))(\sigma)$. We observe that the second part of this composition, $\uparrow \circ \Sigma$, is the operation that simply applies the suspension to each term in the substitution as well as the type of the substitution. This motivates the final definition of this section.

**Definition 2.1.3.** Let the *restricted suspension* of a substitution $\sigma : \Delta \to_A \Gamma$ be a substitution

$$\Sigma'(\sigma) : \Delta \to_{\Sigma(A)} \Sigma(\Gamma)$$

defined inductively by the equations:

$$\Sigma'(\langle A \rangle) = \langle \Sigma(A) \rangle$$
$$\Sigma'(\langle \sigma', t \rangle) = \langle \Sigma'(\sigma'), \Sigma(t) \rangle$$

The suspension of a substitution $\tau : \Delta \to_\star \Gamma$ can be defined by $\Sigma(\tau) = \downarrow \Sigma'(\tau)$.

For the rest of the thesis and the formalisation, the suspension on a substitution is defined as the composition of unrestriction and restricted suspension.

## 2.2 Catt$_\mathcal{R}$: Catt with equality

This section will define the type theory Catt$_\mathcal{R}$, a variation of Catt with specified equality rules. This section, in addition to the following sections in this chapter, will be used to motivate certain choices in the formalisation. All the preliminary definitions as well as syntax, typing, and equality rules are assembled in Figure 2.1.

### 2.2.1 Syntax

The syntax of Catt$_\mathcal{R}$ is based on the syntax of Catt with the changes specified in Section 2.1. This creates a dependence chain of needing to define the base syntax before suspension can be defined, and needing to define suspension before application of substitution can be defined. In the formalisation these are defined in the following files:

- The core syntax is defined in Catt.Syntax.Base.
- Suspension is defined in Catt.Suspension.
- Other syntactic operations are defined in Catt.Syntax, which re-exports the core syntax.

To avoid any issues with $\alpha$-equivalence, especially as we have terms that contain contexts, we work with de Bruijn indices throughout the formalisation. This means that a context is simply a vector of types, a fixed length list, which are given a nicer syntax. Variables are then simply bounded natural numbers, represented by the sets $\mathsf{Fin}_n$, where $\mathsf{Fin}_n$ is the set $\{0, \dots, n-1\}$. Given a context $A, B, C$, the variables over this context are simply var 0, which has type $C$, var 1, which has type $B$, and var 2, with type $A$. We note that $3$ is not in $\mathsf{Fin}_3$, and so var 3 is not a term of this context. Hence, we do not need to deal with unknown variables when applying substitutions. We will still make use of variable names in this text to aid readability, and will ignore any potential problems that could arise from this, knowing that the results are formalised in a setting where they do not appear.

The formalisation also differs from the presentation in the texts by the way that the various notions of syntax are indexed. We index contexts by a natural number representing their length, and then index terms, types, and substitutions over these lengths instead of indexing them by their context. We then get the following 4 syntactic classes defined as mutually inductive families, where $\mathcal{U}$ is a type universe:

$$\mathsf{Ctx} : \mathbb{N} \to \mathcal{U} \quad \mathsf{Type} : \mathbb{N} \to \mathcal{U} \quad \mathsf{Term} : \mathbb{N} \to \mathcal{U} \quad \mathsf{Sub} : (n\ m : \mathbb{N}) \to \mathsf{Type}_m \to \mathcal{U}$$

This decision was made purely for convenience, by indexing by natural numbers instead of contexts, we sometimes avoid the need for providing more explicit arguments to syntactic constructions. It comes with drawback that the context must be provided for certain operations, such as the support of a piece of syntax, or the dimension of a term.

One place an explicit argument can be avoided is when defining the weakening of a piece of syntax, an operation witnessing that for a piece of syntax living in a context $\Gamma$, there is a copy living in $\Gamma, A$ for any $A$. These operations are defined in Catt.Syntax and take the following form, where we re-use the name wk here as an abuse of notation:

$$\mathrm{wk} : \mathsf{Term}_\Gamma \to \mathsf{Term}_{\Gamma,A} \quad \mathrm{wk} : \mathsf{Type}_\Gamma \to \mathsf{Type}_{\Gamma,A} \quad \mathrm{wk} : (\Gamma \to_B \Delta) \to (\Gamma \to_{\mathrm{wk}(B)} \Delta, A)$$

If terms are indexed by contexts then this type $A$ must often be specified, though if they are instead indexed by context length then this is no longer necessary. When using de Bruijn indices, this operation is no longer the identity on terms, as each variable must be incremented due to the index in a variable counting from the end of the context. One might ask why de Bruijn levels (which index from the start of the context) were not used instead, but this would not solve our problem as $\mathsf{Fin}_n$ is not a subtype of $\mathsf{Fin}_{n+1}$ in Agda. Furthermore, using de Bruijn levels would cause the substitution application introduced in Section 1.2.1 (and expanded in Section 2.1) to compute poorly, due to the way substitutions are defined. The definition of weakening is given in Figure 2.1i.

Weakening can be used to give a short inductive definition of the identity substitution, a substitution $\mathsf{id}_\Gamma : \Gamma \to \Gamma$ which sends every variable to itself. On the inductive case $\mathsf{id}_{\Gamma,(x:A)}$, it is clear that the variable $x$ should be sent to $x$, but the constructor for substitutions also requires a substitution $\Gamma \to \Gamma, (x : A)$. This can be obtained by weakening a recursive call to the identity on $\Gamma$. Similarly, an inclusion $\Gamma \to \Gamma, (x : A)$ can be defined as $\mathrm{wk}(\mathsf{id}_\Gamma)$, and applying this substitution is the same operation as weakening.

To begin proving syntactic properties of $\textsc{Catt}_\mathcal{R}$, we need a notion of syntactic equality. This will be written $\Gamma \equiv \Delta$ for contexts $\Gamma$ and $\Delta$, and similarly for terms $s$ and $t$, types $A$ and $B$, and substitutions $\sigma$ and $\tau$. It is given by $\alpha$-equivalence, and so we would hope that the formalisation could leverage the use of de Bruijn indices to use the in-built equality type for syntactic equality. This is too restrictive however, there will be many times when we want to compare two terms of differing context length (in practice this context length will be propositionally equal, instead of definitionally equal).

Therefore, four syntactic equality relations are defined mutually inductively on the constructors of each piece of syntax in Catt.Syntax.Properties. These definitions can easily be heterogeneous, allowing two terms $s : \mathsf{Term}_n$ and $t : \mathsf{Term}_m$ to be compared. Unfortunately, using these comes at the cost of large amounts of boilerplate, as these inductively defined equalities do not come equipped with the J-rule, and so it must be manually proved that each operation respects syntactic equality. An example of such a function is wk-tm-$\simeq$, which states that the weakenings of two syntactically equal terms are syntactically equal.

Catt.Syntax.Properties contains many of the basic properties about the syntax of $\textsc{Catt}_\mathcal{R}$, including:

- Syntactic equality is decidable.
- Syntactic equality is propositional, there is at most one proof of $s \equiv t$.
- Functoriality of suspension.

- Interaction of weakening with substitution application. We have $\mathrm{wk}(s)[\![\langle\sigma, t\rangle]\!] \equiv s[\![\sigma]\!]$ and $s[\![\mathrm{wk}(\sigma)]\!] \equiv \mathrm{wk}(s[\![\sigma]\!])$ and equivalent lemmas for the application of substitution to types and substitutions.

It also contains the following proposition.

**Proposition 2.2.1.** *Application of substitution is associative and unital with respect to the identity substitution. More precisely, given substitutions $\sigma : \Theta \to_A \Delta$ and $\tau : \Delta \to_B \Gamma$, the following equalities hold:*

$$A[\![\sigma]\!][\![\tau]\!] \equiv A[\![\sigma \bullet \tau]\!] \qquad A[\![\mathsf{id}]\!]_\Theta \equiv A$$

$$t[\![\sigma]\!][\![\tau]\!] \equiv t[\![\sigma \bullet \tau]\!] \qquad t[\![\mathsf{id}]\!]_\Theta \equiv t$$

$$(\mu \bullet \sigma) \bullet \tau \equiv \mu \bullet (\sigma \bullet \tau) \qquad \mu \bullet \mathsf{id}_\Theta \equiv \mu \qquad \mathsf{id}_\Xi \bullet \mu \equiv \mu$$

*for types $A \in \mathsf{Type}_\Theta$, terms $t \in \mathsf{Term}_\Theta$, and substitutions $\mu : \Xi \to_C \Theta$.*

*Proof.* The last equation is a simple induction on $\mu$ (and the context $\Xi$). Both the unitality equations and associativity equations, as with the vast majority of syntactic proofs, are given by mutual induction on types, terms, and substitutions. The only difficult case is:

$$\mathsf{Coh}_{(\Theta\,;\,C)}[\mu][\![\sigma]\!][\![\tau]\!] \equiv t[\![\sigma \bullet \tau]\!]$$

where the type part of $\sigma : \Theta \to_A \Delta$ or $\tau : \Delta \to_B \Gamma$ is not $\star$. First suppose $B = s \to_{B'} t$ but $A = \star$:

$$\begin{aligned}
\mathsf{Coh}_{(\Theta\,;\,C)}[\mu][\![\sigma]\!][\![\tau]\!] &\equiv \mathsf{Coh}_{(\Theta\,;\,C)}[\mu \bullet \sigma][\![\tau]\!] \\
&\equiv \mathsf{Coh}_{(\Sigma(\Theta)\,;\,\Sigma(C))}[\Sigma(\mu \bullet \sigma)][\![{\downarrow}\tau]\!] \\
&\equiv \mathsf{Coh}_{(\Sigma(\Theta)\,;\,\Sigma(C))}[\Sigma(\mu) \bullet \Sigma(\sigma)][\![{\downarrow}\tau]\!] \\
&\equiv \mathsf{Coh}_{(\Sigma(\Theta)\,;\,\Sigma(C))}[\Sigma(\mu)][\![\Sigma(\sigma) \bullet {\downarrow}\tau]\!] \\
&\equiv \mathsf{Coh}_{(\Sigma(\Theta)\,;\,\Sigma(C))}[\Sigma(\mu)][\![{\downarrow}(\sigma \bullet \tau)]\!] \\
&\equiv \mathsf{Coh}_{(\Theta\,;\,C)}[\mu][\![\sigma \bullet \tau]\!]
\end{aligned}$$

where the second to last line is given by property

$${\downarrow}(\sigma \bullet \tau) \equiv \Sigma(\sigma) \bullet {\downarrow}\tau$$

which holds for all $\sigma : \Theta \to_\star \Delta$ and is proven in ${\downarrow}$-comp, and the line before is given by the inductive hypothesis.

If instead we had $A = s \to_{A'} t$, then:

$$\begin{aligned}
\mathsf{Coh}_{(\Theta\,;\,C)}[\mu][\![\sigma]\!][\![\tau]\!] &\equiv \mathsf{Coh}_{(\Sigma(\Theta)\,;\,\Sigma(C))}[\Sigma(\mu)][\![{\downarrow}\sigma]\!][\![\tau]\!] \\
&\equiv \mathsf{Coh}_{(\Sigma(\Theta)\,;\,\Sigma(C))}[\Sigma(\mu)][\![{\downarrow}\sigma \bullet \tau]\!] \\
&\equiv \mathsf{Coh}_{(\Sigma(\Theta)\,;\,\Sigma(C))}[\Sigma(\mu)][\![{\downarrow}(\sigma \bullet \tau)]\!] \\
&\equiv \mathsf{Coh}_{(\Theta\,;\,C)}[\mu][\![\sigma \bullet \tau]\!]
\end{aligned}$$

where we use the inductive hypothesis after applying the equality

$${\downarrow}(\sigma \bullet \tau) \equiv {\downarrow}\sigma \bullet \tau$$

which holds for all $\sigma : \Theta \to_{s \to_{A'} t} \Delta$ by ${\downarrow}$-comp-higher. □

This proposition proves that the syntax of $\textsc{Catt}_{\mathcal{R}}$ forms a category, which we will not name as we will work instead with the subcategory containing well-formed contexts and substitutions, introduced in the following sections.

**Discs** We finish our discussion of the syntax of $\textsc{Catt}_{\mathcal{R}}$ by giving formal definitions of disc and sphere contexts, some constructions on these, and their properties. This will allow these to be used as examples in following sections, and pre-empts the use of discs in the first two equality rules that we will introduce, disc removal and endo-coherence removal.

We begin with the definitions of discs, spheres, and sphere types, which can be found in Catt.Discs as Disc, Sphere, and sphere-type. We write the sphere type as $U^n$, which is intentionally close to the notation of the standard type $\mathcal{U}^n_\Delta$, as it will turn out that these coincide.

**Definition 2.2.2.** We mutually define the disc contexts $D^n$, sphere contexts $S^n$, and sphere type $U^n \in \mathsf{Type}_{S^n}$.

$$D^n = S^n, (d_n^- : U^n) \qquad S^0 = \emptyset \qquad S^{n+1} = D^n, (d_n^+ : \mathrm{wk}(U^n))$$

$$U^0 = \star \qquad U^{n+1} = d_n^- \to_{\mathrm{wk}(\mathrm{wk}(U^{n+1}))} d_n^+$$

We will sometimes refer to the last variable of $D^n$ as $d_n$ instead of $d_n^-$, given that there is no $d_n^+$ in the context. We note that the index on the sphere $U^n$ is offset by one compared to the standard definition of the $n$-sphere in topology. This ensures that the index matches the dimension of the sphere type and allows these constructions to be indexed by natural numbers in the formalisation.

We also characterise the substitutions from a sphere or disc. These are given by sub-from-sphere and sub-from-disc in the formalisation.

**Definition 2.2.3.** Let $A : \mathsf{Type}_\Gamma$ be a type and suppose $n = \dim(A)$. Define the substitution $\{A\} : S^n \to \Gamma$ inductively by:

$$\{\star\} = \langle\rangle \qquad \{s \to_A t\} = \langle \{A\}, s, t\rangle$$

Further, given a term $t : \mathsf{Term}_\Gamma$, define the substitution $\{A, t\} : D^n \to \Gamma$ by $\{A, t\} = \langle \{A\}, t\rangle$.

In Catt.Discs.Properties, various facts about these constructions are proved which we list below.

**Lemma 2.2.4.** *The following hold:*

*(i)* $\dim(D^n) = \dim(U^n) = n$ *and* $\dim(S^n) = \max(n-1, 0)$.

*(ii)* $\Sigma(D^n) \equiv D^{n+1}$, $\Sigma(S^n) \equiv S^{n+1}$, *and* $\Sigma(U^n) \equiv U^{n+1}$.

*(iii)* $\{\mathrm{wk}(A)\} \equiv \mathrm{wk}(\{A\})$ *and* $\{\mathrm{wk}(A), \mathrm{wk}(t)\} \equiv \mathrm{wk}(\{A, t\})$.

*(iv)* $\{\Sigma(A)\} \equiv \Sigma(\{A\})$ *and* $\{\Sigma(A), \Sigma(t)\} \equiv \Sigma(\{A, t\})$.

*(v)* $\{A[\![\sigma]\!]\} \equiv \{A\} \bullet \sigma$ *and* $\{A[\![\sigma]\!], t[\![\sigma]\!]\} \equiv \{A, t\} \bullet \sigma$.

*(vi)* $U^n[\![\{A\}]\!] \equiv A$ *and hence* $\mathrm{wk}(U^n)[\![\{A, t\}]\!] \equiv A$.

*(vii) For* $\tau : S^n \to \Gamma$, $\tau \equiv \{U^n[\![\tau]\!]\}$.

*(viii) For* $\tau : D^n \to \Gamma$, $\tau \equiv \{\mathrm{wk}(U^n)[\![\tau]\!], d_n[\![\tau]\!]\}$.

*for all* $n \in \mathbb{N}$ *and appropriate* $A$, $t$, *and* $\sigma$.

The last two statements finish the characterisation of substitutions from spheres and discs as all such substitutions are of the form $\{A\}$ or $\{A, t\}$ respectively.

In Catt.Discs.Pasting, it is shown that $D^n$ is a ps-context for each $n$. Therefore, as in Section 1.2.4, the identity on a term $t$ of type $A$ can be defined as:

$$\mathsf{id}(A, t) = \mathsf{Coh}_{(D^n\,;\,d_n \to_{\mathrm{wk}(U^n)} d_n)}[\{A, t\}]$$

where $n = \dim(A)$. Many properties of identity terms can be easily derived from Lemma 2.2.4.

### 2.2.2 Typing and equality

The typing rules for Catt$_{\mathcal{R}}$ differ from those from Catt in three key ways:

1. The fixed conditions on the support of the types in a coherence have been replaced by a set of operations $\mathcal{O}$. Instead of having two typing rules for coherences, one for equivalences and one for composites, we simply have one typing rule and specify that a coherence $\mathsf{Coh}_{(\Delta\,;\,s \to_A t)}[\sigma]$ can be well-formed when:

   $$(\Delta, \mathrm{Supp}(s), \mathrm{Supp}(t)) \in \mathcal{O}$$

   This will be further motivated and explained in Section 2.3.

2. A definitional equality is added to the system, generated by a set of equality rules $\mathcal{R}$ which specifies pairs of terms which should be equated. The equality takes the form of three new judgements:

   | | |
   |---|---|
   | $\Gamma \vdash A = B$ | $A, B \in \mathsf{Type}_\Gamma$ are equal in context $\Gamma$. |
   | $\Gamma \vdash s = t$ | $s, t \in \mathsf{Term}_\Gamma$ are equal in context $\Gamma$. |
   | $\Gamma \vdash \tau = \sigma$ | $\tau : \Theta \to \Gamma$ and $\sigma : \Delta \to \Gamma$ are equal. |

   These judgements are all mutually defined (and are in fact mutually defined with the typing judgements). We may sometimes abbreviate these judgements to $A = B$, $s = t$, and $\tau = \sigma$ when the contexts of each piece of syntax is clear.

3. The typing rules are adjusted to account for this definitional equality, via the addition of a conversion rule.

The conversion rule is the only additional typing rule that must be added to Catt$_{\mathcal{R}}$, and takes the following form:

$$\frac{\Gamma \vdash s : A \qquad \Gamma \vdash A = B}{\Gamma \vdash s : B}\textsc{conv}$$

allowing the type of any term to vary up to the definitional equality. This rule accounts for all the semistrict behaviour in the theories we introduce in Chapter 4.

By adding this rule, and allowing the type of a term to vary up to definitional equality instead of syntactic equality, we allow more terms in the theory to become composable. Suppose we have terms $f : x \to y$ and $g : y' \to z$. In Catt, we would not be able to form the vertical composition of these terms, as $y$ and $y'$ are not the same. If we now suppose that $\Gamma \vdash y = y'$, then it will follow that $\Gamma \vdash (x \to y) = (x \to y')$, and so using the conversion rule we get:

$$\dfrac{\dfrac{\Gamma \vdash f : x \to y \qquad \dfrac{\Gamma \vdash y = y'}{\Gamma \vdash (x \to y) = (x \to y')}}{\Gamma \vdash f : x \to y'} \qquad \Gamma \vdash g : y' \to z}{\Gamma \vdash f * g : x \to z}$$

We remark that adding definitional equality does not simply quotient the terms of the theory, but also allows new terms to be well-formed as above.

The definitional equality judgements are given by the rules in Figure 2.1c and appear in the formalisation alongside the typing rules in Catt.Typing. These are generated by the set of *equality rules* $\mathcal{R}$, which is a set of triples of the form $(\Gamma, s, t)$ where $\Gamma$ is a context and $s, t \in \mathsf{Term}_\Gamma$. The key inference rule for equality is then:

$$\dfrac{\Gamma \vdash s : A \qquad (\Gamma, s, t) \in \mathcal{R}}{\Gamma \vdash s = t}\textsc{rule}$$

which says that if a triple $(\Gamma, s, t)$ is in $\mathcal{R}$, then $\Gamma \vdash s = t$ if $s$ is well-formed in $\Gamma$. The typing prerequisite forces the definitions of equality and typing to be mutually defined, and ensures that we only apply our equality rules to well-behaved terms.

We note the asymmetry of this rule, in that only the left-hand side is required to be well-formed. Every rule introduced in this thesis will take the form of some reduction from the left-hand side to the right-hand side, and we will be able to prove that typing for the right-hand side follows from typing for the left-hand side for every equality we consider. The converse may not hold in general, necessitating the condition on the left-hand side. This is similar to $\beta$-reduction in the $\lambda$-calculus, where an untyped term can reduce to a simply typed term.

The remainder of the inference rules for equality simply close under each constructor, reflexivity, symmetry, and transitivity. It is only necessary to give symmetry and transitivity rules for terms, and a reflexivity rule for variables, with these properties following for the other judgements by simple induction.

**Lemma 2.2.5.** *The definitional equality relations on terms, types, and substitutions are equivalence relations, for any* $\mathcal{R}$.

*Proof.* Proofs of these are found in Catt.Typing.Properties.Base. □

It is also possible to prove that each term has a canonical type.

**Definition 2.2.6.** The *canonical type* of a term $t : \mathsf{Term}_\Gamma$, $\mathrm{Ty}(t)$, is defined by a case split on $t$. If $t$ is a variable then the canonical type is the corresponding type in the context $\Gamma$. Otherwise, if $t \equiv \mathsf{Coh}_{(\Delta\,;\,A)}[\sigma]$ then the canonical type is $A[\![\sigma]\!]$.

This can be used to show that the type of a well-formed term is unique up to definitional equality, and is equal to this canonical type.

**Lemma 2.2.7.** *If* $\Gamma \vdash s : A$*, then* $\Gamma \vdash s : ty(s)$ *and* $\Gamma \vdash A = \mathrm{Ty}(s)$*. Further, if* $\Gamma \vdash s : A$ *and* $\Gamma \vdash s : B$ *then* $\Gamma \vdash A = B$*.*

*Proof.* We prove the first part by induction on the derivation $\Gamma \vdash s : A$. If the derivation is derived from the conversion rule applied to $\Gamma \vdash s : B$ and $\Gamma \vdash A = B$, then by inductive hypothesis we have $\Gamma \vdash s : \mathrm{Ty}(s)$ and $\Gamma \vdash B = \mathrm{Ty}(s)$. By transitivity, we obtain $\Gamma \vdash A = \mathrm{Ty}(s)$ as required. The second part follows directly from the applying the first part to both derivations. □

Using the canonical type, we can define the canonical identity on a term.

**Definition 2.2.8.** Given a term $t : \mathsf{Term}_\Gamma$, let its *canonical identity* be given by:

$$\mathsf{id}(t) \equiv \mathsf{id}(\mathrm{Ty}(t), t)$$

This construction can be iterated, and we say that a term is an *iterated canonical identity* if it is on the form $\mathsf{id}^k(t)$ for some $k$.

There is not much more that can be proved about the definitional equality at this point without knowing more about the rule set $\mathcal{R}$. In Section 2.4, certain conditions will be imposed on the set of equality rules, that will allow further lemmas to be proved in large generality.

**Disc removal** We now give our first example of an equality rule, *disc removal.* Disc removal removes unary composites, replacing them with the underlying term. We recall that for every $n$, there exists the $n$-dimensional disc context $D^n$, and that given a term $t \in \mathsf{Term}_\Gamma$ and $n$-dimensional type $A \in \mathsf{Type}_\Gamma$, there exists a substitution $\{A, t\} : D^n \to \Gamma$. The unary composite of a term $t$ of type $A$ of dimension $n$ is then the coherence:

$$\mathsf{Coh}_{(D^n\,;\,\mathrm{wk}(U^n))}[\{A, t\}]$$

Disc removal equates this with the term $t$, making the following rule admissible:

$$\frac{\Gamma \vdash t : A \qquad \Gamma \vdash A}{\Gamma \vdash \mathsf{Coh}_{(D^n\,;\,\mathrm{wk}(U^n))}[\{A, t\}] = t}\,\textsc{dr}$$

with the removal of the disc coherence giving the name to this equality rule.

Assembling disc removal into a rule set $\mathcal{R}$ is simple, as it is possible to simply give a syntactic condition with no need to refer to typing.

**Definition 2.2.9.** The *disc removal rule set*, dr, is the set consisting of the triples:

$$(\Gamma, \mathsf{Coh}_{(D^n\,;\,\mathrm{wk}(U^n))}[\{A, t\}], t)$$

for each context $\Gamma$, type $A : \mathsf{Type}_\Gamma$, and term $t : \mathsf{Term}_\Gamma$ where $n = \dim(A)$.

A set of rules $\mathcal{R}$ *contains disc removal* if $\mathsf{dr} \subseteq \mathcal{R}$. Further we say that $\mathcal{R}$ *has disc removal* if the rule DR holds in the generated theory.

The inference rule DR follows the RULE and typing properties about discs which will be given in Section 2.4.

We draw attention to the typing premise of RULE. If we know that the unary composite of a term $t$ is well-formed, then it follows that $t$ itself must have been well-formed, but we cannot infer that the term $\mathsf{Coh}_{(D^n\,;\,\mathrm{wk}(U^n))}[\{A, t\}]$ is well-formed from $t$ being well-formed. In particular, knowing that $t$ is well-formed does not constrain $A$ at all without knowing that the given type $A$ is the type of $t$. We must therefore include an additional typing premise if we want to avoid well-formed and non-well-formed terms being equated.

## 2.3 The set of operations $\mathcal{O}$

In Section 2.2.2, we introduced a set of operations $\mathcal{O}$, which allows us to vary the operations available in the theory, much like the set $\mathcal{R}$ allows us to vary the equality rules of the theory. The set $\mathcal{O}$ replaces the conditions on the support of the type contained in a coherence, and consists of a set of triples of a context $\Delta$, along with two sets $x, y \subseteq \mathrm{Var}(\Delta)$. A certain type $s \to_A t : \mathsf{Type}_\Delta$ is permitted to appear in a coherence exactly when $(\Delta, \mathrm{Supp}(s), \mathrm{Supp}(t))$ is an element of $\mathcal{O}$.

There are two key advantages to setting up the theory this way.

- A clear separation is introduced in the metatheory and formalisation between properties that are specific to the support conditions in Catt and those that are independent of the specific support conditions present.
- The results in the following sections can be proven generically for different variants of Catt.

In particular, the main utility we extract in this thesis is the ability to define groupoidal versions of the various semistrict theories we define in Chapter 4. By letting $\mathcal{O}$ consists of all possible triples, the support condition is effectively removed, producing a version of Catt closer to Grothendieck's definition of $\infty$-groupoid (see Section 1.1.2).

### 2.3.1 Operation sets

As previously mentioned, an operation set $\mathcal{O}$ consists of a collection of triples of a context $\Delta$ and two subsets of the variables of $\Delta$.

We call a subset of the variables of a context a *variable set*. In the formalisation, these variable sets are given as a list of booleans, one boolean for each variable of the context. These are given in Catt.Support, which also contains many constructions on them, including unions of these sets, subset relations, and the free variables of each piece of syntax. The variable sets of $\Delta$ form a lattice with top element $\mathrm{Var}(\Delta)$ and bottom element $\emptyset$. The free variable constructions commute with weakening, as is proved in Catt.Support.Properties by mutual induction.

We recall the function DC on these variable sets, given by DC in the formalisation, which produces the downwards closure of a variable set. This admits the following properties:

**Proposition 2.3.1.** DC *is an idempotent join-semilattice homomorphism. It preserves binary joins (unions), subset inclusions, and preserves the top and bottom element of the lattice.*

We further define the application of a substitution to a variable set below.

**Definition 2.3.2.** Given a variable set $V$ of $\Delta$ and (regular) substitution $\sigma : \Delta \to \Gamma$, we define the application of $\sigma$ to $V$, written $V[\![\sigma]\!]$ to be a variable set of $\Gamma$ given by:

$$V[\![\langle\rangle]\!] = \emptyset$$
$$V[\![\langle\sigma, t\rangle]\!] = \begin{cases} (V \setminus \{x\})[\![\sigma]\!] \cup \mathrm{FV}(t) & \text{if } x \in V \\ V[\![\sigma]\!] & \text{otherwise} \end{cases}$$

Where $x$ is assumed to be the last variable of $\Delta$ in the second case.

We note that when representing variable sets as a list of booleans, these definitions are given by simple inductions on the length of the context. These constructions admit the following properties.

**Proposition 2.3.3.** *Let $\Delta$ be a context. Then the function taking a variable set $V$ of $\Delta$ to $V[\![\sigma]\!]$ is a join-semilattice homomorphism for any substitution $\sigma : \Delta \to \Gamma$. Further, for a term $t : \mathsf{Term}_\Delta$, a type $A : \mathsf{Type}_\Delta$, or a substitution $\tau : \Theta \to_A \Delta$, the following equalities hold:*

$$\mathrm{FV}(t[\![\sigma]\!]) = \mathrm{FV}(t)[\![\sigma]\!]$$
$$\mathrm{FV}(A[\![\sigma]\!]) = \mathrm{FV}(A)[\![\sigma]\!]$$
$$\mathrm{FV}(\tau \bullet \sigma) = \mathrm{FV}(\tau)[\![\sigma]\!]$$

*and hence* $\mathrm{Var}(\Delta)[\![\sigma]\!] = \mathrm{FV}(\mathrm{id}_\Delta)[\![\sigma]\!] = \mathrm{FV}(\mathrm{id}_\Delta \bullet \sigma) = \mathrm{FV}(\sigma)$. *For any variable set* $V \subseteq \mathrm{Var}(\Theta)$ *we have:*

$$V[\![\mathrm{id}_\Theta]\!] = V \qquad V[\![\tau \bullet \sigma]\!] = V[\![\tau]\!][\![\sigma]\!]$$

*for* $\tau : \Theta \to \Delta$ *and* $\sigma : \Delta \to \Gamma$.

*Proof.* All proofs proceed by induction on the length of the context $\Delta$ and are given in Catt.Support.Properties. □

An operation set is then a collection of triples of type:

$$\Sigma_{\Delta:\mathsf{Ctx}} \mathcal{P}(\mathrm{Var}(\Delta)) \times \mathcal{P}(\mathrm{Var}(\Delta))$$

In the formalisation this is defined in Catt.Ops to be a function from a context and two variable sets of that context to a universe.

*Remark* 2.3.4. The definition of an operation set in the formalisation deviates from the presentation given here, as the version in the formalisation is proof relevant. The proof relevant definition allows us to give any type as the type of witnesses that a certain triple appears in $\mathcal{O}$, including a type containing many distinct witnesses.

If we wished to recover a definition closer to the classical set-based definition, we could enforce that this function has a universe of propositions as its codomain, instead of a universe of types, and use propositional truncations to define various versions of $\mathcal{O}$. This is however unnecessary for any of the proofs appearing in this thesis, hence the choice of the proof relevant definition for simplicity. A similar observation will apply to the definition of equality rule sets introduced in Section 2.4.

We can now introduce our first operation set, the operation set for groupoidal operations, which imposes no support conditions and allows all operations.

**Definition 2.3.5.** We define the *groupoidal operation set* Group as:

$$\mathsf{Group} = \{(\Delta, U, V) \mid \Delta : \mathsf{Ctx}, U \subseteq \mathrm{Var}(\Delta), V \subseteq \mathrm{Var}(\Delta)\}$$

We will refer to $\textsc{Catt}_{\mathcal{R}}$ with the operation set Group as *groupoidal* $\textsc{Catt}_{\mathcal{R}}$ or *groupoidal* $\textsc{Catt}$ (when $\mathcal{R} = \emptyset$).

To recover the standard definition of Catt, we must define the boundary sets of a pasting diagram. In Section 1.2.3, these are given as the free variables of the boundary inclusion substitutions of pasting diagrams. Here we will instead give a direct definition of the variable sets corresponding to the free variables of the substitutions, delaying the definition of boundary inclusions of pasting diagrams until Section 3.2.

**Definition 2.3.6.** Let $\Delta$ be a ps-context. Define the $n$-boundary variable sets $\partial_n^-(\Delta)$ and $\partial_n^+(\Delta)$ by induction on $\Delta$:

$$\partial_i^\epsilon((x : \star)) = \{x\}$$

$$\partial_i^\epsilon(\Gamma, (y : A), (f : x \to_A y)) = \begin{cases} \partial_i^\epsilon(\Gamma) & \text{if } i < \dim(A) \\ \partial_i^-(\Gamma) & \text{if } i = \dim(A) \text{ and } \epsilon = - \\ (\partial_i^+(\Gamma) \cup \{y\}) \setminus \{x\} & \text{if } i = \dim(A) \text{ and } \epsilon = + \\ \partial_i^\epsilon(\Gamma) \cup \{y, f\} & \text{otherwise} \end{cases}$$

These boundary sets appear in the formalisation as pd-bd-vs.

*Example* 2.3.7. We consider the boundaries of the pasting context $\Delta$ given by the following diagram:

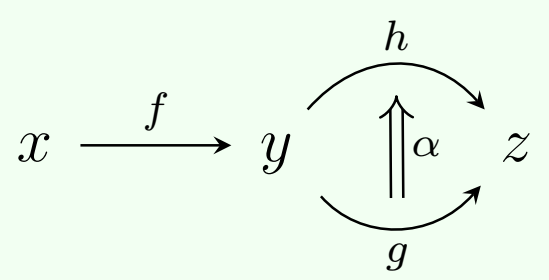

Letting $\Gamma$ be the prefix of $\Delta$ without $h$ and $\alpha$ (such that $\Delta = \Gamma, (h : y \to z), (\alpha : g \to h)$),

$$
\frac{}{\star : \mathsf{Type}_\Gamma} \qquad \frac{x \in \mathsf{Var}(\Gamma)}{x : \mathsf{Term}_\Gamma} \qquad \frac{A : \mathsf{Type}_\Gamma}{\langle A \rangle : \emptyset \to \Gamma}
$$

$$
\frac{}{\emptyset : \mathsf{Ctx}} \qquad \frac{\Gamma : \mathsf{Ctx} \qquad A : \mathsf{Type}_\Gamma}{\Gamma, (x : A) : \mathsf{Ctx}}
$$

$$
\frac{\sigma : \Delta \to_A \Gamma \qquad t : \mathsf{Term}_\Gamma \qquad B : \mathsf{Type}_\Delta}{\langle \sigma, t \rangle : \Delta, (x : B) \to_A \Gamma}
$$

$$
\frac{A : \mathsf{Type}_\Gamma \qquad s : \mathsf{Term}_\Gamma \qquad t : \mathsf{Term}_\Gamma}{s \to_A t : \mathsf{Type}_\Gamma}
$$

$$
\frac{\Delta : \mathsf{Ctx} \qquad A : \mathsf{Type}_\Delta \qquad \sigma : \Delta \to_\star \Gamma}{\mathsf{Coh}_{(\Delta\,;\,A)}[\sigma] : \mathsf{Term}_\Gamma}
$$

(a) Syntax.

$$
\frac{}{\emptyset \vdash} \qquad \frac{\Gamma \vdash \qquad \Gamma \vdash A}{\Gamma, (x : A) \vdash} \qquad \frac{}{\Gamma \vdash \star}
$$

$$
\frac{\Gamma \vdash s : A \qquad \Gamma \vdash A \qquad \Gamma \vdash t : A}{\Gamma \vdash s \to_A t}
$$

$$
\frac{\Gamma \vdash A}{\Gamma \vdash \langle A \rangle : \emptyset} \qquad \frac{\Gamma \vdash \sigma : \Delta \qquad \Gamma \vdash t : A[\![\sigma]\!]}{\Gamma \vdash \langle \sigma, t \rangle : \Delta, (x : A)}
$$

$$
\frac{(x : A) \in \Gamma}{\Gamma \vdash x : A} \qquad \frac{\Gamma \vdash t : A \qquad \Gamma \vdash A = B}{\Gamma \vdash t : B}
$$

$$
\frac{\begin{array}{c}\Delta \vdash_{\mathsf{ps}} \qquad \Delta \vdash s \to_A t \\ \Gamma \vdash \sigma : \Delta \qquad (\Delta, \mathsf{Supp}(s), \mathsf{Supp}(t)) \in \mathcal{O}\end{array}}{\Gamma \vdash \mathsf{Coh}_{(\Delta\,;\,s \to_A t)}[\sigma] : s[\![\sigma]\!] \to_{A[\![\sigma]\!]} t[\![\sigma]\!]}
$$

(b) Typing.

$$
\frac{\Gamma \vdash s : A \qquad (\Gamma, s, t) \in \mathcal{R}}{\Gamma \vdash s = t}\textsc{rule} \qquad \frac{x \in \mathsf{Var}(\Gamma)}{\Gamma \vdash x = x} \qquad \frac{\Gamma \vdash s = t}{\Gamma \vdash t = s}
$$

$$
\frac{\Gamma \vdash s = t \qquad \Gamma \vdash t = u}{\Gamma \vdash s = u} \qquad \frac{\Delta \vdash A = B \qquad \Gamma \vdash \sigma = \tau}{\Gamma \vdash \mathsf{Coh}_{(\Delta\,;\,A)}[\sigma] = \mathsf{Coh}_{(\Delta\,;\,B)}[\tau]} \qquad \frac{}{\Gamma \vdash \star = \star}
$$

$$
\frac{\Gamma \vdash s = s' \qquad \Gamma \vdash t = t' \qquad \Gamma \vdash A = A'}{\Gamma \vdash s \to_A t = s' \to_{A'} t'} \qquad \frac{\Gamma \vdash A = B}{\Gamma \vdash \langle A \rangle = \langle B \rangle}
$$

$$
\frac{\Gamma \vdash \sigma = \tau \qquad \Gamma \vdash s = t}{\Gamma \vdash \langle \sigma, s \rangle = \langle \tau, t \rangle}
$$

(c) Equality.

$$
\frac{}{(x : \star) \vdash_{\mathsf{ps}} x : \star}
$$

$$
\frac{\Gamma \vdash_{\mathsf{ps}} x : A}{\Gamma, (y : A), (f : x \to_A y)}
$$

$$
\frac{\Gamma \vdash_{\mathsf{ps}} x : s \to_A t}{\Gamma \vdash_{\mathsf{ps}} t : A} \qquad \frac{\Gamma \vdash_{\mathsf{ps}} x : \star}{\Gamma \vdash_{\mathsf{ps}}}
$$

(d) Ps-contexts.

$$
\mathrm{FV}(\star) = \{\} \qquad \mathrm{FV}(\langle A \rangle) = \mathrm{FV}(A)
$$

$$
\mathrm{FV}(x) = \{x\} \text{ for } x \in \mathrm{Var}
$$

$$
\mathrm{FV}(\mathsf{Coh}_{(\Delta\,;\,A)}[\sigma]) = \mathrm{FV}(\sigma)
$$

$$
\mathrm{FV}(s \to_A t) = \mathrm{FV}(s) \cup \mathrm{FV}(A) \cup \mathrm{FV}(t)
$$

$$
\mathrm{FV}(\langle \sigma, t \rangle) = \mathrm{FV}(\sigma) \cup \mathrm{FV}(t)
$$

(e) Free variables.

Figure 2.1: $\textsc{Catt}_\mathcal{R}$: syntax, typing, and operations.

$$\begin{aligned}
\mathrm{DC}_\emptyset(\emptyset) &= \emptyset \\
\mathrm{DC}_{\Gamma,x:A}(V) &= \begin{cases} \mathrm{DC}_\Gamma(V) & \text{if } x \notin V \\ \{x\} \cup \mathrm{DC}_\Gamma(V \setminus \{x\} \cup \mathsf{FV}(A)) & \text{if } x \in V \end{cases} \\
\mathsf{Supp}(t) &= \mathrm{DC}_\Gamma(\mathsf{FV}(t)) \text{ for } t \in \mathsf{Term}_\Gamma \\
\mathsf{Supp}(A) &= \mathrm{DC}_\Gamma(\mathsf{FV}(A)) \text{ for } A \in \mathsf{Type}_\Gamma \\
\mathsf{Supp}(\sigma) &= \mathrm{DC}_\Gamma(\mathsf{FV}(\sigma)) \text{ for } \sigma : \Delta \to_A \Gamma
\end{aligned}$$

(f) Support.

$$\begin{aligned}
x[\![\sigma]\!] &= t \text{ if } (x \mapsto t) \in \sigma \\
\mathsf{Coh}_{(\Theta\,;\,A)}[\tau][\![\sigma]\!] &= \begin{cases} \mathsf{Coh}_{(\Theta\,;\,A)}[\tau \bullet \sigma] & \text{if } \dim(\mathsf{Ty}(\sigma)) = 0 \\ \mathsf{Coh}_{(\Sigma(\Theta)\,;\,\Sigma(A))}[\Sigma(\tau)][\![\downarrow \sigma]\!] & \text{otherwise} \end{cases} \\
\star[\![\sigma]\!] &= \mathsf{Ty}(\sigma) \\
(s \to_A t)[\![\sigma]\!] &= s[\![\sigma]\!] \to_{A[\![\sigma]\!]} t[\![\sigma]\!] \\
\langle A \rangle \bullet \sigma &= \langle A[\![\sigma]\!] \rangle \\
\langle \tau, t \rangle \bullet \sigma &= \langle \tau \bullet \sigma, t[\![\sigma]\!] \rangle
\end{aligned}$$

(g) Substitution application.

$$\begin{aligned}
\Sigma(\emptyset) &= (N : \star), (S : \star) \\
\Sigma(\Gamma, (x : A)) &= \Sigma\Gamma, (x : \Sigma A) \\
\Sigma(\star) &= N \to_\star S \\
\Sigma(s \to_A t) &= \Sigma s \to_{\Sigma A} \Sigma t \\
\Sigma(x) &= x \\
\Sigma(\mathsf{Coh}_{(\Delta\,;\,A)}[\sigma]) &= \mathsf{Coh}_{(\Sigma(\Delta)\,;\,\Sigma(A))}[\Sigma(\sigma)] \\
\Sigma(\sigma) &= \downarrow(\Sigma'(\sigma))
\end{aligned}$$

$$\begin{aligned}
\Sigma'(\langle A \rangle) &= \langle \Sigma(A) \rangle \\
\Sigma'(\langle \sigma, x \rangle) &= \langle \Sigma'(\sigma), \Sigma(t) \rangle \\
\downarrow\langle s \to_A t \rangle &= \langle A, s, t \rangle \\
\downarrow\langle \sigma, t \rangle &= \langle \downarrow \sigma, t \rangle
\end{aligned}$$

(h) Suspension.

$$\begin{aligned}
\mathrm{wk}(\star) &= \star \\
\mathrm{wk}(s \to_A t) &= \mathrm{wk}(s) \to_{\mathrm{wk}(A)} \mathrm{wk}(t) \\
\mathrm{wk}(x) &= x \\
\mathrm{wk}(\mathsf{Coh}_{(\Delta\,;\,A)}[\sigma]) &= \mathsf{Coh}_{(\Delta\,;\,A)}[\mathrm{wk}(\sigma)] \\
\mathrm{wk}(\langle A \rangle) &= \langle \mathrm{wk}(A) \rangle \\
\mathrm{wk}(\langle \sigma, t \rangle) &= \langle \mathrm{wk}(\sigma), \mathrm{wk}(t) \rangle
\end{aligned}$$

(i) Weakening.

$$\begin{aligned}
\mathsf{id}_\emptyset &= \langle \star \rangle \\
\mathsf{id}_{\Gamma,(x:A)} &= \langle \mathrm{wk}(\mathsf{id}_\Gamma), x \rangle
\end{aligned}$$

(j) Identity substitution.

Figure 2.1: CATT$_\mathcal{R}$: syntax, typing, and operations.

we immediately see that:

$$\begin{aligned}\partial_1^\epsilon(\Gamma) &= \partial_1^\epsilon((x:\star),(y:\star),(f:x\to y)) \cup \{z,g\}\\ &= \partial_1^\epsilon((x:\star)) \cup \{y,f,z,g\}\\ &= \{x,y,f,z,g\}\end{aligned}$$

and therefore we have that:

$$\begin{aligned}\partial_1^-(\Delta) &= \partial_1^-(\Gamma) = \mathrm{Var}(\Gamma)\\ \partial_1^+(\Delta) &= (\partial_1^-(\Gamma) \cup \{h\}) \setminus \{g\} = \{x,y,f,z,h\}\end{aligned}$$

Similar calculations show that $\partial_2^\epsilon(\Delta) = \mathrm{Var}(\Delta)$, $\partial_0^-(\Delta) = \{x\}$, and $\partial_0^+(\Delta) = \{z\}$.

The following lemma is immediate:

**Lemma 2.3.8.** *If* $n \geq \dim(\Delta)$*, then* $\partial_n^\epsilon(\Delta) = \mathrm{Var}(\Delta)$.

*Proof.* A simple induction on the definition. A formalised proof appears as pd-bd-vs-full in the module Catt.Support.Properties. □

With this definition we can introduce the regular operation set, which recovers the regular support conditions used in the definition of Catt.

**Definition 2.3.9.** The *regular operation set* Reg is defined to be:

$$\mathrm{Reg} = \{(\Delta, \mathrm{Var}(\Delta), \mathrm{Var}(\Delta)) \mid \Delta \vdash_{\mathrm{ps}}\} \cup \{(\Delta, \partial^-_{\dim(\Delta)-1}(\Delta), \partial^+_{\dim(\Delta)-1}(\Delta)) \mid \Delta \vdash_{\mathrm{ps}}\}$$

The first component allows equivalences to be well-formed, and the second gives the support condition for composites.

The regular operation set has more standard presentation.

**Proposition 2.3.10.** *Let the set Std of standard operations be defined as:*

$$\mathit{Std} = \{(\Delta, \partial_n^-(\Delta), \partial_n^+(\Delta)) \mid \Delta \vdash_{\mathrm{ps}}, n \geq \dim(\Delta) - 1\}$$

*Then Std = Reg.*

*Proof.* Suppose $(\Delta, U, V) \in \mathrm{Reg}$. If $U = \partial^-_{\dim(\Delta)-1}(\Delta)$ and $V = \partial^+_{\dim(\Delta)-1}(\Delta)$, then $(\Delta, U, V)$ is trivially in Std by letting $n = \dim(\Delta) - 1$. If instead $U = V = \mathrm{Var}(\Delta)$, then $(\Delta, U, V) \in \mathrm{Std}$ by letting $n = \dim(\Delta)$ and applying Lemma 2.3.8.

Conversely, assume $(\Delta, U, V) \in \mathrm{Std}$. Then there is $n \geq \dim(\Delta) - 1$ with $U = \partial_n^-(\Delta)$ and $V = \partial_n^+(\Delta)$. If $n = \dim(\Delta) - 1$ then $(\Delta, U, V)$ is trivially in Reg, and otherwise by Lemma 2.3.8 we have $U = V = \mathrm{Var}(\Delta)$, and so $(\Delta, U, V)$ is again an element of Reg. Hence, $\mathrm{Reg} = \mathrm{Std}$. □

This more uniform presentation is sometimes easier to work with, and will be used to prove properties of Reg in Section 2.3.2.

*Remark* 2.3.11. By letting $\mathcal{O} = \emptyset$, we recover the type theory GSeTT [BFM24], a type theory for globular sets.

It would be possible to generalise the notion of operation set presented here by instead letting the set $\mathcal{O}$ consist of triples $(\Delta, s, t)$ where $s$ and $t$ are terms over $\Delta$ instead of variable sets over $\Delta$. This would allow more control over which operations were allowed in the theory. As an example, we would be able to restrict the class of composites to contain only the standard composites, or even further restrict it to binary composites.

This is however unnecessary to present the regular and groupoidal versions of $\textsc{Catt}_{\mathcal{R}}$. By only allowing the set of available operations to be specified up to the support of the contained terms, it is possible to show that a coherence being an operation is closed under equality by proving that equality preserves the support of a term.

### 2.3.2 Operation properties

Currently, our set of operations is completely unconstrained, and we will be limited in the constructions that can be made in $\textsc{Catt}_{\mathcal{R}}$. We therefore constrain these sets in two ways. The first enforces that our set of operations is closed under suspension, for which we need to be able to suspend variable sets. This is defined in the formalisation as susp-vs.

**Definition 2.3.12.** Let $\Delta$ be a context. The suspension of a variable set $V$ over $\Delta$ is defined to be:

$$\Sigma(V) = \{N, S\} \cup V$$

where $\Sigma(V)$, the suspension of $V$ is a variable set over $\Sigma(\Delta)$.

The suspension of a variable set commutes with taking the support of a piece of syntax, as shown in the next lemma.

**Lemma 2.3.13.** *The following equalities hold:*

$$\text{Supp}(\Sigma(s)) = \Sigma(\text{Supp}(s)) \qquad \text{Supp}(\Sigma(A)) = \Sigma(\text{Supp}(A)) \qquad \text{Supp}(\Sigma(\sigma)) = \Sigma(\text{Supp}(\sigma))$$

*for term* $s : \mathsf{Term}_\Gamma$, *type* $A : \mathsf{Type}_\Gamma$, *and substitution* $\sigma : \Delta \to_\star \Gamma$.

*Proof.* All equalities hold by a mutual induction on terms, types, and substitutions, with a secondary induction on the context $\Gamma$ for the case of the variables and the base type $\star$. These calculations are given in Catt.Suspension.Support. □

We can then define our first property on operation sets.

**Definition 2.3.14.** An operation set $\mathcal{O}$ is *suspendable* if:

$$(\Delta, U, V) \in \mathcal{O} \implies (\Sigma(\Delta), \Sigma(U), \Sigma(V)) \in \mathcal{O}$$

For $\Delta : \mathsf{Ctx}$ and $U, V \subseteq \mathsf{Var}(\Delta)$.

The groupoidal operation set is trivially suspendable. To show that the regular operation set is suspendable, we prove the following proposition.

**Proposition 2.3.15.** *Let $\Delta$ be a ps-context. Then:*

$$\Sigma(\partial_n^\epsilon(\Delta)) = \partial_{n+1}^\epsilon(\Sigma(\Delta))$$

*for $n \in \mathbb{N}$ and $\epsilon \in \{-, +\}$.*

*Proof.* We proceed by induction on $\Delta$. First suppose $\Delta = (x : \star)$. We then have:

$$\Sigma(\partial_n^\epsilon((x : \star))) = \Sigma(\{x\}) = \{N, S, x\} = \partial_{n+1}^\epsilon(\Sigma((x : \star)))$$

Now suppose that $\Delta = \Delta', (y : A), (f : x \to_A y)$. We split into cases on $n$, $\dim(A)$, and $\epsilon$:

- If $n < \dim(A)$ then

$$\begin{aligned} \Sigma(\partial_n^\epsilon(\Delta)) &= \Sigma(\partial_n^\epsilon(\Delta')) \\ &= \partial_{n+1}^\epsilon(\Sigma(\Delta')) && \text{by inductive hypothesis} \\ &= \partial_{n+1}^\epsilon(\Sigma(\Delta)) && \text{as } n+1 < \dim(\Sigma(A)) \end{aligned}$$

- If $n = \dim(A)$ and $\epsilon = -$ then the proof is similar to the preceding case.
- If $n = \dim(A)$ and $\epsilon = +$ then:

$$\begin{aligned} \Sigma(\partial_n^+(\Delta)) &= \Sigma((\partial_n^+(\Delta') \cup \{y\}) \setminus \{x\}) \\ &= (\Sigma(\partial_n^+(\Delta')) \cup \{y\}) \setminus \{x\} \\ &= (\partial_{n+1}^+(\Sigma(\Delta')) \cup \{y\}) \setminus \{x\} && \text{by inductive hypothesis} \\ &= \partial_{n+1}^+(\Sigma(\Delta)) && \text{as } n+1 = \dim(\Sigma(A)) \end{aligned}$$

- If $n > \dim(A)$ then

$$\begin{aligned} \Sigma(\partial_n^\epsilon(\Delta)) &= \Sigma((\partial_n^\epsilon(\Delta') \cup \{y, f\}) \\ &= \Sigma(\partial_n^+(\Delta')) \cup \{y, f\} \\ &= \partial_{n+1}^+(\Sigma(\Delta')) \cup \{y, f\} && \text{by inductive hypothesis} \\ &= \partial_{n+1}^+(\Sigma(\Delta)) && \text{as } n+1 > \dim(\Sigma(A)) \end{aligned}$$

Hence, the desired equality holds in all cases. □

**Corollary 2.3.16.** *The regular operation set is suspendable.*

*Proof.* By Proposition 2.3.10, it suffices to show that the standard operation set is suspendable, which is clear from the above proposition. □

The second restriction we put on operation sets is that there are enough operations to create the standard coherences presented in Section 1.2.4.

**Definition 2.3.17.** An operation set $\mathcal{O}$ *contains the standard operations* if $\mathsf{Std} \subseteq \mathcal{O}$.

The groupoidal operation set clearly contains the standard operations, and the regular operation set also does due to Proposition 2.3.10. The empty operation set does not contain the standard operations. We end this section with the following proposition about the support of terms in a disc.

**Proposition 2.3.18.** *For $n \in \mathbb{N}$ the following two equations hold:*

$$\partial_n^-(D^{n+1}) = \mathrm{Var}(S^n) \cup \{d_n^-\} = \mathrm{Var}(D^n) \qquad \partial_n^+(D^{n+1}) = \mathrm{Var}(S^n) \cup \{d_{n+1}^+\}$$

*Further, the following equations hold:*

$$\mathrm{FV}(U^n) = \mathrm{Var}(S^n) \qquad \mathrm{Supp}(d_n^-) = \mathrm{Var}(D^n) = \partial_n^-(D^{n+1}) \qquad \mathrm{Supp}(d_n^+) = \partial_n^+(D^{n+1})$$

*again for any $n \in \mathbb{N}$.*

*Proof.* The first equations follow by a simple case analysis, using that $\partial_n^-(D^n) = \mathrm{Var}(D^n)$ by Lemmas 2.2.4(i) and 2.3.8. The free variables of $U^n$ are easily calculated inductively, and the support of $d_n^-$ and $d_n^+$ are easy to compute using the first parts of the proposition, and that $FV(U^n) \subseteq \mathrm{Supp}(d_n^-)$ and $\mathrm{FV}(U^n) \subseteq \mathrm{Supp}(d_n^+)$ as the support of a term is downwards closed.

These proofs are formalised in Catt.Discs.Support. □

**Corollary 2.3.19.** *Both $(D^{n+1}, d_n^-, d_n^+)$ and $(D^n, d_n, d_n)$ are in Std for each $n$.*

## 2.4 The set of equality rules $\mathcal{R}$

In Catt$_\mathcal{R}$, the definitional equality relation is generated by a set of rules $\mathcal{R}$ formed of triples containing a context and two terms in the context which should be made equal. In this section we discuss some operations on these equality sets and properties that they may have.

*Remark* 2.4.1. In the formalisation the set of equality rules is defined similarly to the set of operations $\mathcal{O}$. It is defined as a function that takes a context and two terms over that context and returns a type. It is therefore proof relevant in the same way as the operation sets.

The equality rule sets inherit some operations and relations just by being sets. We can easily form the empty equality set, which allows us to recover the weak type theory Catt, and given two equality sets we can take their union, to get a type theory with equalities from both sets (we note that the equality generated by a union is in general coarser than the union of the equalities generated by the individual sets).

To aid readability when reasoning about typing and equality with multiple distinct operations, we may subscript the turnstile symbol in various judgements with the set of equality rules being used. For example, we may write the judgements for typing of a term $t$ in the type theory generated from rules $\mathcal{R}$ as

$$\Gamma \vdash_\mathcal{R} t : A$$

and the corresponding judgement for the equality of two terms $s$ and $t$ as

$$\Gamma \vdash_{\mathcal{R}} s = t$$

Equality rule sets can also be subsets of each other, leading to the following lemma.

**Lemma 2.4.2.** *Let $\mathcal{R}$ and $\mathcal{S}$ be two equality rule sets and suppose that*

$$\Gamma \vdash_{\mathcal{S}} s = t$$

*for all $(\Gamma, s, t) \in \mathcal{R}$ with $\Gamma \vdash_{\mathcal{S}} s : A$ for some $A : \mathsf{Type}_\Gamma$. Then the following inference rules hold:*

$$\frac{\Gamma \vdash_{\mathcal{R}}}{\Gamma \vdash_{\mathcal{S}}} \qquad \frac{\Gamma \vdash_{\mathcal{R}} t : A}{\Gamma \vdash_{\mathcal{S}} t : A} \qquad \frac{\Gamma \vdash_{\mathcal{R}} A}{\Gamma \vdash_{\mathcal{S}} A} \qquad \frac{\Gamma \vdash_{\mathcal{R}} \sigma : \Delta}{\Gamma \vdash_{\mathcal{S}} \sigma : \Delta}$$

$$\frac{\Gamma \vdash_{\mathcal{R}} s = t}{\Gamma \vdash_{\mathcal{S}} s = t} \qquad \frac{\Gamma \vdash_{\mathcal{R}} A = B}{\Gamma \vdash_{\mathcal{S}} A = B} \qquad \frac{\Gamma \vdash_{\mathcal{R}} \sigma = \tau}{\Gamma \vdash_{\mathcal{S}} \sigma = \tau}$$

*In particular these inference rules hold when $\mathcal{R} \subseteq \mathcal{S}$.*

*Proof.* Follows from a simple induction. Details are given in the formalisation in module Catt.Typing.Rule.Properties. □

**Corollary 2.4.3.** *Any context, term, type, or substitution that is well-formed in* Catt *is also well-formed in* $\text{Catt}_{\mathcal{R}}$*, for any equality set $\mathcal{R}$.*

Furthermore, we can immediately show that the application of a substitution to a piece of syntax that is well-formed in Catt is well-formed.

**Lemma 2.4.4.** *Let $\mathcal{R}$ be any equality rule set. Then the following inference rules hold for $\sigma : \Delta \to_\star \Gamma$:*

$$\frac{\Delta \vdash_{\emptyset} A \qquad \Gamma \vdash_{\mathcal{R}} \sigma : \Delta}{\Gamma \vdash_{\mathcal{R}} A[\![\sigma]\!]} \qquad \frac{\Delta \vdash_{\emptyset} s : A \qquad \Gamma \vdash_{\mathcal{R}} \sigma : \Delta}{\Gamma \vdash_{\mathcal{R}} s[\![\sigma]\!] : A[\![\sigma]\!]} \qquad \frac{\Delta \vdash_{\emptyset} \tau : \Theta \qquad \Gamma \vdash_{\mathcal{R}} \sigma : \Delta}{\Gamma \vdash_{\mathcal{R}} \tau \bullet \sigma : \Theta}$$

*where the judgements with a subscript empty set are judgements in the theory generated by the empty rule sets (judgements in* Catt*).*

*Proof.* Follows immediately from a mutual induction, using that any equality in Catt is syntactic. The proof is formalised in Catt.Typing.Properties.Base. □

An arbitrary set $\mathcal{R}$ has very few restrictions on the equality relation it generates, and the terms that are well-formed because of it. A rule set $\mathcal{R}$ could identify terms of different types, or identify two different variables (or even identify all variables or terms). This makes it difficult to prove much about the theory generated by an arbitrary set $\mathcal{R}$.

To this end, we introduce certain conditions that these equality rule sets can satisfy. The first three of these conditions put certain closure properties on the set of rules $\mathcal{R}$, and each allow various constructions to be well-formed. We call theories that satisfy these three properties *tame theories* and introduce these in Section 2.4.1. In Section 2.4.2, we introduce two more conditions which take the form of a property that the generated equality must satisfy.

By introducing these conditions, we can prove various metatheoretic properties about $\textsc{Catt}_{\mathcal{R}}$ in a modular and generic way. This will allow the re-use of many constructions and proofs about the properties of these constructions in Chapter 4, where two distinct type theories for semistrict $\infty$-categories are given.

In the following subsections, we will also show that the rule set for disc removal satisfies all these conditions. For all these conditions, we will have that if the condition holds on $\mathcal{R}$ and on $\mathcal{S}$ then it also holds on $\mathcal{R} \cup \mathcal{S}$, and so these conditions can be proved individually for each rule set that is introduced. Further, the empty set will satisfy all of these conditions vacuously, and so all proofs and constructions in the section apply to Catt.

### 2.4.1 Tame theories

Here we introduce the three core conditions on the equality rule set $\mathcal{R}$ which we expect hold for any reasonable choice of rule set:

- The *weakening condition*, which allows weakening to be well-formed.
- The *suspension condition*, which allows suspension to be well-formed.
- The *substitution condition*, which implies that the application of substitution to terms, types, and other substitutions (as substitution composition) preserves typing and equality.

We call an equality rule set *tame* if it satisfies all three of these conditions, and call the corresponding theory $\textsc{Catt}_{\mathcal{R}}$ a *tame theory*.

**Weakening condition** For the weakening operation to be well-formed, meaning that the weakening of a well-formed piece of syntax is itself well-formed, the following closure property must hold on the set of rules $\mathcal{R}$.

**Definition 2.4.5.** A set of rules $\mathcal{R}$ satisfies the *weakening condition* if for all $(\Gamma, s, t) \in \mathcal{R}$ we have:

$$((\Gamma, (x : A)), \mathrm{wk}(s), \mathrm{wk}(t)) \in \mathcal{R}$$

for all $A : \mathsf{Type}_\Gamma$.

The following proposition is immediately provable by mutual induction on typing and equality. Its proof is given in Catt.Typing.Properties.Weakening.

**Proposition 2.4.6.** *Let $\mathcal{R}$ satisfy the weakening condition. Then the following inference rules are admissible in $\textsc{Catt}_{\mathcal{R}}$.*

$$\frac{\Gamma \vdash B}{\Gamma, (x : A) \vdash \mathrm{wk}(B)} \qquad \frac{\Gamma \vdash s : B}{\Gamma, (x : A) \vdash \mathrm{wk}(s) : \mathrm{wk}(B)} \qquad \frac{\Gamma \vdash \sigma : \Delta}{\Gamma, (x : A) \vdash \mathrm{wk}(\sigma) : \Delta}$$

*for types $A, B : \mathsf{Type}_\Gamma$, term $s : \mathsf{Term}_\Gamma$ and substitution $\sigma : \Delta \to_C \Gamma$.*

**Corollary 2.4.7.** *If $\mathcal{R}$ satisfies the weakening condition then:*

$$\Gamma \vdash \mathsf{id}_\Gamma : \Gamma$$

*for any* $\Gamma : \mathsf{Ctx}$.

Using only the above proposition we can immediately prove typing properties for several constructions using discs.

**Lemma 2.4.8.** *Suppose the weakening condition holds. Then the following judgements hold:*

$$S^n \vdash U^n \qquad S^n \vdash \qquad D^n \vdash$$

*For all* $n \in \mathbb{N}$*. Further, the following inference rules are admissible:*

$$\frac{\Gamma \vdash A \qquad n = \dim(A)}{\Gamma \vdash \{A\} : S^n} \qquad \frac{\Gamma \vdash A \qquad n = \dim(A) \qquad \Gamma \vdash s : A}{\Gamma \vdash \{A, s\} : D^n}$$

$$\frac{\Gamma \vdash \{A\} : S^n}{\Gamma \vdash A} \qquad \frac{\Gamma \vdash \{A, s\} : D^n}{\Gamma \vdash A} \qquad \frac{\Gamma \vdash \{A, s\} : D^n}{\Gamma \vdash s : A}$$

*For* $A : \mathsf{Type}_\Gamma$ *and* $s : \mathsf{Term}_\Gamma$.

*Proof.* The first three typing judgements follow from a simple mutual induction, making use of the typing of weakening. We prove that $\Gamma \vdash \{A\} : S^n$ by induction on $n$ and $A$. The base case is trivial. For the inductive step we assume that $\Gamma \vdash s \to_A t$, with $n = \dim(A)$, and want to show that:

$$\Gamma \vdash \langle \{A\}, s, t \rangle : S^n, (d^-_{n+1} : U^n), (d^+_{n+1} : \mathrm{wk}(U^n))$$

The judgement $\Gamma \vdash \{A\} : S^n$ holds by inductive hypothesis, and so it remains to show that the following two judgements hold:

$$\Gamma \vdash s : U^n[\![\{A\}]\!] \qquad \Gamma \vdash t : \mathrm{wk}(U^n)[\![\langle \{A\}, s \rangle]\!]$$

As $\Gamma \vdash s \to_A t$, we know (by case analysis on the typing derivation) that $\Gamma \vdash s : A$ and $\Gamma \vdash t : A$. These judgements are sufficient to finish the proof, since $A \equiv U^n[\![\{A\}]\!] \equiv \mathrm{wk}(U^n)[\![\langle \{A\}, s \rangle]\!]$ by Lemma 2.2.4(vi) and the interaction of weakening with substitution application.

To show that $\Gamma \vdash A$ follows from $\Gamma \vdash \{A\} : S^n$, we instead show that $\Gamma \vdash U^n[\![\{A\}]\!]$, leveraging that typing is invariant under syntactic equality. The typing of $U^n[\![\{A\}]\!]$ follows from $U^n$ being well-formed in Catt (as it is well-formed in any theory with the weakening property), and Lemma 2.4.4. The second to last inference rule follows trivially from the preceding one. For the last rule, we get that $\Gamma \vdash s : U^n[\![\{A\}]\!]$ by case analysis on $\Gamma \vdash \{A, s\} : D^n$, and so we are finished by the invariance of typing rules under syntactic equality. □

If we further have that the set of operations includes the standard operations then we get the following corollary.

**Corollary 2.4.9.** *Suppose that $\mathcal{O}$ contains the standard operations in addition to $\mathcal{R}$ satisfying the weakening condition. Then the following are equivalent:*

- $\Gamma \vdash A$ *and* $\Gamma \vdash t : A$,
- *There exists some* $B : \mathsf{Type}_\Gamma$ *such that* $\Gamma \vdash \mathsf{id}(A, t) : B$,
- $\Gamma \vdash \mathsf{id}(A, t) : t \to_A t$.

*If we further have that* $\dim(A) \neq 0$ *then the following two conditions are also equivalent:*

- *There exists some* $B : \mathsf{Type}_\Gamma$ *such that* $\Gamma \vdash \mathsf{Coh}_{(D^n\,;\,\mathrm{wk}(U^n))}[\{A, t\}] : B$,
- $\Gamma \vdash \mathsf{Coh}_{(D^n\,;\,\mathrm{wk}(U^n))}[\{A, t\}] : A$.

*where* $n = \dim(A)$.

*Proof.* The proof follows from Lemmas 2.4.8 and 2.2.4(vi) and Corollary 2.3.19. □

We end this discussion with the following proposition.

**Proposition 2.4.10.** *The set dr satisfies the weakening condition.*

*Proof.* It suffices to show that for all $\Gamma$ : Ctx, $A, B$ : $\mathsf{Type}_\Gamma$, and $t$ : $\mathsf{Term}_\Gamma$ that:

$$((\Gamma, (x : B)), \mathsf{Coh}_{(D^n\,;\,\mathrm{wk}(U^n))}[\mathrm{wk}(\{A, t\})], \mathrm{wk}(t)) \in \mathsf{dr}$$

when $n = \dim(A)$. By Lemma 2.2.4(iii), $\mathrm{wk}(\{A, t\}) \equiv \{\mathrm{wk}(A), \mathrm{wk}(t)\}$ and so the triple above is clearly contained in dr. □

The semistrict type theories $\textsc{Catt}_{\mathsf{su}}$ and $\textsc{Catt}_{\mathsf{sua}}$ (which will be introduced in Sections 4.2 and 4.3) will be generated by equality rule sets that are the union of multiple smaller rule sets (including disc removal). Since the weakening condition is clearly preserved under unions, we will be able to show that the rule sets generating $\textsc{Catt}_{\mathsf{su}}$ and $\textsc{Catt}_{\mathsf{sua}}$ satisfy the weakening condition by showing that it is satisfied by each individual component.

**Suspension condition** For suspension, we introduce the following condition, which is similar to the corresponding condition for weakening.

**Definition 2.4.11.** A set of equality rules $\mathcal{R}$ satisfies the *suspension condition* if

$$(\Sigma(\Gamma), \Sigma(s), \Sigma(t)) \in \mathcal{R}$$

for all $(\Gamma, s, t) \in \mathcal{R}$.

If the set of operations $\mathcal{O}$ is suspendable, then this condition is sufficient to show that the suspension of a well-formed piece of syntax is well-formed.

**Proposition 2.4.12.** *Suppose $\mathcal{O}$ is suspendable and $\mathcal{R}$ satisfies the suspension condition. Then the following inference rules are admissible for* $\Gamma, \Delta, \Delta'$ : Ctx, $A, B, C, D$ : $\mathsf{Type}_\Gamma$, $s, t$ :

$\mathsf{Term}_\Gamma$, $\sigma : \Delta \to_C \Gamma$, *and* $\tau : \Delta' \to_D \Gamma$.

$$\frac{\Gamma \vdash}{\Sigma(\Gamma) \vdash} \qquad \frac{\Gamma \vdash A}{\Sigma(\Gamma) \vdash \Sigma(A)} \qquad \frac{\Gamma \vdash s : A}{\Sigma(\Gamma) \vdash \Sigma(s) : \Sigma(A)} \qquad \frac{\Gamma \vdash \sigma : \Delta}{\Sigma(\Gamma) \vdash \Sigma'(\sigma) : \Delta}$$

$$\frac{\Gamma \vdash A = B}{\Sigma(\Gamma) \vdash \Sigma(A) = \Sigma(B)} \qquad \frac{\Gamma \vdash s = t}{\Sigma(\Gamma) \vdash \Sigma(s) = \Sigma(t)} \qquad \frac{\Gamma \vdash \sigma = \tau}{\Sigma'(\Gamma) \vdash \Sigma'(\sigma) = \Sigma(\tau)}$$

*For all* $\mu : \Delta \to_{s \to_A t} \Gamma$ *and* $\mu' : \Delta' \to_{s' \to_{A'} t'} \Gamma'$ *the following two rules are admissible:*

$$\frac{\Gamma \vdash \mu : \Delta}{\Gamma \vdash \downarrow \mu : \Sigma(\Delta)} \qquad \frac{\Gamma \vdash \mu = \mu'}{\Gamma \vdash \downarrow \mu = \downarrow \mu'}$$

*and so the inference rules*

$$\frac{\Gamma \vdash \sigma : \Delta}{\Sigma(\Gamma) \vdash \Sigma(\sigma) : \Sigma(\Delta)} \qquad \frac{\Gamma \vdash \sigma = \tau}{\Sigma(\Gamma) \vdash \Sigma(\sigma) = \Sigma(\tau)}$$

*hold for* $\sigma : \Delta \to_\star \Gamma$ *and* $\tau : \Delta' \to_\star \Gamma$.

*Proof.* The rules concerning the unrestriction operation follow by simple induction on the typing judgement or equality in the premise, and in fact do not need the suspension condition.

The remainder of the rules follow from a routine mutual induction on all typing and equality rules, which can be found in Catt.Suspension.Typing. The suspendability of the operation set is used for the case involving the typing rule for coherences, which also makes use of Lemma 2.3.13. In this case, the functoriality of suspension is used to show that the coherence has the correct type. The suspension condition is used for the rule constructor of the equality of terms. □

Similarly to the weakening condition, the suspension condition is closed under unions of rule sets, and we can show it is satisfied by dr, with a similar proof to the proof for weakening.

**Proposition 2.4.13.** *The set dr satisfies the suspension condition.*

*Proof.* It is sufficient to prove that for all $\Gamma : \mathsf{Ctx}$, $A : \mathsf{Type}_\Gamma$, and $t : \mathsf{Term}_\Gamma$ that:

$$(\Sigma(\Gamma), \mathsf{Coh}_{(\Sigma(D^n)\,;\,\Sigma(\mathrm{wk}(U^n)))}[\Sigma(\{A, t\})], \Sigma(t)) \in \mathsf{dr}$$

when $n = \dim(A)$. By Lemma 2.2.4(ii), we get that $\Sigma(D^n) \equiv D^{n+1}$ and $\Sigma(\mathrm{wk}(U^n)) \equiv \mathrm{wk}(\Sigma(U^n) \equiv \mathrm{wk}(U^{n+1})$. By Lemma 2.2.4(iv), $\Sigma(\{A, t\}) \equiv \{\Sigma(A), \Sigma(t)\}$. Therefore, it is sufficient to show that:

$$(\Sigma(\Gamma), \mathsf{Coh}_{(D^{n+1}\,;\,\mathrm{wk}(U^{n+1}))}[\{\Sigma(A), \Sigma(t)\}], \Sigma(t)) \in \mathsf{dr}$$

which is clear as $\dim(\Sigma(A)) = \dim(A) + 1 = n + 1$. □

**Substitution condition** The substitution condition takes a slightly different form to the previous two conditions. Instead of requiring that the rule set is closed under application of any arbitrary substitution $\sigma$, we instead only ensure it is closed under well-formed substitutions. This will not prevent us proving that typing is closed under the application of substitutions, but will be critical in proving that the supported rules construction, which will be given in Definition 2.4.27 and is used for proving the support condition, satisfies the substitution condition.

**Definition 2.4.14.** An equality rule set $\mathcal{R}$ satisfies the *$\mathcal{R}'$-substitution condition* if:

$$(\Gamma, s[\![\sigma]\!], t[\![\sigma]\!]) \in \mathcal{R}$$

whenever $(\Delta, s, t) \in \mathcal{R}$ and $\sigma : \Delta \to_\star \Gamma$ with $\Gamma \vdash_{\mathcal{R}'} \sigma : \Delta$. We say the set $\mathcal{R}$ satisfies the *substitution condition* if it satisfies the $\mathcal{R}$-substitution condition.

We make two comments about this definition:

- We only close under substitutions with type part $\star$. It will still be possible that typing is preserved by arbitrary (well-formed) substitutions when combined with the suspension condition.
- We introduce a second rule set $\mathcal{R}'$ in the definition, which is only used for the typing premise of the substitution $\sigma$. The reason for this is that the substitution condition is not closed under unions, and so we will instead prove that certain rule sets satisfy the $\mathcal{R}'$-substitution condition for an arbitrary $\mathcal{R}'$, a condition which is closed under unions.

The substitution condition allows us to give the next proposition.

**Proposition 2.4.15.** *Suppose $\mathcal{R}$ satisfies the substitution condition. For any $\sigma : \Delta \to_\star \Gamma$, the following rules are admissible:*

$$\frac{\Delta \vdash A \qquad \Gamma \vdash \sigma : \Delta}{\Gamma \vdash A[\![\sigma]\!]} \qquad \frac{\Delta \vdash s : A \qquad \Gamma \vdash \sigma : \Delta}{\Gamma \vdash s[\![\sigma]\!] : A[\![\sigma]\!]} \qquad \frac{\Delta \vdash \tau : \Theta \qquad \Gamma \vdash \sigma : \Delta}{\Gamma \vdash \tau \bullet \sigma : \Theta}$$

$$\frac{\Delta \vdash A = B \qquad \Gamma \vdash \sigma : \Delta}{\Gamma \vdash A[\![\sigma]\!] = B[\![\sigma]\!]} \qquad \frac{\Delta \vdash s = t \qquad \Gamma \vdash \sigma : \Delta}{\Gamma \vdash s[\![\sigma]\!] = t[\![\sigma]\!]} \qquad \frac{\Delta \vdash \tau = \mu \qquad \Gamma \vdash \sigma : \Delta}{\Gamma \vdash \tau \bullet \sigma = \mu \bullet \sigma}$$

*If $\mathcal{R}$ additionally satisfies the suspension conditions, then all the above rules are admissible for any substitution $\sigma : \Delta \to_B \Gamma$.*

*Proof.* The proof for a non-extended substitution is given by another routine mutual induction in Catt.Typing.Properties.Substitution. For an arbitrary substitution $\sigma : \Delta \to_B \Gamma$, we also proceed by mutual induction, but for the application of the substitution to an equality of terms $s$ and $t$ we further split on $B$. If $B = \star$, then the proof for non-extended substitutions can be used. Otherwise, we have:

$$\begin{aligned} s[\![\sigma]\!] &\equiv \Sigma s[\![\downarrow \sigma]\!] \\ &= \Sigma t[\![\downarrow \sigma]\!] \\ &\equiv t[\![\sigma]\!] \end{aligned}$$

with the non-syntactic equality following from the preservation of equality by suspension and inductive hypothesis. The proofs that the extended versions of these rules are admissible are found in Catt.Typing.Properties.Substitution.Suspended. □

We also prove that application of substitution respects equality in its second argument, which does not in fact need the substitution condition. This is also proved by a simple mutual induction in Catt.Typing.Properties.Substitution.

**Proposition 2.4.16.** *The following inference rules are admissible:*

$$\frac{\Gamma \vdash \sigma = \tau}{\Gamma \vdash s[\![\sigma]\!] = s[\![\tau]\!]} \qquad \frac{\Gamma \vdash \sigma = \tau}{\Gamma \vdash A[\![\sigma]\!] = A[\![\tau]\!]} \qquad \frac{\Gamma \vdash \sigma = \tau}{\Gamma \vdash \mu \bullet \sigma = \mu \bullet \tau}$$

*for substitutions* $\sigma : \Delta \to_A \Gamma$, $\tau : \Delta \to_B \Gamma$, *and* $\mu : \Theta \to_C \Delta$, *term* $s : \mathsf{Term}_\Delta$, *and type* $A : \mathsf{Type}_\Delta$.

This allows us to define a category of well-formed syntax in $\text{C}\scriptstyle{\text{ATT}}_\mathcal{R}$, which is well-defined by the two preceding definitions.

**Definition 2.4.17.** Suppose $\mathcal{R}$ satisfies the substitution and weakening conditions. Then we can define the *syntactic category* of $\text{C}\scriptstyle{\text{ATT}}_\mathcal{R}$, which by an abuse of notation we call $\mathsf{Catt}_\mathcal{R}$, to have:

- Objects given by contexts $\Gamma$ where $\Gamma \vdash$.
- Morphisms $\Delta \to \Gamma$ given by substitutions $\sigma : \Delta \to_\star \Gamma$ where $\Gamma \vdash \sigma : \Delta$ quotiented by the relation which equates substitutions $\sigma$ and $\tau$ when $\Gamma \vdash \sigma = \tau$.
- The identity morphism $\Gamma \to \Gamma$ given by $\mathrm{id}_\Gamma$.
- Composition is given by $\tau \circ \sigma = \sigma \bullet \tau$.

By Corollary 2.4.7, the identity substitution is a well-defined morphism, and the above two propositions prove that composition is well-defined. Composition satisfies associativity and unitality by Proposition 2.2.1. We remind the reader that the direction of morphisms in this category is opposite to the direction of morphisms in most categorical models of type theory, as explained in Remark 1.2.1.

By taking the weakening of the identity substitution $\mathrm{id}_\Gamma : \Gamma \to \Gamma$, we get a substitution:

$$\mathsf{proj}_\Gamma = \mathrm{wk}(\mathrm{id}_\Gamma) : \Gamma \to \Gamma, (x : A)$$

which includes $\Gamma$ into $\Gamma, x : A$. It can be checked (and is given by apply-project-is-wk-tm in the formalisation) that applying this substitution to a term is the same operation as weakening the term. Using this, the following can be proved:

**Lemma 2.4.18.** *Suppose* $\mathcal{R}$ *satisfies the substitution condition. Then it also satisfies the weakening condition.*

*Proof.* For $(\Gamma, s, t) \in \mathcal{R}$ and $A : \mathsf{Type}_\Gamma$, we must prove that:

$$((\Gamma, (x : A)), \mathrm{wk}(s), \mathrm{wk}(t)) \equiv ((\Gamma, (x : A)), s[\![\mathsf{proj}_\Gamma]\!], t[\![\mathsf{proj}_\Gamma]\!]) \in \mathcal{R}$$

which will follow from the substitution condition if it can be proved that

$$\Gamma, x : A \vdash_{\mathcal{R}} \mathsf{proj}_\Gamma : \Gamma$$

holds. This judgement is easy to derive when $\mathcal{R}$ satisfies the weakening condition, but this is what we are trying to prove. Instead, since $\emptyset$ trivially satisfies the weakening condition, $\mathsf{proj}_\Gamma$ is well-formed in Catt, and so the derivation above follows from Corollary 2.4.3. □

We lastly show that dr also satisfies the substitution condition.

**Proposition 2.4.19.** *The set dr satisfies the $\mathcal{R}$-substitution condition for any equality set $\mathcal{R}$.*

*Proof.* The proof is similar to Propositions 2.4.10 and 2.4.13, and follows from the equality $\{A, t\} \bullet \sigma \equiv \{A[\![\sigma]\!], t[\![\sigma]\!]\}$ which holds by Lemma 2.2.4(v) . □

*Remark* 2.4.20. The proof of the substitution condition for dr makes no use of the typing of $\sigma$. In fact this premise is only necessary for the supported rules construction which will be given in Definition 2.4.27

**Tameness** We can now define tameness.

**Definition 2.4.21.** An equality rule set $\mathcal{R}$ is tame if it satisfies the weakening, substitution, and suspension conditions. An operation set $\mathcal{O}$ is tame if it is suspendable and contains the standard operations. A theory generated by $\mathcal{R}$ and $\mathcal{O}$ is tame if both $\mathcal{R}$ and $\mathcal{O}$ are.

**Proposition 2.4.22.** *The set dr is tame.*

In the formalisation, each module is parameterised by the various conditions that the module needs, and where possible we avoid using extra unnecessary conditions. Given that every theory we will consider in this thesis is tame, and that it is hard to imagine a sensible theory that isn't tame, the argument could be made that the effort put into making distinctions between these conditions is wasted or at least unnecessary.

The case for including the weakening condition is especially unconvincing as it is implied by the substitution condition which likely holds in any theory of significant interest. It is however included here as it is used in the formalisation, where its introduction is an artefact of the natural progression of this research.

To this end, from Chapter 3, we will assume that the theory we are working over is tame, and build a library of constructions and results that work in any tame theory, even when some results may not need all the conditions above.

Since we have limited use for proving properties about theories that do not satisfy the substitution condition, we could have instead enforced that all theories respect substitution by adding a constructor to the (term) equality relation that takes an equality $\Delta \vdash s = t$ and typing relation $\Gamma \vdash \sigma : \Delta$ to an equality $\Gamma \vdash s[\![\sigma]\!] = t[\![\sigma]\!]$. This may remove some overhead of setting up the weakening and substitution conditions. It would also allow more minimal equality rule sets to be given, as a rule set such as disc removal could be given by

$$\{(D^n, \mathsf{Coh}_{(D^n\,;\,\mathrm{wk}(U^n))}[\mathsf{id}_{D^n}], d_n) \mid n \in \mathbb{N}\}$$

On the other hand, including the extra constructor would effectively add an extra case to each inductive proof, and it is less clear how to minimise some of the equality rules that will be introduced in Chapter 3. Taking either approach would likely lead to a similar development of the theory.

### 2.4.2 Further conditions

Knowing that the theory we are working in is tame will be sufficient for giving most of the constructions and proofs in Chapter 3. Here we introduce some extra conditions that instead serve to aid in the proof of metatheoretic properties of the generated theory. These conditions take the form of predicates on each rule in the equality rule sets, rather than being closure properties as the conditions for tameness were.

**Support condition** The support of a term plays a central role in classifying the operations of the theory (see Section 2.3). Although it is known that support is respected by syntactic equality, we have not yet shown it is preserved by definitional equality. The following condition allows this to be proved.

**Definition 2.4.23.** A set $\mathcal{R}$ satisfies the *$\mathcal{R}'$-support condition* for an equality set $\mathcal{R}'$ when:

$$\Gamma \vdash_{\mathcal{R}'} s : A \implies \mathrm{Supp}(s) = \mathrm{Supp}(t)$$

for each $(\Gamma, s, t) \in \mathcal{R}$ and $A : \mathsf{Type}_\Gamma$. A set $\mathcal{R}$ satisfies the *support condition* if it satisfies the $\mathcal{R}$-support condition.

The use of support instead of free variables in this definition is critical, as we do not expect the free variables of a piece of syntax to be preserved by equality in general. As an example, we would like to have the equality:

$$D^1 \vdash \mathsf{Coh}_{(D^1\,;\,U^1)}[\mathsf{id}_{D^1}] = d_1$$

given by disc removal, yet the free variables of each side are not equal (though the support of each side is).

We also draw attention to the typing premise. Without this, the left-hand side of each equality rule is too unconstrained (at least with how the equality rules are currently presented), and this condition would fail to hold on the equality sets we introduce in this thesis. Having this typing premise come from a separate rule set $\mathcal{R}'$ allows the support condition to be preserved by unions of equality sets, similar to the substitution condition.

From the support condition, we immediately get the following proposition, proved by mutual induction.

**Proposition 2.4.24.** *Let $\mathcal{R}$ satisfy the support condition. Then the following rules are admissible:*

$$\frac{\Gamma \vdash s = t}{\mathsf{Supp}(s) = \mathsf{Supp}(t)} \qquad \frac{\Gamma \vdash A = B}{\mathsf{Supp}(A) = \mathsf{Supp}(B)} \qquad \frac{\Gamma \vdash \sigma = \tau}{\mathsf{Supp}(\sigma) = \mathsf{Supp}(\tau)}$$

*For $s, t : \mathsf{Term}_\Gamma$, $A, B : \mathsf{Type}_\Gamma$ and substitutions $\sigma : \Delta \to_A \Gamma$ and $\tau : \Theta \to_B \Gamma$.*

In traditional presentations of Catt, $\mathsf{FV}(t) \cup \mathsf{FV}(A)$ is used instead of $\mathsf{Supp}(t)$ for a term $t$ of type $A$. Equipped with the support condition we can now show that these are the same.

**Lemma 2.4.25.** *The following hold when $\mathcal{R}$ satisfies the support condition:*

*(i)* $\mathsf{Supp}(A) = \mathsf{FV}(A)$ *when* $\Gamma \vdash A$,

*(ii)* $\mathsf{Supp}(\sigma) = \mathsf{FV}(\sigma)$ *when* $\Gamma \vdash \sigma : \Delta$,

*(iii)* $\mathsf{Supp}(t) = \mathsf{Supp}(A) \cup \mathsf{FV}(t)$ *when* $\Gamma \vdash t : A$,

*(iv)* $\mathsf{Supp}(t) = \mathsf{FV}(A) \cup \mathsf{FV}(t) = \mathsf{Supp}(A) \cup \mathsf{Supp}(t)$ *when* $\Gamma \vdash t : A$ *and* $\Gamma \vdash A$.

*Proof.* All properties are proven by a single mutual induction on the typing derivations in the premises.

(i) Suppose $\Gamma \vdash A$. If $A \equiv \star$ then $\mathsf{Supp}(A) = \mathsf{FV}(A) = \emptyset$. Instead, suppose $A \equiv s \to_B t$. Then we have that $\Gamma \vdash B$, $\Gamma \vdash s : B$, and $\Gamma \vdash t : B$ and so:

$$\begin{aligned} \mathsf{Supp}(A) &= \mathsf{Supp}(B) \cup \mathsf{Supp}(s) \cup \mathsf{Supp}(t) \\ &= \mathsf{FV}(B) \cup (\mathsf{FV}(B) \cup \mathsf{FV}(s)) \cup (\mathsf{FV}(B) \cup \mathsf{FV}(t)) && (*) \\ &= \mathsf{FV}(B) \cup \mathsf{FV}(s) \cup \mathsf{FV}(t) \\ &= \mathsf{FV}(A) \end{aligned}$$

where the equality $(*)$ is derived from the inductive hypothesis for (i) applied to $B$ and the inductive hypothesis for (iv) applied to $s$ and $t$.

(ii) Suppose $\Gamma \vdash \sigma : \Delta$. If $\sigma \equiv \langle A \rangle$ then $\Gamma \vdash A$ and so:

$$\mathsf{Supp}(\sigma) = \mathsf{Supp}(A) = \mathsf{FV}(A) = \mathsf{FV}(\sigma)$$

If instead $\sigma \equiv \langle \tau, t \rangle$ and $\Delta = \Theta, (x : A)$ then $\Gamma \vdash \tau : \Theta$ and $\Gamma \vdash t : A[\![\tau]\!]$ and so:

$$\begin{aligned} \mathsf{Supp}(\sigma) &= \mathsf{Supp}(\tau) \cup \mathsf{Supp}(t) \\ &= \mathsf{Supp}(\tau) \cup (\mathsf{Supp}(A[\![\tau]\!]) \cup \mathsf{FV}(t)) && (*) \\ &= \mathsf{DC}_\Gamma(\mathsf{FV}(\tau) \cup \mathsf{FV}(A[\![\tau]\!])) \cup \mathsf{FV}(t) \\ &= \mathsf{Supp}(\tau) \cup \mathsf{FV}(t) && \text{as } \mathsf{FV}(A[\![\tau]\!]) \subseteq \mathsf{FV}(\tau) \\ &= \mathsf{FV}(\tau) \cup \mathsf{FV}(t) && (\dagger) \\ &= \mathsf{FV}(\sigma) \end{aligned}$$

where the equality $(*)$ is derived from the inductive hypothesis for (iii) applied to $t$ and the equality $(\dagger)$ is derived from the inductive hypothesis for (ii) applied to $\tau$.

(iii) Suppose $\Gamma \vdash t : A$. We then split on the constructor used for the typing derivation:

If the derivation is the result of a conversion rule applied to $\Gamma \vdash t : B$ and $\Gamma \vdash A = B$, then inductive hypothesis gives $\mathrm{Supp}(t) = \mathrm{Supp}(B) \cup \mathrm{FV}(t)$ and Proposition 2.4.24 gives $\mathrm{Supp}(A) = \mathrm{Supp}(B)$ and so $\mathrm{Supp}(t) = \mathrm{Supp}(A) \cup \mathrm{FV}(t)$ as required.

If the derivation is derived from the typing rule for variables, then a simple induction on the context $\Gamma$, using that $\mathrm{Supp}(\mathrm{wk}(A)) = \mathrm{Supp}(A)$, gives the required result.

If the derivation is given by the typing rule for coherences then $t \equiv \mathsf{Coh}_{(\Delta\,;\,B)}[\sigma]$, $\Gamma \vdash \sigma : \Delta$, and $A \equiv B[\![\sigma]\!]$. Therefore,

$$\begin{aligned}
\mathrm{Supp}(t) &= \mathrm{Supp}(\sigma) \\
&= \mathrm{DC}_\Gamma(\mathrm{FV}(B[\![\sigma]\!]) \cup \mathrm{FV}(\sigma)) && \text{as } \mathrm{FV}(B[\![\sigma]\!]) \subseteq \mathrm{FV}(\sigma) \\
&= \mathrm{Supp}(A) \cup \mathrm{Supp}(\sigma) \\
&= \mathrm{Supp}(A) \cup \mathrm{FV}(\sigma) && (*) \\
&= \mathrm{Supp}(A) \cup \mathrm{FV}(t)
\end{aligned}$$

where the equality $(*)$ is the result of applying the inductive hypothesis for (ii) to $\sigma$.

(iv) If $\Gamma \vdash t : A$ and $\Gamma \vdash A$ then:

$$\mathrm{Supp}(t) = \mathrm{Supp}(A) \cup \mathrm{FV}(t) = \mathrm{FV}(A) \cup \mathrm{FV}(t)$$

trivially follows from (i) and (iii) and:

$$\mathrm{Supp}(t) = \mathrm{DC}_\Gamma(\mathrm{Supp}(t)) = \mathrm{DC}_\Gamma(\mathrm{FV}(A) \cup \mathrm{FV}(t)) = \mathrm{Supp}(A) \cup \mathrm{Supp}(t)$$

with the first equality resulting from the idempotency of the downwards closure operator.

This proof is formalised in Catt.Typing.Properties.Support. □

**Corollary 2.4.26.** *Let $\mathcal{R}$ satisfy the support condition and suppose $\Gamma \vdash \sigma : \Delta$. Then the following equality holds:*

$$\mathrm{DC}_\Gamma(V[\![\sigma]\!]) = \mathrm{DC}_\Delta(V)[\![\sigma]\!]$$

*for all $V \subseteq \mathrm{Var}(\Delta)$; downwards closure commutes with the application of $\sigma$ to variable sets.*

*Proof.* Proceed by induction on $\Delta$. If $\Delta \equiv \emptyset$ then the equation is trivial. Therefore, assume $\Delta \equiv \Theta, (x : A)$ and so $\sigma \equiv \langle \tau, t\rangle$ with $\Gamma \vdash \tau : \Theta$ and $\Gamma \vdash t : A[\![\tau]\!]$ by case analysis. We now split on whether $x \in V$.

If $x \notin V$ then $\mathrm{DC}_\Gamma(V[\![\sigma]\!]) = \mathrm{DC}_\Gamma(V[\![\tau]\!]) = \mathrm{DC}_\Theta(V)[\![\tau]\!] = \mathrm{DC}_\Delta(V)[\![\tau]\!]$ with the second equality due to inductive hypothesis. Otherwise, $x \in V$ and so letting $U = V \setminus \{x\}$ we get

the equality:

$$
\begin{aligned}
\mathrm{DC}_\Gamma(V\llbracket\sigma\rrbracket) &= \mathrm{DC}_\Gamma(U\llbracket\tau\rrbracket \cup \mathrm{FV}(t)) \\
&= \mathrm{DC}_\Gamma(U\llbracket\tau\rrbracket) \cup \mathrm{Supp}(t) \\
&= \mathrm{DC}_\Gamma(U\llbracket\tau\rrbracket) \cup \mathrm{Supp}(A\llbracket\tau\rrbracket) \cup \mathrm{FV}(t) && (\dagger) \\
&= \mathrm{DC}_\Gamma(U\llbracket\tau\rrbracket) \cup \mathrm{DC}_\Gamma(\mathrm{FV}(A)\llbracket\tau\rrbracket) \cup \mathrm{FV}(t) \\
&= \mathrm{DC}_\Gamma(U\llbracket\tau\rrbracket \cup \mathrm{FV}(A)\llbracket\tau\rrbracket) \cup \mathrm{FV}(t) \\
&= \mathrm{DC}_\Gamma((U \cup \mathrm{FV}(A))\llbracket\tau\rrbracket) \cup \mathrm{FV}(t) \\
&= \mathrm{DC}_\Theta(U \cup \mathrm{FV}(A))\llbracket\tau\rrbracket \cup \mathrm{FV}(t) && (*) \\
&= (\{x\} \cup \mathrm{DC}_\Theta(U \cup \mathrm{FV}(A)))\llbracket\sigma\rrbracket \\
&= \mathrm{DC}_\Delta(V)\llbracket\sigma\rrbracket
\end{aligned}
$$

where equality $(*)$ is by inductive hypothesis and equality $(\dagger)$ is by Lemma 2.4.25(iii). □

Unfortunately, proving that the support condition holds for most equality rule sets is not as trivial as the proofs for the tameness properties. Consider the case for disc removal, which gives rise to the equality

$$\Gamma \vdash \mathsf{Coh}_{(D^n\,;\,\mathrm{wk}(U^n))}[\{A, t\}] = t$$

To prove the support condition for this case we need to show that:

$$\mathrm{Supp}(\{A, t\}) = \mathrm{Supp}(t)$$

where we can assume that $\Gamma \vdash t : A$. Intuitively this should hold, as the support of a substitution should be equal to the support of the locally maximal arguments, and if the derivation $\Gamma \vdash t : A$ held in Catt, we would be able to prove this. However, this proof (and intuition) relies on the derivation $\Gamma \vdash_{\mathcal{R}} t : A$ holding in a theory generated by $\mathcal{R}$ where $\mathcal{R}$ already satisfies the support condition, without which the typing derivation offers little utility.

We therefore introduce a proof strategy for showing that the support condition holds. The key insight of this strategy is to prove by induction that every equality and every typing derivation in the system is well-behaved with respect to support. Then, for the case of an equality $\Gamma \vdash s = t$ arising from a rule $(\Gamma, s, t)$, we have $\Gamma \vdash s : A$ as a premise and so by inductive hypothesis can assume that this typing derivation is well-behaved with respect to support.

We formalise this with the following definition, called the *supported rules* construction:

**Definition 2.4.27.** Let $\mathcal{R}$ be some equality rule set. The *supported rules* construction applied to $\mathcal{R}$ produces the equality rule set $\mathcal{R}_{\mathsf{S}}$, given by:

$$\mathcal{R}_{\mathsf{S}} = \{(\Gamma, s, t) \in \mathcal{R} \mid \mathrm{Supp}(s) = \mathrm{Supp}(t)\}$$

The rule set $\mathcal{R}_{\mathsf{S}}$ satisfies the support condition by construction.

The proof strategy then proceeds as follows: to prove that $\mathcal{R}$ satisfies the support condition, we instead prove that $\mathcal{R}$ satisfies the $\mathcal{R}_{\mathsf{S}}$-support condition, leveraging that $\mathcal{R}_{\mathsf{S}}$ itself satisfies the support condition. The proof is then completed by the following lemma:

**Lemma 2.4.28.** *Let $\mathcal{R}$ be an equality rule set that satisfies the $\mathcal{R}_\mathsf{S}$-support condition. Then the following inference rules are admissible:*

$$\frac{\Gamma \vdash_{\mathcal{R}} A}{\Gamma \vdash_{\mathcal{R}_\mathsf{S}} A} \qquad \frac{\Gamma \vdash_{\mathcal{R}} s : A}{\Gamma \vdash_{\mathcal{R}_\mathsf{S}} s : A} \qquad \frac{\Gamma \vdash_{\mathcal{R}} \sigma : \Delta}{\Gamma \vdash_{\mathcal{R}_\mathsf{S}} \sigma : \Delta} \qquad \frac{\Gamma \vdash_{\mathcal{R}} A = B}{\Gamma \vdash_{\mathcal{R}_\mathsf{S}} A = B} \qquad \frac{\Gamma \vdash_{\mathcal{R}} s = t}{\Gamma \vdash_{\mathcal{R}_\mathsf{S}} s = t}$$

$$\frac{\Gamma \vdash_{\mathcal{R}} \sigma = \tau}{\Gamma \vdash_{\mathcal{R}_\mathsf{S}} \sigma = \tau}$$

*and hence $\mathcal{R}$ satisfies the support condition.*

*Proof.* The inference rules are all proven using a mutual induction on all typing and equality rules, using that $\mathcal{R}$ satisfies the $\mathcal{R}_\mathsf{S}$-support condition in the case where the equality $\Gamma \vdash s = t$ is derived from a rule $(\Gamma, s, t) \in \mathcal{R}$. This induction is formalised in Catt.Support.Typing.

The set $\mathcal{R}$ then satisfies the support condition as if $(\Gamma, s, t) \in \mathcal{R}$ and $\Gamma \vdash_{\mathcal{R}} s : A$, then $\Gamma \vdash_{\mathcal{R}_\mathsf{S}} s : A$ holds by the first part of the lemma and so $\mathrm{Supp}(s) = \mathrm{Supp}(t)$ as $\mathcal{R}$ is already known to satisfy the $\mathcal{R}_\mathsf{S}$-support condition. ∎

*Remark* 2.4.29. The original motivation for parameterising Catt by an arbitrary set of equality rules $\mathcal{R}$ was not to share proofs between $\text{Catt}_{\text{su}}$ and $\text{Catt}_{\text{sua}}$ but was to be able to state the supported rules construction.

To be able to prove that $\mathcal{R}$ satisfies the $\mathcal{R}_\mathsf{S}$-support condition, we will commonly need to know that $\mathcal{R}_\mathsf{S}$ satisfies various tameness conditions, which are given by the next lemma.

**Lemma 2.4.30.** *Let $\mathcal{R}$ be any equality set. Then $\mathcal{R}_\mathsf{S}$ satisfies the weakening, suspension, and substitution conditions if $\mathcal{R}$ respects the corresponding condition.*

*Proof.* Let $(\Gamma, s, t) \in \mathcal{R}$ be an arbitrary rule. To show $\mathcal{R}_\mathsf{S}$ satisfies the weakening condition we need to show that:

$$(\Gamma, s, t) \in \mathcal{R}_\mathsf{S} \implies ((\Gamma, (x : A)), \mathrm{wk}(s), \mathrm{wk}(t)) \in \mathcal{R}_\mathsf{S}$$

By assumption, $(\Gamma, \mathrm{wk}(s), \mathrm{wk}(t)) \in \mathcal{R}$ and by the premise of the implication we have $\mathrm{Supp}(s) = \mathrm{Supp}(t)$. From this it follows that $\mathrm{Supp}(\mathrm{wk}(s)) = \mathrm{Supp}(\mathrm{wk}(t))$ and so the conclusion of the implication holds.

The case for suspension is similar except we need to use the equality:

$$\mathrm{Supp}(\Sigma(s)) = \Sigma(\mathrm{Supp}(s)) = \Sigma(\mathrm{Supp}(t)) = \mathrm{Supp}(\Sigma(t))$$

derived from Lemma 2.3.13 and $\mathrm{Supp}(s) = \mathrm{Supp}(t)$ from the premise of the implication.

For the substitution condition we need to show that:

$$\mathrm{Supp}(s) = \mathrm{Supp}(t) \implies \mathrm{Supp}(s[\![\sigma]\!]) = \mathrm{Supp}(t[\![\sigma]\!])$$

under the assumption that $\Delta \vdash_{\mathcal{R}_\mathsf{S}} \sigma : \Gamma$. Since $\mathcal{R}_\mathsf{S}$ satisfies the support rule, we can use Corollary 2.4.26 to get:

$$\mathrm{Supp}(s[\![\sigma]\!]) = \mathrm{DC}_\Gamma(\mathrm{FV}(s)[\![\sigma]\!]) = \mathrm{Supp}(s)[\![\sigma]\!] = \mathrm{Supp}(t)[\![\sigma]\!] = \mathrm{DC}_\Gamma(\mathrm{FV}(t)[\![\sigma]\!]) = \mathrm{Supp}(t[\![\sigma]\!])$$

as required. ∎

We now prove the appropriate support condition for disc removal.

**Proposition 2.4.31.** *Let $\mathcal{R}$ satisfy the support and weakening conditions. Then the set dr satisfies the $\mathcal{R}$-support condition.*

*Proof.* It is sufficient to prove that given $s : \mathsf{Term}_\Gamma$, $A : \mathsf{Type}_\Gamma$, and $n = \dim(A)$ that:

$$\Gamma \vdash_{\mathcal{R}} \mathsf{Coh}_{(D^n\,;\,\mathrm{wk}(U^n))}[\{A,t\}] : B \implies \mathrm{Supp}(\{A,t\}) = \mathrm{Supp}(t)$$

Assume the premise of the implication. Then $\Gamma \vdash_{\mathcal{R}} \{A,t\} : D^n$ by case analysis on the typing derivation and so $\Gamma \vdash_{\mathcal{R}} A$ and $\Gamma \vdash_{\mathcal{R}} t : A$ by Lemma 2.4.8 as $\mathcal{R}$ satisfies the weakening condition.

By a simple induction, it can be shown that $\mathrm{Supp}(\{A,t\}) = \mathrm{Supp}(A) \cup \mathrm{Supp}(t)$. By Lemma 2.4.25(iv) we have $\mathrm{Supp}(t) = \mathrm{Supp}(A) \cup \mathrm{Supp}(t)$ as $\mathcal{R}$ satisfies the support condition and so $\mathrm{Supp}(\{A,t\}) = \mathrm{Supp}(t)$ as required. ∎

**Preservation condition** Our last condition allows us to prove preservation, the property that typing is preserved by equality.

**Definition 2.4.32.** A set $\mathcal{R}$ satisfies the $\mathcal{R}'$*-preservation condition* for an equality set $\mathcal{R}'$ when:

$$\Gamma \vdash_{\mathcal{R}'} s : A \implies \Gamma \vdash_{\mathcal{R}'} t : A$$

for each $(\Gamma, s, t) \in \mathcal{R}$ and $A : \mathsf{Type}_\Gamma$. The set $\mathcal{R}$ satisfies the *preservation condition* if it satisfies the $\mathcal{R}$-preservation condition.

When a rule set $\mathcal{R}$ has all the properties presented in this section, we are able to show preservation for the generated theory.

**Proposition 2.4.33.** *Let $\mathcal{R}$ satisfy the support condition and preservation condition, as well as being tame. Then the following inference rules are admissible:*

$$\frac{\Gamma \vdash A \qquad \Gamma \vdash A = B}{\Gamma \vdash B} \qquad\qquad \frac{\Gamma \vdash s : A \qquad \Gamma \vdash s = t \qquad \Gamma \vdash A = B}{\Gamma \vdash t : B}$$

$$\frac{\Gamma \vdash \sigma : \Delta \qquad \Gamma \vdash \sigma = \tau}{\Gamma \vdash \tau : \Delta}$$

*for $A, B : \mathsf{Type}_\Gamma$, $s, t : \mathsf{Term}_\Gamma$, $\sigma : \Delta \to_A \Gamma$, and $\tau : \Delta \to_B \Gamma$.*

*Proof.* We prove the following bidirectional versions of the inference rules by mutual induc-tion on the equality derivation:

$$\begin{aligned}
\Gamma \vdash A = B &\implies (\Gamma \vdash A \iff \Gamma \vdash B)\\
\Gamma \vdash s = t &\implies (\forall A.\, \Gamma \vdash s : A \iff \Gamma \vdash t : A)\\
\Gamma \vdash \sigma = \tau &\implies (\Gamma \vdash \sigma : \Delta \iff \Gamma \vdash \tau : \Delta)
\end{aligned}$$

which imply the inference rules of the proposition are admissible (using the conversion rule for the second rule).

The only non-trivial cases are for the statement for terms. We split on the equality derivation $\Gamma \vdash s = t$. The cases for reflexivity on variables and transitivity are also trivial. The case for symmetry follows from the symmetry of the "if and only if" relation.

Now suppose the equality is of the form $\mathsf{Coh}_{(\Delta\,;\,A)}[\sigma] = \mathsf{Coh}_{(\Delta\,;\,B)}[\tau]$ and is derived from the equality rule for coherences from equalities $\Delta \vdash A = B$ and $\Gamma \vdash \sigma = \tau$. We prove the first direction, with the second following symmetrically. We therefore assume we have a typing derivation $\Gamma \vdash \mathsf{Coh}_{(\Delta\,;\,A)}[\sigma] : C$, and will induct on this derivation to construction a derivation of $\Gamma \vdash \mathsf{Coh}_{(\Delta\,;\,B)}[\tau] : C$.

- If the derivation is constructed with the conversion rule from $\Gamma \vdash \mathsf{Coh}_{(\Delta\,;\,A)}[\sigma] : D$ and $\Gamma \vdash D = C$, then we get a derivation $\Gamma \vdash \mathsf{Coh}_{(\Delta\,;\,B)}[\tau] : D$ by inductive hypothesis and can apply the conversion rule to get a derivation $\Gamma \vdash \mathsf{Coh}_{(\Delta\,;\,B)}[\tau] : C$.
- If instead the derivation is constructed with the coherence rule then $C \equiv A[\![\sigma]\!]$ and $A \equiv s \to_{A'} t$ and therefore $B \equiv u \to_{B'} v$ with $\Delta \vdash s = u$ and $\Delta \vdash t = v$. We also have that $\Delta \vdash_{\mathsf{ps}}$, $(\Delta, \mathsf{Supp}(s), \mathsf{Supp}(t)) \in \mathcal{O}$, $\Delta \vdash A$, and $\Gamma \vdash \sigma : \Delta$. By the inductive hypothesis on the equality, we have $\Delta \vdash B$ and $\Gamma \vdash \tau : \Delta$. By Proposition 2.4.24, $\mathsf{Supp}(s) = \mathsf{Supp}(u)$ and $\mathsf{Supp}(t) = \mathsf{Supp}(v)$ and so $(\Delta, \mathsf{Supp}(u), \mathsf{Supp}(v)) \in \mathcal{O}$. Hence, by the coherence rule we have $\Gamma \vdash \mathsf{Coh}_{(\Delta\,;\,B)}[\tau] : B[\![\tau]\!]$. By Propositions 2.4.15 and 2.4.16, $\Gamma \vdash A[\![\sigma]\!] = B[\![\tau]\!]$ and so by the conversion rule we obtain a derivation $\Gamma \vdash \mathsf{Coh}_{(\Delta\,;\,B)}[\tau] : C$.

Finally, suppose the equality is derived from RULE, such that $(\Gamma, s, t) \in \mathcal{R}$ and $\Gamma \vdash s : A$. If $\Gamma \vdash s : B$, then the preservation condition gives a derivation $\Gamma \vdash t : B$. Conversely, if $\Gamma \vdash t : B$, then we need to show that $\Gamma \vdash A = B$. By applying the preservation condition to the derivation $\Gamma \vdash s : A$, we get a derivation $\Gamma \vdash t : A$. Then by Lemma 2.2.7, we have $\Gamma \vdash A = B$ and so the proof is complete by applying the conversion rule to the derivation $\Gamma \vdash s : A$. □

As with the other conditions, we end this section by showing that dr satisfies the preservation condition.

**Proposition 2.4.34.** *Suppose $\mathcal{R}$ satisfies the weakening condition, and the set of operations $\mathcal{O}$ contains the standard operations. Then dr satisfies the $\mathcal{R}$-preservation condition.*

*Proof.* Take $(\Gamma, \mathsf{Coh}_{(D^n\,;\,\mathrm{wk}(U^n))}[\{A, t\}], t) \in \mathsf{dr}$ and suppose $\Gamma \vdash \mathsf{Coh}_{(D^n\,;\,U^n)}[\{A, t\}] : B$. Then by Lemma 2.2.7:

$$\Gamma \vdash B = \mathrm{wk}(U^n)[\![\{A, t\}]\!] \equiv A$$

By Lemma 2.4.8, $\Gamma \vdash t : A$ and so by the conversion rule $\Gamma \vdash t : B$ as required. □

### 2.4.3 Endo-coherence removal

We conclude this chapter with a second example of a family of equality rules called *endo-coherence removal.* As suggested by the name, these equalities simplify a class of terms known as endo-coherences.

**Definition 2.4.35.** An *endo-coherence* is a coherence term $\mathsf{Coh}_{(\Delta\,;\,s \to_A s)}[\sigma]$.

If we consider the (ps-context):

$$\Delta = (x : \star)(y : \star)(f : x \to_\star y)(z : \star)(g : y \to_\star z)$$

then we see that there are two distinct endo-coherences with source and target $f * g$, the identity on $f * g$ and the "fake identity" $\mathsf{Coh}_{(\Delta\,;\,f*g \to f*g)}[\mathsf{id}_\Delta]$. In the type theories $\text{Catt}_{\mathsf{su}}$ and $\text{Catt}_{\mathsf{sua}}$ introduced in Sections 4.2 and 4.3, identities will be privileged, and these fake identities will be reduced to the true identity.

More generally, for each term $t$ there is a canonical endo-coherence with source and target $t$, the identity on $t$. Endo-coherence removal simplifies any other endo-coherence on that term to an identity. It makes the following rule admissible:

$$\frac{\Delta \vdash_{\mathsf{ps}} \qquad \Delta \vdash A \qquad \Delta \vdash s : A \qquad \mathsf{Supp}(s) = \mathsf{Var}(\Delta) \qquad \Gamma \vdash \sigma : \Delta}{\Gamma \vdash \mathsf{Coh}_{(\Delta\,;\,s \to_A s)}[\sigma] = \mathsf{id}(A[\![\sigma]\!], s[\![\sigma]\!])}\ \textsc{ecr}$$

Endo-coherence removal can be assembled into the following equality rule set.

**Definition 2.4.36.** The *endo-coherence removal set*, ecr, is the set consisting of the triples:

$$\Gamma, \mathsf{Coh}_{(\Delta\,;\,s \to_A s)}[\sigma], \mathsf{id}(A[\![\sigma]\!], s[\![\sigma]\!]))$$

for contexts $\Gamma$ and $\Delta$, $A : \mathsf{Type}_\Delta$, $s : \mathsf{Term}_\Delta$, and $\sigma : \Delta \to_\star \Gamma$.

A set of rules $\mathcal{R}$ *contains endo-coherence removal* if ecr $\subseteq \mathcal{R}$. We say that $\mathcal{R}$ *has endo-coherence removal* if the rule ECR holds in the generated theory.

The set ecr satisfies all the conditions introduced in this chapter, as proven in the next proposition, which concludes this chapter.

**Proposition 2.4.37.** *Suppose the set of operations $\mathcal{O}$ contains the standard operations. Then the set ecr satisfies the following properties:*

*(i) The set ecr satisfies the weakening condition.*

*(ii) The set ecr satisfies the suspension condition.*

*(iii) The set ecr satisfies the $\mathcal{R}$-substitution condition, for any equality set $\mathcal{R}$.*

*(iv) The set ecr satisfies the $\mathcal{R}$-support condition, for any equality set $\mathcal{R}$ satisfying the support condition.*

*(v) The set ecr satisfies the $\mathcal{R}$-preservation condition, for any equality set $\mathcal{R}$ satisfying the weakening and substitution conditions.*

*Proof.* Suppose $(\Gamma, \mathsf{Coh}_{(\Delta\,;\,s \to_A s)}[\sigma], \mathsf{id}(A[\![\sigma]\!], s[\![\sigma]\!])) \in \mathsf{ecr}$. To show that the substitution holds, we suppose that $\tau : \Gamma \to_\star \Theta$, and then must prove that:

$$(\Theta, \mathsf{Coh}_{(\Delta\,;\,s \to_A s)}[\sigma \bullet \tau], \mathsf{id}(A[\![\sigma]\!], s[\![\sigma]\!])[\![\tau]\!]) \in \mathsf{ecr}$$

It is immediate that:

$$(\Theta, \mathsf{Coh}_{(\Delta\,;\,s \to_A s)}[\sigma \bullet \tau], \mathsf{id}(A[\![\sigma \bullet \tau]\!], s[\![\sigma \bullet \tau]\!])) \in \mathsf{ecr}$$

and so it suffices to prove that $\mathsf{id}(A[\![\sigma]\!], s[\![\sigma]\!])[\![\tau]\!] \equiv \mathsf{id}(A[\![\sigma \bullet \tau]\!], s[\![\sigma \bullet \tau]\!])$, but this follows from Lemma 2.2.4(v) and Proposition 2.2.1. The weakening condition then follows from the substitution condition.

For the suspension condition, it must be shown that:

$$(\Sigma(\Gamma), \mathsf{Coh}_{(\Sigma(\Delta)\,;\,\Sigma(s) \to_{\Sigma(A)} \Sigma(s))}[\Sigma(\sigma)], \Sigma(\mathsf{id}(A[\![\sigma]\!], s[\![\sigma]\!]))) \in \mathsf{ecr}$$

and so it suffices to show that $\mathrm{Supp}(\Sigma(s)) = \mathrm{Var}(\Sigma(\Delta))$, which follows from $\mathrm{Supp}(\Sigma(s)) = \Sigma(\mathrm{Supp}(s))$, and

$$\Sigma(\mathsf{id}(A[\![\sigma]\!], s[\![\sigma]\!])) \equiv \mathsf{id}(\Sigma(A)[\![\Sigma(\sigma)]\!], \Sigma(s)[\![\Sigma(\sigma)]\!])$$

which follows from the functoriality of suspension and Lemmas 2.2.4(ii) and 2.2.4(iv).

For the support condition, assume that $\Gamma \vdash_{\mathcal{R}} \mathsf{Coh}_{(\Delta\,;\,s \to_A s)}[\sigma] : B$ for some $B : \mathsf{Type}_\Gamma$ and that $\mathcal{R}$ satisfies the support condition. Then:

$$\begin{aligned}
\mathrm{Supp}(\mathsf{Coh}_{(\Delta\,;\,s \to_A s)}[\sigma]) &= \mathrm{Supp}(\sigma) \\
&= \mathrm{FV}(\sigma) && \text{by Lemma 2.4.25(ii)} \\
&= \mathrm{Var}(\Delta)[\![\sigma]\!] \\
&= \mathrm{Supp}(s)[\![\sigma]\!] && \text{by assumption} \\
&= (\mathrm{Supp}(A) \cup \mathrm{Supp}(s))[\![\sigma]\!] && \text{by Lemma 2.4.25(iv)} \\
&= \mathrm{DC}_\Delta(\mathrm{FV}(A) \cup \mathrm{FV}(s))[\![\sigma]\!] \\
&= \mathrm{DC}_\Gamma(\mathrm{FV}(A)[\![\sigma]\!] \cup \mathrm{FV}(s)[\![\sigma]\!]) && \text{by Corollary 2.4.26} \\
&= \mathrm{DC}_\Gamma(\mathrm{FV}(A[\![\sigma]\!]) \cup \mathrm{FV}(s[\![\sigma]\!])) && \text{by Proposition 2.3.3} \\
&= \mathrm{Supp}(A[\![\sigma]\!]) \cup \mathrm{Supp}(s[\![\sigma]\!]) \\
&= \mathrm{Supp}(\mathsf{id}(A[\![\sigma]\!], s[\![\sigma]\!]))
\end{aligned}$$

as required.

Lastly for the preservation condition, let $\mathcal{R}$ satisfy the weakening and substitution conditions, and assume $\Gamma \vdash \mathsf{Coh}_{(\Delta\,;\,s \to_A s)}[\sigma] : B$. By deconstructing the typing derivation, we must have that $\Delta \vdash A$, $\Delta \vdash s : A$, and $\Gamma \vdash \sigma : \Delta$. Therefore, by Proposition 2.4.15, $\Gamma \vdash A[\![\sigma]\!]$ and $\Gamma \vdash s[\![\sigma]\!] : A[\![\sigma]\!]$. Hence, by Corollary 2.4.9, $\Gamma \vdash \mathsf{id}(A[\![\sigma]\!], s[\![\sigma]\!]) : (s \to_A s)[\![\sigma]\!]$. It remains to prove that $\Gamma \vdash (s \to_A s)[\![\sigma]\!] = B$, but this is immediate from Lemma 2.2.7, applied to the derivation $\Gamma \vdash \mathsf{Coh}_{(\Delta\,;\,s \to_A s)}[\sigma] : B$. □

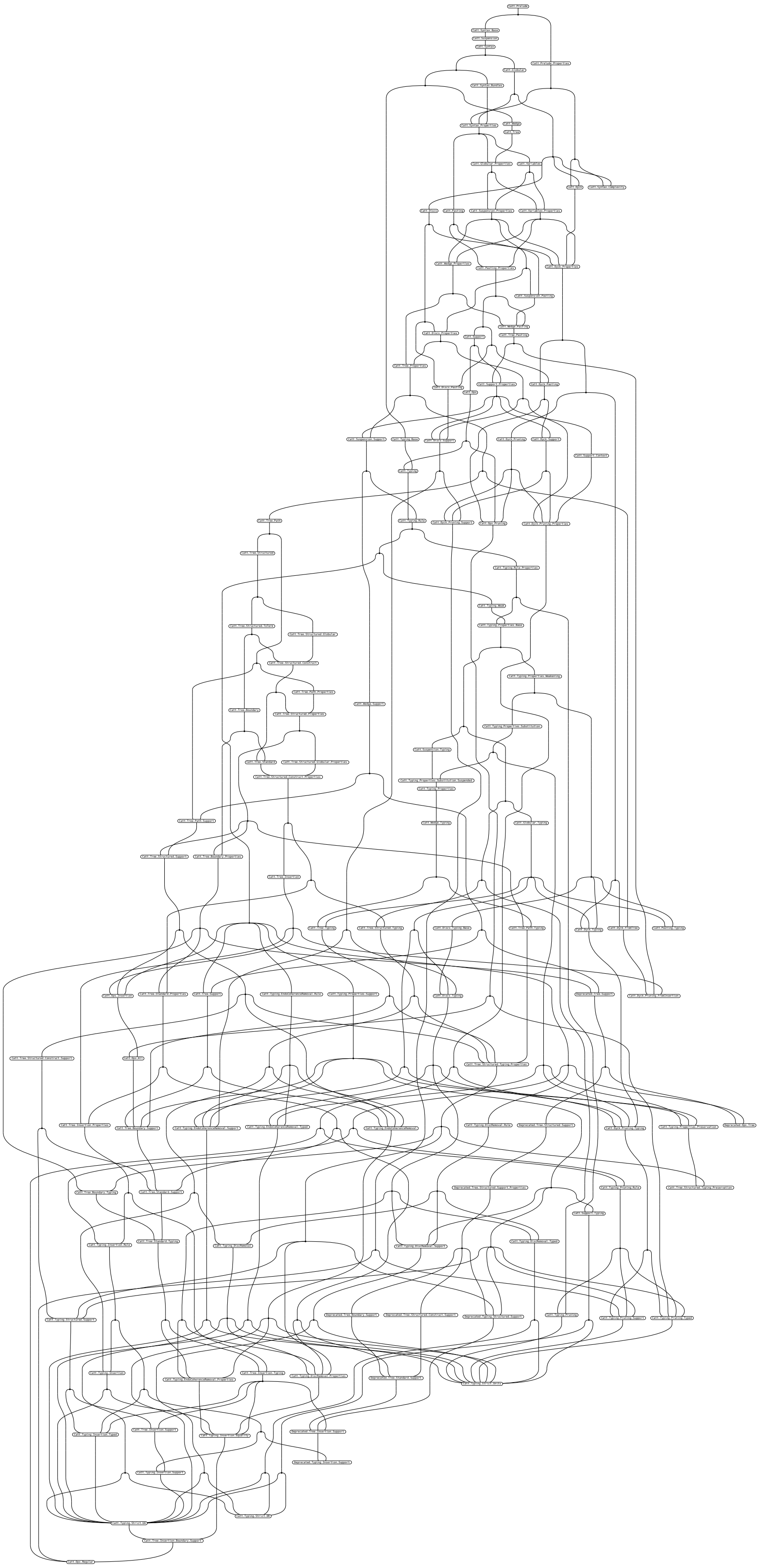

Figure 2.2: Dependency graph of Agda formalisation.

# Chapter 3

# Constructions in $\textsc{Catt}_{\mathcal{R}}$

This chapter will investigate some more involved constructions that can be given in the type theory $\textsc{Catt}_{\mathcal{R}}$. These constructions will be central to defining the reductions that underpin the type theories $\textsc{Catt}_{\text{su}}$ and $\textsc{Catt}_{\text{sua}}$ which appear in Chapter 4. We will give a definition of each construction, describe under what conditions it is well-formed, and state various properties describing the behaviour of the construction and its interaction with other constructions.

For this chapter we will assume that we are working in a tame theory, as described in Section 2.4.1. This means that all proofs in this section will hold in any variant of $\textsc{Catt}_{\mathcal{R}}$ such that the equality set $\mathcal{R}$ satisfies the weakening, substitution, and suspension conditions, and the set of operations $\mathcal{O}$ is suspendable and contains the standard operations. We will also use all the relevant proofs from Section 2.2, without explaining exactly what condition of the set $\mathcal{R}$ is being used.

The formalisation is commonly more specific when specifying which conditions are necessary for each module, for example omitting the suspension condition when it is not needed for a specific construction, but for the body of this text we ignore these distinctions and simply assume that every theory we work with will be tame, as will be case for all theories introduced in Chapter 4.

This chapter builds up to the following constructions, that can be viewed as meta-operations on $\textsc{Catt}_{\mathcal{R}}$.

- The *pruning* operation will be introduced in Section 3.1 and is the main component of the type theory $\textsc{Catt}_{\text{su}}$, defined in Section 4.2, a type theory for strictly unital $\infty$-categories. Pruning removes unnecessary identities from a term, simplifying the resulting term in the process.
- The *insertion* operation will be introduced in Section 3.4. It powers the type theory $\textsc{Catt}_{\text{sua}}$, a type theory for strictly unital and associative $\infty$-categories. Insertion merges certain arguments to a coherence into the body of the coherence itself, effectively "inserting" the argument into the head term. It can be viewed as a generalisation of pruning, but is a more complex construction.

Both pruning and insertion perform more radical modifications to the structure of a term than disc removal and endo-coherence removal, the equality rules we have seen so far. Pruning and insertion modify the pasting diagram in the coherence at the head of the term they act on. In this chapter, more combinatorial descriptions of pasting diagrams will be introduced to enable the pasting diagrams involved in these constructions to be constructed by induction.

The pruning construction identifies locally maximal arguments of a coherence that are syntactically identities, and removes these arguments from the term, while also removing the component of the pasting diagram in the coherence which corresponds to this argument. Pruning could be applied to the term $f * g * \mathsf{id}$, a ternary composite, to remove the identity argument and convert the ternary composite to a binary composite, returning the term $f * g$.

Insertion does not just simply remove parts of a term, but flattens the structure of a term, moving data from a locally maximal argument into the head term. The motivating example for insertion is the term $f * (g * h)$, a binary composite where one of the locally maximal arguments is itself a binary composite. Under insertion, the inner composite $g * h$ is merged with the outer binary composite to form a single ternary composite $f * g * h$.

When a locally maximal argument is an identity, it will always be insertable, and the result of inserting the identity into the head term will be similar to pruning the same argument, motivating the viewpoint that insertion is a generalisation of pruning. At the end of this chapter, this relationship will be made precise.

Insertion again performs more radical changes to the head coherence of the term than pruning, and needs to be able to merge two pasting diagrams into one along a locally maximal argument. The operation on pasting diagrams is best understood as an operation on *trees*, an alternative characterisation of pasting diagrams which will be introduced in Section 3.2.

Although the definition of these trees is simple, to be able to use them effectively we must be able to describe their relationship to the Catt contexts they represent. It will also be necessary to describe the morphisms between these trees, which correspond to substitutions between the underlying contexts, and the composition of such morphisms.

Certain constructions on trees will not compute nicely with the syntax in Catt. We therefore introduce a new notion of *structured term*, an alternative syntax for Catt which allows more complex representations of terms over contexts derived from trees. Structured terms effectively retain more information about how they are constructed, allowing constructions to compute on them in ways that are not possible on the raw syntax of Catt. This representation of terms will be crucial in the formalisation, as it aids the proof assistant in simplifying various constructions. These structured terms are defined in Section 3.3.

Finally, Section 3.4 defines the constructions used in the insertion operation, using the structured syntax from the preceding section. In this section, many properties of insertion are stated, including a universal property that it satisfies.

## 3.1 Pruning

Pruning drives the strictly unital behaviour of $\text{Catt}_{\text{su}}$. Unitality in $\infty$-categories is the property that the identity acts as a unit with respect to composition, so that composing with the unit is equivalent to the original term. If an $\infty$-category is strictly unital, then it exhibits this behaviour up to equality rather than equivalence.

For Catt, strict unitality means that a composition containing an identity as one of its arguments should be definitionally equal to the term with this argument removed. Pruning is the operation that removes an argument from a composition, taking a term such as $f * g * \mathsf{id}$ to $f * g$, or $\mathsf{id} * f$ to the unary composite on $f$. In the presence of strict units, it is also desirable to simplify the higher-dimensional data that witnessed the (weak) unitality in Catt. For example, the left unitor on $f$, given by the term:

$$\mathsf{Coh}_{((x:\star),(y:\star),(f:x\to_\star y)\,;\,\mathsf{id}(x)*f\to f)}[\mathsf{id}]$$

which witnesses that composing on the left with an identity is equivalent to the original term, can be simplified to the identity on $f$, and the triangle equations which govern the coherence laws for the unitors can also trivialise. For this reason, pruning is defined to be able to apply to any term which has identities as a locally maximal argument. We review the definition of a locally maximal argument below.

**Definition 3.1.1.** In a context $\Gamma$, a *locally maximal variable* is a variable $x$ of $\Gamma$ that does not appear in the source or target of any other variable of $\Gamma$. Equivalently, $x$ is locally maximal when:

$$x \notin \mathrm{Supp}(y)$$

for any $y \neq x \in \mathrm{Var}(\Gamma)$. Given a substitution $\sigma : \Delta \to \Gamma$, a *locally maximal argument* of $\sigma$ is a term $x[\![\sigma]\!]$ where $x$ is a locally maximal variable of $\Delta$.

*Example* 3.1.2. Consider the pasting diagram given by the following diagram:

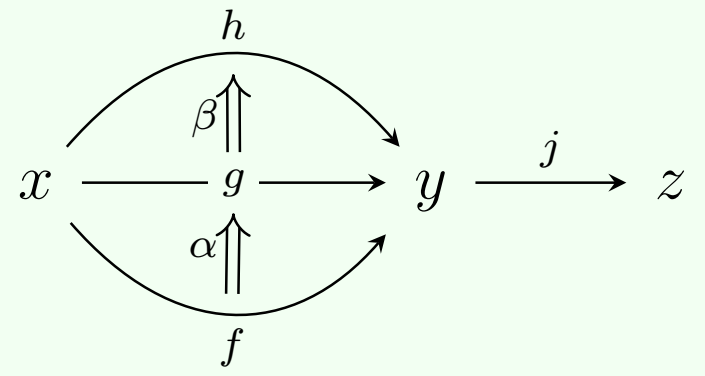

which corresponds to the Catt context (written to highlight the dimension of each term):

$$\begin{aligned}\Theta = (x:\star),&\\ (y:\star),(f:x\to y),&\\ (g:x\to y),(\alpha:f\to g),&\\ (h:x\to y),(\beta:g\to h),&\\ (z:\star),(j:y\to z)&\end{aligned}$$

The locally maximal variables of $\Theta$ are $\alpha$, $\beta$, and $j$. Note that $j$ is locally maximal, despite not being of maximal dimension in the context. Pruning the context $\Theta$ along locally maximal variable $\alpha$ removes the variables $\alpha$ and $g$ from the context, and must amend the type of $\beta$ so that its source is $f$.

To perform the pruning construction, we start with a coherence term $\mathsf{Coh}_{(\Delta\,;\,A)}[\sigma] : \mathsf{Term}_\Gamma$, and assume that some locally maximal argument of $\sigma$ is an identity, that is $x[\![\sigma]\!] \equiv \mathsf{id}(B,t)$ for some locally maximal variable $x$, type $B : \mathsf{Type}_\Gamma$, and term $t : \mathsf{Term}_\Gamma$. We then construct the following:

- A new pasting diagram $\Delta /\!\!/ x$, corresponding to $\Delta$ with the variable $x$ and its target removed.

- A new set of arguments $\sigma \sslash x$, consisting of the same terms as $\sigma$ except those corresponding to $x$ and its target.
- A projection substitution $\pi_x : \Delta \to \Delta \sslash x$, from which a type $A[\![\pi_x]\!] : \mathsf{Type}_{\Delta \sslash x}$ can be obtained. This projection sends $x$ to the identity on its source, the target of $x$ to the source of $x$, and every other variable to itself.

We note that the source and target of the locally maximal variable $x$ are well-defined as $x$ must be sent by $\sigma$ to an identity, which cannot be zero dimensional.

### 3.1.1 Dyck words

To be able to easily reason about the structures involved in pruning, we wish to define them by induction. To do this we introduce a different presentation of pasting diagrams called *Dyck words*, which have a simpler inductive structure. Dyck words more directly encode the structure of the pasting diagram, and will allow us to give an inductive characterisation of the locally maximal variables of the associated context.

**Definition 3.1.3.** The set of *Dyck words*, $\mathsf{Dyck}_d$ of trailing dimension $d$ consists of lists of "up" and "down" moves according to the following rules.

$$\frac{}{\circleddash : \mathsf{Dyck}_0} \qquad \frac{d : \mathbb{N} \qquad \mathcal{D} : \mathsf{Dyck}_d}{\mathcal{D} \Uparrow : \mathsf{Dyck}_{d+1}} \qquad \frac{d : \mathbb{N} \qquad \mathcal{D} : \mathsf{Dyck}_{d+1}}{\mathcal{D} \Downarrow : \mathsf{Dyck}_d}$$

In any prefix of a Dyck word $D : \mathsf{Dyck}_d$, the number of "up" moves (given by constructor $\Uparrow$) must be greater than or equal to the number of "down" moves (given by constructor $\Downarrow$). The difference between the number of each move is given by the trailing dimension $d$.

Dyck words can be given a visual interpretation as a *mountain diagram*. To obtain such a diagram we start on the left-hand side, and draw a continuous line by drawing an upwards sloping segment for each $\Uparrow$ in the word, and a downwards sloping line for each $\Downarrow$ in the word. An example of such a diagram is given in Figure 3.1.

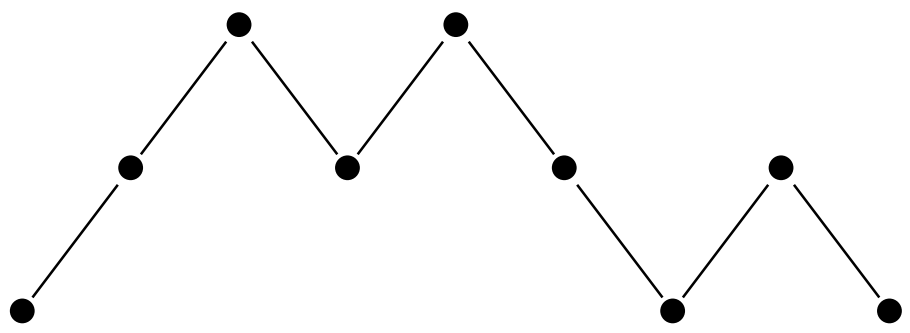

Figure 3.1: Mountain diagram for $\circleddash \Uparrow \Uparrow \Downarrow \Uparrow \Downarrow \Downarrow \Uparrow \Downarrow : \mathsf{Dyck}_0$.

The rules $\circleddash$, $\Uparrow$, and $\Downarrow$ directly correspond to the rules PSS, PSE, and PSD that generate the typing judgement for ps-contexts. From a Dyck word, we can directly construct this context by induction.

**Definition 3.1.4.** For a Dyck word $\mathcal{D} : \mathsf{Dyck}_d$, its associated context $\lfloor \mathcal{D} \rfloor$, associated type $\mathrm{Ty}_{\mathcal{D}} : \mathsf{Type}_{\lfloor \mathcal{D} \rfloor}$, and associated term $\mathrm{Tm}_{\mathcal{D}} : \mathsf{Term}_{\lfloor \mathcal{D} \rfloor}$ are defined by mutual induction on $\mathcal{D}$:

$$\begin{aligned}
\lfloor \ominus \rfloor &= (x : \star) \\
\lfloor \mathcal{D} \Uparrow \rfloor &= \lfloor \mathcal{D} \rfloor, (y_{\mathcal{D}} : \mathrm{Ty}_{\mathcal{D}}), (f_{\mathcal{D}} : \mathrm{wk}(\mathrm{Tm}_{\mathcal{D}}) \to_{\mathrm{wk}(\mathrm{Ty}_{\mathcal{D}})} y_{\mathcal{D}}) \\
\lfloor \mathcal{D} \Downarrow \rfloor &= \lfloor \mathcal{D} \rfloor
\end{aligned}$$

$$\begin{aligned}
\mathrm{Ty}_{\ominus} &= \star \\
\mathrm{Ty}_{\mathcal{D}\Uparrow} &= \mathrm{wk}(\mathrm{wk}(\mathrm{Tm}_{\mathcal{D}})) \to_{\mathrm{wk}(\mathrm{wk}(\mathrm{Ty}_{\mathcal{D}}))} y_{\mathcal{D}} \\
\mathrm{Ty}_{\mathcal{D}\Downarrow} &= \mathrm{base}(\mathrm{Ty}_{\mathcal{D}}) && \text{where } \mathrm{base}(s \to_A t) = A
\end{aligned}$$

$$\begin{aligned}
\mathrm{Tm}_{\ominus} &= x \\
\mathrm{Tm}_{\mathcal{D}\Uparrow} &= f_{\mathcal{D}} \\
\mathrm{Tm}_{\mathcal{D}\Downarrow} &= \mathrm{tgt}(\mathrm{Ty}_{\mathcal{D}}) && \text{where } \mathrm{tgt}(s \to_A t) = t
\end{aligned}$$

The variable names given here are used to avoid ambiguity in the definition. As we consider contexts up to $\alpha$-equality, we may freely change these variable names. The tgt and base operations are well-defined here as it may be checked by a simple induction that $\dim(\mathrm{Ty}_{\mathcal{D}}) = d$ for $\mathcal{D} : \mathsf{Dyck}_d$, ensuring that we only apply tgt and base to types of strictly positive dimension.

The tight correspondence between the rules used to construct Dyck words and ps-contexts allow an easy proof that the contexts associated to Dyck words are in fact pasting diagrams.

**Lemma 3.1.5.** *For a Dyck word $\mathcal{D} : \mathsf{Dyck}_d$, its associated context, type, and term are all well-formed:*

$$\lfloor \mathcal{D} \rfloor \vdash \qquad \lfloor \mathcal{D} \rfloor \vdash \mathrm{Ty}_{\mathcal{D}} \qquad \lfloor \mathcal{D} \rfloor \vdash \mathrm{Tm}_{\mathcal{D}} : \mathrm{Ty}_{\mathcal{D}}$$

*In addition to being a well-formed context, the context associated to a Dyck word is a ps-context; the following judgement holds:*

$$\lfloor \mathcal{D} \rfloor \vdash_{\mathsf{ps}} \mathrm{Tm}_{\mathcal{D}} : \mathrm{Ty}_{\mathcal{D}}$$

*and so if $\mathcal{D} : \mathsf{Dyck}_0$, we have $\lfloor \mathcal{D} \rfloor \vdash_{\mathsf{ps}}$. Further, all ps-contexts are the associated context of a Dyck word.*

*Proof.* Due to the similarity of the rules for ps-contexts and Dyck words, this follows quickly from simple inductions, which are given in the formalisation. The proofs for the typing judgements appear in Catt.Dyck.Typing and the proofs for the ps-context judgements appear in Catt.Dyck.Pasting. □

The locally maximal variables in the context associated to a Dyck word correspond exactly to the points in the word where there is an upwards move followed immediately by a downwards move, creating a peak in the mountain diagram. These peaks can be given an inductive characterisation.

**Definition 3.1.6.** Let $\mathcal{D} : \mathsf{Dyck}_d$ be a Dyck word. A *peak* of $\mathcal{D}$, $p : \mathsf{Peak}_{\mathcal{D}}$ is inductively defined by the following rules:

$$\frac{d \in \mathbb{N} \qquad \mathcal{D} : \mathsf{Dyck}_d}{\mathcal{D} \Updownarrow_{\mathsf{pk}} : \mathsf{Peak}_{\mathcal{D}\Uparrow\Downarrow}} \qquad\qquad \frac{d \in \mathbb{N} \qquad \mathcal{D} : \mathsf{Dyck}_d \qquad p : \mathsf{Peak}_{\mathcal{D}}}{p \Uparrow_{\mathsf{pk}} : \mathsf{Peak}_{\mathsf{D}\Uparrow}}$$

$$\frac{d \in \mathbb{N} \qquad \mathcal{D} : \mathsf{Dyck}_{d+1} \qquad p : \mathsf{Peak}_{\mathcal{D}}}{p \Downarrow_{\mathsf{pk}} : \mathsf{Peak}_{\mathcal{D}\Downarrow}}$$

From each peak $p : \mathsf{Peak}_{\mathcal{D}}$, a term $\lfloor p \rfloor$ of $\lfloor \mathcal{D} \rfloor$ can be inductively defined by:

$$\lfloor \mathcal{D} \Updownarrow_{\mathsf{pk}} \rfloor = f_{\mathcal{D}} \qquad \lfloor p \Uparrow_{\mathsf{pk}} \rfloor = \mathrm{wk}(\mathrm{wk}\lfloor p \rfloor) \qquad \lfloor p \Downarrow_{\mathsf{pk}} \rfloor = \lfloor p \rfloor$$

The term $\lfloor p \rfloor$ is a locally maximal variable of $\lfloor \mathcal{D} \rfloor$.

*Example* 3.1.7. Recall the ps-context $\Theta$ from Example 3.1.2. This context is the associated context of the Dyck word:

$$\circleddash \Uparrow \Uparrow \Downarrow \Uparrow \Downarrow \Downarrow \Uparrow \Downarrow$$

for which the mountain diagram is given in Figure 3.1. The three locally maximal variables $\alpha$, $\beta$, and $j$ correspond to the peaks:

$$\circleddash \Uparrow \Updownarrow_{\mathsf{pk}} \Uparrow_{\mathsf{pk}} \Downarrow_{\mathsf{pk}} \Downarrow_{\mathsf{pk}} \Uparrow_{\mathsf{pk}} \Downarrow_{\mathsf{pk}} \qquad \circleddash \Uparrow \Uparrow \Downarrow \Updownarrow_{\mathsf{pk}} \Downarrow_{\mathsf{pk}} \Uparrow_{\mathsf{pk}} \Downarrow_{\mathsf{pk}} \qquad \circleddash \Uparrow \Uparrow \Downarrow \Uparrow \Downarrow \Downarrow \Updownarrow_{\mathsf{pk}}$$

which themselves correspond to the three peaks of the mountain diagram, with the height of each peak corresponding to the dimension of each locally maximal variable.

All disc contexts are pasting diagrams, and hence are the associated context of a Dyck word.

**Definition 3.1.8.** Let $\mathcal{D}^n$ be the Dyck word with $n$ upwards moves followed by $n$ downwards moves. The equality $\lfloor \mathcal{D}^n \rfloor \equiv D^n$ follows from a trivial induction. If $n > 0$, There is a unique peak of $\mathcal{D}^n$ with associated term $d_n$.

We lastly show that a Dyck word can be suspended, which is expected as ps-contexts are closed under suspension. The various constructions associated to a suspended Dyck word are equal to the same constructions on the unsuspended Dyck word.

**Lemma 3.1.9.** *Dyck words are closed under suspension. We define the suspension of a Dyck word $\mathcal{D} : \mathsf{Dyck}_d$ to be the Dyck word $\Sigma(\mathcal{D}) : \mathsf{Dyck}_{d+1}$ which is obtained by inserting an additional up move to the start of the word, or can alternatively be inductively defined by:*

$$\Sigma(\circleddash) = \circleddash \Uparrow \qquad \Sigma(\mathcal{D} \Uparrow) = \Sigma(\mathcal{D}) \Uparrow \qquad \Sigma(\mathcal{D} \Downarrow) = \Sigma(\mathcal{D}) \Downarrow$$

*The following equalities hold:*

$$\lfloor \Sigma(\mathcal{D}) \rfloor = \Sigma(\lfloor \mathcal{D} \rfloor) \qquad \mathrm{Ty}_{\Sigma(\mathcal{D})} = \Sigma(\mathrm{Ty}_{\mathcal{D}}) \qquad \mathrm{Tm}_{\Sigma(\mathcal{D})} = \Sigma(\mathrm{Tm}_{\mathcal{D}})$$

*for each Dyck word $\mathcal{D}$. For each peak $p : \mathsf{Peak}_{\mathcal{D}}$, there is an associated peak $\Sigma(p) : \mathsf{Peak}_{\Sigma(\mathcal{D})}$ which is defined similarly.*

*Proof.* These properties are all proved by straight forward induction on $\mathcal{D}$. The formalised proofs appear in Catt.Dyck.Properties. □

The Dyck words presented in this section can be viewed as a more direct syntax for pasting contexts, which allow induction to be easily performed. For this reason, most of the properties of Dyck words follow from routine inductions, and hence are relegated to the formalisation. The key contribution of this (sub)section is the characterisation of locally maximal variables as peaks, which have an easy inductive definition due to the simplicity of Dyck words.

*Remark* 3.1.10. All locally maximal variables of ps-contexts are identified with peaks, except for the unique variable of the singleton context. This discrepancy will make no difference for pruning, as a $0$-dimensional variable could never have been sent to an identity and so would never have been a candidate for pruning.

### 3.1.2 The pruning construction

Equipped with Dyck words, and a classification of locally maximal variables as peaks, we are now able to define each of the constructions used in the pruning operation.

**Definition 3.1.11.** Let $\mathcal{D} : \mathsf{Dyck}_d$ be a Dyck word, and $p : \mathsf{Peak}_{\mathcal{D}}$ be a peak of $\mathcal{D}$. The pruned Dyck word $\mathcal{D} \mathbin{/\!\!/} p : \mathsf{Dyck}_d$ and substitution $\pi_p : \lfloor \mathcal{D} \rfloor \to \lfloor \mathcal{D} \mathbin{/\!\!/} p \rfloor$ are then defined inductively on the peak $p$ by the following equations:

$$\begin{aligned}
\mathcal{D} \Uparrow \Downarrow \mathbin{/\!\!/} \mathcal{D} \Updownarrow_{\mathsf{pk}} &= \mathcal{D} \\
\mathcal{D} \Uparrow \mathbin{/\!\!/} p \Uparrow_{\mathsf{pk}} &= (\mathcal{D} \mathbin{/\!\!/} p) \Uparrow \\
\mathcal{D} \Downarrow \mathbin{/\!\!/} p \Downarrow_{\mathsf{pk}} &= (\mathcal{D} \mathbin{/\!\!/} p) \Downarrow
\end{aligned}$$

$$\begin{aligned}
\pi_{\mathcal{D} \Updownarrow_{\mathsf{pk}}} &= \langle \mathsf{id}_{\lfloor \mathcal{D} \rfloor}, \mathrm{Tm}_{\mathcal{D}}, \mathsf{id}(\mathrm{Ty}_{\mathcal{D}}, \mathrm{Tm}_{\mathcal{D}}) \rangle \\
\pi_{p \Uparrow_{\mathsf{pk}}} &= \langle \mathrm{wk}(\mathrm{wk}(\pi_p)), y_{\mathcal{D}}, f_{\mathcal{D}} \rangle \\
\pi_{p \Downarrow_{\mathsf{pk}}} &= \pi_p
\end{aligned}$$

If we further have a substitution $\sigma : \lfloor \mathcal{D} \rfloor \to_\star \Gamma$ for some context $\Gamma$, then the pruned substitution $\sigma \mathbin{/\!\!/} p : \lfloor \mathcal{D} \mathbin{/\!\!/} p \rfloor \to_\star \Gamma$ can be formed:

$$\begin{aligned}
\langle \sigma, s, t \rangle \mathbin{/\!\!/} \mathcal{D} \Updownarrow_{\mathsf{pk}} &= \sigma \\
\langle \sigma, s, t \rangle \mathbin{/\!\!/} p \Uparrow_{\mathsf{pk}} &= \langle \sigma \mathbin{/\!\!/} p, s, t \rangle \\
\sigma \mathbin{/\!\!/} p \Downarrow_{\mathsf{pk}} &= \sigma \mathbin{/\!\!/} p
\end{aligned}$$

Each peak in a Dyck word corresponds to a consecutive upwards arrow and downwards arrow. Pruning this peak corresponds removing these two arrows, which does not change the trailing dimension of the Dyck word. The effect on the mountain diagram representation can be seen in Figure 3.2.

When a peak is pruned the locally maximal variable and its target are removed from the associated context. The substitution $\pi_{\mathcal{D} \Updownarrow_{\mathsf{pk}}}$ simply maps these two variables to $\mathsf{id}(\mathrm{Ty}_{\mathcal{D}}, \mathrm{Tm}_{\mathcal{D}})$ and $\mathrm{Tm}_{\mathcal{D}}$, where the Dyck term $\mathrm{Tm}_{\mathcal{D}}$ is the source of the locally maximal variable. Pruning a substitution simply removes the terms corresponding to the removed variables in the associated

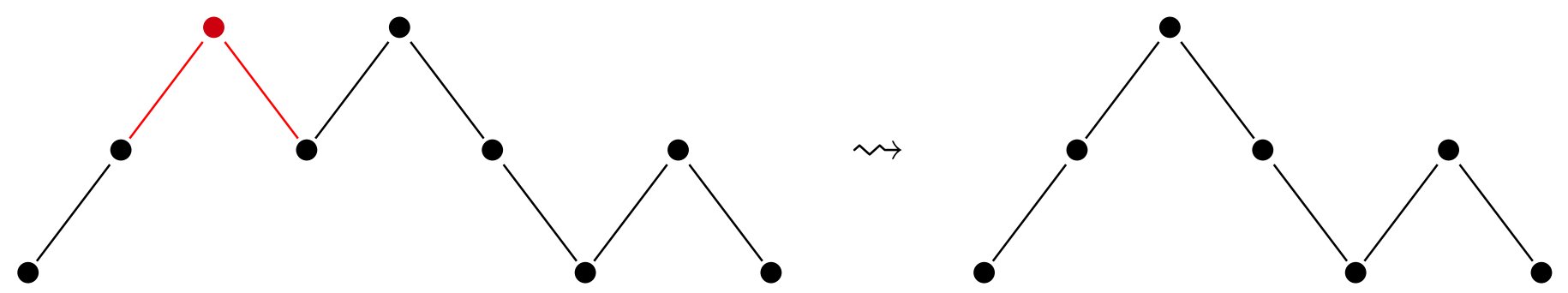

Figure 3.2: Pruning of peak $\circleddash \Uparrow \Updownarrow_{\mathsf{pk}} \Uparrow_{\mathsf{pk}} \Downarrow_{\mathsf{pk}} \Downarrow_{\mathsf{pk}} \Uparrow_{\mathsf{pk}} \Downarrow_{\mathsf{pk}}$.

context.

*Example* 3.1.12. Let $\Gamma = (x : \star), (f : x \to_\star x)$ and consider the term $f * \mathsf{id}(x)$, which is given by:

$$\mathsf{Coh}_{((a:\star),(b:\star),(c:a\to b),(d:\star),(e:b\to d)\,;\,a\to d)}[\langle x, x, f, x, \mathsf{id}(\star, x)\rangle]$$

The context in this coherence is the associated context of the Dyck word $\circleddash \Uparrow \Downarrow \Uparrow \Downarrow$ which has a peak $\circleddash \Uparrow \Downarrow \Updownarrow_{\mathsf{pk}}$, which corresponds to the locally maximal variable $e$. Since $e$ is sent to an identity by the substitution, pruning can be applied to get:

$$\begin{aligned} \circleddash \Uparrow \Downarrow \Uparrow \Downarrow /\!\!/ \circleddash \Uparrow \Downarrow \Updownarrow_{\mathsf{pk}} &= \circleddash \Uparrow \Downarrow \\ \pi_{\circleddash \Uparrow \Downarrow \Updownarrow_{\mathsf{pk}}} &= \langle a, b, c, b, \mathsf{id}(\star, b)\rangle \\ \langle x, x, f, x, \mathsf{id}(\star, x)\rangle /\!\!/ \circleddash \Uparrow \Downarrow \Updownarrow_{\mathsf{pk}} &= \langle x, x, f\rangle \end{aligned}$$

Which results in the term:

$$\mathsf{Coh}_{((a:\star),(b:\star),(c:a\to b)\,;\,(a\to d)[\![\langle a,b,c,b,\mathsf{id}(\star,b)\rangle]\!])}[\langle x, x, f\rangle] \equiv \mathsf{Coh}_{((a:\star),(b:\star),(c:a\to b)\,;\,(a\to b))}[\langle x, x, f\rangle]$$

which is the unary composite on $f$. In the presence of disc removal, this term could further simplify to the variable $f$.

With these constructions, we can define the pruning rule.

**Definition 3.1.13.** A term $t$ *is an identity* if $t \equiv \mathsf{id}(A, s)$ for some type $A$ and some term $s$. The *pruning rule set*, prune, is the set consisting of the triples:

$$(\Gamma, \mathsf{Coh}_{(\lfloor \mathcal{D} \rfloor\,;\,A)}[\sigma], \mathsf{Coh}_{(\lfloor \mathcal{D} /\!\!/ p \rfloor\,;\,A[\![\pi_p]\!])}[\sigma /\!\!/ p])$$

for each Dyck word $\mathcal{D} : \mathsf{Dyck}_0$, peak $p : \mathsf{Peak}_{\mathcal{D}}$, type $A : \mathsf{Type}_{\lfloor \mathcal{D} \rfloor}$, and substitution $\sigma : \lfloor \mathcal{D} \rfloor \to_\star \Gamma$ where $\lfloor p \rfloor [\![\sigma]\!]$ is an identity.

A set of rules $\mathcal{R}$ *contains pruning* if prune $\subseteq \mathcal{R}$. Pruning makes the following rule admissible:

$$\frac{\mathcal{D} : \mathsf{Dyck}_0 \qquad p : \mathsf{Peak}_{\mathcal{D}} \qquad \lfloor \mathcal{D} \rfloor \vdash A \qquad \Gamma \vdash \sigma : \lfloor \mathcal{D} \rfloor \\ (\lfloor \mathcal{D} \rfloor, \mathsf{Supp}(\mathsf{src}(A)), \mathsf{tgt}(A)) \in \mathcal{O} \qquad \lfloor p \rfloor [\![\sigma]\!] \text{ is an identity}}{\Gamma \vdash \mathsf{Coh}_{(\lfloor \mathcal{D} \rfloor\,;\,A)}[\sigma] = \mathsf{Coh}_{(\lfloor \mathcal{D} /\!\!/ p\,;\,A[\![\pi_p]\!])}[\sigma /\!\!/ p]} \textsc{prune}$$

The set $\mathcal{R}$ *has pruning* if the rule PRUNE holds in the generated theory.

### 3.1.3 Properties of pruning

We start with the aim of proving that each construction involved in pruning satisfies the expected typing judgements. To do this the following lemma will be necessary, which describes the interaction of the Dyck word construction with pruning.

**Lemma 3.1.14.** *Let $\mathcal{D}$ : $\mathsf{Dyck}_d$ be a Dyck word. Then the following equations hold:*

$$\begin{aligned}\mathrm{Ty}_{\mathcal{D}}[\![\pi_p]\!] &\equiv \mathrm{Ty}_{\mathcal{D} /\!\!/ p}\\ \mathrm{Tm}_{\mathcal{D}}[\![\pi_p]\!] &\equiv \mathrm{Tm}_{\mathcal{D} /\!\!/ p}\end{aligned}$$

*for any peak $p$ : $\mathsf{Peak}_{\mathcal{D}}$ of $\mathcal{D}$.*

*Proof.* The proof proceeds by an induction on the peak $p$, proving both equations simultaneously. Both equations hold by routine calculations given in Catt.Dyck.Pruning.Properties by the functions dyck-type-prune and dyck-term-prune. □

This allows the main typing properties of this section to be given.

**Proposition 3.1.15.** *Let $\mathcal{D}$ : $\mathsf{Dyck}_d$ be a Dyck word and let $p$ : $\mathsf{Peak}_{\mathcal{D}}$ be a peak of this word. Then:*

$$\lfloor \mathcal{D} /\!\!/ p \rfloor \vdash \pi_p : \lfloor \mathcal{D} \rfloor$$

*Given a substitution $\sigma$ with $\Gamma \vdash \sigma : \lfloor \mathcal{D} \rfloor$, where $\lfloor p \rfloor[\![\sigma]\!]$ is an identity, the equality and typing judgements:*

$$\Gamma \vdash \sigma = \pi_p \bullet (\sigma /\!\!/ p) \qquad \Gamma \vdash \sigma : \lfloor \mathcal{D} /\!\!/ p \rfloor$$

*hold.*

*Proof.* We prove each judgement holds in turn by induction on the peak $p$. For the judgement:

$$\lfloor \mathcal{D} /\!\!/ p \rfloor \vdash \pi_p : \lfloor \mathcal{D} \rfloor$$

the case when the peak is of the form $p \Downarrow_{\mathsf{pk}}$ is trivial. The case for when it is of the form $\mathcal{D} \Updownarrow_{\mathsf{pk}}$ easily follows from Lemma 3.1.5 and Corollary 2.4.9. For the case where the peak is of the form $p \Uparrow_{\mathsf{pk}}$, it must be shown that:

$$\Delta \vdash \langle \mathrm{wk}(\mathrm{wk}(\pi_p)), y, f \rangle : \lfloor \mathcal{D} \rfloor, (y : \mathrm{Ty}_{\mathcal{D}}), (f : \mathrm{wk}(\mathrm{Tm}_{\mathcal{D}}) \to_{\mathrm{wk}(\mathrm{Ty}_{\mathcal{D}})} y)$$

where $\Delta = \lfloor \mathcal{D} /\!\!/ p \rfloor, (y : \mathrm{Ty}_{\mathcal{D} /\!\!/ p}), (f : \mathrm{wk}(\mathrm{Tm}_{\mathcal{D} /\!\!/ p}) \to_{\mathrm{wk}(\mathrm{Ty}_{\mathcal{D} /\!\!/ p})} y)$. This requires proofs of:

$$\begin{aligned}&\Delta \vdash \mathrm{wk}(\mathrm{wk}(\pi_p)) : \lfloor \mathcal{D} \rfloor\\ &\Delta \vdash y : \mathrm{Ty}_{\mathcal{D}}[\![\pi_p]\!]\\ &\Delta \vdash f : (\mathrm{wk}(\mathrm{Tm}_{\mathcal{D}}) \to_{\mathrm{wk}(\mathrm{Ty}_{\mathcal{D}})} y)[\![\langle \mathrm{wk}(\pi_p), y \rangle]\!]\end{aligned}$$

The first part follows from inductive hypothesis (and typing of weakening). The other two judgements follow from some calculation and Lemma 3.1.14.

For the second judgement:

$$\Gamma \vdash \sigma = \pi_p \bullet (\sigma /\!\!/ p)$$

The $p \Downarrow_{\mathsf{pk}}$ case is again trivial. The $p \Uparrow_{\mathsf{pk}}$ case follows easily from properties of weakening and the inductive hypothesis. For the $\mathcal{D} \Updownarrow_{\mathsf{pk}}$ case we suppose the substitution is of the form $\langle \sigma, s, \mathsf{id}(A, t) \rangle$ and are required to show that:

$$\Gamma \vdash \langle \mathsf{id}_{\mathcal{D}}, \mathrm{Tm}_{\mathcal{D}}, \mathsf{id}(\mathrm{Ty}_{\mathcal{D}}, \mathrm{Tm}_{\mathcal{D}}) \rangle \bullet \sigma = \langle \sigma, s, \mathsf{id}(A, t) \rangle$$

It is immediate that $\mathsf{id}_{\mathcal{D}} \bullet \sigma \equiv \sigma$ and so it remains to show that $\Gamma \vdash \mathrm{Tm}_{\mathcal{D}}[\![\sigma]\!] = s$ and $\Gamma \vdash \mathsf{id}(\mathrm{Ty}_{\mathcal{D}}, \mathrm{Tm}_{\mathcal{D}})[\![\sigma]\!] = \mathsf{id}(A, t)$. By deconstructing the typing derivation of $\langle \sigma, s, \mathsf{id}(A, t) \rangle$, we have:

$$\Gamma \vdash \mathsf{id}(A, t) : (\mathrm{wk}(\mathrm{Tm}_{\mathcal{D}}) \to_{\mathrm{wk}(\mathrm{Ty}_{\mathcal{D}})} y)[\![\langle \sigma, s \rangle]\!]$$

By Corollary 2.4.9 and the uniqueness of typing, we must have:

$$\Gamma \vdash t \to_A t = (\mathrm{wk}(\mathrm{Tm}_{\mathcal{D}}) \to_{\mathrm{wk}(\mathrm{Ty}_{\mathcal{D}})} y)[\![\langle \sigma, s \rangle]\!] \equiv \mathrm{Tm}_{\mathcal{D}}[\![\sigma]\!] \to_{\mathrm{Ty}_{\mathcal{D}}[\![\sigma]\!]} s$$

and so $A = \mathrm{Ty}_{\mathcal{D}}[\![\sigma]\!]$ and $s = t = \mathrm{Tm}_{\mathcal{D}}[\![\sigma]\!]$. The equality $\mathsf{id}(\mathrm{Ty}_{\mathcal{D}}, \mathrm{Tm}_{\mathcal{D}}) = \mathsf{id}(A, t)$ follows as equality is respected by the identity construction, which can be proved by a simple induction.

Lastly, we consider the judgement:

$$\Gamma \vdash \sigma /\!\!/ p : \lfloor \mathcal{D} /\!\!/ p \rfloor$$

The only difficult case is for the peak $p \Uparrow_{\mathsf{pk}}$, where we can assume that the substitution is of the form $\langle \sigma, s, t \rangle$, such that:

$$\langle \sigma, s, t \rangle /\!\!/ p \Uparrow_{\mathsf{pk}} \equiv \langle \sigma /\!\!/ p, s, t \rangle$$

Typing for $\sigma /\!\!/ p$ follows from inductive hypothesis, and the typing for $s$ and $t$ follow from applying conversion rules to the corresponding parts of the typing derivation for $\langle \sigma, s, t \rangle$. After some computation, the following equalities are needed for these conversion rules:

$$\Gamma \vdash \mathrm{Tm}_{\mathcal{D}}[\![\sigma]\!] = \mathrm{Tm}_{\mathcal{D} /\!\!/ p}[\![\sigma /\!\!/ p]\!]$$
$$\Gamma \vdash \mathrm{Ty}_{\mathcal{D}}[\![\sigma]\!] = \mathrm{Ty}_{\mathcal{D} /\!\!/ p}[\![\sigma /\!\!/ p]\!]$$

The first is given by:

$$\begin{aligned} \mathrm{Tm}_{\mathcal{D}}[\![\sigma]\!] &= \mathrm{Tm}_{\mathcal{D}}[\![\pi_p \bullet (\sigma /\!\!/ p)]\!] \\ &\equiv \mathrm{Tm}_{\mathcal{D}}[\![\pi_p]\!][\![\sigma /\!\!/ p]\!] \\ &\equiv \mathrm{Tm}_{\mathcal{D} /\!\!/ p}[\![\sigma /\!\!/ p]\!] \end{aligned}$$

and the second follows similarly, completing the proof. □

We next show that pruning has the expected properties on the Dyck words $\mathcal{D}^n$, which correspond to disc contexts.

**Proposition 3.1.16.** *Let $n > 0$, and $p$ be the unique peak of $\mathcal{D}^n$. Then:*

$$\mathcal{D}^n /\!\!/ p \equiv \mathcal{D}^{n-1} \qquad \{s \to_A t, u\} /\!\!/ p \equiv \{A, s\}$$

*for all* $A, s, t, u$ *where* $\dim(A) = n - 1$.

*Proof.* Both properties are immediate. □

We now turn our attention to proving that the pruning equality set satisfies all the conditions from Section 2.4. We begin with the tameness conditions, omitting the weakening condition, as it follows from the substitution condition.

**Proposition 3.1.17.** *For all* $\mathcal{D} : \mathsf{Dyck}_d$ *and peaks* $p : \mathsf{Peak}_{\mathcal{D}}$, *and substitutions* $\sigma : \lfloor \mathcal{D} \rfloor \to \Delta$ *and* $\tau : \Delta \to \Gamma$ *the following equality holds:*

$$(\sigma /\!\!/ p) \bullet \tau \equiv (\sigma \bullet \tau) /\!\!/ p$$

*Hence, the set prune satisfies the* $\mathcal{R}$*-substitution condition for any equality set* $\mathcal{R}$, *and so also satisfies the weakening condition.*

*Furthermore, the following equalities hold:*

$$\Sigma(\mathcal{D}) /\!\!/ \Sigma(p) = \Sigma(\mathcal{D} /\!\!/ p) \qquad \pi_{\Sigma(p)} \equiv \Sigma(\pi_p) \qquad \Sigma(\sigma /\!\!/ p) \equiv \Sigma(\sigma) /\!\!/ \Sigma(p)$$

*Therefore, the set prune also satisfies the suspension condition, making the equality set prune tame.*

*Proof.* The proofs of each syntactic equality are easily proved by induction on the peak $p$. Their proofs are given in the formalisation in Catt.Dyck.Pruning.Properties as //s-sub, prune-susp-peak, susp-$\pi$, and susp-//s. □

To show that the support property holds, we must prove that $\mathsf{Supp}(\sigma) = \mathsf{Supp}(\sigma /\!\!/ p)$. We aim to do this by observing that $\mathsf{Supp}(\sigma) = \mathsf{Supp}(\pi_p \bullet (\sigma /\!\!/ p))$ and that $\mathsf{Supp}(\pi_p \bullet (\sigma /\!\!/ p)) = \mathsf{Supp}(\sigma /\!\!/ p)$. By employing the proof strategy for the support condition introduced in Section 2.4.2, the first will follow from the equality $\sigma = \pi_p \bullet (\sigma /\!\!/ p)$, which we can assume holds in a theory which satisfies the support condition. For the second we need the following lemma.

**Lemma 3.1.18.** *For all* $n : \mathbb{N}$, $\epsilon \in \{-, +\}$, $\mathcal{D} : \mathsf{Dyck}_d$, *and* $p : \mathsf{Peak}_{\mathcal{D}}$*:*

$$\partial_n^\epsilon(\lfloor \mathcal{D} \rfloor)[\![\pi_p]\!] = \partial_n^\epsilon(\lfloor \mathcal{D} /\!\!/ p \rfloor)$$

*and so* $\mathsf{Supp}(\pi_p) = \mathsf{Var}(\lfloor \mathcal{D} /\!\!/ p \rfloor)$.

*Proof.* The main equation in this lemma is given by a long and technical induction on the peak $p$. The details of this induction appear in the formalisation in the function $\pi$-boundary-vs which appears in the module Catt.Dyck.Pruning.Support.

The equation $\mathsf{Supp}(\pi_p) = \mathsf{Var}(\lfloor \mathcal{D} /\!\!/ p \rfloor)$ follows from Proposition 2.3.3 and Lemma 2.3.8, by setting $n = \dim(\lfloor \mathcal{D} \rfloor)$. □

We are now ready to prove that the support condition holds.

**Proposition 3.1.19.** *Let $\mathcal{R}$ be a tame equality rule set that satisfies the support condition. Then the set prune satisfies the $\mathcal{R}$-support condition.*

*Proof.* It suffices to prove that:

$$\mathrm{Supp}(\mathsf{Coh}_{(\lfloor \mathcal{D} \rfloor\,;\,A)}[\sigma]) = \mathrm{Supp}(\mathsf{Coh}_{(\lfloor \mathcal{D} /\!\!/ p \rfloor\,;\,A[\![\pi_p]\!])}[\sigma /\!\!/ p])$$

for $\mathcal{D} : \mathsf{Dyck}_0$, $p : \mathsf{Peak}_{\mathcal{D}}$, type $A$, and substitution $\sigma : \lfloor \mathcal{D} \rfloor \to \Gamma$, where $\lfloor p \rfloor [\![\sigma]\!]$ is an identity and $\Gamma \vdash_{\mathcal{R}} \mathsf{Coh}_{(\lfloor \mathcal{D} \rfloor\,;\,A)}[\sigma] : B$ for some $B$. By inspection of the typing derivation we obtain an instance of the judgement $\Gamma \vdash_{\mathcal{R}} \sigma : \lfloor \mathcal{D} \rfloor$, and so:

$$\begin{aligned}
\mathrm{Supp}(\mathsf{Coh}_{(\lfloor \mathcal{D} \rfloor\,;\,A)}[\sigma]) &= \mathrm{Supp}(\sigma) \\
&= \mathrm{Supp}(\pi_p \bullet (\sigma /\!\!/ p)) && (*) \\
&= \mathrm{Supp}(\pi_p)[\![\sigma /\!\!/ p]\!] \\
&= \mathrm{Var}\ \lfloor \mathcal{D} /\!\!/ p \rfloor [\![\sigma /\!\!/ p]\!] && \text{by Lemma 3.1.18} \\
&= \mathrm{Supp}(\sigma /\!\!/ p) \\
&= \mathrm{Supp}(\mathsf{Coh}_{(\lfloor \mathcal{D} /\!\!/ p \rfloor\,;\,A[\![\pi_p]\!])}[\sigma /\!\!/ p])
\end{aligned}$$

where equality $(*)$ is derived by applying Proposition 2.4.24 to the equality

$$\Gamma \vdash_{\mathcal{R}} \sigma = \pi_p \bullet (\sigma /\!\!/ p)$$

from Proposition 3.1.15. □

To prove that the preservation condition holds, it is necessary to show that the type $A[\![\pi_p]\!]$ created by pruning is a valid operation. This cannot be deduced from any of the conditions that have been imposed on the operation set $\mathcal{O}$ so far. Therefore, we introduce the following additional condition.

**Definition 3.1.20.** An operation set $\mathcal{O}$ *supports pruning* if for all $\mathcal{D} : \mathsf{Dyck}_0$, $p : \mathsf{Peak}_{\mathcal{D}}$, and variable sets $U, V \subseteq \mathrm{Var}(\lfloor \mathcal{D} \rfloor)$ we have:

$$(\lfloor \mathcal{D} /\!\!/ p \rfloor, U[\![\pi_p]\!], V[\![\pi_p]\!]) \in \mathcal{O}$$

whenever $(\lfloor \mathcal{D} \rfloor, U, V) \in \mathcal{O}$.

The globular operation set trivially supports pruning. From Lemma 3.1.18 and Proposition 2.3.10, it can be proved that the regular operation set supports pruning. We can now prove that the preservation condition holds.

**Proposition 3.1.21.** *Let $\mathcal{R}$ be a tame equality rule set and suppose the operation set $\mathcal{O}$ supports pruning. Then the set prune satisfies the $\mathcal{R}$-preservation condition.*

*Proof.* Let $\mathcal{D} : \mathsf{Dyck}_d$ be a Dyck word and $p : \mathsf{Peak}_{\mathcal{D}}$ be a peak of $\mathcal{D}$. Further suppose $s \to_A t : \mathsf{Type}_{\lfloor \mathcal{D} \rfloor}$, and $\sigma : \lfloor \mathcal{D} \rfloor \to \Gamma$ such that $\lfloor p \rfloor [\![\sigma]\!]$ is an identity and:

$$\Gamma \vdash_{\mathcal{R}} \mathsf{Coh}_{(\lfloor \mathcal{D} \rfloor\,;\,s \to_A t)}[\sigma] : B$$

for some type $B$ : Type$_\Gamma$. By inspection on this typing derivation we have:

$$\lfloor \mathcal{D} \rfloor \vdash_{\mathcal{R}} A \qquad \Gamma \vdash_{\mathcal{R}} \sigma \lfloor \mathcal{D} \rfloor \qquad (\lfloor \mathcal{D} \rfloor, \mathsf{Supp}(s), \mathsf{Supp}(t)) \in \mathcal{O} \qquad \Gamma \vdash_{\mathcal{R}} B = (s \to_A t)[\![\sigma]\!]$$

and so by Proposition 3.1.15, we have:

$$\lfloor \mathcal{D} /\!\!/ p \rfloor \vdash_{\mathcal{R}} \pi_p : \lfloor \mathcal{D} \rfloor \qquad \Gamma \vdash_{\mathcal{R}} \sigma /\!\!/ p : \lfloor \mathcal{D} /\!\!/ p \rfloor$$

therefore, as $\mathcal{O}$ supports pruning, the following judgement holds:

$$\Gamma \vdash_{\mathcal{R}} \mathsf{Coh}_{(\lfloor \mathcal{D} /\!\!/ p \rfloor \,;\, (s \to_A t)[\![\pi_p]\!])}[\sigma /\!\!/ p] : (s \to_A t)[\![\pi_p]\!][\![\sigma /\!\!/ p]\!]$$

and so by applying the conversion rule, it suffices to show that

$$\Gamma \vdash_{\mathcal{R}} B = (s \to_A t)[\![\pi_p]\!][\![\sigma /\!\!/ p]\!]$$

but this follows from the equality $B = (s \to_A t)[\![\sigma]\!]$ and the equality $\sigma = \pi_p \bullet (\sigma /\!\!/ p)$ from Proposition 3.1.15. □

We end this section with a property of pruning that will be required to prove confluence. Suppose we have a Dyck word $\mathcal{D}$ and two distinct peaks $p, q$ : Peak$_{\mathcal{D}}$. Then both peaks can be pruned from $\mathcal{D}$ in either order. Consider the example below on the Dyck word from Example 3.1.7.

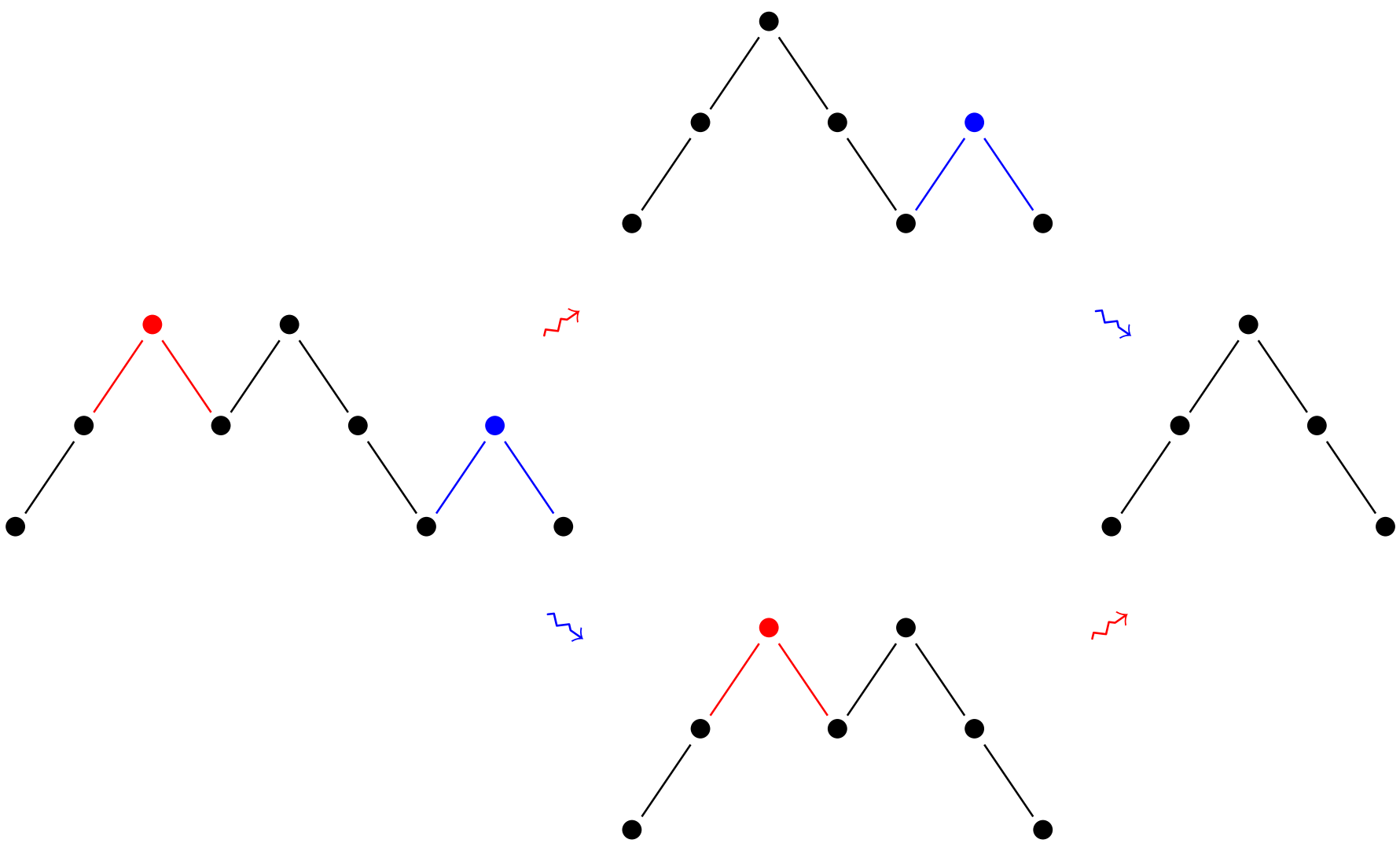

The following proposition proves that both peaks of the Dyck word can be pruned, and that the order in which this is done does not matter.

**Proposition 3.1.22.** *Suppose $\mathcal{D}$ : Dyck$_d$ is a Dyck word and let $p$ and $q$ be two distinct peaks of $\mathcal{D}$. Then there is a peak $q_p$ of $\mathcal{D} /\!\!/ p$ such that:*

$$\lfloor q_p \rfloor \equiv \lfloor q \rfloor [\![\pi_p]\!]$$

*and a similar peak $p_q$ of $\mathcal{D} /\!\!/ q$. Furthermore, the following equations hold syntactically:*

$$(\mathcal{D} /\!\!/ p) /\!\!/ q_p = (\mathcal{D} /\!\!/ q) /\!\!/ p_q \qquad \pi_p \bullet \pi_{q_p} \equiv \pi_q \bullet \pi_{p_q} \qquad (\sigma /\!\!/ p) /\!\!/ q_p = (\sigma /\!\!/ q) /\!\!/ p_q$$

*where the last equation holds for any $\sigma : \lfloor \mathcal{D} \rfloor \to \Gamma$.*

*Proof.* All proofs proceed by a simultaneous induction on both the peaks $p$ and $q$, and are given in Catt.Dyck.Pruning.Properties in the formalisation. The construction of the peak $q_p$ is given by the function prune-peak, the equality $\lfloor q_p \rfloor \equiv \lfloor q \rfloor [\![\pi_p]\!]$ is given by prune-peak-prop, and the remaining three equations are given by prune-conf, $\pi$-conf, and prune-sub-conf. □

## 3.2 Trees

During the next sections we build up to defining the insertion operation. This operation performs larger modifications to pasting diagrams than the pruning operation, and we will again want to represent pasting diagrams differently to make the definition in Section 3.4 as natural as possible. It is well known that pasting diagrams correspond to planar rooted trees [Web04; Lei04; Bat98b], which we will simply refer to as *trees* and can be defined as follows.

**Definition 3.2.1.** A *tree* $T$ : Tree is inductively defined to be a (possibly empty) list of trees.

Throughout this section we will make use of standard operations and notations for lists. A list that contains the elements $x_i$ for $i$ from 0 to $n$ will be written in square bracket notation as $[x_0, x_1, x_2, \dots, x_n]$. Further, we use the notation $[\,]$ for the empty list and $+\!\!+$ for the concatenation of lists, which is associative and has the empty list as its unit. We will use the Agda-like notation of writing $n :: ns$ for a list for which the first element (the head) is $n$ and the rest of the list (the tail) is $ns$. The length of a list will be given by the operation len.

We will use the notation $\Sigma(T) = [T]$, and call $\Sigma(T)$ the suspension of $T$, for reasons that will become immediate once the context generated by a tree has been defined in Section 3.2.2.

We note that it will be common to see expressions of the form $S :: T$ where $S$ and $T$ are both trees. It may seem as if this was an error, and that a concatenation operation should have been given instead, but in this case we are exploiting the identification of trees and lists of trees to treat $S$ as a tree (as an element of the list) and $T$ as a list of trees.

We now define some common operations on trees.

**Definition 3.2.2.** The *height* of a tree $\mathrm{h}(T)$ is 0 if $T$ is empty or $1 + \max_k \mathrm{h}(T_k)$ if $T = [T_0, \dots, T_n]$. For a tree $T$, its *trunk height*, $\mathsf{th}(T)$, is $1 + \mathsf{th}(T_0)$ if $T = [T_0]$ and 0 otherwise. A tree is *linear* if its trunk height equals its height.

Subtrees of a tree can be indexed by a list of natural numbers $P$, giving a subtree $T^P$ by letting $T^{[\,]} = T$ and $T^{k::P} = (T_k)^P$ if $T = [T_0, \dots, T_n]$.

As these trees represent pasting diagrams, a context can be associated to each one. To be able to make effective use of trees we will need to understand this mapping to contexts, and the associated constructions used in this mapping. One of these constructions is suspension, which we have already seen. The second is an operation known as the wedge sum, which will be introduced in Section 3.2.1. Both these operations are mappings from contexts to contexts which preserve ps-context derivations. We will see in Section 3.2.2 that a further result holds, that these two operations (along with the singleton context) are sufficient to generate all ps-contexts.

*Remark* 3.2.3. In the formalisation, trees are binary and defined in Catt.Tree. This exploits an isomorphism between binary trees and trees with arbitrary (finite) branching. The constructors for the trees in the formalisation are called Sing, which stands for "singleton" and takes no arguments, and Join, which takes two trees as arguments. The isomorphism is generated from the following rules:

$$\frac{}{\mathsf{Sing} \simeq [\,]} \qquad\qquad \frac{S \simeq S' \qquad T \simeq T'}{\mathsf{Join}(S, T) \simeq S' :: T'}$$

Presenting trees in this way in the formalisation allows any induction to be done as a single induction over the constructors of a tree, instead of simultaneously inducting on the height of the tree and on lists. We retain the standard notation of trees for this text for simplicity of notation. Under the above isomorphism, this has no effect on the formal development.

### 3.2.1 Wedge sums

The wedge sum, just like suspension, is an operation inspired by a similar operation on topological spaces. Given two spaces $X$ and $Y$ and points $x$ of $X$ and $y$ of $Y$, the space $X \vee Y$ can be formed, by taking the disjoint union of $X$ and $Y$, and identifying the points $x$ and $y$.

This construction satisfies a universal property: it is the colimit of the following diagram:

$$X \xleftarrow{\;x\;} \{*\} \xrightarrow{\;y\;} Y \tag{3.2.4}$$

where the arrows labelled $x$ and $y$ send the unique point $*$ to $x$ and $y$ respectively. Such a universal construction gives rise to two inclusions:

$$\mathrm{inl}_{X;Y} : X \to X \vee Y \qquad \mathrm{inr}_{X;Y} : Y \to X \vee Y$$

A similar colimit can be formed in the syntactic category of $\textsc{Catt}_{\mathcal{R}}$. Leveraging that the variables of a context are ordered, every (non-empty) context in $\textsc{Catt}$ is naturally bipointed. For a context $\Gamma$, the first point is given by the first variable of the context (which must have type $\star$), which we name $\mathrm{fst}(\Gamma)$, and the second point is given by the last $0$-dimensional variable in the context, which we name $\mathrm{snd}(\Gamma)$. We therefore restrict the construction above to when the chosen point for the left context $\Gamma$ is $\mathrm{snd}(\Gamma)$ and the chosen point for the second context is $\mathrm{fst}(\Delta)$. This simplifies the construction, and will be the only case we need for forming trees. We note that $\mathrm{fst}(\Sigma(\Gamma)) \equiv N$ and $\mathrm{snd}(\Sigma(\Gamma)) \equiv S$, as we will commonly take the wedge sums of suspended contexts.

**Definition 3.2.5.** Let $\Gamma$ and $\Delta$ be non-empty contexts. We then mutually define the *wedge sum* $\Gamma \vee \Delta$ and inclusions $\mathrm{inl}_{\Gamma;\Delta} : \Gamma \to_\star \Gamma \vee_t \Delta$ and $\mathrm{inr}_{\Gamma;\Delta} : \Delta \to_\star \Gamma \vee \Delta$ by induction on the context $\Delta$, noting that the base case is $\Delta = (x : A)$ as $\Delta$ is non-empty.

$$\begin{aligned} \Gamma \vee (x : A) &= \Gamma \\ \Gamma \vee \Delta, (x : A) &= \Gamma \vee \Delta, (x : A[\![\mathrm{inr}_{\Gamma;\Delta}]\!]) \end{aligned}$$

$$\mathrm{inl}_{\Gamma;\Delta} = \mathrm{wk}^{n-1}(\mathsf{id}_\Gamma) \qquad\qquad \text{when } \Delta \text{ has length } n$$

$$\begin{aligned} \mathrm{inr}_{\Gamma;(x:A)} &= \langle \mathrm{snd}(\Gamma) \rangle \\ \mathrm{inr}_{\Gamma;\Delta,(x:A)} &= \langle \mathrm{wk}(\mathrm{inr}_{\Gamma;\Delta}), x \rangle \end{aligned}$$

If we further have substitutions $\sigma : \Gamma \to_A \Theta$ and $\tau : \Delta \to_A \Theta$, then we can define the substitution $\sigma \vee \tau : \Gamma \vee \Delta \to_A \Theta$ again by induction on $\Delta$:

$$\begin{aligned} \sigma \vee \langle A, s \rangle &= \sigma \\ \sigma \vee \langle \tau, s \rangle &= \langle \sigma, s \rangle \end{aligned}$$

We note that no extra property is needed to define this universal map, though to show it is well-formed we will need that $\mathrm{snd}(\Gamma)[\![\sigma]\!] = \mathrm{fst}(\Delta)[\![\tau]\!]$.

We firstly prove some basic properties required for $\Gamma \vee \Delta$ to be the colimit of Diagram 3.2.4.

**Lemma 3.2.6.** *Let $\Gamma$ and $\Delta$ be non-empty contexts. Then:*

$$\mathrm{inl}_{\Gamma;\Delta} \vee \mathrm{inr}_{\Gamma;\Delta} \equiv \mathsf{id}_{\Gamma \vee \Delta}$$

*Further, the following equations hold:*

$$\mathrm{inl}_{\Gamma;\Delta} \bullet (\sigma \vee \tau) \equiv \sigma \qquad \mathrm{inr}_{\Gamma;\Delta} \bullet (\sigma \vee \tau) \equiv \tau$$

*for substitutions $\sigma : \Gamma \to_A \Theta$ and $\tau : \Delta \to_A \Theta$ where the second equality requires that* $\mathrm{snd}(\Gamma)[\![\sigma]\!] \equiv \mathrm{fst}(\Delta)[\![\tau]\!]$. *Lastly:*

$$(\sigma \vee \tau) \bullet \mu \equiv (\sigma \bullet \mu) \vee (\tau \bullet \mu)$$

*where $\mu : \Theta \to_B \Theta'$ is another substitution.*

*Proof.* Proofs appear as sub-from-wedge-prop, sub-from-wedge-inc-left, sub-from-wedge-inc-right, and sub-from-wedge-sub in Catt.Wedge.Properties. ☐

To simplify definitions of substitutions between wedge sums of contexts, we will write substitutions diagrammatically by specifying the individual components. Consider the following diagram:

$$\begin{array}{ccccc} \Gamma' & \vee & \Delta' & \vee & \Theta' \\ {\scriptstyle\sigma}\big\uparrow & & {\scriptstyle\tau}\big\uparrow & & \\ \Gamma & \vee & \Delta & & \end{array}$$

which is generated from substitutions $\sigma : \Gamma \to \Gamma'$ and $\tau : \Delta \to \Delta'$. A substitution $\Gamma \vee \Delta \to \Gamma' \vee \Delta' \vee \Theta'$ can be generated by composing each arrow in the diagram with suitable inclusions so that its target is $\Gamma' \vee \Delta' \vee \Theta'$, and then using the universal property of the wedge to map out of the source context. In the diagram above the generated substitution is:

$$((\sigma \bullet \mathsf{inl}_{\Gamma';\Delta'} \bullet \mathsf{inl}_{\Gamma'\vee\Delta';\Theta'}) \vee (\tau \bullet \mathsf{inl}_{\Gamma';\Delta'} \bullet \mathsf{inr}_{\Gamma'\vee\Delta';\Theta'}))$$

To ensure these definitions are unique, the following proposition is needed:

**Proposition 3.2.7.** *The wedge sum* $\vee$ *is associative and has the singleton context* $(x : \star)$ *as its left and right unit. Given a context* $\Gamma$*, the inclusions satisfy the following unitality properties:*

$$\mathsf{inl}_{\Gamma;(x:\star)} \equiv \mathsf{id}_\Gamma \qquad \mathsf{inr}_{(x:\star);\Gamma} \equiv \mathsf{id}_\Gamma$$

*and given substitutions* $\sigma : \Gamma \to_A \Xi$*,* $\tau : \Delta \to_A \Xi$*, and* $\mu : \Theta \to_A \Xi$ *we have:*

$$(\sigma \vee \tau) \vee \mu \equiv \sigma \vee (\tau \vee \mu)$$

*There is a unique way of including each of the contexts* $\Gamma$*,* $\Delta$*, and* $\Theta$ *into* $\Gamma \vee \Delta \vee \Theta$*, that is there is a unique substitution* $\Gamma \to \Gamma \vee \Delta \vee \Theta$ *which is built from a composite of inclusions and similarly for* $\Delta$ *and* $\Theta$*.*

*Proof.* The proofs of these are given in Catt.Wedge.Properties, and are all given by inducting on the right most context. The proof for the right unitality of $\vee$ is omitted from the formalisation as it is immediate from the definitions.

The uniqueness of inclusions substitutions is given by

- wedge-inc-left-assoc, which says:

$$\mathsf{inl}_{\Gamma;\Delta} \bullet \mathsf{inl}_{\Gamma\vee\Delta;\Theta} : \Gamma \to (\Gamma \vee \Delta) \vee \Theta \equiv \mathsf{inl}_{\Gamma;\Delta\vee\Theta} : \Gamma \to \Gamma \vee (\Delta \vee \Theta)$$

- wedge-incs-assoc, which says:

$$\mathsf{inr}_{\Gamma;\Delta} \bullet \mathsf{inl}_{\Gamma\vee\Delta;\Theta} : \Delta \to (\Gamma \vee \Delta) \vee \Theta \equiv \mathsf{inl}_{\Delta;\Theta} \bullet \mathsf{inr}_{\Gamma;\Delta\vee\Theta} : \Delta \to \Gamma \vee (\Delta \vee \Theta)$$

- wedge-inc-right-assoc, which says:

$$\mathsf{inr}_{\Gamma\vee\Delta;\Theta} : \Theta \to (\Gamma \vee \Delta) \vee \Theta \equiv \mathsf{inr}_{\Delta;\Theta} \bullet \mathsf{inr}_{\Gamma;\Delta\vee\Theta} : \Theta \to \Gamma \vee (\Delta \vee \Theta)$$

We note that the definition of the wedge sum differs slightly in the formalisation, specifying a term $t$ in $\Gamma$ which takes the role of $\mathsf{snd}(\Gamma)$, in order to give more computational control. By replacing the terms $t$ in the formalisation by $\mathsf{snd}(\Gamma)$ for the appropriate context $\Gamma$, and noting that $\mathsf{snd}(\Delta)[\![\mathsf{inr}_{\Gamma;\Delta}]\!] \equiv \mathsf{snd}(\Gamma \vee \Delta)$ (which can be proved by an easy induction), the results written here can be recovered. □

The previous proposition ensures that the diagrammatic notation for substitutions between wedge sums uniquely defines a substitution. We next show that all the constructions in this section have the expected typing properties.

**Lemma 3.2.8.** *The following inference rules are admissible in* $\text{CATT}_{\mathcal{R}}$*:*

$$\frac{\Gamma \vdash \quad \Delta \vdash}{\Gamma \vee \Delta \vdash} \qquad \frac{}{\Gamma \vee \Delta \vdash \mathrm{inl}_{\Gamma;\Delta} : \Gamma} \qquad \frac{}{\Gamma \vee \Delta \vdash \mathrm{inr}_{\Gamma;\Delta} : \Delta}$$

$$\frac{\Theta \vdash \mathrm{snd}(\Gamma)[\![\sigma]\!] = \mathrm{fst}(\Delta)[\![\tau]\!]}{\Theta \vdash \mathrm{inr}_{\Gamma;\Delta} \bullet (\sigma \vee \tau) = \tau} \qquad \frac{\Theta \vdash \sigma : \Gamma \quad \Theta \vdash \tau : \Gamma \quad \Theta \vdash \mathrm{snd}(\Gamma)[\![\sigma]\!] = \mathrm{fst}(\Delta)[\![\tau]\!]}{\Theta \vdash \sigma \vee \tau : \Gamma \vee \Delta}$$

$$\frac{\Theta \vdash \sigma = \sigma' \quad \Theta \vdash \tau = \tau'}{\Theta \vdash \sigma \vee \tau = \sigma' \vee \tau'}$$

*Proof.* All proofs are given in Catt.Wedge.Typing. □

We finally show that the wedge sum preserves pasting diagrams, the property that wedge sums were initially introduced for.

**Proposition 3.2.9.** *The wedge sum of two ps-contexts is a ps-context: If* $\Gamma \vdash_{\mathrm{ps}}$ *and* $\Delta \vdash_{\mathrm{ps}}$*, then* $\Gamma \vee \Delta \vdash_{\mathrm{ps}}$

*Proof.* It can first be proven that if the derivation $\Gamma \vdash_{\mathrm{ps}}$ is generated by $\Gamma \vdash_{\mathrm{ps}} x : \star$, then $x \equiv \mathrm{snd}(\Gamma)$, by showing for all derivations $\Gamma \vdash_{\mathrm{ps}} x : A$, where $\dim(A) > 0$ that the 0-target of the type $A$ is $\mathrm{snd}(\Gamma)$ by induction, and then case splitting on the original derivation. Then $\Gamma \vdash_{\mathrm{ps}}$ implies that $\Gamma \vdash_{\mathrm{ps}} \mathrm{snd}(\Gamma) : \star$.

The statement of the proposition is then proven by induction on the following statement: If $\Gamma \vdash_{\mathrm{ps}}$ and $\Delta \vdash_{\mathrm{ps}} x : A$, then:

$$\Gamma \vee \Delta \vdash_{\mathrm{ps}} x[\![\mathrm{inr}_{\Gamma;\Delta}]\!] : A[\![\mathrm{inr}_{\Gamma;\Delta}]\!]$$

The base case is given by the preceding paragraph, and the other cases follow from routine calculation.

These proofs are given in Catt.Wedge.Pasting. □

We lastly give a version of the wedge sum construction for variable sets.

**Definition 3.2.10.** Let $\Gamma$ and $\Delta$ be two non-empty contexts, and let $U \subseteq \mathrm{Var}(\Gamma)$ and $V \subseteq \mathrm{Var}(\Delta)$ be variable sets. Then define:

$$U \vee V = U[\![\mathrm{inl}_{\Gamma;\Delta}]\!] \cup V[\![\mathrm{inr}_{\Gamma;\Delta}]\!]$$

to be a variable set of $\Gamma \vee \Delta$.

### 3.2.2 Tree contexts

We have now defined suspensions and wedge sums, and shown that both operations preserve ps-contexts. This allows us to define the context generated by a tree.

**Definition 3.2.11.** For a tree $T$, the context $\lfloor T \rfloor$ generated from it is defined recursively by:

$$\lfloor [\,] \rfloor = D^0 \qquad \lfloor [T_0, \dots, T_n] \rfloor = \bigvee_{i=0}^{n} \Sigma \lfloor T_i \rfloor$$

It is immediate from this definition that $\lfloor \Sigma(T) \rfloor \equiv \Sigma(\lfloor T \rfloor)$, $\lfloor S \mathbin{+\!\!+} T \rfloor \equiv \lfloor S \rfloor \vee \lfloor T \rfloor$, and that $\dim(\lfloor T \rfloor) = \mathrm{h}(T)$.

We can immediately give some examples of trees and their associated contexts. The context $D^0$ is defined to be the context associated to $[\,]$, and so as $D^{n+1} \equiv \Sigma(D^n)$, all the disc contexts can easily be recovered from trees as $D^n \equiv \lfloor \Sigma^n([\,]) \rfloor$. Each tree $\Sigma^n([\,])$ is linear and has height $n$.

Trees can also be drawn graphically as follows: For a tree $[T_0, \dots, T_n]$, first recursively draw the trees $T_i$ and lay these out in a horizontal line. Then a single point is drawn underneath these subtrees which we call the root of the tree, and a line is and drawn between the root of the tree and the root of each subtree. An example is given in Figure 3.3.

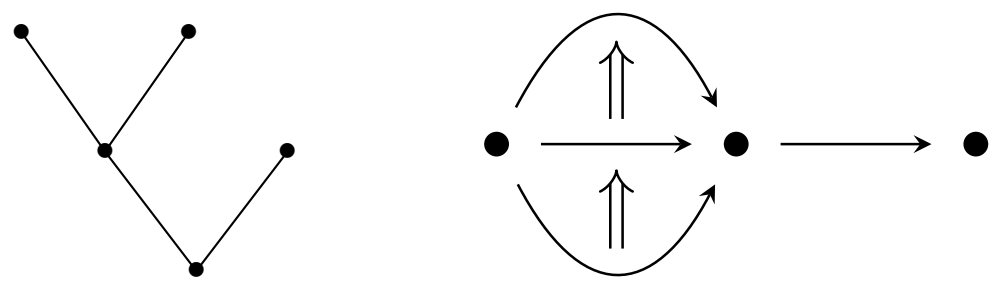

Figure 3.3: The tree $[[[\,],[\,]],[\,]]$ and generated context.

The context associated to a tree is clearly a pasting diagram, as the context is built only using the singleton context, wedge sums, and suspension. In fact, the set of contexts generated by trees is exactly the set containing the singleton context, and closed under wedge sums and suspensions. Further, it is proven in the formalisation module Catt.Dyck.FromTree that all pasting diagrams are generated by some tree, though this will not be needed for any formal development of our type theories.

We next introduces *paths*, which can be thought of as the variables in a tree.

**Definition 3.2.12.** Let $T$ be a tree. *Paths* $p : \mathsf{Path}_T$ are non-empty lists of natural numbers of the form $q \mathbin{+\!\!+} [n]$ such that $q$ indexes a subtree $T^q$ of $T$ and $0 \le n \le \mathrm{len}(T^q)$.

For path $p : \mathsf{Path}_T$, we obtain a variable of $\lfloor T \rfloor$ by recursion on $p$ as follows:

- Suppose $p = [n]$. Let $T = [T_0, \dots, T_k]$. It is clear that $\lfloor T \rfloor$ has exactly $k+2$ variables of dimension 0, corresponding to (inclusion of) the first variable of each context $\Sigma(\lfloor T_i \rfloor)$ as well as the variable corresponding to the inclusion of $\mathrm{snd}(\Sigma(T_i))$. We then define $\lfloor [n] \rfloor$ to be the $n^{\text{th}}$ such variable, indexing from 0.
- Let $p = k :: q$ and $T = [T_0, \dots, T_k, \dots]$, where $q$ is a path of $T_k$. Then by recursion we have a variable $\lfloor q \rfloor$ of $\lfloor T_k \rfloor$. This gives a variable $\Sigma(\lfloor q \rfloor)$ of $\Sigma(\lfloor T_k \rfloor)$ which can be included into $\lfloor T \rfloor$ by the appropriate inclusion to get $\lfloor p \rfloor$.

We lastly define the set of *maximal paths* $\mathsf{MaxPath}_T$ of $T$ to be paths $p \mathbin{+\!\!+} [0]$ such that $T^p = [\,]$. Such paths correspond to locally maximal variables of $\lfloor T \rfloor$.

We now turn our attention to substitutions from a tree context $\sigma : \lfloor T \rfloor \to \Gamma$. A substitution can be viewed as a function from the variables of its source context to terms of the target context. Therefore, a substitution $\sigma : \lfloor T \rfloor \to \Gamma$ acts on variables of $\lfloor T \rfloor$. However, we have seen that the more natural notion of a variable in a tree context is a path. This motivates the following definition.

**Definition 3.2.13.** A term-labelling $L : T \to \Gamma$ from a tree $T$ to context $\Gamma$ is a pair containing a function $\mathsf{Path}_T \to \mathsf{Term}_\Gamma$ and a type of $\Gamma$. To apply the function component of a labelling to a path $p$, we write $L(p)$ or $L[x_0, x_1, \dots]$ for a path $[x_0, x_1, \dots]$. The type component of the labelling is given by $\mathrm{Ty}(L)$.

If $T = [T_0, \dots, T_n]$, then there are natural projections $L_i : T_i \to \Gamma$ given by $L_i(p) = L(i :: p)$ and $\mathrm{Ty}(L) = L[i] \to_{\mathrm{Ty}(L)} L[i+1]$ for $0 \leq i \leq n$.

For labellings to play the role of substitutions, a substitution $\lfloor L \rfloor : \lfloor T \rfloor \to_{\mathrm{Ty}(L)} \Gamma$ will be defined for each term-labelling $L : T \to \Gamma$. A natural way to define this substitution is by induction on the tree $T$, which motivates the use of extended substitutions. Suppose we start with a labelling $L : [T_0, \dots, T_n] \to \Gamma$. To proceed, we will apply the inductive hypothesis to obtain the substitutions:

$$\lfloor L_i \rfloor : \lfloor T_i \rfloor \to_{L[i] \to_{\mathrm{Ty}(L)} L[i+1]} \Gamma$$

These substitutions are not regular (non-extended) substitutions, even if $L$ has associated type $\star$, and so corresponds to a regular substitution.

**Definition 3.2.14.** Let $L : T \to \Gamma$ be a term-labelling. We define the substitution:

$$\lfloor L \rfloor : \lfloor T \rfloor \to_{\mathrm{Ty}(L)} \Gamma$$

by induction on the tree $T$ as $\langle \mathrm{Ty}(L), L[0] \rangle$ if $T = [\,]$ and:

$$\downarrow \lfloor L_0 \rfloor \vee \downarrow \lfloor L_1 \rfloor \vee \cdots \vee \downarrow \lfloor L_n \rfloor$$

if $T = [T_0, \dots, T_n]$. Although it looks like the 0-dimensional terms in the labelling are not used to generate the substitution, they appear in the types of the labellings $L_i$, and so appear in the unrestricted substitutions.

There are many ways of giving a more syntactic presentation of labellings. Given a tree $T = [T_0, \dots, T_n]$, a labelling $L : T \to \Gamma$ can be written as:

$$t_0\{L_0\}t_1\{L_1\}t_2 \cdots t_n\{L_n\}t_{n+1} : \mathrm{Ty}(L)$$

where each $t_i$ is the term $L[i]$ and the sublabellings $L_i$ have been recursively put in this syntactic bracketing format (omitting the type). The syntactic presentation contains all the information of the original labelling, which can be recovered by letting $L[i] = t_i$ for each $i$, $L[i :: p] = L_i(p)$.

As an example, take the tree $T = [[[\,],[\,]],[\,]]$ from Figure 3.3, and let:

$$\Gamma = (x : \star), (f : x \to x), (\alpha : f * f \to f)$$

Then we can define the labelling $L : T \to \Gamma$ by:

$$L = x\big\{f * f\{\alpha\}f\{\mathsf{id}(f)\}f\big\}x\{f\}x : \star$$

which sends the (maximal) paths $[0,0,0]$ to $\alpha$, $[0,1,0]$ to $\mathsf{id}(f)$, and $[1,0]$ to $f$, and has associated substitution:

$$\lfloor L \rfloor = \langle x, x, f * f, f, \alpha, f, \mathsf{id}(f), x, f \rangle$$

The curly brackets notation for labellings is used instead of a typical round bracket notation to avoid clashes with notations that already use round brackets, such as $\mathsf{id}(f)$.

We finish this section by examining a boundary operation for trees. We have already seen that for every ps-context $\Gamma$ and $n \in \mathbb{N}$, there are the boundary variable sets:

$$\partial_n^-(\Gamma) \qquad \partial_n^+(\Gamma)$$

Since $\lfloor T \rfloor$ is a ps-context for any tree $T$, we immediately obtain such boundary variable sets for $\lfloor T \rfloor$. However, by recalling the definitions for the wedge sum of variable sets given in Section 3.2.1 and the suspension of a variable set given in Section 2.3.2, a more natural definition can be given.

**Definition 3.2.15.** For any tree $T$ : Tree, dimension $n \in \mathbb{N}$, and $\epsilon \in \{-,+\}$, we define the boundary set:

$$\partial_n^\epsilon(T)$$

by induction on $n$ and $T$. If $n = 0$, then we define:

$$\partial_0^-(T) = \mathrm{FV}(\mathrm{fst}(\lfloor T \rfloor)) \qquad \partial_0^+(T) = \mathrm{FV}(\mathrm{snd}(\lfloor T \rfloor))$$

Now suppose $n$ is not 0. If the tree $T$ is the singleton tree, then $\partial_n^\epsilon(T) = \mathrm{Var}(\lfloor T \rfloor)$. Now suppose that $T = [T_0, \dots, T_n]$. We then define:

$$\partial_n^\epsilon(T) = \partial_{n-1}^\epsilon(T_0) \vee \cdots \vee \partial_{n-1}^\epsilon(T_n)$$

with the boundary sets $\partial_n^\epsilon(T_i)$ obtained by inductive hypothesis.

In the formalisation module Catt.Tree.Support, we prove that the boundary sets $\partial_n^\epsilon(T)$, the tree boundary, and $\partial_n^\epsilon(\lfloor T \rfloor)$, the ps-context boundary, coincide. Therefore:

$$(\lfloor S \rfloor, \partial_n^-(T), \partial_n^+(T)) \in \mathsf{Std}$$

for each $n \geq \mathrm{h}(S) - 1$.

## 3.3 Structured syntax

We now introduce a new class of syntax named *structured syntax*. Terms over tree contexts are commonly built using several of the standard constructions we have seen so far, such as paths, labellings, suspensions, and inclusions. By recording which of these constructions was used in the formation of a term, these terms can compute more usefully, which we will exploit to prove more involved lemmas about insertion in Section 3.4. Structured syntax will be our variation on the base syntax of Catt which records these constructions.

The key problem with the base syntax for Catt is that term-labellings are difficult to compose. We have so far considered term-labellings of the form $L : T \to \Gamma$, where $\Gamma$ is any arbitrary context, but there is no reason a labelling couldn't be of the form $M : S \to \lfloor T \rfloor$ for trees $S$ and $T$. We would then hope to be able to compose these labellings to get a labelling of the form:

$$M \bullet L : S \to \Gamma$$

Such a labelling would need to send a path $p : \mathsf{Path}_S$ to a term of $\Gamma$. The only reasonable way forward is to apply $M$ to $p$ to get a term of $\lfloor T \rfloor$, and then apply $\lfloor L \rfloor$ to this term to get a term of $\Gamma$. Unfortunately, for an arbitrary term $t : \mathsf{Term}_{\lfloor T \rfloor}$ and labelling $L : T \to \Gamma$, the term:

$$t[\![\lfloor L \rfloor]\!]$$

does not have nice computational properties. We examine two examples:

- Suppose $t$ was of the form $\lfloor p \rfloor$ for some path $p$. We then have:

  $$\lfloor p \rfloor[\![\lfloor L \rfloor]\!] \equiv L(p)$$

  and would hope that this syntactic equality would fall out immediately, and that the left-hand side would reduce to the right-hand side in the formalisation. This is not the case however, and proving that such a syntactic equality holds is non-trivial.

- Suppose $t \equiv \Sigma(s)$ and $L = a\{L_1\}b : A$. Similar to the above case we would hope that the syntactic equality:

  $$\Sigma(s)[\![\lfloor a\{L_1\}b : A \rfloor]\!] \equiv s[\![\lfloor L_1 \rfloor]\!]$$

  holds "on the nose". This however is not the case.

Structured terms alleviate these problems by recording that such a term $t$ was generated from a path or generated using suspension. This allows the application of a labelling to a structured term to use this information, for example letting the two syntactic equalities above to hold by definition. If a labelling is the "correct" notion of substitution from a tree, then a structured term is the "correct" notion of term in a tree.

**Definition 3.3.1.** Let $\mathbf{U}$ be a member of $\mathsf{Ctx} \uplus \mathsf{Tree}$, either some context $\Gamma$ or some tree $T$. We then define the *structured syntax* classes $\mathsf{STerm}_{\mathbf{U}}$ of *structured terms*, $\mathsf{SType}_{\mathbf{U}}$ of *structured types*, and *(*STerm*-)labellings* $S \to \mathbf{U}$ for some tree $S$. The syntax classes for structured terms and types are generated by the following rules:

$$\frac{p : \mathsf{Path}_T}{\mathsf{SPath}(p) : \mathsf{STerm}_T} \qquad \frac{s : \mathsf{STerm}_{T_i} \qquad 0 \leq i \leq n}{\mathsf{Inc}_i(s) : \mathsf{STerm}_{[T_0,\dots,T_n]}}$$

$$\frac{S : \mathsf{Tree} \qquad A : \mathsf{SType}_S \qquad L : S \to \mathbf{U}}{\mathsf{SCoh}_{(S\,;\,A)}[L] : \mathsf{STerm}_{\mathbf{U}}} \qquad \frac{t : \mathsf{Term}_\Gamma}{\mathsf{SOther}(t) : \mathsf{STerm}_\Gamma}$$

$$\frac{}{\star : \mathsf{SType}_{\mathbf{U}}} \qquad \frac{s : \mathsf{STerm}_{\mathbf{U}} \qquad A : \mathsf{SType}_{\mathbf{U}} \qquad t : \mathsf{STerm}_{\mathbf{U}}}{s \to_A t : \mathsf{SType}_{\mathbf{U}}}$$

Labellings $L : S \to \mathbf{U}$ are defined as pairs of a function $\mathsf{Path}_S \to \mathsf{STerm}_{\mathbf{U}}$ and structured type, similarly to term-labellings in Section 3.2.2. We note that the syntax for structured types is shared with the syntax for Catt types, and will be careful to make it clear which syntax we are using when necessary.

Each piece of structured syntax can be converted back into the base syntax of CATT, using many of the constructions already introduced.

**Definition 3.3.2.** Suppose $\mathbf{U} : \mathsf{Ctx} \uplus \mathsf{Tree}$. Define $\lfloor \mathbf{U} \rfloor$ to be $\Gamma$ if $\mathbf{U} = \Gamma$ for some context $\Gamma$ or $\lfloor T \rfloor$ if $\mathbf{U} = T$ for some tree $T$. Now, for a structured term $s : \mathsf{STerm}_{\mathbf{U}}$, a structured type $A : \mathsf{SType}_{\mathbf{U}}$, or a labelling $L : S \to \mathbf{U}$, we define:

$$\lfloor s \rfloor : \mathsf{Term}_{\lfloor \mathbf{U} \rfloor} \qquad \lfloor A \rfloor : \mathsf{Type}_{\lfloor \mathbf{U} \rfloor} \qquad \lfloor L \rfloor : \lfloor S \rfloor \to_{\lfloor \mathrm{Ty}(L) \rfloor} \lfloor \mathbf{U} \rfloor$$

by the equations:

$$\begin{aligned} \lfloor \mathsf{SPath}(p) \rfloor &= \lfloor p \rfloor \\ \lfloor \mathsf{Inc}_i(s) \rfloor &= \Sigma(\lfloor s \rfloor)[\![\mathsf{inc}_i]\!] \\ \lfloor \mathsf{SCoh}_{(S\,;\,A)}[L] \rfloor &= \mathsf{Coh}_{(\lfloor S \rfloor\,;\,\lfloor A \rfloor)}[\mathsf{id}_{\lfloor S \rfloor}][\![\lfloor L \rfloor]\!] \\ \lfloor \mathsf{SOther}(t) \rfloor &= t \end{aligned}$$

$$\begin{aligned} \lfloor \star \rfloor &= \star \\ \lfloor s \to_A t \rfloor &= \lfloor s \rfloor \to_{\lfloor A \rfloor} \lfloor t \rfloor \end{aligned}$$

defining $\mathrm{inc}_i$ be the unique (by Proposition 3.2.7) inclusion:

$$\mathrm{inc}_i : \Sigma\lfloor T_i \rfloor \to \lfloor T \rfloor$$

Further define $\lfloor L \rfloor$ similarly to term labellings except $\lfloor L \rfloor = \langle \lfloor \mathrm{Ty}(L) \rfloor, \lfloor L[0] \rfloor \rangle$ for labellings $L : [\,] \to \mathbf{U}$ from the singleton tree. We refer to $\lfloor a \rfloor$, $\lfloor A \rfloor$ and $\lfloor L \rfloor$ as the term, type, or substitution generated by $a$,$A$, or $L$.

For any tree $T$, there is an *identity labelling* $\mathsf{id}_T$ given by:

$$\mathsf{id}_T(p) = \mathsf{SPath}(p) \qquad \mathrm{Ty}(\mathsf{id}_T) = \star$$

The function id-label-to-sub in the formalisation (see Catt.Tree.Structured.Properties) shows that:

$$\lfloor \mathsf{id}_T \rfloor = \mathsf{id}_{\lfloor T \rfloor}$$

The main motivation for introducing structured syntax was to be able to define a composition of labellings, which we do now by defining the application of a labelling to a structured term, structured type, or another labelling.

**Definition 3.3.3.** Let $L : T \to \mathbf{U}$ be a labelling (with $\mathbf{U} : \mathsf{Ctx} \uplus \mathsf{Tree}$). We define the application of $L$ to a structured term $s : \mathsf{STerm}_T$, a structured type $A : \mathsf{SType}_T$, and a labelling $M : S \to T$ to give:

$$s[\![L]\!] : \mathsf{STerm}_{\mathbf{U}} \qquad A[\![L]\!] : \mathsf{SType}_{\mathbf{U}} \qquad M \bullet L : S \to \mathbf{U}$$

These definitions are given by mutual recursion:

$$\begin{aligned}
\mathsf{SPath}(p)[\![L]\!] &= L(p)\\
\mathsf{Inc}_i(s)[\![L]\!] &= s[\![L_i]\!]\\
\mathsf{SCoh}_{(S\,;\,A)}[M][\![L]\!] &= \mathsf{SCoh}_{(S\,;\,A)}[M \bullet L]\\
\mathsf{SOther}(t)[\![L]\!] &= t[\![\lfloor L \rfloor]\!]
\end{aligned}$$

$$\begin{aligned}
\star[\![L]\!] &= B\\
(s \to_A t)[\![L]\!] &= s[\![L]\!] \to_{A[\![L]\!]} t[\![L]\!]
\end{aligned}$$

$$\begin{aligned}
(M \bullet L)(p) &= M(p)[\![L]\!]\\
\mathsf{Ty}(M \bullet L) &= \mathsf{Ty}(M)[\![L]\!]
\end{aligned}$$

It can easily be seen that these definitions satisfy the computational properties given at the start of the section.

The main theorem of this section is that the application of a labelling to a structured term is compatible with the map from structured syntax to Catt syntax.

**Theorem 3.3.4.** *For any labelling $L : T \to \mathbf{U}$ and structured term $s : \mathsf{STerm}_T$, structured type $A : \mathsf{SType}_T$, or labelling $M : S \to T$, we have:*

$$\lfloor s[\![L]\!] \rfloor \equiv \lfloor s \rfloor[\![\lfloor L \rfloor]\!] \qquad \lfloor A[\![L]\!] \rfloor \equiv \lfloor A \rfloor[\![\lfloor L \rfloor]\!] \qquad \lfloor M \bullet L \rfloor \equiv \lfloor M \rfloor \bullet \lfloor L \rfloor$$

*Proof.* We proceed by proving all statements by mutual induction. Suppose $s : \mathsf{STerm}_T$ is a structured term. We split on the form of $s$:

- Suppose $s$ is of the form $\mathsf{SCoh}_{(S\,;\,A)}[M]$. Then $s[\![L]\!]$ is $\mathsf{SCoh}_{(S\,;\,A)}[M \bullet L]$ and so the required statement follows from the inductive hypothesis for labellings.
- Suppose $s$ is of the form $\mathsf{SOther}(t)$. Then $\lfloor s[\![L]\!] \rfloor \equiv \lfloor \mathsf{SOther}(t[\![\lfloor L \rfloor]\!]) \rfloor \equiv t[\![\lfloor L \rfloor]\!] \equiv \lfloor s \rfloor[\![\lfloor L \rfloor]\!]$.
- Suppose $T = [T_0, \dots, T_n]$ and $s$ is of the form $\mathsf{Inc}_i(t)$. Then:

$$\begin{aligned}
\lfloor \mathsf{Inc}_i(t) \rfloor[\![\lfloor L \rfloor]\!] &\equiv \Sigma(t)[\![\mathsf{inc}_i]\!][\![{\downarrow}\lfloor L_0 \rfloor \vee \cdots \vee {\downarrow}\lfloor L_n \rfloor]\!]\\
&\equiv \Sigma(t)[\![\mathsf{inc}_i \bullet ({\downarrow}\lfloor L_0 \rfloor \vee \cdots \vee {\downarrow}\lfloor L_n \rfloor)]\!]\\
&\equiv \Sigma(t)[\![{\downarrow}\lfloor L_i \rfloor]\!] && \text{by Lemma 3.2.6}\\
&\equiv \lfloor t \rfloor[\![\lfloor L_i \rfloor]\!]\\
&\equiv \lfloor t[\![L_i]\!] \rfloor && \text{by inductive hypothesis}\\
&\equiv \lfloor \mathsf{Inc}_i(t)[\![L]\!] \rfloor
\end{aligned}$$

- Suppose $s$ is of the form $\mathsf{SPath}(p)$. Then if $\lfloor p \rfloor$ is not a $0$-dimensional variable, then an argument similar to the preceding case can be made. If instead $\lfloor p \rfloor$ is of the form $[k]$ and $T = [T_0, \dots, T_n]$ then first suppose that $k < n + 1$ such that

$\lfloor [k] \rfloor \equiv \mathsf{fst}(\lfloor T_k \rfloor)[\![\mathrm{inc}_k]\!]$. Then:

$$\begin{aligned}\lfloor [k] \rfloor \lfloor L \rfloor &\equiv \mathsf{fst}(\lfloor T_k \rfloor)[\![\mathrm{inc}_k]\!][\![\downarrow\lfloor L_0 \rfloor \vee \cdots \vee \downarrow\lfloor L_n \rfloor \rfloor]\!] \\ &\equiv \mathsf{fst}(\lfloor T_k \rfloor)[\![\downarrow\lfloor L_k \rfloor]\!] \\ &= \lfloor L[k] \rfloor\end{aligned}$$

where the last equality follows from the labelling $L_k$ having type component $\mathsf{Ty}(L_k) \equiv \lfloor L[k] \rfloor \to_B \lfloor L[k+1] \rfloor$. The case where $k = n+1$ is similar to above using $\mathsf{snd}(T_n)$ instead of $\mathsf{fst}(T_k)$ (as there is no tree $T_k$ in this case).

The case for structured types follows by a simple induction using the case for terms. We now consider the case for a label $M : S \to T$. Suppose $S = [S_0, \dots, S_n]$. Then:

$$\begin{aligned}\lfloor M \rfloor \bullet \lfloor L \rfloor &\equiv \left(\bigvee_i \downarrow\lfloor M_i \rfloor\right) \bullet \lfloor L \rfloor \\ &\equiv \bigvee_i \downarrow\lfloor M_i \rfloor \bullet \lfloor L \rfloor && \text{by Lemma 3.2.6} \\ &\equiv \bigvee_i \downarrow(\lfloor M_i \rfloor \bullet \lfloor L \rfloor) \\ &\equiv \bigvee_i \downarrow\lfloor M_i \bullet L \rfloor && \text{by inductive hypothesis} \\ &\equiv \lfloor M \bullet L \rfloor\end{aligned}$$

with the last line following from $(M \bullet L)_i$ and $M_i \bullet L$ being the same labelling. This concludes all cases. □

Structured syntax is only used as computational aid for reasoning about the base syntax of Catt, and therefore the desired notion of "syntactic" equality of structured syntax is syntactic equality of the underlying Catt terms, that is we say $s \equiv t$ for structured terms $s$ and $t$ exactly when $\lfloor s \rfloor \equiv \lfloor t \rfloor$. On labellings $L, M : T \to \mathbf{U}$ we can instead provide the equality:

$$L \equiv M \iff \mathsf{Ty}(L) \equiv \mathsf{Ty}(M) \wedge \forall (p : \mathsf{Path}_T).\ L(p) \equiv M(p)$$

and by observing the proof of Theorem 3.3.4, we see that this equality implies equality of the generated substitutions.

It is therefore possible to derive many properties for this equality of structured terms simply by reducing all constructions used to the corresponding Catt constructions, and using the corresponding result for the syntax of Catt.

**Proposition 3.3.5.** *Composition of labellings is associative and has a left and right unit given by the identity labelling.*

*Proof.* Follows immediately from Theorem 3.3.4, the identity labelling generating the identity substitution, and the corresponding results for Catt. □

Using this technique, every syntactic result about Catt can be transported to structured syntax. Further, it is easy to prove that the equality relation is preserved by each constructor, for example if $L \equiv M$ and $A \equiv B$, then $\mathsf{SCoh}_{(S\,;\,A)}[L] \equiv \mathsf{SCoh}_{(A\,;\,B)}[M]$.

To extend this, we redefine some constructions we have seen for CATT in the previous sections, this time for structured terms.

**Definition 3.3.6.** We define the suspension for a structured term $a : \mathsf{STerm}_{\mathbf{U}}$, structured type $A : \mathsf{STerm}_{\mathbf{U}}$, and restricted substitution for a labelling $L : T \to \mathbf{U}$, giving structured term $\Sigma(a) : \mathsf{STerm}_{\Sigma(\mathbf{U})}$, structured type $\Sigma(A) : \mathsf{STerm}_{\Sigma(\mathbf{U})}$, and labelling $\Sigma'(L) : T \to \Sigma(\mathbf{U})$. These are all defined by mutual induction as follows:

$$\begin{aligned}
\Sigma(a) &\equiv \mathsf{Inc}_0(a) && \text{if } \mathbf{U} \text{ is a tree}\\
\Sigma(\mathsf{SCoh}_{(S\,;\,A)}[M]) &\equiv \mathsf{SCoh}_{(S\,;\,A)}[\Sigma'(M)] && \text{if } \mathbf{U} \text{ is a context}\\
\Sigma(\mathsf{SOther}(t)) &\equiv \mathsf{SOther}(\Sigma(t))
\end{aligned}$$

$$\begin{aligned}
\Sigma(\star) &= N \to_\star S && \text{if } \mathbf{U} \text{ is a context}\\
\Sigma(\star) &= \mathsf{SPath}[0] \to_\star \mathsf{SPath}[1] && \text{otherwise}\\
\Sigma(s \to_A t) &= \Sigma(s) \to_{\Sigma(A)} \Sigma(t)
\end{aligned}$$

$$\begin{aligned}
\Sigma'(L)(p) &= \Sigma(L(p))\\
\mathsf{Ty}(\Sigma'(L)) &= \Sigma(\mathsf{Ty}(L))
\end{aligned}$$

We further define an unrestriction operation that takes a labelling of the form $M : T \to \mathbf{U}$ with $\mathsf{Ty}(M) \equiv s \to_A t$ and produces a labelling

$$\downarrow M : \Sigma(T) \to \mathbf{U} \equiv s\{M\}t : A$$

This can be used to define the full suspension of a labelling as with CATT substitutions by defining $\Sigma(L)$ to be $\downarrow \Sigma'(L)$.

A simple case analysis demonstrates that these constructions commute with $\lfloor _ \rfloor$. They therefore inherit the properties of the suspension on CATT terms, types, and substitutions. We lastly recover wedge sums for structured syntax.

**Definition 3.3.7.** We have seen that the wedge sum of trees $S$ and $T$ is given by $S \mathbin{+\!\!+} T$. Letting $S = [S_0, \dots, S_m]$ and $T = [T_0, \dots, T_n]$, we further define inclusion labellings:

$$\mathrm{inl}_{S;T} : S \to S \mathbin{+\!\!+} T \qquad \mathrm{inr}_{S;T} : T \to S \mathbin{+\!\!+} T$$

by the equations:

$$\begin{aligned}
&\mathrm{inl}_{S;T}([k]) \equiv \mathsf{SPath}[k] && \mathrm{inl}_{S;T}(k :: p) \equiv \mathsf{SPath}(k :: p) && \mathsf{Ty}(\mathrm{inl}_{S;T}) \equiv \star\\
&\mathrm{inr}_{S;T}([k]) \equiv \mathsf{SPath}[m+k] && \mathrm{inr}_{S;T}(k :: p) \equiv \mathsf{SPath}(m+k :: p) && \mathsf{Ty}(\mathrm{inr}_{S;T}) \equiv \star
\end{aligned}$$

and finally, we suppose $L : S \to \mathbf{U}$ and $M : T \to \mathbf{U}$ are labellings of the form:

$$L \equiv s_0\{L_0\}s_1 \cdots s_n\{L_n\}t_0 : A \qquad M \equiv t_0\{M_0\}t_1 \cdots t_n\{M_n\}t_{n+1} : A$$

and define their concatenation to be the labelling:

$$L \mathbin{+\!\!+} M \equiv s_0\{L_0\}s_1 \cdots s_n\{L_n\}t_0\{M_0\}t_1 \cdots t_n\{M_n\}t_{n+1} : A$$

where $L \mathbin{+\!\!+} M : S \mathbin{+\!\!+} T \to \mathbf{U}$.

Many properties of these constructions among others are given in the formalisation module Catt.Tree.Structured.Construct.Properties. In particular, the diagrammatic notation for substitutions between wedge sums can be repurposed to define labellings, which will be used to define certain labellings in Section 3.4.

It will be useful to be able to interpret all Catt syntax as structured syntax. For terms such a mapping is trivially given by the SOther constructor. For a type $A$, a structured type $\lceil A \rceil$ can be formed by a simple induction, applying the SOther constructor to each term in the type. For substitutions, we give the following definition.

**Definition 3.3.8.** Let $\sigma : \lfloor S \rfloor \to_A \Gamma$ be a substitution. We then define the labelling:

$$\lceil \sigma \rceil : S \to \Gamma$$

by $\lceil \sigma \rceil(p) = \mathsf{SOther}(\lfloor p \rfloor [\![ \sigma ]\!])$ and $\mathsf{Ty}(\lceil \sigma \rceil) = \lceil A \rceil$.

This construction is an inverse to taking generating a substitution from a labelling.

**Proposition 3.3.9.** *Let $\sigma : \lfloor S \rfloor \to_A \Gamma$ be a substitution. Then $\lfloor \lceil \sigma \rceil \rfloor \equiv \sigma$. Further, for any labelling $L : S \to \Gamma$, $\lceil \lfloor L \rfloor \rceil \equiv L$.*

*Proof.* We note that every variable of $\lfloor S \rfloor$ is given by $\lfloor p \rfloor$ for some path $p$. We then have the equality:

$$\lfloor p \rfloor [\![ \lfloor \lceil \sigma \rceil \rfloor ]\!] \equiv \lfloor p [\![ \lceil \sigma \rceil ]\!] \rfloor \equiv \lfloor \mathsf{SOther}(\lfloor p \rfloor [\![ \sigma ]\!]) \rfloor \equiv \lfloor p \rfloor [\![ \sigma ]\!]$$

and so $\sigma$ and $\lfloor \lceil \sigma \rceil \rfloor$ have the same action on each variable and so are equal.

Letting $L : S \to \Gamma$ be a labelling. Then for any path $p$:

$$\lceil \lfloor L \rfloor \rceil(p) \equiv \mathsf{SOther}(\lfloor p \rfloor [\![ \lfloor L \rfloor ]\!]) \equiv \mathsf{SOther}(\lfloor L(p) \rfloor)$$

and so $\lfloor \lceil \lfloor L \rfloor \rceil(p) \rfloor \equiv \lfloor L(p) \rfloor$. Therefore, $L \equiv \lceil \lfloor L \rfloor \rceil$ by definition. □

### 3.3.1 Typing and equality

Similarly to the definition of syntactic equality for structured syntax, we also want the equality rules for structured terms and structured types to be inherited from the equality relations on their generated terms, and so define:

$$\mathbf{U} \vdash s = t \iff \lfloor \mathbf{U} \rfloor \vdash \lfloor s \rfloor = \lfloor t \rfloor \qquad \mathbf{U} \vdash A = B \iff \lfloor \mathbf{U} \rfloor \vdash \lfloor A \rfloor = \lfloor B \rfloor$$

For labellings, (definitional) equality can be defined similarly to the syntactic equality relation:

$$\mathbf{U} \vdash L = M \iff \mathbf{U} \vdash \mathsf{Ty}(L) = \mathsf{Ty}(M) \wedge \forall (p : \mathsf{Path}_T).\, \mathbf{U} \vdash L(p) = M(p)$$

Using Lemma 3.2.8, it can be proven by a simple induction that equality of labellings (along with equality of their associated types) induces equality of the generated substitutions.

We also want the typing rules for $s : \mathsf{STerm}_{\mathbf{U}}$ and $A : \mathsf{SType}_{\mathbf{U}}$ to be inherited from the typing rules for $\lfloor s \rfloor$ and $\lfloor A \rfloor$. We re-use the notation for each typing judgement. For labellings, we introduce the following more natural typing judgement:

**Definition 3.3.10.** For a labelling $L : T \to \mathbf{U}$, where $\mathbf{U} : \mathsf{Ctx} \uplus \mathsf{Tree}$, we define the judgement:

$$\mathbf{U} \vdash L : T$$

to mean that the labelling $L$ is well-formed. This judgement is generated by the following rule:

$$\frac{\mathbf{U} \vdash L[0] : \mathsf{Ty}(L) \quad \cdots \quad \mathbf{U} \vdash L[n+1] : \mathsf{Ty}(L) \qquad \mathbf{U} \vdash L_0 : T_0 \quad \cdots \quad \mathbf{U} \vdash L_n : T_n}{\mathbf{U} \vdash L : [T_0, \dots, T_n]}$$

Paths $p$ can be equipped with a canonical structured type, $\mathsf{Ty}(p)$, as follows:

- For paths $[k]$, $\mathsf{Ty}([k]) = \star$,
- For paths $k :: p$ where $p$ is a path, the type $\mathsf{Ty}(k :: p)$ is obtained by taking the type $\mathsf{Ty}(p)$, applying $\mathsf{Inc}_k$ to each term, and replacing the $\star$ type at its base by the type $\mathsf{SPath}[k] \to_\star \mathsf{SPath}[k+1]$.

This can be used to prove that the identity labelling is well-formed.

**Proposition 3.3.11.** *Let $S$ be a tree. Then $S \vdash \mathsf{id}_S : S$.*

*Proof.* Let $x$ be a list that indexes a subtree of $S$, and define the labelling $\mathsf{subtree}(x) : S^x \to x$ by $\mathsf{Ty}(\mathsf{subtree}(x)) = \mathsf{Ty}(x +\!\!+ [0])$ and $\mathsf{subtree}(x)(p) = \mathsf{SPath}(x +\!\!+ p)$.

We then prove the more general result that $S \vdash \mathsf{subtree}(x) : S^x$ for each $x$, with the desired result following from the case $x = [\,]$. If $S^x = [\,]$, then the typing judgement follows from $S \vdash S^x[0] : \mathsf{Ty}(S^x[0])$.

If $S^x = [T_0, \dots, T_n]$ then we must show that $S \vdash S^x[k] : \mathsf{Ty}(S^x[0])$, which follows from the observation that $\mathsf{Ty}(S^x[0]) \equiv \mathsf{Ty}(S^x[i])$ for any $i$ as the definition does not use the last element of the path. We are also required to show that $S \vdash S^x_i : T_i$, but $T_i \equiv S^{x +\!\!+ [i]}$ and $S^x_i \equiv S^{x +\!\!+ [i]}$, and so this follows from inductive hypothesis. □

From this typing judgement for labellings, one can obtain a derivation of the typing judgement for the generated substitution.

**Proposition 3.3.12.** *Let $L : T \to \mathbf{U}$, and suppose $\mathbf{U} \vdash L : T$ and $\mathbf{U} \vdash \mathsf{Ty}(L)$. Then:*

$$\lfloor \mathbf{U} \rfloor \vdash \lfloor L \rfloor : \lfloor T \rfloor$$

*Proof.* We induct on the tree $T$, splitting into cases on whether it is empty. If it is, then by case analysis on the judgement for label typing we get:

$$\mathbf{U} \vdash L[0] : \mathsf{Ty}(L)$$

Then, $\lfloor L \rfloor \equiv \langle \lfloor A \rfloor, \lfloor L[0] \rfloor \rangle$, and so the following derivation can be obtained:

$$\dfrac{\dfrac{\dfrac{\dfrac{}{\mathbf{U} \vdash A}}{\lfloor \mathbf{U} \rfloor \vdash \lfloor A \rfloor}}{\lfloor \mathbf{U} \rfloor \vdash \langle \lfloor A \rfloor \rangle : \emptyset} \qquad \dfrac{\dfrac{}{\mathbf{U} \vdash L[0] : A}}{\lfloor \mathbf{U} \rfloor \vdash \lfloor L[0] \rfloor : \lfloor A \rfloor}}{\lfloor \mathbf{U} \rfloor \vdash \langle \lfloor A \rfloor, \lfloor L[0] \rfloor \rangle : \lfloor [\,] \rfloor}$$

Suppose instead that $T = [T_0, \dots, T_n]$, such that:

$$\lfloor L \rfloor \equiv {\downarrow}\lfloor L_0 \rfloor \vee \cdots \vee {\downarrow}\lfloor L_n \rfloor$$

From $\mathbf{U} \vdash L : T$, we obtain $\mathbf{U} \vdash L_i : T_i$ for each $i \in \{0, \dots, n\}$. We further obtain $\mathbf{U} \vdash L[k] : \mathrm{Ty}(L)$ for $0 \leq k \leq n+1$ and so:

$$\mathrm{Ty}(L_i) \equiv L[i] \to_{\mathrm{Ty}(L)} L[i+1]$$

is well-formed and so by inductive hypothesis we have $\lfloor \mathbf{U} \rfloor \vdash \lfloor L_i \rfloor : \lfloor T_i \rfloor$. We have for each $i$ that $\lfloor \mathrm{Ty}(L) \rfloor$ is not the type $\star$ and so the unrestriction ${\downarrow}\lfloor L_i \rfloor$ is well-formed. Furthermore, by construction of the unrestriction we have:

$$\mathrm{fst}(\lfloor T_i \rfloor)[\![\lfloor L_i \rfloor]\!] \equiv \lfloor L[i] \rfloor \qquad \mathrm{snd}(\lfloor T_i \rfloor)[\![\lfloor L_i \rfloor]\!] \equiv \lfloor L[i+1] \rfloor$$

and so by Lemma 3.2.8, the wedge sums are well-formed, completing the proof. □

It can be shown that the reverse implication also holds: if $\lfloor \mathbf{U} \rfloor \vdash \lfloor L \rfloor : \lfloor T \rfloor$ then $\mathbf{U} \vdash L : T$. This follows as a corollary from the following proposition.

**Proposition 3.3.13.** *Let $\sigma : \lfloor T \rfloor \to_A \Gamma$ be a substitution with $\Gamma \vdash \sigma : \lfloor S \rfloor$. Then for any $L : S \to T$ we have:*

$$T \vdash L : S \implies \Gamma \vdash L \bullet \lceil \sigma \rceil : S$$

*and hence $\Gamma \vdash \lceil \sigma \rceil : T$ follows from letting $L$ be the identity labelling.*

*Proof.* Let $S = [S_0, \dots, S_n]$ (where we allow this list to be empty). By the definition of the typing for a labelling, it suffices to show that for each $0 \leq i \leq n$ and $0 \leq k \leq n+1$ that:

$$S \vdash L[k] \bullet \lceil \sigma \rceil : \mathrm{Ty}(L)[\![\lceil \sigma \rceil]\!] \qquad S \vdash (L \bullet \lceil \sigma \rceil)_i : S_i$$

The second typing judgement follows directly from inductive hypothesis, as $(L \bullet \lceil \sigma \rceil)_i \equiv L_i \bullet \lceil \sigma \rceil$. By definition of typing for structured terms, the first judgement requires us to prove that:

$$\lfloor S \rfloor \vdash \lfloor L[k] \bullet \lceil \sigma \rceil \rfloor : \lfloor \mathrm{Ty}(L)[\![\lceil \sigma \rceil]\!] \rfloor$$

which is equivalent to:

$$\lfloor S \rfloor \vdash \lfloor L[k] \rfloor [\![\sigma]\!] : \lfloor \mathrm{Ty}(L) \rfloor [\![\sigma]\!]$$

and so follows from typing being preserved by substitution. □

By these results, many of the properties enjoyed by the typing judgements in $\textsc{Catt}_{\mathcal{R}}$ with a tame rule set $\mathcal{R}$ also apply to the typing judgements for structured terms.

The module Catt.Tree.Structured.Typing.Properties also introduces many functions for constructing the typing judgements for structured syntax. One such function is TySCoh, which represents the admissibility of the following rule:

$$\frac{S \vdash s \to_A t \qquad \mathbf{U} \vdash L : S \qquad \mathbf{U} \vdash \mathrm{Ty}(L) \qquad (\lfloor S \rfloor, \mathrm{Supp}(s), \mathrm{Supp}(t)) \in \mathcal{O}}{\mathbf{U} \vdash \mathsf{SCoh}_{(S\,;\,s \to_A t)}[L]} \tag{3.3.14}$$

In keeping with the theme of this section, one could define $\mathrm{Supp}(s)$ as $\mathrm{Supp}(\lfloor s \rfloor)$ for a structured term $s : \mathsf{STerm}_{\mathbf{U}}$. However, we choose not to do this, instead giving a definition of support for structured syntax that leverages the extra information available in the syntax.

**Definition 3.3.15.** For a path $p : \mathsf{Path}_T$, a structured term $s : \mathsf{STerm}_{\mathbf{U}}$, a structured type $A : \mathsf{SType}_{\mathbf{U}}$, and a labelling $L : S \to \mathbf{U}$, we define their supports $\mathrm{Supp}(p)$, $\mathrm{Supp}(s)$, $\mathrm{Supp}(A)$, and $\mathrm{Supp}(L)$ by mutual recursion:

$$\begin{aligned} \mathrm{Supp}([n]) &= \{\lfloor [0] \rfloor\} \\ \mathrm{Supp}(k :: p) &= \Sigma(\mathrm{Supp}(p))[\![\mathrm{inc}_k]\!] && \text{where } T = [T_1, \dots, T_n] \end{aligned}$$

$$\begin{aligned} \mathrm{Supp}(\mathsf{SPath}(p)) &= \mathrm{Supp}(p) \\ \mathrm{Supp}(\mathsf{Inc}_i(s)) &= \Sigma(\mathrm{Supp}(s))[\![\mathrm{inc}_i]\!] && \text{where } T = [T_1, \dots, T_n] \\ \mathrm{Supp}(\mathsf{SCoh}_{(S\,;\,A)}[L]) &= \mathrm{Supp}(L) \cup \mathrm{Supp}(\mathrm{Ty}(L)) \\ \mathrm{Supp}(\mathsf{SOther}(t)) &= \mathrm{Supp}(t) \end{aligned}$$

$$\begin{aligned} \mathrm{Supp}(\star) &= \emptyset \\ \mathrm{Supp}(s \to_A t) &= \mathrm{Supp}(s) \cup \mathrm{Supp}(A) \cup \mathrm{Supp}(t) \\ \mathrm{Supp}(L) &= \bigcup_{i=0}^{n+1} \mathrm{Supp}(L[i]) \cup \bigcup_{i=0}^{n} \mathrm{Supp}(L_i) \end{aligned}$$

We note that each of these support definitions is naturally downwards closed, and there is no need to apply a downwards closure operator as was necessary for the support of Catt syntax. By some routine calculations given in the formalisation module Catt.Tree.Structured.Support, these support definitions are equivalent to taking the support of the generated piece of syntax. More precisely, the equations:

$$\mathrm{Supp}(p) = \mathrm{Supp}(\lfloor p \rfloor) \qquad \mathrm{Supp}(s) = \mathrm{Supp}(\lfloor s \rfloor) \qquad \mathrm{Supp}(A) = \mathrm{Supp}(\lfloor A \rfloor)$$

$$\mathrm{Supp}(L) \cup \mathrm{Supp}(\mathrm{Ty}(L)) = \mathrm{Supp}(\lfloor L \rfloor)$$

for path $p$, structured term $s$, structured type $A$, and labelling $L$.

By using this notion of support, we are able to avoid a lot of "boilerplate" proof. The above definition of support more closely resembles the format of structured terms, and without this definition, most proofs concerning the support of a structured term would begin by simplifying a variable set similar to $\mathrm{Supp}(\lfloor s \rfloor)$ to one more similar to $\mathrm{Supp}(s)$. Here, we instead give this equivalence proof once.

We end this section by giving alternative equality relations for labellings, which encapsulate the idea that a substitution is fully determined by where it sends locally maximal variables.

These equalities are defined as follows for labellings $L : T \to \mathbf{U}$ and $M : T \to \mathbf{U}$:

$$L \equiv^{\max} M \iff \forall (p : \mathsf{MaxPath}_T).\ L(p) \equiv M(p)$$
$$\mathbf{U} \vdash L =^{\max} M \iff \forall (p : \mathsf{MaxPath}_T).\ \mathbf{U} \vdash L(p) = M(p)$$

and define two labels to be equal exactly when their action on maximal paths is equal. The following theorem gives conditions for when the standard equality relation can be recovered from these.

**Theorem 3.3.16.** *Let $L : S \to \mathbf{U}$ and $M : S \to \mathbf{U}$ be labellings. Then the following rules are admissible:*

$$\frac{\mathbf{U} \vdash L : S \qquad \mathbf{U} \vdash M : S \qquad L \equiv^{\max} M}{\mathbf{U} \vdash L = M} \qquad \frac{\mathbf{U} \vdash L : S \qquad \mathbf{U} \vdash M : S \qquad L \equiv^{\max} M}{\mathbf{U} \vdash \mathrm{Ty}(L) = \mathrm{Ty}(M)}$$

*If the equality rule set $\mathcal{R}$ satisfies the preservation and support conditions, then the rules above are still admissible with $\mathbf{U} \vdash L =^{\max} M$ replacing the syntactic equalities.*

*Proof.* We prove the results for the syntactic equality, with the results for the definitional equality following similarly, but using the preservation property instead of uniqueness of typing. We proceed by induction on the tree $S$, proving the admissibility of both rules simultaneously.

First suppose that $S = [\,]$. Then the path $[0] : \mathsf{Path}_{[\,]}$ is maximal and so $\mathbf{U} \vdash L = M$ follows by the reflexivity of equality. The second rule follows from the uniqueness of typing, as we get $\mathbf{U} \vdash L[0] : \mathrm{Ty}(L)$ and $\mathbf{U} \vdash M[0] : \mathrm{Ty}(M)$ from the premises.

Now suppose that $S = [S_0, \dots, S_n]$. By inductive hypothesis, the following judgements hold for each $i \in \{0, \dots, n\}$:

$$\mathbf{U} \vdash L_i = M_i \qquad \mathbf{U} \vdash L[i] \to_{\mathrm{Ty}(L)} L[i+1] = M[i] \to_{\mathrm{Ty}(M)} M[i+1]$$

From the equalities on types, we immediately get that $\mathbf{U} \vdash \mathrm{Ty}(L) = \mathrm{Ty}(M)$ as is required for the admissibility of the second rule, and also get that $\mathbf{U} \vdash L[i] = M[i]$ for each $0 \le i \le n+1$, which along with equality on (sub)labellings above is sufficient to prove that:

$$\mathbf{U} \vdash L = M$$

which witnesses the admissibility of the first rule. □

### 3.3.2 Standard coherences

In Chapter 1, we gave a preliminary definition of standard coherences, a definition of a canonical coherence over a given pasting diagram. This diagram relies on inclusion substitutions from the boundary of a pasting diagram into its source and target variables, whose definition for ps-contexts can be unpleasant to work with.

In contrast, the $n$-boundary of a tree and its associated source and target inclusions have a natural definition by induction on the tree, where the source and target inclusions are given by labellings. We give this definition below.

**Definition 3.3.17.** Given dimension $n \in \mathbb{N}$ and $T$ : Tree, we define the *$n$-boundary* of the tree $\partial_n(T)$ : Tree by induction on $n$ and $T$:

$$\partial_0(T) = [\,] \qquad \partial_{n+1}([T_0, \dots, T_n]) = [\partial_n(T_0), \dots, \partial_n(T_n)]$$

We further define path-to-path functions $\mathrm{I}_n^\epsilon(T) : \partial_n(T) \to T$ for $\epsilon \in \{-,+\}$ by induction:

$$\begin{aligned} \mathrm{I}_0^-(T)([0]) &= [0] \\ \mathrm{I}_0^+([T_0, \dots, T_m])([0]) &= [m+1] \\ \mathrm{I}_{n+1}^\epsilon([T_0, \dots, T_m])([k]) &= [k] \\ \mathrm{I}_{n+1}^\epsilon([T_0, \dots, T_m])(k :: p) &= [k :: \mathrm{I}_{n+1}^\epsilon(T_k)(p)] \end{aligned}$$

and then can define the *source inclusion labelling* $\delta_n^+(T) : \partial_n(T) \to T$ and *target inclusion labelling* $\delta_n^+(T) : \partial_n(T) \to T$ by:

$$\delta_n^\epsilon(T)(p) = \mathsf{SPath}(\mathrm{I}_n^\epsilon(T)(p)) \qquad \mathsf{Ty}(\delta_n^\epsilon(T)) = \star$$

for each $n$ and $\epsilon \in \{-,+\}$.

In the module Catt.Tree.Boundary.Typing, it is proven that:

$$T \vdash \delta_n^\epsilon(T) : \partial_n(T)$$

for all trees $T$, $n \in \mathbb{N}$, and $\epsilon \in \{-,+\}$.

In Chapter 1, the source and target variable sets were defined to be support of the source and target inclusions. This can now be justified by the following lemma.

**Lemma 3.3.18.** *For a dimension $n \in \mathbb{N}$, $T$ : Tree, and $\epsilon \in \{-,+\}$ we have:*

$$\mathsf{Supp}(\delta_n^\epsilon(T)) = \partial_n^\epsilon(T)$$

*Proof.* The proof is given by the function tree-inc-label-supp in the formalisation module Catt.Tree.Boundary.Support and proceeds by induction on $n$ and $T$. □

This definition also allows simple inductive proofs that the boundary inclusions satisfy the globularity conditions, which we state in the following proposition. These proofs are given in the formalisation module Catt.Tree.Boundary.Properties.

**Proposition 3.3.19.** *Let $n \leq m$ and let $T$ be a tree. Then:*

$$\partial_n(\partial_m(T)) \equiv \partial_n(T)$$

*Further, for $\epsilon, \omega \in \{-,+\}$ we have:*

$$\delta_n^\epsilon(\partial_m(T)) \bullet \delta_m^\omega(T) \equiv \delta_n^\epsilon(T)$$

*If instead $n \geq \mathrm{h}(T)$, then $\partial_n(T) \equiv T$ and $\delta_n^\epsilon(T) \equiv \mathsf{id}_T$.*

Further, these constructions commute with suspension: The equalities $\Sigma(\partial_n(T)) \equiv \partial_{n+1}(\Sigma(T))$ and $\Sigma(\delta_n^\epsilon(T)) \equiv \delta_{n+1}^\epsilon(\Sigma(T))$ hold by definition.

We now recall the definitions of standard type, standard coherence, and standard term for a tree $T$, which are given by mutual induction:

- The *standard type*, $\mathcal{U}_T^n$, is an $n$-dimensional type where each component of the type is given by the standard term over the appropriate boundary of the tree $T$, and then included back into $T$ by applying the inclusion labelling.
- The *standard coherence*, $\mathcal{C}_T^n$, is the canonical dimension $n$ coherence term over a tree $T$. It is formed by a single coherence constructor over $T$ with type given by the standard type, $\mathcal{U}_T^n$.
- The *standard term*, $\mathcal{T}_T^n$, is a variation on the standard coherence which does not introduce unnecessary unary composites. If $T$ is linear (and so represents a disc context), and $n = \mathrm{h}(T)$, then $\mathcal{T}_T^n$ is simply given by the unique maximal path in $T$. Otherwise, it is given by the standard coherence $\mathcal{C}_T^n$.

At the end of Chapter 1 it was stated that $\Sigma(\mathcal{T}_T^n) \equiv \mathcal{T}_{\Sigma(T)}^{n+1}$. Using this, the standard term can instead be defined by letting $\mathcal{T}_{[\,]}^0$ be $\mathsf{SPath}([0])$, $\mathcal{T}_{\Sigma(T)}^{n+1}$ be $\Sigma(\mathcal{T}_T^n)$, and $\mathcal{T}_T^n$ be $\mathcal{C}_T^n$ otherwise, which avoids the case split on the linearity of $T$. We now define all three constructions formally using structured syntax.

**Definition 3.3.20.** We define the $n$-dimensional *standard type* over a tree $T$ as a structured type $\mathcal{U}_T^n : \mathsf{SType}_T$, and the $n$-dimensional *standard coherence* and *standard term* over a tree $T$ as structured terms $\mathcal{C}_T^n, \mathcal{T}_T^n : \mathsf{STerm}_T$ by mutual induction:

$$\begin{aligned} \mathcal{U}_T^0 &= \star \\ \mathcal{U}_T^{n+1} &= \mathcal{T}_{\partial_n(T)}^n[\![\delta_{n+1}^-(T)]\!] \to_{\mathcal{U}_T^n} \mathcal{T}_{\partial_n(T)}^n[\![\delta_{n+1}^+(T)]\!] \end{aligned}$$

$$\mathcal{C}_T^n = \mathsf{SCoh}_{(T\,;\,\mathcal{U}_T^n)}[\mathsf{id}_T]$$

$$\mathcal{T}_T^n = \begin{cases} \mathsf{SPath}([0]) & \text{if } T = [\,] \text{ and } n = 0 \\ \mathsf{Inc}_0(\mathcal{T}_{T_0}^{n-1}) & \text{if } n \neq 0 \text{ and } T = [T_0] \\ \mathcal{C}_T^n & \text{otherwise} \end{cases}$$

when $n = \mathrm{h}(T)$, we call the standard coherence $\mathcal{C}_T^n$ the *standard composite* of $T$.

We can immediately show that these standard construct commute with suspension.

**Lemma 3.3.21.** *For tree $T$ and $n \in \mathbb{N}$, $\Sigma(\mathcal{U}_T^n) \equiv \mathcal{U}_{\Sigma(T)}^{n+1}$ and $\Sigma(\mathcal{C}_T^n) \equiv \mathcal{C}_{\Sigma(T)}^{n+1}$.*

*Proof.* We first consider the standard type. The case for $n = 0$ follows immediately, so we let $n > 0$. We then get for $\epsilon \in \{-, +\}$:

$$\begin{aligned}
\Sigma\left(\mathcal{T}^{n-1}_{\partial_{n-1}(T)}[\![\delta^{\epsilon}_{n-1}(T)]\!]\right) &\equiv \Sigma(\mathcal{T}^{n-1}_{\partial_{n-1}(T)})[\![\Sigma(\delta^{\epsilon}_{n-1}(T))]\!] && \text{by functoriality of suspension}\\
&\equiv \mathcal{T}^{n}_{\Sigma(\partial_{n-1}(T))}[\![\Sigma(\delta^{\epsilon}_{n-1}(T))]\!]\\
&\equiv \mathcal{T}^{n}_{\partial_n(\Sigma(T))}[\![\delta^{\epsilon}_{n}(\Sigma(T))]\!]
\end{aligned}$$

By inductive hypothesis $\Sigma(\mathcal{U}^{n-1}_T) \equiv \mathcal{U}^n_{\Sigma(T)}$ and so

$$\begin{aligned}
\Sigma(\mathcal{U}^n_T) &\equiv \Sigma\left(\mathcal{T}^{n-1}_{\partial_{n-1}(T)}[\![\delta^{-}_{n-1}(T)]\!]\right) \to_{\Sigma(\mathcal{U}^{n-1}_T)} \Sigma\left(\mathcal{T}^{n-1}_{\partial_{n-1}(T)}[\![\delta^{+}_{n-1}(T)]\!]\right)\\
&\equiv \mathcal{T}^{n}_{\partial_n(\Sigma(T))}[\![\delta^{-}_{n}(\Sigma(T))]\!] \to_{\mathcal{U}^n_{\Sigma(T)}} \mathcal{T}^{n}_{\partial_n(\Sigma(T))}[\![\delta^{+}_{n}(\Sigma(T))]\!]\\
&\equiv \mathcal{U}^{n+1}_{\Sigma(T)}
\end{aligned}$$

as required.

For the standard coherence we have:

$$\Sigma(\mathcal{C}^n_T) \equiv \mathsf{SCoh}_{(\Sigma(T)\,;\,\Sigma(\mathcal{U}^n_T))}[\Sigma(\mathsf{id}_T)] \equiv \mathsf{SCoh}_{(\Sigma(T)\,;\,\mathcal{U}^{n+1}_{\Sigma(T)})}[\mathsf{id}_{\Sigma(T)}] \equiv \mathcal{C}^{n+1}_{\Sigma(T)}$$

following from the case for types. □

To prove that the standard constructions are well-formed, we give a couple of lemmas. The first concerns the support of the standard term and standard coherence.

**Lemma 3.3.22.** *For a tree $T$, dimension $n \in \mathbb{N}$, and $\epsilon \in \{-, +\}$, we have:*

$$\mathrm{Supp}\left(\mathcal{T}^n_{\partial_n(T)}[\![\delta^{\epsilon}_n(T)]\!]\right) = \partial^{\epsilon}_n(T) \qquad \mathrm{Supp}\left(\mathcal{C}^n_{\partial_n(T)}[\![\delta^{\epsilon}_n(T)]\!]\right) = \partial^{\epsilon}_n(T)$$

*Proof.* The case for coherences follows from the definition and the equality

$$\mathrm{Supp}(\delta^{\epsilon}_n(T)) = \partial^{\epsilon}_n(T)$$

For the standard term, it suffices to consider cases where the standard term and standard coherence are not equal. If $n = 0$, then $\partial_n(T) \equiv [\,]$, and it suffices to prove that $\mathrm{Supp}([m]) = \mathrm{FV}(\lfloor [m] \rfloor)$, but this is immediate because $\mathrm{Supp}([m]) = \mathrm{Supp}(\lfloor [m] \rfloor)$ and $\lfloor [m] \rfloor$ is a variable of type $\star$ so its support is equal to its free variables.

We therefore consider the case where $n > 0$ and $\mathrm{len}(\partial_n(T)) = 1$. The only case where this happens is if $\mathrm{len}(T) = 1$ too, so assume $T \equiv [T_0]$

$$\begin{aligned}
\mathrm{Supp}\left(\mathcal{T}^n_{\partial_n(T)}[\![\delta^{\epsilon}_n(T)]\!]\right) &= \mathrm{Supp}\left(\mathcal{T}^n_{\Sigma(\partial_{n-1}(T_0))}[\![\Sigma\left(\delta^{\epsilon}_{n-1}(T_0)\right)]\!]\right)\\
&= \mathrm{Supp}\left(\Sigma\left(\mathcal{T}^{n-1}_{\partial_{n-1}(T_0)}\right)[\![\Sigma\left(\delta^{\epsilon}_{n-1}(T_0)\right)]\!]\right)\\
&= \mathrm{Supp}\left(\Sigma\left(\mathcal{T}^{n-1}_{\partial_{n-1}(T_0)}[\![\delta^{\epsilon}_{n-1}(T_0)]\!]\right)\right)\\
&= \Sigma\left(\mathrm{Supp}\left(\mathcal{T}^{n-1}_{\partial_{n-1}(T_0)}[\![\delta^{\epsilon}_{n-1}(T_0)]\!]\right)\right)\\
&= \Sigma\left(\partial^{\epsilon}_{n-1}(T_0)\right)\\
&= \partial^{\epsilon}_n(T)
\end{aligned}$$

as required. □

The second lemma gives a globularity condition for the standard type.

**Lemma 3.3.23.** *Let $T$ be a tree. Then:*

$$\mathcal{U}_T^n \equiv \mathcal{U}_{\partial_m(T)}^n[\![\delta_m^\epsilon(T)]\!]$$

*for $n \leq m$ and $\epsilon \in \{-, +\}$.*

*Proof.* We induct on $n$. If $n = 0$ then both sides of the equation are the type $\star$. We therefore consider the case for $n + 1$ and so we must prove:

$$\begin{aligned}
\mathcal{U}_T^{n+1} &\equiv \mathcal{T}_{\partial_n(T)}^n[\![\delta_n^-(T)]\!] \to_{\mathcal{U}_T^k} \mathcal{T}_{\partial_n(T)}^n[\![\delta_n^+(T)]\!] \\
&\equiv \mathcal{T}_{\partial_n(\partial_m(T))}^n[\![\delta_n^-(\partial_m(T))]\!][\![\delta_m^\epsilon(T)]\!] \to_{\mathcal{U}_{\partial_m(T)}^n[\![\delta_m^\epsilon(T)]\!]} \mathcal{T}_{\partial_n(\partial_m(T))}^n[\![\delta_n^+(\partial_m(T))]\!][\![\delta_m^\epsilon(T)]\!] \\
&\equiv \mathcal{U}_{\partial_m(T)}^{n+1}[\![\delta_m^\epsilon(T)]\!]
\end{aligned}$$

The equality $\mathcal{U}_T^n \equiv \mathcal{U}_{\partial_m(T)}^n[\![\delta_m^\epsilon(T)]\!]$ follows by inductive hypothesis. Further, for $\omega \in \{-, +\}$ we have by Proposition 3.3.19:

$$\begin{aligned}
\mathcal{T}_{\partial_n(\partial_m(T))}^n[\![\delta_n^\omega(\partial_m(T))]\!][\![\delta_n^\epsilon(T)]\!] &\equiv \mathcal{T}_{\partial_n(\partial_m(T))}^n[\![\delta_n^\omega(\partial_m(T)) \bullet \delta_m^\epsilon(T)]\!] \\
&\equiv \mathcal{T}_{\partial_n(T)}^n[\![\delta_n^-(T)]\!]
\end{aligned}$$

which completes the proof. □

We can now state and prove the typing properties of standard constructions.

**Proposition 3.3.24.** *Suppose that $\mathcal{O}$ contains the standard operations. Then the following rules are admissible:*

$$\frac{T : \mathsf{Tree} \qquad n \in \mathbb{N}}{T \vdash \mathcal{U}_T^n} \qquad \frac{T : \mathsf{Tree} \qquad n \neq 0 \qquad n \geq \mathrm{h}(T)}{T \vdash \mathcal{C}_T^n : \mathcal{U}_T^n} \qquad \frac{T : \mathsf{Tree} \qquad n \geq \mathrm{h}(T)}{T \vdash \mathcal{T}_T^n : \mathcal{U}_T^n}$$

*Proof.* We prove that all three rules are admissible by mutual induction. First consider the cases for types. The case when $n = 0$ is trivial, so we consider the case for $n + 1$. We need to show that:

$$T \vdash \mathcal{T}_{\partial_n(T)}^n[\![\delta_n^-(T)]\!] \to_{\mathcal{U}_n^T} \mathcal{T}_{\partial_n(T)}^n[\![\delta_n^+(T)]\!]$$

The inductive hypothesis on types gives that $T \vdash \mathcal{U}_n^T$ and so we must show that:

$$T \vdash \mathcal{T}_{\partial_n(T)}^n[\![\delta_n^\epsilon(T)]\!] : \mathcal{U}_n^T$$

for $\epsilon \in \{-, +\}$. By inductive hypothesis for terms, we have $\partial_n(T) \vdash \mathcal{T}_{\partial_n(T)}^n : \mathcal{U}_{\partial_n(T)}^n$ as we have $\mathrm{h}(\partial_n(T)) \leq n$. As $T \vdash \delta_n^\epsilon(T) : \partial_n(T)$ we have that:

$$T \vdash \mathcal{T}_{\partial_n(T)}^n[\![\delta_n^\epsilon(T)]\!] : \mathcal{U}_{\partial_n(T)}^n[\![\delta_n^\epsilon(T)]\!]$$

and so by Lemma 3.3.23, this case is complete.

For the standard coherence, we apply Rule 3.3.14, using the inductive hypothesis for types. To show that $(T, \mathrm{src}(\mathcal{U}_T^n), \mathrm{tgt}(\mathcal{U}_T^n)) \in \mathcal{O}$, we apply Lemma 3.3.22.

For the standard term, like previous proofs it is sufficient to consider the cases where it is defined differently to the standard coherence. For $n = 0$ we must have $T = [\,]$ by the condition on the height of $T$. Hence, $\mathcal{T}_T^n \equiv [0]$ which is well-formed as has type $\star \equiv \mathcal{U}_T^n$ as required.

We now consider $\mathcal{T}_{\Sigma(T)}^{n+1} \equiv \Sigma(\mathcal{T}_T^n)$. By inductive hypothesis on dimension, $T \vdash \mathcal{T}_T^n : \mathcal{U}_T^n$ and so we immediately have that:

$$\Sigma(T) \vdash \mathcal{T}_{\Sigma(T)}^{n+1} : \Sigma(\mathcal{U}_T^n)$$

and so the proof is complete by Lemma 3.3.22. □

The equality relations we have seen so far make heavy use of disc contexts and associated terms and types. We therefore pause to consider the form of these as structured syntax and to relate them to the standard constructions presented in this section.

All disc contexts are the result of applying iterated suspensions to the singleton context, and so it follows that disc contexts correspond exactly to linear trees. By an abuse of notation we write:

$$D^n = \Sigma^n([\,])$$

As we further have that $\Sigma(U^n) \equiv U^{n+1}$ for the sphere type $U^n$, it can be proved for a simple induction that:

$$U^n \equiv \lfloor \mathcal{U}_{D^n}^n \rfloor$$

As we have already noted, the maximal dimension term $d_n : \mathsf{Term}_{D^n}$ is given by $\lfloor \mathcal{T}_{D^n}^n \rfloor$. It is also equal to the unique maximal path, $p^n = \Sigma^n[0]$, which is the list containing $n + 1$ zeros.

The only missing construction is an equivalent for the substitution from a disc context. From a structured term $s : \mathsf{STerm}_{\mathbf{U}}$ of type $A : \mathsf{SType}_{\mathbf{U}}$, there should be a labelling $\{A, s\}$ from $D^n$ to $\mathbf{U}$. This however proves more challenging to define as trees and types have opposite inductive structure. For a labelling, it is natural to specify the lower-dimensional terms first and fill in higher-dimensional terms by induction, though when deconstructing a type, we first receive the highest dimensional terms, only receiving the lower-dimensional terms by further deconstructing the type.

To define the labelling $\{A, t\}$, we define the extension of labelling from a linear tree, which allows us to add higher-dimensional terms to the labelling, and use this to define the labelling from a linear tree.

**Definition 3.3.25.** Let $L : D^n \to \mathbf{U}$ be a labelling from a linear tree, and let $s, t : \mathsf{STerm}_{\mathbf{U}}$ be structured terms. The *extension* of $L$ by $s$ and $t$, $\mathrm{ext}(L, s, t)$, is defined inductively on $n$ by:

$$\mathrm{Ty}(\mathrm{ext}(L, s, t)) = \mathrm{Ty}(L) \qquad \mathrm{ext}(L, s, t) = \begin{cases} L[0]\ \{t\}\ s & \text{if } n = 0 \\ L[0]\ \{\mathrm{ext}(L_0, s, t)\}\ L[1] & \text{otherwise} \end{cases}$$

We then define the labelling $\{A, t\}$ by induction on $A$:

$$\{\star, t\} = (p \mapsto t) \qquad \{s \to_A t, u\} = \mathrm{ext}(\{A, s\}, t, u) \qquad \mathrm{Ty}(\{A, t\}) = \star$$

These constructions all satisfy the expected typing judgements. More precisely the following inference rules are admissible:

$$\frac{\mathbf{U} \vdash L : D^n \qquad \mathbf{U} \vdash s : \mathcal{U}^n_{D^n}[\![L]\!] \qquad \mathbf{U} \vdash t : p^n[\![L]\!] \to_{\mathcal{U}^n_{D^n}[\![L]\!]} s}{\mathbf{U} \vdash \mathsf{ext}(L, s, t) : D^{n+1}}$$

$$\frac{\mathbf{U} \vdash A \qquad \mathbf{U} \vdash t : A}{\mathbf{U} \vdash \{A, t\} : D^{\dim(A)}}$$

The admissibility of the above rules is routine to verify.

Using these constructions, we can recover structured term definitions of the unary composite of a (structured) term $t$ of type $A$ of dimension $n$ as $\mathcal{C}^n_{D^n}[\![\{A, t\}]\!]$ and can define the identity of the same term $t$ as $\mathcal{C}^{n+1}_{D^n}[\![\{A, t\}]\!]$. Therefore, the rules for disc removal and endo-coherence removal can be rephrased in terms of structured syntax to get the following rules:

$$\frac{\mathbf{U} : \mathsf{Ctx} \uplus \mathsf{Tree} \qquad \mathbf{U} \vdash A \qquad \mathbf{U} \vdash t : A \qquad \dim(A) = n > 0}{\mathbf{U} \vdash \mathcal{C}^n_{D^n}[\![\{A, t\}]\!] = t}\textsc{dr},$$

$$\frac{\begin{array}{c}\mathbf{U} : \mathsf{Ctx} \uplus \mathsf{Tree} \qquad T : \mathsf{Tree} \qquad L : S \to_\star \mathbf{U} \qquad n = \dim(A) \\ T \vdash A \qquad T \vdash s : A \qquad \mathsf{Supp}(s) = \mathsf{Var}(T) \qquad \mathbf{U} \vdash L : T\end{array}}{\mathbf{U} \vdash \mathsf{SCoh}_{(T\,;\,s \to_A s)}[L] = \mathcal{C}^{n+1}_{D^n}[\![\{A, s\} \bullet L]\!]}\textsc{ecr},$$

which are admissible if the equality rule set $\mathcal{R}$ has disc removal or endo-coherence removal respectively.

We end this section with two further results that can be proven in the presence of disc removal and endo-coherence removal. The first states that disc removal is sufficient (and necessary) to unify standard coherences and standard terms.

**Theorem 3.3.26.** *The tame equality rule set $\mathcal{R}$ has disc removal if and only if the rule:*

$$\frac{T : \mathsf{Tree} \qquad n \in \mathbb{N} \qquad n \geq \mathrm{h}(T) > 0}{T \vdash \mathcal{C}^n_T = \mathcal{T}^n_T}$$

*is admissible.*

*Proof.* We note that $\mathcal{C}^n_T$ and $\mathcal{T}^n_T$ only differ when $T = D^n$. If $\mathcal{R}$ has disc removal, then for each $n \neq 0$ we have $\mathcal{C}^n_{D^n} = \mathsf{SPath}(p^n) \equiv \mathcal{T}^n_{D^n}$. Conversely, if $\mathcal{C}^n_T = \mathcal{T}^n_T$ when $n > 0$ or $\mathrm{h}(T) > 0$, then $\mathcal{C}^n_{D^n} = \mathcal{T}^n_{D^n}$ for any $n > 0$. Then as $\mathcal{R}$ is tame, we can apply the substitution $\{A, t\}$ to both sides of the equation to get the statement of disc removal. □

Lastly, under the presence of endo-coherence removal, the standard coherences $\mathcal{T}^n_T$ for which $n > \mathrm{h}(T)$ can be shown to be equal to identities.

**Theorem 3.3.27.** *Suppose the equality rule set $\mathcal{R}$ has endo-coherence removal. Let $T$ be a tree and suppose $n \geq \mathrm{h}(T)$. Then:*

$$T \vdash \mathcal{C}^{n+1}_T = \mathcal{C}^{n+1}_{D^n}[\![\{\mathcal{U}^n_T, \mathcal{T}^n_T\}]\!]$$

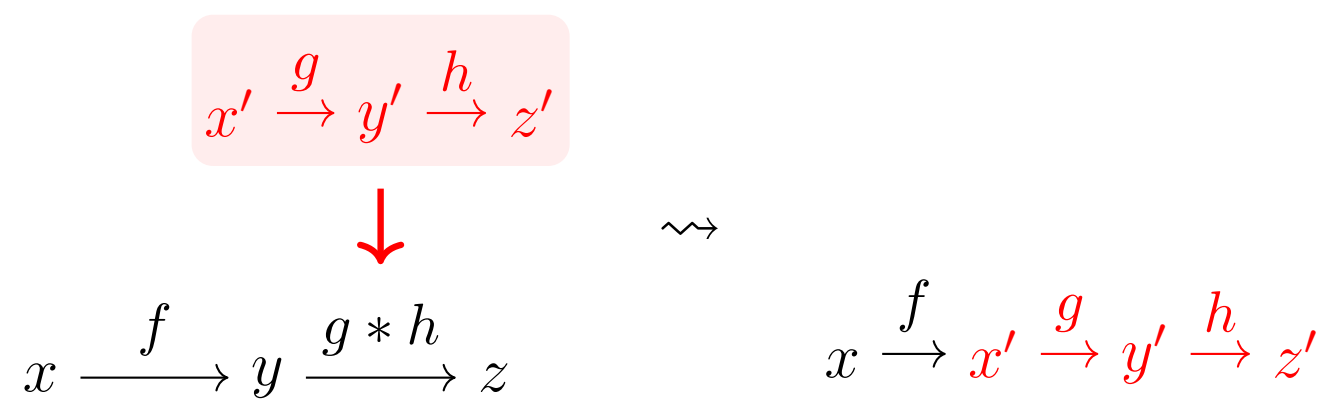

Figure 3.4: Insertion acting on the composite $f * (g * h)$.

*Proof.* The following chain of equalities hold:

$$\begin{aligned}
\mathcal{C}_T^{n+1} &\equiv \mathsf{SCoh}_{(T\,;\,\mathcal{T}^n_{\partial_n(T)}[\![\delta_n^T(-)]\!] \to_{\mathcal{U}_T^n} \mathcal{T}^n_{\partial_n(T)}[\![\delta_n^T(+)]\!])}[\mathsf{id}_S] \\
&\equiv \mathsf{SCoh}_{(T\,;\,\mathcal{T}_T^n \to_{\mathcal{U}_T^n} \mathcal{T}_T^n)}[\mathsf{id}_S] && \text{by Proposition 3.3.19} \\
&= \mathcal{C}_{D^n}^{n+1}[\![\{\mathcal{U}_T^n, \mathcal{T}_T^n\}]\!] && \text{by ECR'}
\end{aligned}$$

where ECR' can be applied as $\mathrm{Supp}(\mathcal{T}_T^n) = \mathrm{Var}(\lfloor T \rfloor)$ by Lemma 3.3.22. □

Due to these two theorems, every standard term $\mathcal{T}_T^n$ with $n \geq \mathrm{h}(T)$ is equal to either the unique variable of the singleton context (when $n = \mathrm{h}(T) = 0$), a standard composite (when $n = \mathrm{h}(T) > 0$) or an identity (when $n > \mathrm{h}(T)$), hence completely classifying the well-formed standard terms.

## 3.4 Insertion

We now introduce *insertion*, the construction that powers the strictly associative behaviour of $\textsc{Catt}_{\mathsf{sua}}$. Insertion incorporates part of the structure of a locally maximal argument term into the head coherence, simplifying the overall syntax of the term.

Consider the composite $f * (g * h)$. This term has two locally maximal arguments, $f$ and $g * h$, the second of which is a (standard) coherence. Insertion allows us to merge these two composites into one by "inserting" the pasting diagram of the inner coherence into the pasting diagram of the outer coherence. In the case above we will get that the term $f * (g * h)$ is equal to the ternary composite $f * g * h$, a term with a single coherence. As the term $(f * g) * h$ also reduces by insertion to the ternary composite, we see that both sides of the associator become equal under insertion. The action of insertion on these contexts is shown in Figure 3.4.

Insertion is an operation that is best understood with respect to trees instead of ps-contexts. Insertion merges the structure of two trees along a *branch* of the first tree.

**Definition 3.4.1.** Let $S$ be a tree. A *branch* of $S$ is a non-empty list of natural numbers $P$ which indexes a subtree $S^P$ which is linear. From each branch $P$, a maximal path $\overline{P}$ can be obtained by concatenating $P$ with $p^{\mathrm{h}(S^P)}$, the unique maximal path of $S^P$.

For a branch $P$, we further define the *branch height*, $\mathsf{bh}(P)$, to be one less than the length of $P$ (noting that branches are non-empty lists), and the *leaf height*, $\mathsf{lh}(P)$, to be one less than the length of $\overline{P}$, which is equal to the dimension of $\lfloor \hat{P} \rfloor$.

While each branch $P$ uniquely determines a maximal path $\overline{P}$, the converse does not hold.

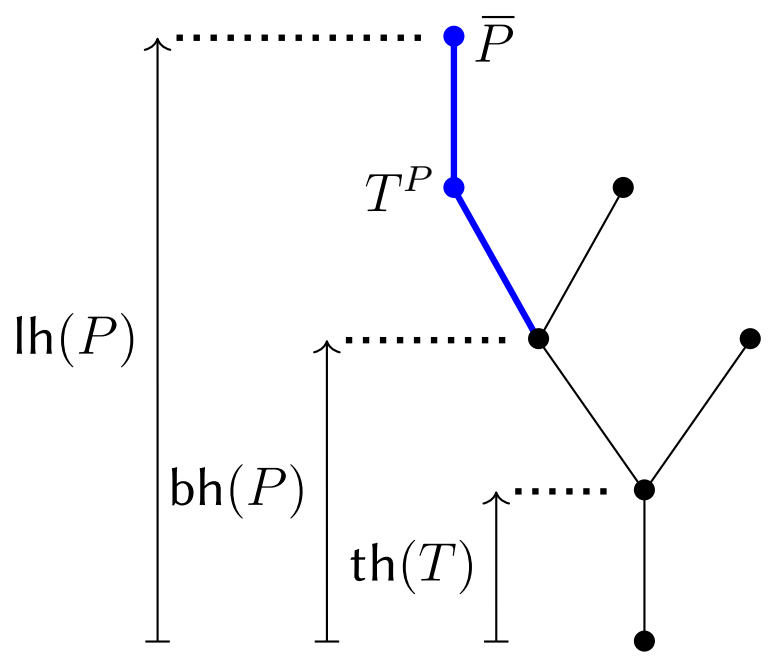

Figure 3.5: Leaf height, branch height and trunk height.

There may be multiple branches of a tree which correspond to the same maximal path. Consider the tree $T = [[[[[\,]], [\,]], [\,]]]$. This has two distinct branches $P = [0, 0, 0]$ and $Q = [0, 0, 0, 0]$ which both correspond to the maximal path $[0, 0, 0, 0, 0]$. We graphically depict these branches below by drawing them in blue.

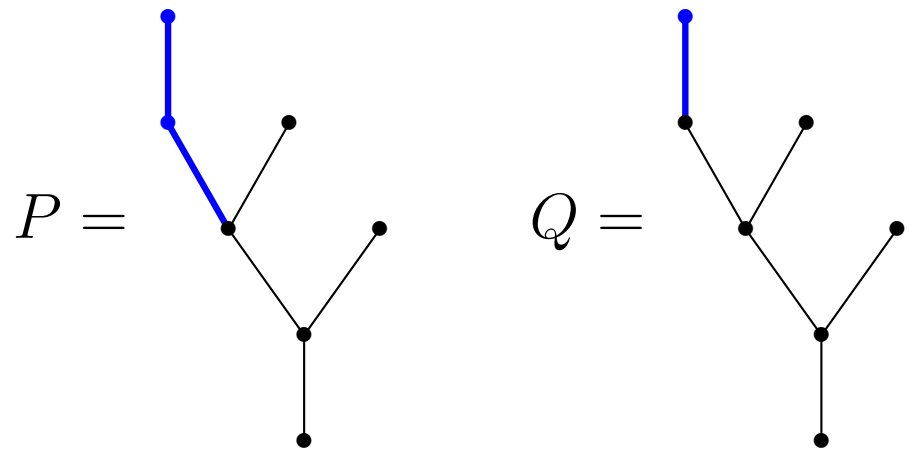

While $P$ and $Q$ represent the same path, they have different branch heights: the branch height of $P$ is 2 while the branch height of $Q$ is 3. This will cause insertions along these two branches to proceed differently (though we will see later in Lemma 3.4.27 that if both insertions are valid then the results are equivalent). The leaf height and branch height of the branch $P$ is demonstrated in Figure 3.5, where we also depict the trunk height of $T$, which was defined in Section 3.2.

Let us again consider the tree $S = [[[\,], [\,]], [\,]]$ from Figure 3.3. This tree has three branches, corresponding to the maximal paths $[0, 0, 0]$, $[0, 1, 0]$, and $[1, 0]$. We consider the action of insertion of three trees $T_1, T_2, T_3$, given below, into branch $P = [0, 0]$, which corresponds to the first of these maximal paths.

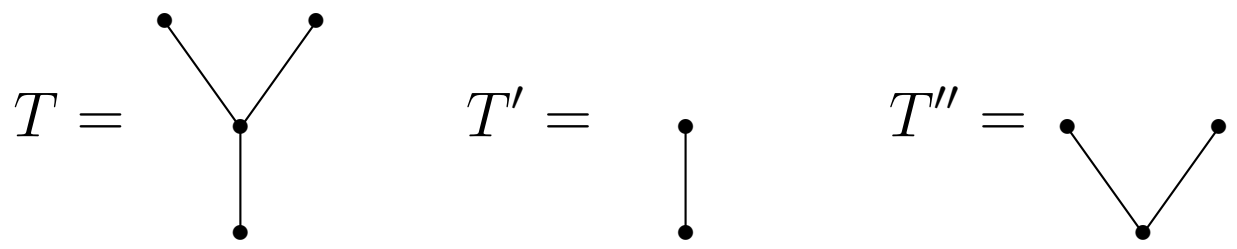

We first consider the insertion of $T$ into $S$, which returns the inserted tree $S \ll_P T$, where $P$ is drawn in blue on the diagram.

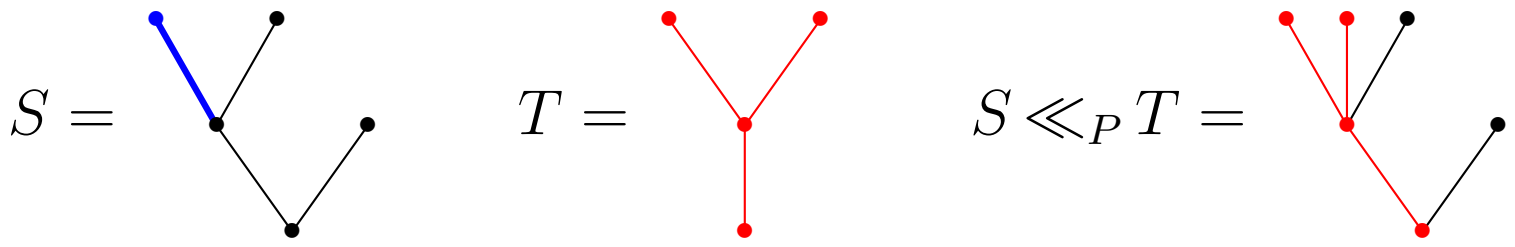

In this case the structure of $T$ is compatible with the point of insertion $P$ and $T$ can be inserted into $S$, replacing the branch $P$ with the appropriate part of $T$, where this appropriate part is obtained by removing the trunk of $T$.

We now consider the insertion of $T'$ into $S$. Despite $T'$ having a lower height than $S$, it is still insertable, forming the following tree $S \ll_P T'$.

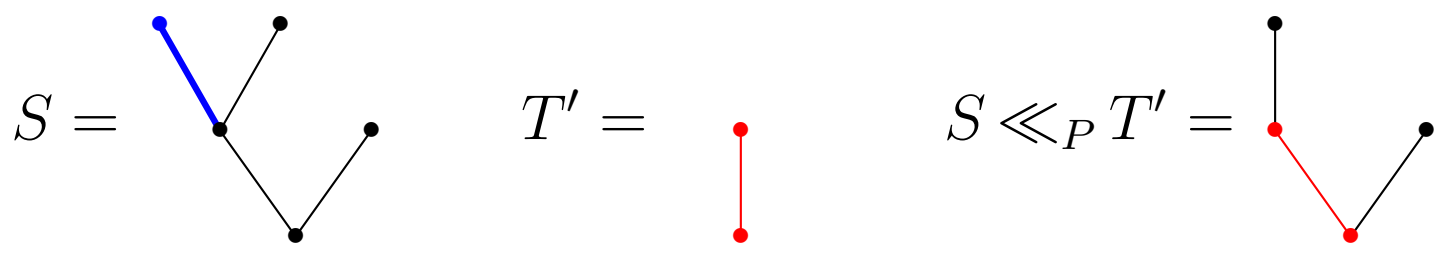

Here, the branch $P$ is replaced by a singleton tree, which is the remaining part of $T'$ after removing its trunk. We note that this operation is the same as pruning the locally maximal variable $\lfloor \overline{P} \rfloor$ from $\lfloor T \rfloor$. We will see in Section 3.4.1 that all instances of pruning can be represented as an instance of insertion.

When we consider the insertion of $T''$ into $S$, it is not clear how to proceed, as there is no "corresponding part" of $T''$ to replace the branch $P$ with. In the other two cases this is obtained by removing the trunk of the tree, but $T''$ has no trunk to remove. In this case we say that the insertion is not possible to perform as $\mathsf{bh}(P) > \mathsf{th}(T'')$, a condition necessary for insertion.

More generally we consider a (structured) coherence term $\mathsf{SCoh}_{(S\,;\,A)}[L] : \mathsf{STerm}_{\mathbf{U}}$. To apply insertion to this term, we must first identify a branch $P$ of $S$ such that $\overline{P}[\![L]\!] \equiv \mathcal{C}_T^{\mathsf{lh}(P)}[\![M]\!]$, that is there is a locally maximal argument of $L$ which is a standard coherence. We then must construct the following data as part of the insertion operation:

- The *inserted tree* $S \ll_P T$, obtained by inserting $T$ into $S$ along the branch $P$. We have already given some examples of this operation.
- The *interior labelling* $\iota : T \to S \ll_P T$, the inclusion of $T$ into a copy of $T$ living in the inserted tree.
- The *exterior labelling* $\kappa : S \to S \ll_P T$, which maps $\overline{P}$ to standard coherence over the copy of $T$, or more specifically $\mathcal{C}_\Theta^{\mathsf{lh}(P)}[\![\iota]\!]$, and other maximal paths to their copy in the inserted tree.
- The *inserted labelling* $L \ll_P M : S \ll_P T \to \mathbf{U}$, which collects the appropriate parts of $L$ and $M$.

Using this notation, insertion yields the following equality:

$$\mathsf{Coh}_{(S\,;\,A)}[L] = \mathsf{Coh}_{(S \ll_P T\,;\,A[\![\kappa]\!])}[L \ll_P M]$$

These constructions can be assembled into the following diagram, where $n = \mathsf{lh}(P)$:

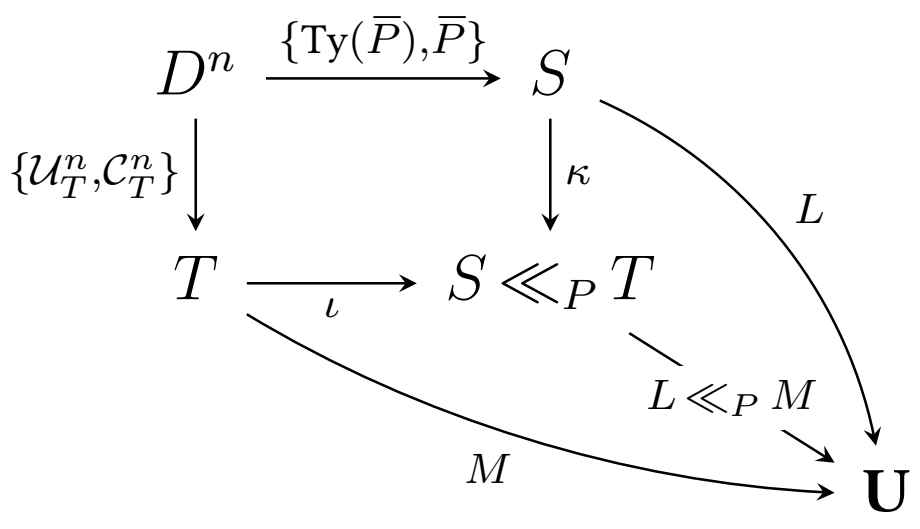

It will be proven in Section 3.4.1 that the square above is cocartesian, and so $S \ll_P T$ is the pushout of $S$ and $T$.

We now begin to define each of these constructions in turn. As we need a lot of data to perform an insertion, we will package it up to avoid repetition.

**Definition 3.4.2.** An *insertion point* is a triple $(S, P, T)$ such that $S$ and $T$ are trees and $P$ is a branch of $S$ with $\mathsf{bh}(P) \leq \mathsf{th}(T)$ and $\mathsf{lh}(S) \geq \dim(T)$.

An *insertion redex* is a sextuple $(S, P, T, \mathbf{U}, L, M)$ such that $(S, P, T)$ is an insertion point, $L : S \to \mathbf{U}$ and $M : T \to \mathbf{U}$ are labellings with $\mathrm{Ty}(L) \equiv \mathrm{Ty}(M) \equiv \star$, and $L(\underline{\overline{P}}) \equiv \mathcal{C}_T^{\mathsf{lh}(P)}[\![M]\!]$.

We can now define the insertion operation on trees.

**Definition 3.4.3** (Inserted tree)**.** Let $(S, P, T)$ be an insertion point. Define the *inserted tree* $S \ll_P T$ by induction on the branch $P$, noting that $P$ is always non-empty.

- Suppose $P = [k]$ and $S = [S_0, \dots, S_k, \dots, S_n]$. Then:

$$S \ll_P T = [S_0, \dots, S_{k-1}] \mathbin{+\!\!+} T \mathbin{+\!\!+} [S_{k+1}, \dots, S_n]$$

- Suppose $P = k :: Q$ and again $S = [S_0, \dots, S_k, \dots, S_n]$. We note that $Q$ is a branch of $S_k$ and by the condition on trunk height of $T$ we have $T = \Sigma(T_0)$. Then:

$$S \ll_P T = [S_0, \dots, S_{k-1}, (S_k \ll_Q T_0), S_{k+1}, \dots, S_n]$$

We draw attention to the condition of the trunk height of $T$ being at least the branch height of $P$, which is necessary for the induction to proceed. We recall that a tree is identified with a list of trees, and that in the first case of insertion $T$ is treated as a list, and in the second case $S_k \ll_Q T_0$ is treated as a single tree which forms one of the subtrees of $S \ll_P T$.

We now proceed to define the interior and exterior labellings, which will be done using the diagrammatic notation introduced in Section 3.2.1.

**Definition 3.4.4** (Interior labelling)**.** Given an insertion point $(S, P, T)$ we define the interior labelling $\iota_{S,P,T} : T \to S \ll_P T$ by induction on $P$.

- When $P = [k]$ and $S = [S_0, \dots, S_k, \dots, S_n]$ we define $\iota$ by $\mathrm{Ty}(\iota) = \star$ and:

$$\begin{array}{ccccc} [S_0, \dots, S_{k-1}] & \mathbin{+\!\!+} & T & \mathbin{+\!\!+} & [S_{k+1}, \dots, S_n] \\ & & \uparrow{\scriptstyle \mathsf{id}} & & \\ & & T & & \end{array}$$

- When $P = k :: Q$, $S = [S_0, \dots, S_k, \dots, S_n]$, and $T = [T_0]$ (by the trunk height condition) we define $\iota$ by $\mathrm{Ty}(\iota) = \star$ and:

$$\begin{array}{ccccc} [S_0, \dots, S_{k-1}] & \vee & \Sigma S_k \ll_Q T_0 & \vee & [S_{k+1}, \dots, S_n] \\ & & \uparrow{\scriptstyle \Sigma \iota_{S_k,Q,T_0}} & & \\ & & \Sigma T_0 & & \end{array}$$

We may drop the subscripts on $\iota$ when they are easily inferred.

**Definition 3.4.5** (Exterior labelling). Given an insertion point $(S, P, T)$, we define the exterior labelling $\kappa_{S,P,T} : S \to S \ll_P T$ by induction on $P$.

- When $P = [k]$ and $S = [S_0, \dots, S_k, \dots, S_n]$ we define $\kappa$ by $\mathrm{Ty}(\kappa) = \star$ and:

$$\begin{array}{ccccc} [S_0, \dots, S_{k-1}] & +\!\!+ & T & +\!\!+ & [S_{k+1}, \dots, S_n] \\ \mathsf{id}\uparrow & & \{\mathcal{U}_T^m, \mathcal{C}_T^m\}\uparrow & & \mathsf{id}\uparrow \\ [S_0, \dots, S_{k-1}] & \vee & \Sigma S_k & \vee & [S_{k+1}, \dots, S_n] \end{array}$$

  Where we note that by the condition of $P$ being a branch we have that $S_k$ is linear and so $\Sigma\lfloor S_k \rfloor$ is a some disc $D^m$ where $m = \mathrm{h}(S_k) + 1$.

- When $P = k :: Q$, $S = [S_0, \dots, S_k, \dots, S_n]$, and $T = [T_0]$ (by the trunk height condition) we define $\kappa$ by $\mathrm{Ty}(\kappa) = \star$ and:

$$\begin{array}{ccccc} [S_0, \dots, S_{k-1}] & \vee & \Sigma S_k \ll_Q T_0 & \vee & [S_{k+1}, \dots, S_n] \\ \mathsf{id}\uparrow & & \Sigma\kappa_{S_k,Q,T_0}\uparrow & & \mathsf{id}\uparrow \\ [S_0, \dots, S_{k-1}] & \vee & \Sigma S_k & \vee & [S_{k+1}, \dots, S_n] \end{array}$$

Again the subscripts on $\kappa$ may be dropped where they can be inferred.

Lastly we define the inserted labelling, the labelling out of the inserted tree.

**Definition 3.4.6** (Inserted labelling). Given an insertion point $(S, P, T)$ with $L : S \to \mathbf{U}$ and $M : T \to \mathbf{U}$, we define the *inserted labelling* $L \ll_P M : S \ll_P T \to \mathbf{U}$. Let

$$S = [S_0, \dots, S_n] \qquad L = s_0\{L_0\}s_1 \cdots \{L_n\}s_{n+1} : A$$

and then proceed by induction on $P$.

- Let $P = [k]$, and

$$T = [T_0, \dots, T_m] \qquad M = t_0\{M_0\}t_1 \cdots \{M_m\}t_{m+1} : B$$

  Then define $L \ll_{[k]} M$ to be:

$$s_0\{L_0\}s_1 \cdots \{L_{k-1}\}t_0\{M_0\}t_1 \cdots \{M_m\}t_{m+1}\{L_{k+1}\}s_{k+2} \cdots \{L_n\}s_{n+1} : A$$

- Suppose $P = k :: Q$ so that

$$T = [T_0] \qquad M = t_0\{M_0\}t_1 : B$$

  Define $L \ll_P M$ as:

$$s_0\{L_0\}s_1 \cdots \{L_{k-1}\}t_0\{L_k \ll_Q M_0\}t_1\{L_{k+1}\}s_{k+2} \cdots \{L_n\}s_{n+1} : A$$

We now proceed to prove that each of these constructions used to generate insertion is well-formed. We begin with the following small lemma.

**Lemma 3.4.7.** *Let $(S, P, T, \mathbf{U}, L, M)$ be an insertion redex. If we further suppose that $\mathbf{U} \vdash L : S$ and $\mathbf{U} \vdash M : T$, then:*

$$\mathbf{U} \vdash L[k] \to_{\mathtt{Ty}(L)} L[k+1] = M[0] \to_{\mathtt{Ty}(M)} M[m+1]$$

*where $k$ is the first element of $P$ (as $P$ is non-empty) and $T$ has length $m$.*

*Proof.* From the insertion redex, we have $L(\overline{P}) \equiv \mathcal{C}_T^{\mathsf{lh}(P)}[\![M]\!]$. By assumption, $P$ is of the form $k :: p$, where $p$ is a path and $S = [S_0, \dots, S_n]$ and so

$$\mathsf{SPath}(\overline{P}) \equiv \mathsf{Inc}_k(\mathsf{SPath}(p))$$

and so supposing that $S_k \vdash \mathsf{SPath}(p) : A$ (as every path is well-formed), we can obtain:

$$\mathbf{U} \vdash \mathsf{SPath}(\overline{P})[\![L]\!] : \Sigma(A)[\![\mathrm{inc}_k]\!][\![L]\!]$$

By Proposition 3.3.24, $\mathbf{U} \vdash \mathcal{C}_T^{\mathsf{lh}(P)}[\![M]\!] : \mathcal{U}_T^{\mathsf{lh}(P)}[\![M]\!]$. Therefore, by uniqueness of types (using the syntactic equality from the insertion redex), we have:

$$\mathbf{U} \vdash \Sigma(A)[\![\mathrm{inc}_k \bullet L]\!] = \mathcal{U}_T^{\mathsf{lh}(P)}[\![M]\!]$$

By truncating both sides of this equality $\mathsf{lh}(P) - 1$ times we get:

$$\mathbf{U} \vdash \Sigma(\star)[\![\mathrm{inc}_k \bullet L]\!] = \mathcal{U}_T^{1}[\![M]\!]$$

which after expanding definitions on both sides gives the required equality. □

The typing properties of each of the constructions involved in insertion are given in the following proposition.

**Proposition 3.4.8.** *Let $(S, P, T)$ be an insertion point. Then:*

$$S \ll_P T \vdash \iota_{S,P,T} : T \qquad S \ll_P T \vdash \kappa_{S,P,T} : S$$

*If we further have $\mathbf{U} \vdash L : S$ and $\mathbf{U} \vdash M : S$ with $L(\overline{P}) \equiv \mathcal{C}_T^{\mathsf{lh}(P)}[\![M]\!]$ then:*

$$\mathbf{U} \vdash L \ll_P M : S \ll_P T$$

*Proof.* The labellings $\iota$ and $\kappa$ are formed using constructions that have already been shown to be well-formed. We therefore focus on the typing judgement for the inserted labelling. As in the definition of the inserted labelling, we let

$$S = [S_0, \dots, S_n] \qquad L = s_0\{L_0\}s_1 \cdots \{L_n\}s_{n+1} : A$$

By inspection of the typing derivation $\mathbf{U} \vdash L : S$ we have that $\mathbf{U} \vdash s_i : A$ and $\mathbf{U} \vdash L_i : S_i$ for each $i$.

We then proceed by induction on $P$.

- Let $P = [k]$ and

$$T = [T_0, \dots, T_m] \qquad M = t_0\{M_0\}t_1 \cdots \{M_m\}t_{m+1} : B$$

By $\mathbf{U} \vdash M : T$, we have that $\mathbf{U} \vdash t_i : B$ and $\mathbf{U} \vdash M_i : T_i$ for each $i$. Applying Lemma 3.4.7, we have $\mathbf{U} \vdash A = B$, $\mathbf{U} \vdash s_k = t_0$, and $\mathbf{U} \vdash s_{k+1} = t_{m+1}$. Therefore, by applying the conversion rule, $\mathbf{U} \vdash t_i : A$. To complete this case, we must show that for each $i$:

$$\mathbf{U} \vdash (L \ll_P M)_i : (S \ll_P T)_i$$

For most $i$ this is trivial, however there is a subtlety for $i = k-1$ that $(L \ll_P M)_{k-1} \not\equiv L_{k-1}$, as:

$$\mathsf{Ty}((L \ll_P M)_{k-1}) \equiv s_{k-1} \to_A t_0 \not\equiv s_{k-1} \to_A s_k \equiv \mathsf{Ty}(L_{k-1})$$

However, the equality $\mathbf{U} \vdash s_k = t_0$ means that these two types are definitionally equal, and so the required typing derivation follows from $\mathbf{U} \vdash L_{k-1} : S_k$. A similar argument is needed to prove that $\mathbf{U} \vdash L_{k+1} : S_{k+1}$, completing this case.

- Suppose $P = k :: Q$ so that

$$T = [T_0] \qquad M = t_0\{M_0\}t_1 : B$$

with $\mathbf{U} \vdash M_0 : T_0$ and $\mathbf{U} \vdash t_i : B$ for $i \in \{0, 1\}$. Then:

$$\begin{aligned} L_k(\overline{Q}) &\equiv L(\overline{P}) \\ &\equiv \mathcal{C}_T^{\mathsf{lh}(P)}[\![M]\!] \\ &\equiv \Sigma\left(\mathcal{C}_{T_0}^{\mathsf{lh}(Q)}\right)[\![M]\!] \\ &\equiv \mathcal{C}_{T_0}^{\mathsf{lh}(Q)}[\![M_0]\!] \end{aligned}$$

and so by inductive hypothesis, we have $\mathbf{U} \vdash L_k \ll_Q M_0 : S_k \ll_Q T_0$. Then by a similar argument to above it can be shown that $L \ll_P M$ is well-formed.

Hence, $\mathbf{U} \vdash L \ll_P M : S \ll_P T$ for all branches $P$. □

We now end this section by formally giving the equality rule set for insertion.

**Definition 3.4.9.** The *insertion rule set*, insert, is the set consisting of the triples:

$$(\Gamma, \lfloor \mathsf{SCoh}_{(S\,;\,A)}[L] \rfloor, \lfloor \mathsf{SCoh}_{(S \ll_P T\,;\,A[\![\kappa_{S,P,T}]\!])}[L \ll_P M] \rfloor)$$

for each insertion redex $(S, P, T, \Gamma, L, M)$, and structured type $A$.

A set of rules $\mathcal{R}$ *contains insertion* if insert $\subseteq \mathcal{R}$. Insertion makes the following rule admissible:

$$\frac{(S, P, T, \Gamma, L, M) \text{ is an insertion redex} \qquad S \vdash A \qquad \Gamma \vdash L : S}{\Gamma \vdash \mathsf{SCoh}_{(S\,;\,A)}[L] = \mathsf{SCoh}_{(S \ll_P T\,;\,A[\![\kappa_{S,P,T}]\!])}[L \ll_P M]}$$

The set $\mathcal{R}$ *has insertion* if the rule INSERT holds in the generated theory.

### 3.4.1 Universal property of insertion

As stated in the previous section, the constructions involved in insertion arise as a pushout square. In this section, we prove this result, which we state below. Throughout this section we assume that we are working in a tame theory for which the support and preservation conditions hold. Further, we only give the maximal arguments of substitutions from a disc, as we only work with well-formed syntax up to definitional equality and so the type will always be inferable.

**Theorem 3.4.10.** *Let $(S, P, T)$ be an insertion point. Then the following commutative square of* $\mathsf{Catt}_{\mathcal{R}}$ *is cocartesian:*

$$\begin{array}{ccc} D^{\mathsf{lh}(P)} & \xrightarrow{\{\lfloor \overline{P} \rfloor\}} & \lfloor S \rfloor \\ {\scriptstyle \{\lfloor \mathcal{C}_T^{\mathsf{lh}(P)} \rfloor\}} \downarrow & & \downarrow {\scriptstyle \lfloor \kappa \rfloor} \\ \lfloor T \rfloor & \xrightarrow[\lfloor \iota \rfloor]{} & \lfloor S \ll_P T \rfloor \end{array}$$

*The context $\lfloor S \ll_P T \rfloor$ is the pushout of $\lfloor S \rfloor$ and $\lfloor T \rfloor$ along the maps that send the maximal variable of $D^n$ to the locally maximal variable corresponding to the branch $P$ and the standard coherence of over $T$ of dimension equal to the leaf height of $P$.*

This theorem allows an intuitive understanding of the insertion operation; the inserted tree $S \ll_P T$ is the result of taking the disjoint union of $S$ and $T$ and gluing the locally maximal variable of $S$ corresponding to the branch $P$ to the composite of $T$. The original motivation for insertion was to take a term where one of the locally maximal arguments was a standard composition and flatten the structure, which aligns with the intuition given by the universal property.

*Remark* 3.4.11. As contexts have an interpretation as freely generated $\infty$-categories, and the category of $\infty$-categories is cocomplete, there is an $\infty$-category pushout of this square. It however may be surprising that this pushout is freely generated and happens to be freely generated by a pasting diagram.

We work towards Theorem 3.4.10 by introducing a couple of lemmas. These lemmas will mostly be proven by deferring to the formalisation, using the machinery of structured terms introduced in Section 3.3 to simplify the computations involved. We first show that the square is commutative, while also justifying the description of the exterior labelling given at the start of the section.

**Lemma 3.4.12.** *Let $(S, P, T)$ be an insertion point. Then $\kappa(\overline{P}) \equiv \mathcal{C}_T^{\mathsf{lh}(P)}[\![\iota]\!]$.*

*Proof.* See κ-branch-path in Catt.Tree.Insertion.Properties. □

We next state two factorisation properties for the interior and exterior labellings.

**Lemma 3.4.13.** *For insertion redex $(S, P, T, \mathbf{U}, L, M)$, the following hold:*

$$\iota_{S,P,T} \circ (L \ll_P M) \equiv M \qquad \kappa_{S,P,T} \circ (L \ll_P M) \equiv^{\max} L$$

*Hence, the maps $L$ and $M$ factor through the labellings $\kappa$ and $\iota$ respectively.*

*Proof.* See ι-comm and κ-comm in Catt.Tree.Insertion.Properties. □

We can now proceed with the proof of Theorem 3.4.10.

*Proof of Theorem 3.4.10.* Let $(S, P, T)$ be an insertion point. We must first show that the candidate pushout square is in fact commutative, for which it is sufficient to show:

$$\{\mathrm{Ty}(\overline{P}), \overline{P}\} \bullet \kappa \equiv^{\max} \{\mathcal{U}_T^{\mathsf{lh}(P)}, \mathcal{C}_T^{\mathsf{lh}(P)}\} \bullet \iota$$

which follows from Lemma 3.4.12. To prove that this square is cocartesian, we take two substitutions $\sigma : \lfloor S \rfloor \to \Gamma$ and $\tau : \lfloor T \rfloor \to \Gamma$ such that the following diagram is commutative:

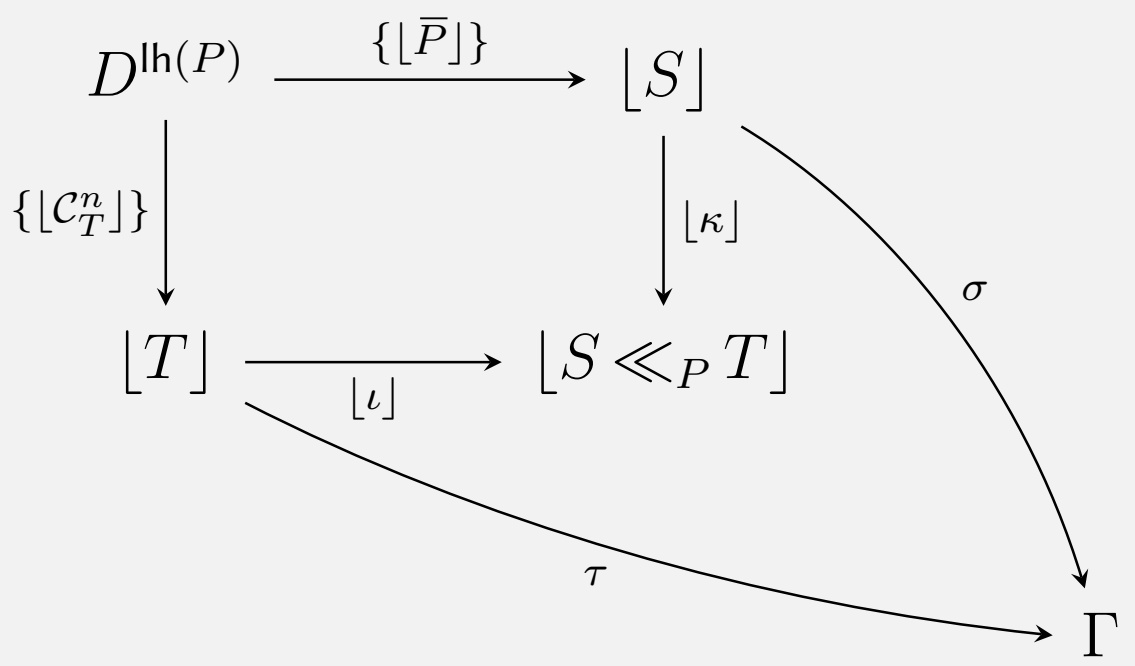

We therefore have that $\lceil \sigma \rceil$ is a labelling $S \to \Gamma$ and $\lceil \tau \rceil$ is a labelling $T \to \Gamma$ with

$$\Gamma \vdash \lceil \sigma \rceil(\overline{P}) = \mathcal{C}_T^{\mathsf{lh}(P)}[\![\lceil \tau \rceil]\!]$$

To apply Lemma 3.4.13, we need this to be a syntactic equality. We therefore define $M = \lceil \tau \rceil$ and $L$ to be given by:

$$L(p) = \begin{cases} \mathcal{C}_T^{\mathsf{lh}(P)}[\![M]\!] & \text{if } p = \overline{P} \\ \lceil \sigma \rceil(p) & \text{otherwise} \end{cases}$$

by the equality above, $L$ is well-formed and $\lfloor L \rfloor = \sigma$. We then get a well-formed map $\lfloor L \ll_P M \rfloor$ from $\lfloor S \ll_P T \rfloor$ to $\Gamma$ such that the following diagram is commutative by Lemma 3.4.13:

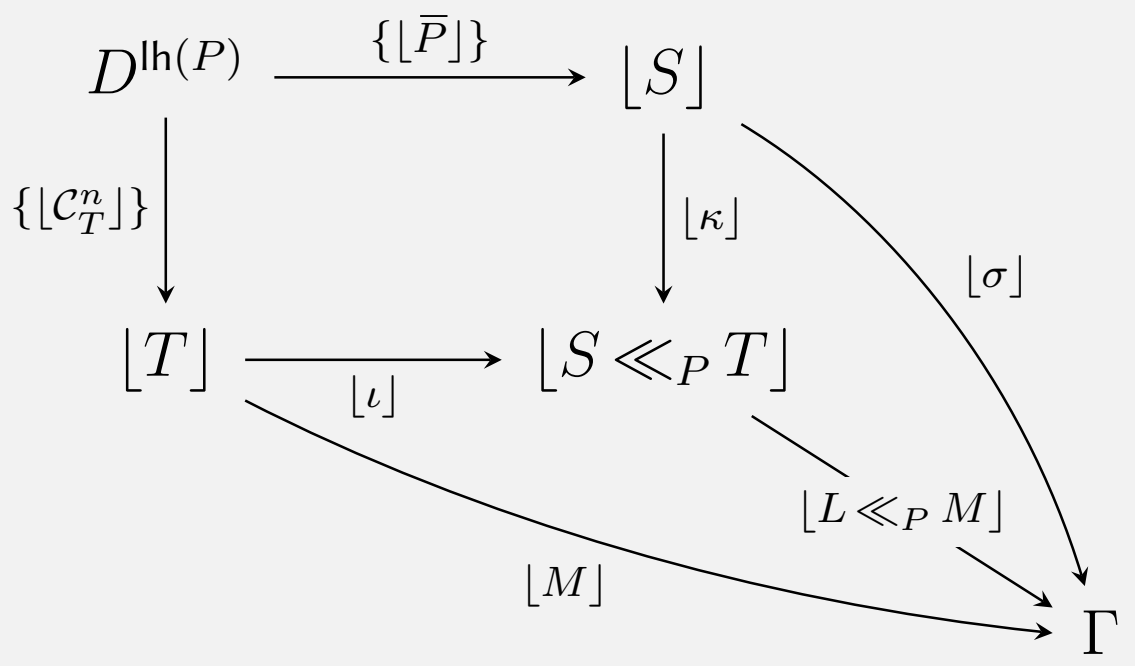

The uniqueness of this morphism follows from the observation that every path of $S \ll_P T$ is either of the form $\iota(p)$ for some $p : \mathsf{Path}_T$ or $\kappa(q)$ for some $q : \mathsf{Path}_S$. □

From this result we will be able to show that having insertion in a theory implies the existence of pruning. The plan will be to show that pruning satisfies a similar universal property.

**Proposition 3.4.14.** *Let $\mathcal{D}$ : $\mathsf{Dyck}_0$ be a Dyck word, and let $p$ be a peak of $\mathcal{D}$. Then the following square is a pushout square:*

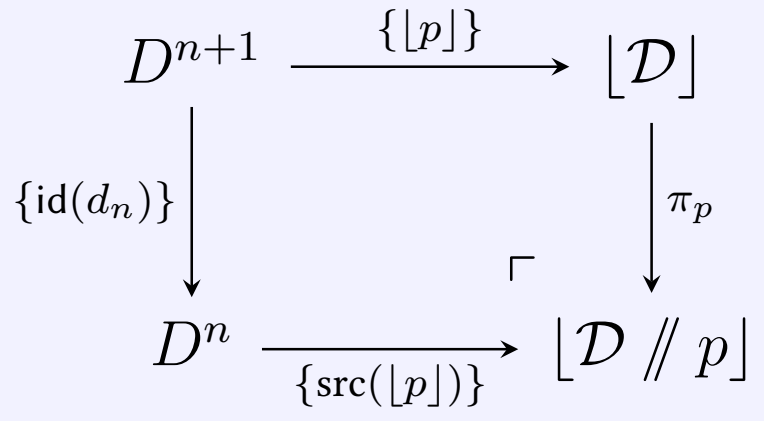

*where* $\dim(A) = n$*, and each substitution from a disc is given only by its maximal element.*

*Proof.* As discussed in Section 3.1.2, the substitution $\pi_p$ sends $\lfloor p \rfloor$ to the identity on the source of $\lfloor p \rfloor$, which makes the square commute, as it suffices to consider the action of each substitution on $d_{n+1}$, the maximal variable of $D^{n+1}$. We now assume that we have substitutions $\sigma : \lfloor \mathcal{D} \rfloor \to \Gamma$ and $\{t\} : D^n \to \Gamma$ such that the following diagram commutes:

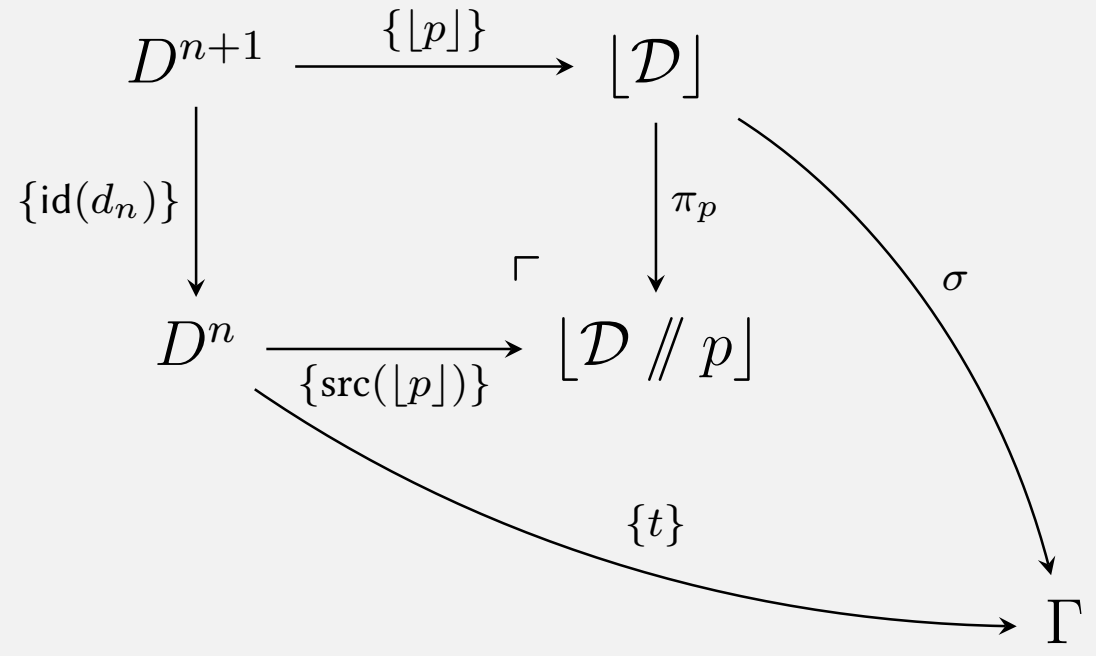

We immediately have that $\lfloor p \rfloor [\![\sigma]\!] = \mathsf{id}(\{t\})$. We can therefore let $\sigma'$ the same substitution as $\sigma$ but with $\lfloor p \rfloor [\![\sigma]\!]$ replaced by $\mathsf{id}(\{t\})$, and then can form the substitution:

$$\sigma /\!\!/ p \equiv \sigma' /\!\!/ p : \lfloor \mathcal{D} /\!\!/ p \rfloor \to \Gamma$$

By Proposition 3.1.15, we immediately have $\sigma = \sigma' = \pi_p \bullet \sigma /\!\!/ p$. The other equality follows from a diagram chase, noting that $d_n^-$ in $D^{n+1}$ is sent to the variable $d^n$ in $D^n$ by the map $\{\mathsf{id}(d_n)\}$.

It remains to show that the chosen universal map $\sigma /\!\!/ p$ is unique, but this is trivial as every variable of $\lfloor \mathcal{D} /\!\!/ p \rfloor$ is also a variable of $\lfloor \mathcal{D} \rfloor$, and so the universal map is fully determined by the substitution $\sigma$. □

**Corollary 3.4.15.** *Let $\mathcal{R}$ have insertion. Then $\mathcal{R}$ has pruning.*

*Proof.* Assume $\mathcal{R}$ has insertion. Then take a term $\mathsf{Coh}_{(\lfloor \mathcal{D} \rfloor \,;\, A)}[\sigma]$ : $\mathsf{Term}_\Gamma$ with a peak $p$ : $\mathsf{Peak}_\mathcal{D}$ such that:

$$\lfloor p \rfloor [\![\sigma]\!] \equiv \mathsf{id}(A, t)$$

for some term $t$ and type $A$ of $\Gamma$. We then need to show that:

$$\Gamma \vdash \mathsf{Coh}_{(\lfloor \mathcal{D} \rfloor\, ;\, A)}[\sigma] = \mathsf{Coh}_{(\lfloor \mathcal{D} /\!\!/ p \rfloor\, ;\, A[\![\pi_p]\!])}[\sigma /\!\!/ p]$$

From $\lfloor \mathcal{D} \rfloor$ we can obtain a tree $S$ with $\lfloor S \rfloor \equiv \lfloor \mathcal{D} \rfloor$. Further, $\lfloor p \rfloor$ is a locally maximal variable of $\lfloor \mathcal{D} \rfloor$, and so there exists a branch $P$ such that $\lfloor \bar{P} \rfloor$ is this locally maximal variable, and $\mathsf{bh}(P) = \mathsf{lh}(P) - 1$. Then the diagram:

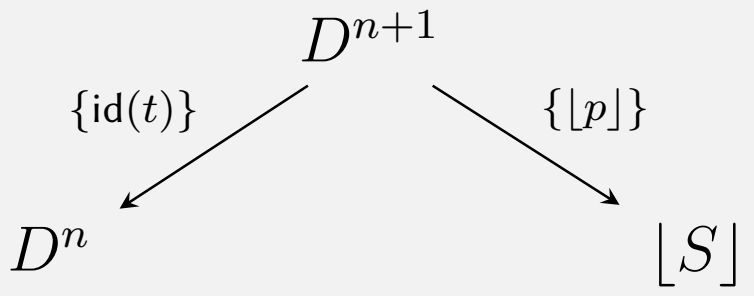

has two pushouts, the one given by insertion, and the one given by pruning. Therefore, we obtain an isomorphism $\lfloor S \ll_P D^n \rfloor \cong \lfloor \mathcal{D} /\!\!/ p \rfloor$. By Proposition 1.2.6, this isomorphism must be the identity (as both pushouts exist in Catt), and so we can deduce that $\pi_p = \kappa_{S,P,D^n}$ and $\sigma /\!\!/ p = \lfloor \lceil \sigma \rceil \ll_P \{\lceil t \rceil\} \rfloor$. Therefore, the above equality is given by an insertion along $P$. □

### 3.4.2 The insertion rule

We now prove that the insertion rule set given in Section 3.4.2 satisfies the various conditions presented in Section 2.4. We begin with the following lemma.

**Lemma 3.4.16.** *Let $(S, P, T)$ be an insertion point and let $L : S \to \mathbf{U}$ and $M : T \to \mathbf{U}$ be labellings. Let $f : \mathsf{STerm}_{\mathbf{U}} \to \mathsf{STerm}_{\mathbf{U}'}$ be any function from structured terms of $\mathbf{U}$ to structured terms of $\mathbf{U}'$. Then for any path $p$ of $S \ll_P T$ we have:*

$$f((S \ll_P T)(p)) \equiv ((f \circ L) \ll_P (f \circ M))(p)$$

*where $f \circ L$ is the result of composing $f$ to the function component of $L$.*

*Proof.* The proof of this follows by a simple induction on $P$ and is given in the formalisation module Catt.Tree.Insertion.Properties by function label-from-insertion-map. □

**Proposition 3.4.17.** *The insertion rule set, insert, satisfies the suspension condition. It further satisfies the $\mathcal{R}$-substitution condition for any rule set $\mathcal{R}$, and so also satisfies the weakening condition.*

*Proof.* Let $(S, P, T, \Gamma, L, M)$ be an insertion redex and let $A$ be a structured type of $S$, such that:

$$s \equiv \mathsf{SCoh}_{(S\,;\,A)}[L] \qquad t \equiv \mathsf{SCoh}_{(S \ll_P T\,;\,A[\![\kappa_{S,P,T}]\!])}[L \ll_P T] \qquad (\Gamma, \lfloor s \rfloor, \lfloor t \rfloor) \in \mathsf{insert}$$

To prove the suspension condition, we observe that $0 :: P$ is a branch of $\Sigma(S)$ such that $\Sigma(S) \ll_{0::P} \Sigma(T) \equiv \Sigma(S \ll_P T)$ and $\kappa_{\Sigma(S),0::P,\Sigma(T)} \equiv \Sigma(\kappa_{S,P,T})$ by definition. By applying Lemma 3.4.16 with $f = \Sigma$, we get:

$$\Sigma'(L) \ll_P \Sigma'(M) \equiv \Sigma'(L \ll_P M)$$

and so by unwrapping definitions we obtain $\Sigma(L) \ll_{0::P} \Sigma(M) \equiv \Sigma(L \ll_P M)$. Therefore, we have:

$$\begin{aligned}\Sigma(s) &\equiv \mathsf{SCoh}_{(\Sigma(S)\,;\,\Sigma(A))}[\Sigma(L)]\\ \Sigma(t) &\equiv \mathsf{SCoh}_{(\Sigma(S) \ll_{0::P} \Sigma(T)\,;\,\Sigma(A)[\![\kappa_{\Sigma(S),0::P,\Sigma(T)}]\!])}[\Sigma(L) \ll_{0::P} \Sigma(M)]\end{aligned}$$

and so as

$$\Sigma(L)(0 :: \bar{P}) \equiv \Sigma'(L)(\bar{P}) \equiv \Sigma(\mathcal{C}_T^{\mathsf{lh}(P)}[\![M]\!]) \equiv \mathcal{C}_{\Sigma(T)}^{\mathsf{lh}(0::P)}[\![\Sigma(M)]\!]$$

we get $(\Sigma(\Gamma), \Sigma(\lfloor s \rfloor), \Sigma(\lfloor t \rfloor)) \in \mathsf{insert}$ as required.

For the substitution condition we let $\sigma : \Gamma \to_\star \Delta$ be any substitution. Then:

$$\lfloor s \rfloor[\![\sigma]\!] \equiv \lfloor \mathsf{SCoh}_{(S\,;\,A)}[L \bullet \lceil \sigma \rceil] \rfloor \qquad \lfloor t \rfloor[\![\sigma]\!] \equiv \lfloor \mathsf{SCoh}_{(S \ll_P T\,;\,A[\![\kappa_{S,P,T}]\!])}[(L \ll_P T) \bullet \lceil \sigma \rceil] \rfloor$$

Again using Lemma 3.4.16, this time with $f = u \mapsto u[\![\lceil \sigma \rceil]\!]$, we have:

$$(L \ll_P M) \bullet \lceil \sigma \rceil \equiv L \bullet \lceil \sigma \rceil \ll_P M \bullet \lceil \sigma \rceil$$

Further, we have the equality:

$$(L \bullet \lceil \sigma \rceil)(\bar{P}) \equiv L(p)[\![\lceil \sigma \rceil]\!] \equiv \mathcal{C}_T^{\mathsf{lh}(P)}[\![M \bullet \lceil \sigma \rceil]\!]$$

and so $(\Gamma, \lfloor s \rfloor[\![\sigma]\!], \lfloor t \rfloor[\![\sigma]\!]) \in \mathsf{insert}$ and so $\mathsf{insert}$ satisfies the $\mathcal{R}$-substitution condition for any $\mathcal{R}$, as we made no assumption on $\sigma$ being well-formed. □

We next prove the support condition for the insertion rule set. We start with the following support lemma for the exterior labelling.

**Lemma 3.4.18.** *Let $(S \ll_P T)$ be an insertion point. Then:*

$$\mathrm{Supp}(\kappa_{S,P,T}) = \mathrm{Var}(S \ll_P T)$$

*The exterior labelling is full.*

*Proof.* Proof proceeds by induction on $P$, the only non-trivial case is $P = [k]$, where we rely on $\mathrm{Supp}(\{\mathcal{U}_T^{\mathsf{lh}P}, \mathcal{C}_T^{\mathsf{lh}P}\})$ being $\mathrm{Var}(T)$. A full proof is given in the formalisation module Catt.Tree.Insertion.Support. □

Similar to the other rule sets introduced so far, to prove the support condition for the insertion rule set, we will take an arbitrary rule set $\mathcal{R}$ that is tame and satisfies the support condition, and prove instead that the insertion set satisfies the $\mathcal{R}$-support condition. This result can be then used as part of the strategy for proving the support condition outlined in Lemma 2.4.28.

**Proposition 3.4.19.** *Let $\mathcal{R}$ be a tame equality rule set that satisfies the support condition. Then $\mathsf{insert}$ satisfies the $\mathcal{R}$-support condition.*

*Proof.* As in the previous proposition, let $(S, P, T, \Gamma, L, M)$ be an insertion redex and $A$ a structured type of $S$, such that:

$$s \equiv \mathsf{SCoh}_{(S\,;\,A)}[L] \qquad t \equiv \mathsf{SCoh}_{(S \ll_P T\,;\,A[\![\kappa_{S,P,T}]\!])}[L \ll_P T] \qquad (\Gamma, \lfloor s \rfloor, \lfloor t \rfloor) \in \mathsf{insert}$$

We now assume that $\Gamma \vdash_{\mathcal{R}} \lfloor s \rfloor : B$ for some $B$ and must prove that $\mathrm{Supp}(s) = \mathrm{Supp}(t)$. By inspecting the typing judgement, we can obtain proofs of the following typing judgements:

$$\Gamma \vdash L : S \qquad S \vdash A \qquad \Gamma \vdash M : T$$

where the typing of $M$ is obtained by transporting the typing of $L(\overline{P})$ along the syntactic equality $L(\overline{P}) \equiv \mathcal{C}_T^{\mathsf{lh}(P)}[\![M]\!]$. By Lemma 3.4.13, we have:

$$\kappa_{S,P,T} \bullet L \ll_P M \equiv^{\max} L$$

By Proposition 3.4.8, both sides of this equation are well-formed and so by Theorem 3.3.16, we obtain the equality:

$$\Gamma \vdash_{\mathcal{R}} \kappa_{S,P,T} \bullet L \ll_P M = L$$

As $\mathcal{R}$ satisfies the support property, we get:

$$\begin{aligned}
\mathrm{Supp}(s) &= \mathrm{Supp}(L) \\
&= \mathrm{Supp}(\kappa_{S,P,T} \bullet L \ll_P M) \\
&= \mathrm{Supp}(\kappa_{S,P,T})[\![L \ll_P M]\!] \\
&= \mathrm{Var}(S \ll_P T)[\![L \ll_P M]\!] && \text{by Lemma 3.4.18} \\
&= \mathrm{Supp}(L \ll_P M) \\
&= \mathrm{Supp}(t)
\end{aligned}$$

and so $\mathrm{Supp}(\lfloor s \rfloor) = \mathrm{Supp}(\lfloor t \rfloor)$ as required. □

Similarly to the situation in pruning, we are not able to show that the type $A[\![\kappa]\!]$ is a valid operation without knowing more about the set of operations $\mathcal{O}$. We therefore introduce the following additional condition on the set of operations.

**Definition 3.4.20.** An operation $\mathcal{O}$ *supports insertion* if for all insertion points $(S, P, T)$ and variable sets $U, V \subseteq \mathrm{Var}(S)$ we have:

$$(\lfloor S \ll_P T \rfloor, \lfloor U[\![\kappa_{S,P,T}]\!] \rfloor, \lfloor V[\![\kappa_{S,P,T}]\!] \rfloor) \in \mathcal{O}$$

whenever $(\lfloor S \rfloor, U, V) \in \mathcal{O}$

Using this property, we can give the preservation condition for the insertion rule set.

**Proposition 3.4.21.** *Let $\mathcal{R}$ be a tame equality rule set and suppose the operation set $\mathcal{O}$ supports insertion. Then the set* insert *satisfies the $\mathcal{R}$-preservation condition.*

*Proof.* Let $(S, P, T, \Gamma, L, M)$ be an insertion redex and let $a \to_A b$ be a structured type such that:

$$s \equiv \mathsf{SCoh}_{(S\,;\,a \to_A b)}[L] \qquad t \equiv \mathsf{SCoh}_{(S \ll_P T\,;\,(a \to_A b)[\![\kappa_{S,P,T}]\!])}[L \ll_P T] \qquad (\Gamma, \lfloor s \rfloor, \lfloor t \rfloor) \in \mathsf{insert}$$

We now suppose that $\Gamma \vdash \lfloor s \rfloor : B$ and aim to prove that $\Gamma \vdash \lfloor t \rfloor : B$. By inspecting the typing derivation we get:

$$S \vdash a \to_A b \qquad \Gamma \vdash L : S \qquad \Gamma \vdash M : T \qquad (\lfloor S \rfloor, \mathrm{Supp}(a), \mathrm{Supp}(b)) \in \mathcal{O}$$
$$\Gamma \vdash (a \to_A b)[\![L]\!] = A$$

and so by Proposition 3.4.8 we have:

$$S \ll_P T \vdash \kappa_{S,P,T} : S \qquad \Gamma \vdash L \ll_P M : S \ll_P T$$

As the operation set supports insertion with $\mathrm{Supp}(a[\![\kappa]\!]) = \mathrm{Supp}(a)[\![\kappa]\!]$ and $\mathrm{Supp}(b[\![\kappa]\!]) = \mathrm{Supp}(b)[\![\kappa]\!]$ we get:

$$(\lfloor S \ll_P T \rfloor, \mathrm{Supp}(a[\![\kappa]\!]), \mathrm{Supp}(b[\![\kappa]\!]))$$

and so we obtain:

$$\Gamma \vdash \mathsf{SCoh}_{(S \ll_P T\,;\,(a \to_A b)[\![\kappa]\!])}[L \ll_P M] : (a \to_A b)[\![\kappa]\!][\![L \ll_P M]\!]$$

By Lemma 3.4.13 and Theorem 3.3.16, $\Gamma \vdash \kappa \bullet L \ll_P M = L$, and so:

$$\begin{aligned}(a \to_A b)[\![\kappa]\!][\![L \ll_P M]\!] &\equiv (a \to_A b)[\![\kappa \bullet (L \ll_P M)]\!] \\ &= (a \to_A b)[\![L]\!] \\ &= B\end{aligned}$$

and so by applying the conversion rule we obtain $\Gamma \vdash \lfloor t \rfloor : B$ as required. □

### 3.4.3 Further properties

It has now been proved that insertion can form part of a reasonable type theory. We now proceed to prove further properties of the insertion construction that will be critical for proving the confluence of $\textsc{Catt}_{\mathsf{sua}}$ in Section 4.3. The majority of these properties will therefore concern the interaction of insertion with other constructions and itself. We will justify each property with up to three of the following methods:

- For each property, we will give a graphical depiction of the constructions involved, similar to the diagram for Proposition 3.1.22, which should help build intuition for the constructions at play.
- Where applicable, each combination of constructions will be described using the universal property from Section 3.4.1. This can be used to classify these constructions up to definitional equality.
- As these properties are used in a confluence proof, we will need a more syntactic form than can be offered by the universal property approach. To do this we fall back to the formalisation, using the computation power of structured terms to brute force each property.

The first two properties we consider concern the interaction of insertion with disc contexts, and will be crucial for proving confluence cases involving insertion and disc removal. Disc contexts often admit insertions, and the disc acts as a left and right unit for the insertion operation.

**Insertion into a disc** We begin by considering insertions into a disc. A disc context has a branch of height $0$, and so if the locally maximal variable is sent to a standard coherence, then insertion can always be preformed. Inserting into a disc effectively performs disc removal, replacing the entire disc with the entirety of the inner context. We illustrate this by the following diagram, where we take the branch $[0,0]$ of $D^4$ (which we note is not the minimal branch).

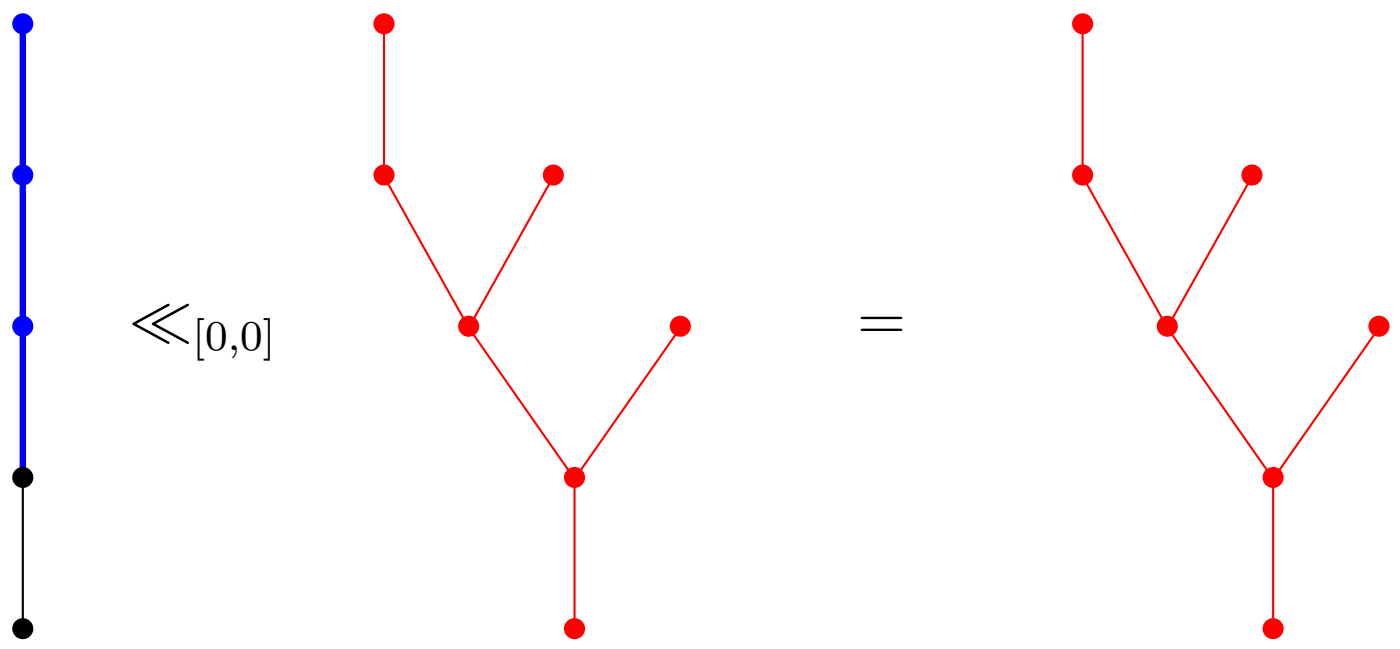

This property of insertion also has a simple proof by universal property. Suppose we have disc $D^n$ with a branch $P$ and we insert tree $T$. Then the inserted tree is given by the following pushout.

$$\begin{array}{ccc} D^n & \xrightarrow{\mathsf{id}} & D^n \\ {\scriptstyle \{\mathcal{C}^n_T\}}\downarrow & & \downarrow{\scriptstyle \kappa} \\ T & \xrightarrow[\iota]{} & D^n \ll_P T \end{array}$$

By standard properties of pushouts, we have that $D^n \ll_P T$ is isomorphic to $T$. As this pushout holds in Catt, we have a Catt isomorphism between pasting contexts and so by Proposition 1.2.6, $T = D^n \ll_P T$, $\iota = \mathsf{id}$. The following lemma gives syntactic versions of these properties.

**Lemma 3.4.22.** *Let $T$ be a tree, $n \geq \mathsf{dim}(T)$, and $P$ a branch of $D^n$ with $\mathsf{bh}(P) \leq \mathsf{th}(T)$. Then $D^n \ll_P T = T$ and $\iota_{D^n,P,T} \equiv \mathsf{id}$. Suppose further that $(D^n, P, T, \Gamma, L, M)$ is an insertion redex. Then $L \ll_P M \equiv M$.*

*Proof.* See the functions disc-insertion, disc-ι, and disc-label-from in formalisation module Catt.Tree.Insertion.Properties. □

**Insertion of a disc** We now consider the opposite situation, where a disc context is inserted into an arbitrary tree. For a tree $T$, with a branch $P$, we can always insert the disc context $D^{\mathsf{lh}(P)}$, as the trunk height condition will be satisfied by the linearity of the disc context. Inserting such a disc context makes no change to the tree $T$, as the operation effectively replaces a branch of $T$ (which is linear by construction) by a disc. The diagram below depicts this construction.

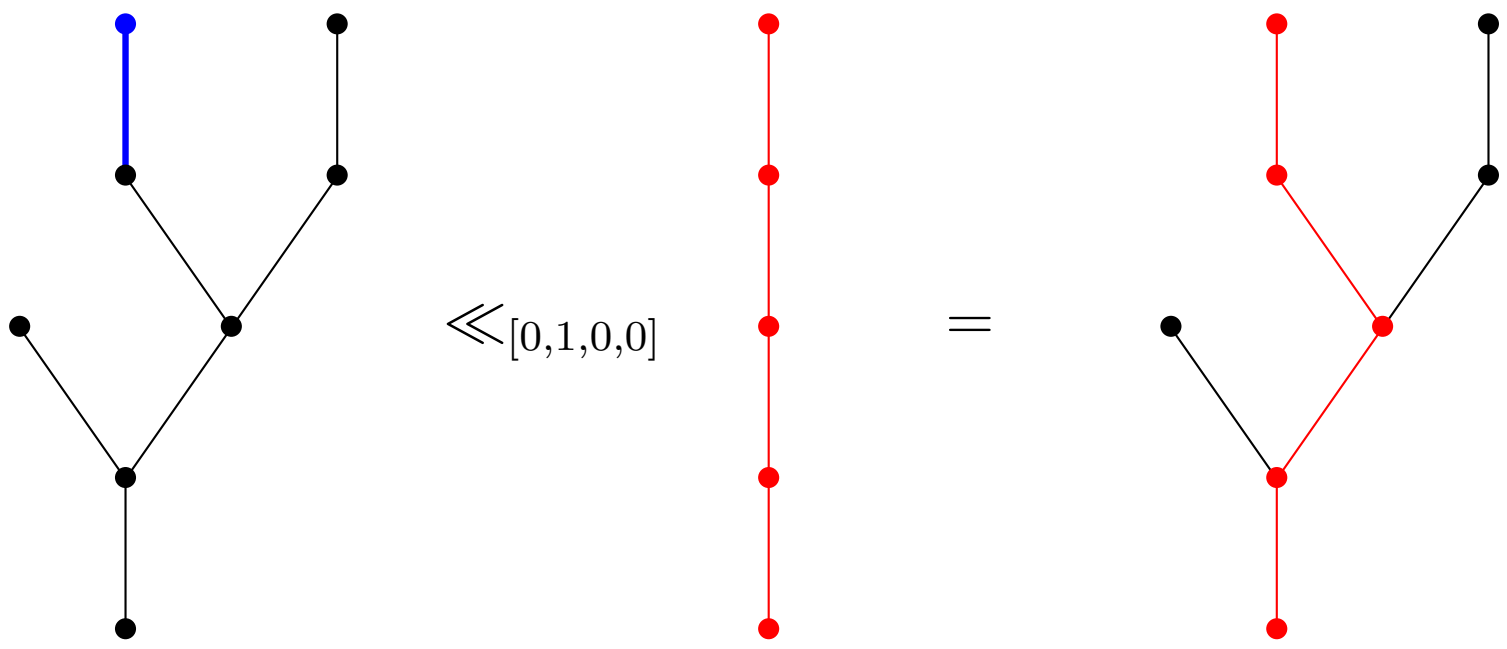

Similar to the insertion into the disc, the insertion of a disc can be characterised by universal property. Take any tree $T$ with a branch $P$. Then the tree $T \ll_P D^{\mathsf{lh}(P)}$ is the following pushout:

$$\begin{array}{ccc} D^n & \xrightarrow{\{\bar{P}\}} & T \\ {\scriptstyle \{\mathcal{C}^n_{D^n}\}}\downarrow & & \downarrow{\scriptstyle \kappa} \\ D^n & \xrightarrow[\iota]{} & T \ll_P D^n \end{array}$$

The situation here is less clear than before, as the map $D^n \to D^n$ is not the identity. However, in the presence of disc removal this map becomes equal to the identity, and in this case a similar argument can be made to determine that $\kappa$ should be the identity and $T \ll_P D^{\mathsf{lh}(P)}$ should be equal to the tree $T$. The results are given in the lemma below:

**Lemma 3.4.23.** *Let* $(T, P, D^{\mathsf{lh}(P)}, \Gamma, L, M)$ *be an insertion redex. Then:*

$$T \ll_P D^{\mathsf{lh}(P)} \equiv S \qquad L \ll_P M \equiv^{\mathsf{max}} L$$

*We further have:*

$$S \vdash_{\mathcal{R}} \kappa_{S,P,D^{\mathsf{lh}(P)}} =^{\mathsf{max}} \mathsf{id}_S$$

*if* $\mathcal{R}$ *is a (tame) equality rule set which has disc removal.*

*Proof.* See the functions insertion-disc and disc-label-from-2 in the formalisation module Catt.Tree.Insertion.Properties and κ-disc in Catt.Typing.Insertion.Equality. □

**Insertion of an endo-coherence** We now turn our attention to the interaction between insertion and endo-coherence removal. Unlike in $\text{Catt}_{\text{su}}$, the locally maximal argument in an insertion redex need not be in normal form. In particular, since the only condition on the locally maximal argument is that it is a standard coherence, it may be an endo-coherence. In such a situation there are two distinct ways of applying equalities:

- The endo-coherence could be directly inserted into the head term.

- The endo-coherence could be transformed into an identity on a standard coherence (see Theorem 3.3.27) after which the head term could undergo two insertions, the first of which "prunes" the identity, and the second of which inserts the locally maximal argument of the pruned identity.

As the insertion of an identity acts in a similar way to pruning (see Corollary 3.4.15), we re-use the notation.

**Definition 3.4.24.** Let $S$ be a tree, and $P$ be a branch of $S$. Then define:

$$S \mathbin{/\!\!/} P = S \ll_P D^{\mathsf{lh}(P)-1} \qquad \pi_P = \kappa_{S,P,D^{\mathsf{lh}(P)-1}}$$

where we note that $(S, P, D^{\mathsf{lh}(P)-1})$ is always an insertion point.

To perform the second equality path of pruning an identity followed by inserting the maximal argument of that identity, we must obtain a branch of the pruned context $S \mathbin{/\!\!/} P$. This can be done when $\mathsf{lh}(P) - \mathsf{bh}(P) \geq 2$ by taking the same list as $P$, as depicted in Figure 3.6. We name such a branch the *pruned branch*.

**Definition 3.4.25.** Let $S$ be a tree, and $P$ be a branch of $S$ with $\mathsf{lh}(P) - \mathsf{bh}(P) \geq 2$. We then define the *pruned branch* $P'$ of $S \mathbin{/\!\!/} P$ to be given by the same list as $P$.

If $\mathsf{lh}(P) - \mathsf{bh}(P) = 1$ (noting that $\mathsf{lh}(P) - \mathsf{bh}(P)$ cannot be zero) then pruning the branch $P$ removes the branch entirely, and so the condition $\mathsf{lh}(P) - \mathsf{bh}(P) \geq 2$ is necessary to form the pruned branch. It is clear that $\mathsf{bh}(P') = \mathsf{bh}(P)$ and $\mathsf{lh}(P') = \mathsf{lh}(P) - 1$.

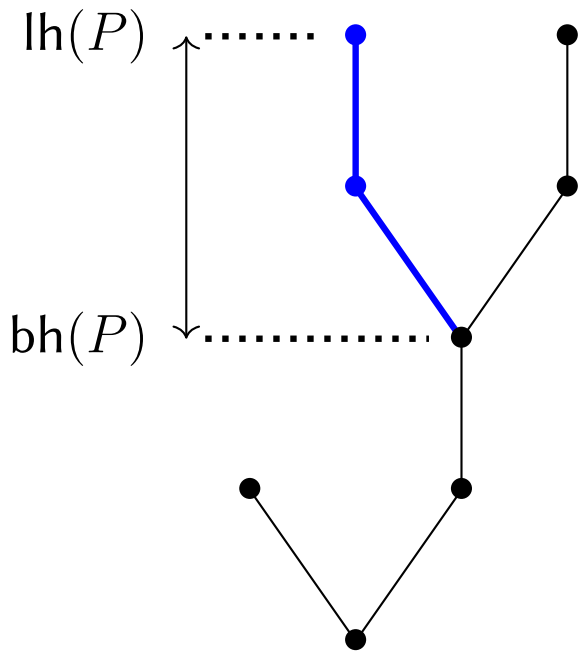

(a) Tree $S$ and branch $P = [1, 0, 0]$.

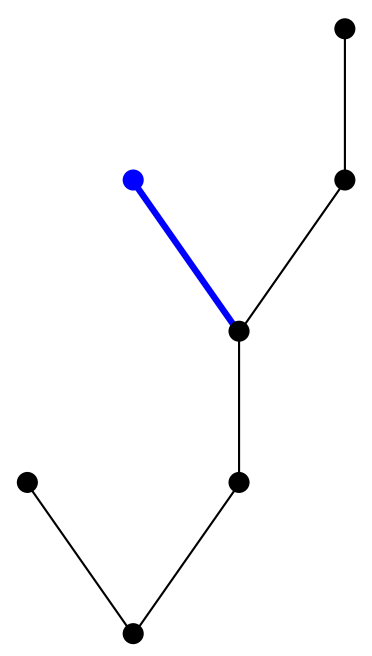

(b) Tree $S \mathbin{/\!\!/} P$ and branch $P' = [1, 0, 0]$.

Figure 3.6: The pruned branch.

We also note that the path $\overline{P'}$ is the maximal argument of the labelling $\iota_{S,P,D^{\mathsf{lh}P-1}}$, the inclusion of $D^{\mathsf{lh}(P)-1}$ into $S \mathbin{/\!\!/} P$. Insertion along the pruned branch is then characterised by the following pushout.

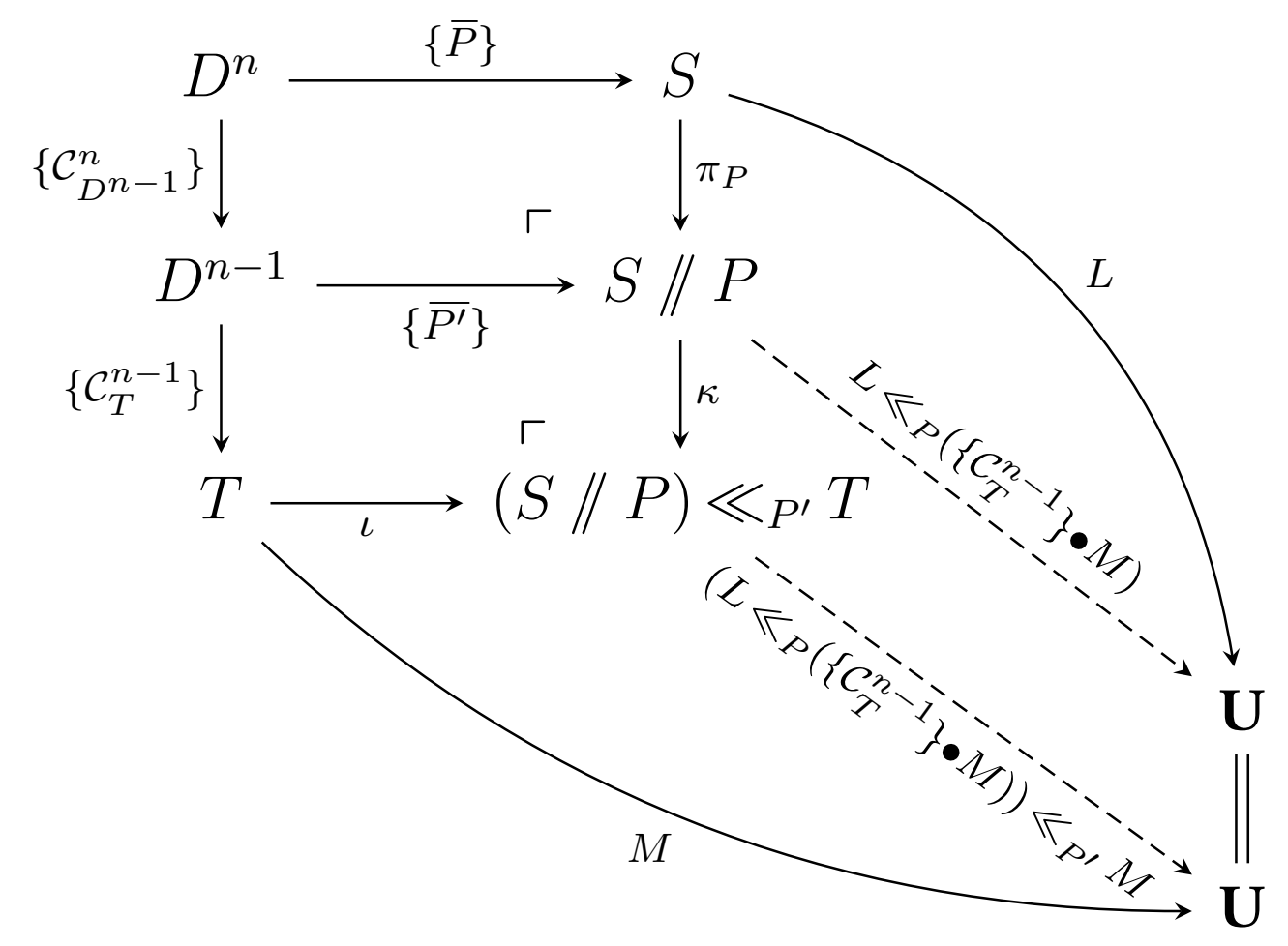

The top pushout is from the construction of $S /\!\!/ P$, noting that $\iota_{S,P,D^{\mathsf{lh}(P)-1}} = \{\overline{P'}\}$. The bottom pushout is from the construction of the insertion along the pruned branch. By the pasting lemma for pushouts, the whole outer rectangle is also a pushout along the maps $\{\hat{P}\}$ and $\{\mathcal{C}^n_{D^{n-1}}\} \bullet \{\mathcal{C}^{n-1}_T\}$. In the presence of endo-coherence removal we have:

$$\{\mathcal{C}^n_{D^{n-1}}\} \bullet \{\mathcal{C}^{n-1}_T\} = \{\mathcal{C}^n_T\}$$

by Theorem 3.3.27 and so the outer pushout rectangle is the pushout generated by directly inserting the endo-coherence. There are two ways to form the unique map $(S /\!\!/ P) \ll_{P'} T \to \mathbf{U}$, one by the outer pushout rectangle that gives the map $L \ll_P M$, and one by first using the top pushout square with the maps $L$ and $\{\mathcal{C}^{n-1}_T\} \bullet M$ to get a map $S /\!\!/ P \to \mathbf{U}$, and then using this map with the bottom pushout square and $M$ to get the morphisms depicted in the commutative diagram.

These results appear in the next lemma.

**Lemma 3.4.26.** *Suppose $S$ has branch $P$ with $\mathsf{lh}(P) - \mathsf{bh}(P) \geq 2$. Then $\iota_{S,P,D^{\mathsf{lh}(P)-1}} \equiv \{\overline{P'}\}$. Further suppose that $(S, P, T)$ is an insertion point. Then if the (tame) rule set $\mathcal{R}$ has disc removal and endo-coherence removal we get:*

$$(S /\!\!/ P) \ll_{P'} T = S \ll_P T \qquad \mathbf{U} \vdash_{\mathcal{R}} \pi_P \bullet \kappa_{S /\!\!/ P, P', T} =^{\max} \kappa_{S,P,T}$$

*If we further have that $(S, P, T, \mathbf{U}, L, M)$ is an insertion redex then:*

$$(L \ll_P (\{\mathcal{C}^{\mathsf{lh}(P)-1}_T\} \bullet M)) \ll_{P'} M \equiv^{\max} L \ll_P M$$

*Proof.* See the functions insertion-tree-pruned-branch, pruned-branch-prop, and label-from-pruned-branch in formalisation module Catt.Tree.Insertion.Properties, and pruned-branch-κ in Catt.Typing.Insertion.Equality. □

**Branch irrelevance** As has already been noted, a tree $S$ may admit multiple branches $P$ and $Q$ which represent the same locally maximal variable, that is $\overline{P} \equiv \overline{Q}$. If there is an insertion that can be applied along either branch $P$ or $Q$ then it does not matter which branch we choose.

This can be immediately seen by the universal property: The pushout square for an insertion point $(S, P, T)$ only mentions the path $\overline{P}$ and never uses the actual branch $P$.

**Lemma 3.4.27.** *Suppose $(S, P, T)$ and $(S, Q, T)$ are insertion points with $\overline{P} \equiv \overline{Q}$. Then $S \ll_P T \equiv S \ll_Q T$ and $\kappa_{S,P,T} \equiv^{\max} \kappa_{S,Q,T}$. If we further have $L : S \to \Gamma$ and $M : T \to \Gamma$, then $L \ll_P M \equiv^{\max} L \ll_Q M$.*

*Proof.* See the functions insertion-irrel, κ-irrel, and irrel-label-from in formalisation module Catt.Tree.Insertion.Properties. □

It is natural to ask why we define branches at all, and don't identify points where insertion can be performed by a maximal path, implicitly taking the branch of minimal branching height. While this could be done, it would make other confluence cases more difficult, as the branch associated to a maximal path could significantly change if a different branch is pruned from the tree.

**Parallel insertion** We now begin to consider the interaction between insertion and itself. In contrast to the previous case, we now consider two branches $P$ and $Q$ such that $\overline{P}$ and $\overline{Q}$ are not the same maximal path, in which case we say the branches $P$ and $Q$ are *parallel*. Assume we have a tree $S$ such that $(S, P, T)$ and $(S, Q, U)$ are insertion points. We then aim to perform both insertions, and prove that the order they occur in is irrelevant. To do this we must form a branch of the inserted tree $S \ll_P T$, which is intuitively given by the branch $Q$, but such a branch must be adapted to the new inserted tree.

**Definition 3.4.28.** Let $(S, P, T)$ be an insertion point and let $Q$ be a branch of $S$ such that $\overline{P} \neq \overline{Q}$. Then we define the branch $Q \ll_P T$ of $S \ll_P T$ by induction on $P$ and $Q$.

- Suppose $P = [k]$ and $Q = j :: x$. Then if $j < k$ we let $Q \ll_P T = Q$. Otherwise, we let:
$$Q \ll_P T = (j + \text{len}(T) - 1) :: x$$
- Suppose $P = k :: P_2$ and $Q = j :: x$. If $j \neq k$ then let $Q \ll_P T = Q$. Otherwise, both $P_2$ and $x$ are branches of $S_k$ and so we let
$$Q \ll_P T = k :: x \ll_{P_2} T$$

It is clear that $Q \ll_P T$ satisfies the condition for being a branch.

The maximal path associated to the branch $Q \ll_P T$ is obtained by applying the labelling $\kappa$ to the maximal path associated to $Q$. That is:

$$\overline{Q \ll_P T} \equiv \overline{Q}[\![\kappa_{S,P,T}]\!]$$

A graphical example of such a situation is given in Figure 3.7 where we note how the right branch changes after the left-hand insertion is performed. We also note that the final trees at the bottom of the diagram are coloured slightly differently, which corresponds to the inserted labellings from these trees being different. To remedy this, we introduce a variant of the inserted labelling, which takes arguments from the head labelling instead of the argument labelling wherever possible.

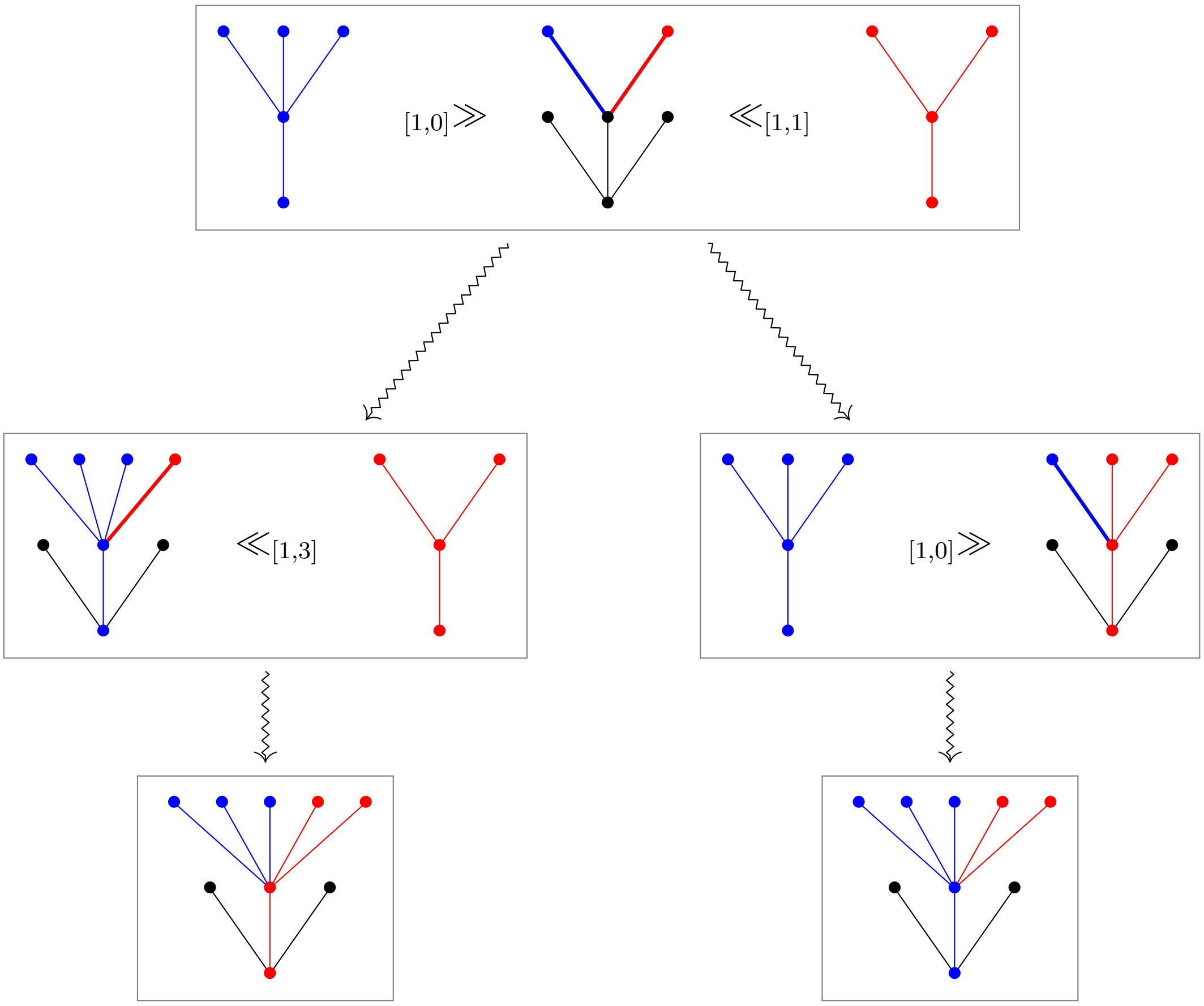

Figure 3.7: Parallel insertions.

**Definition 3.4.29.** We define an alternative to the inserted labelling as follows: Given an insertion point $(S, P, T)$ with $L : S \to \mathbf{U}$ and $M : T \to \mathbf{U}$ we define this *alternative inserted labelling* $L \ll'_P M : S \ll_P T \to \mathbf{U}$. Let

$$S = [S_0, \dots, S_n] \qquad L = s_0\{L_0\}s_1 \cdots \{L_n\}s_{n+1} : A$$

and then proceed by induction on $P$.

- Let $P = [k]$, and

  $$T = [T_0, \dots, T_m] \qquad M = t_0\{M_0\}t_1 \cdots \{M_m\}t_{m+1} : B$$

  Then define $L \ll_{[k]} M$ to be:

  $$s_0\{L_0\}s_1 \cdots \{L_{k-1}\}\mathbf{s_k}\{M_0\}t_1 \cdots \{M_m\}\mathbf{s_{k+1}}\{L_{k+1}\}s_{k+2} \cdots \{L_n\}s_{n+1} : A$$

- Suppose $P = k :: Q$ so that

  $$T = [T_0] \qquad M = t_0\{M_0\}t_1 : B$$

  Define $L \ll_P M$ as:

  $$s_0\{L_0\}s_1 \cdots \{L_{k-1}\}\mathbf{s_k}\{L_k \ll_Q M_0\}\mathbf{s_{k+1}}\{L_{k+1}\}s_{k+2} \cdots \{L_n\}s_{n+1} : A$$

The terms that differ from the regular inserted labelling are written in bold. In the edge case where $M = [\,]$, we arbitrarily use $s_k$ instead of $s_{k+1}$ for the definition of $L \ll'_{[k]} M$.

It is immediate that the alternative inserted labelling only differs up to definitional equality.

**Proposition 3.4.30.** *Let $(S, P, T, \mathbf{U}, L, M)$ be an insertion redex. Then:*

$$L \ll'_P M = L \ll_P M$$

*Proof.* See function label-from-insertion-eq in the module Catt.Tree.Insertion.Typing. □

We now examine the universal property of parallel insertion. This is given by the following diagram, where we insert along $P$ first, followed by $Q$, letting $n = \mathsf{lh}(P)$ and $m = \mathsf{lh}(Q)$.

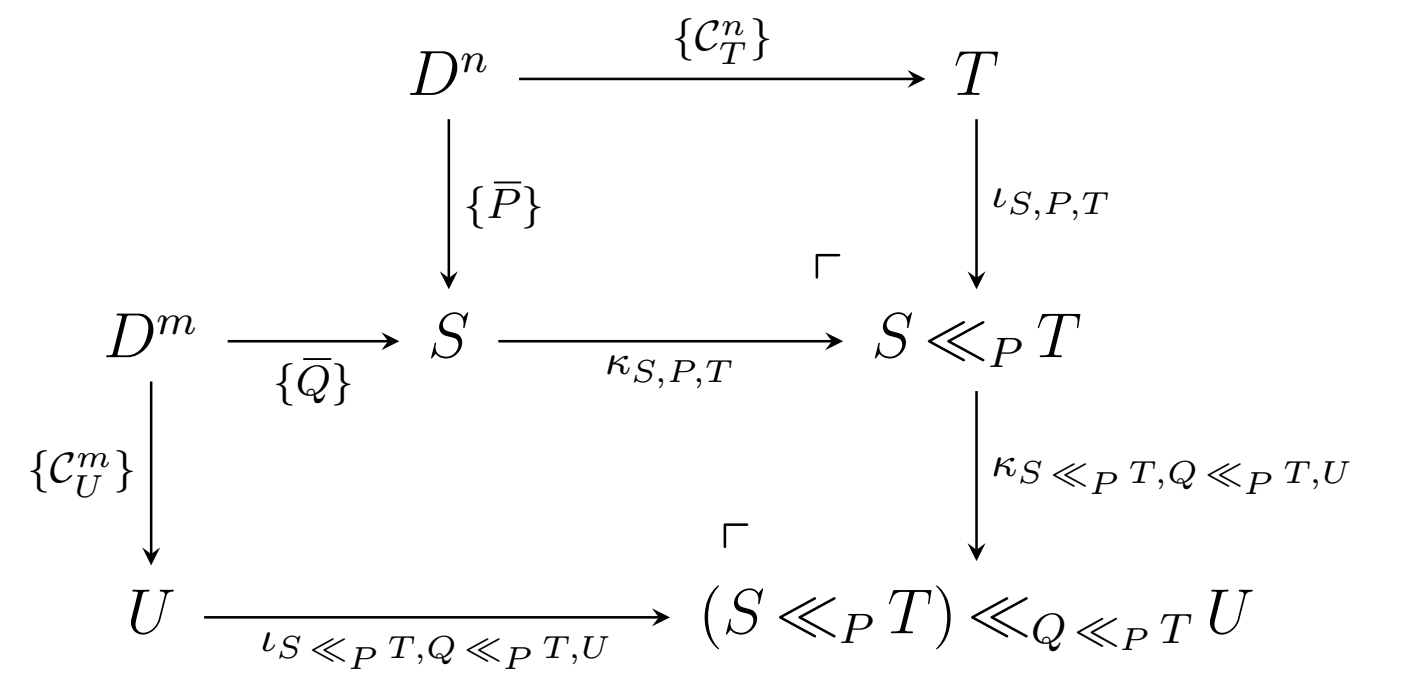

Here, the top pushout square is given by the insertion along $P$, and the bottom square is given by the insertion along $Q \ll_P T$, noting that:

$$\{\bar{Q}\} \bullet \kappa_{S,P,T} \equiv \{\overline{Q \ll_P T}\}$$

The construction is therefore given by the colimit of the top-left border of the diagram. By a symmetric argument, it can be seen that performing the insertions in the opposite order also leads to a colimit of the same diagram. We state the lemma that formally states these ideas.

**Lemma 3.4.31.** *Let $(S, P, T)$ and $(S, Q, U)$ be insertion points such that $\bar{P} \not\equiv \bar{Q}$. Then we have:*

$$(S \ll_P T) \ll_{Q \ll_P T} U \equiv (S \ll_Q U) \ll_{P \ll_Q U} T$$
$$\kappa_{S,P,T} \circ \kappa_{S \ll_P T, Q \ll_P T, U} \equiv^{\max} \kappa_{S,Q,U} \circ \kappa_{S \ll_Q U, P \ll_Q U, T}$$

*Further:*

$$(L \ll_P M) \ll'_{Q \ll_P T} N \equiv^{\max} (L \ll_Q N) \ll'_{P \ll_Q U} M$$

*for any insertion redexes $(S, P, T, \mathbf{U}, L, M)$ and $(S, P, T, \mathbf{U}, L, N)$.*

*Proof.* See functions insertion-parallel, κ-parallel, and label-from-parallel in formalisation module Catt.Tree.Insertion.Properties. □

**Boundaries of inserted trees** We now work towards the most complex property of insertion, the action of insertion on an insertable argument. To do this, we must first understand the action of insertion on standard coherences, which itself requires an understanding of how insertion interacts with the boundary inclusion maps of trees.

There are two fundamental cases for the boundary of an inserted tree:

- The boundary has low enough dimension such that it is unaffected by the insertion. In this case applying the boundary to the inserted tree is the same as applying the boundary to the original tree.
- The boundary has sufficient dimension such that the boundary of the original tree still contains the insertion branch. In this case applying the boundary to the inserted tree is the same as inserting into the boundary of the original tree along this branch.

We begin with the first case. Suppose we have an insertion point $(S, P, T)$ and a dimension $n \in \mathbb{N}$. The main criterion for the boundary having no interaction with the insertion is that:

$$n \leq \mathsf{th}(T)$$

When this condition holds, taking the $n$-boundary of $T$ returns a linear tree, and we have already seen that inserting linear trees has no effect on the head tree. We illustrate this case in the diagram below, where the tree $T$ has trunk height 3 and we set $n = 2$. The dashed line represents taking the boundary operation, and it is easy to see that the two boundary of $S$ and the insertion tree $S \ll_P T$ are the same.

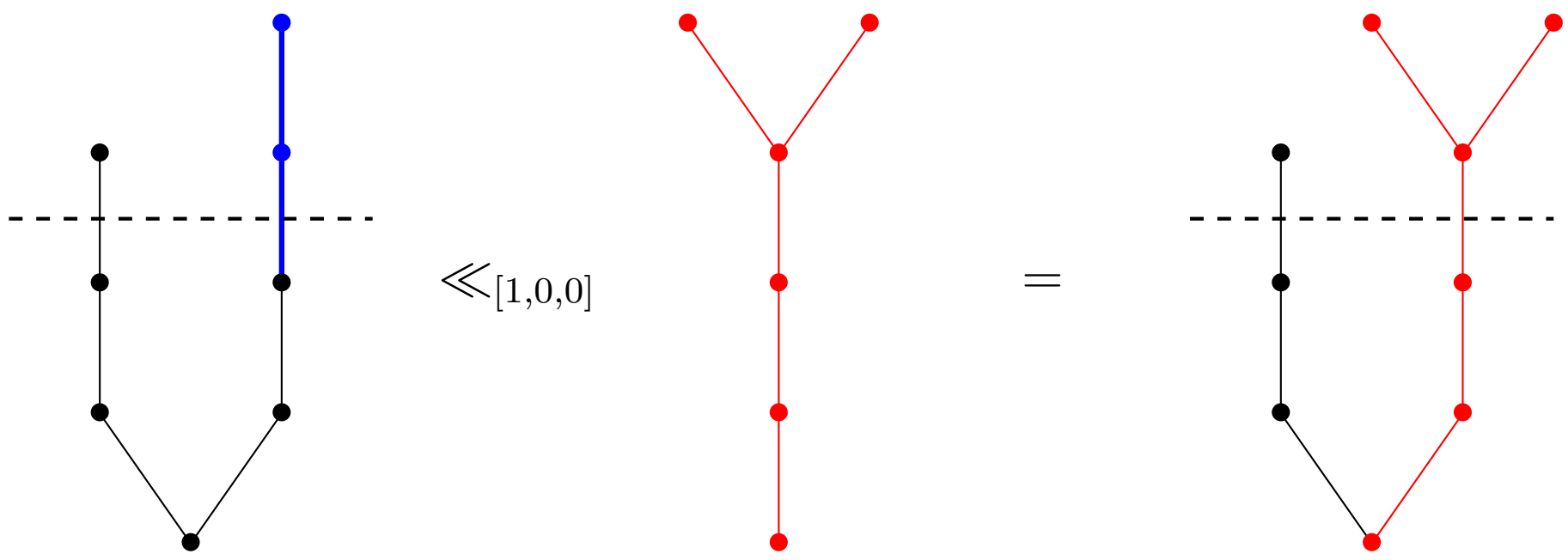

As well as knowing about the interaction of the boundary with the inserted tree, we also need to investigate the interaction of the inclusion maps with the exterior labelling. In this first case, we would hope to prove that:

$$\delta_d^-(S) \bullet \kappa_{S,P,T} \equiv \delta_d^-(S \ll_P T)$$

Now that $\partial_n(S \ll_P T) \equiv \partial_n(S)$, there are two ways to encode the source inclusion $\partial_d(S)$ into $S \ll_P T$. The right-hand side of the above equation directly includes $\partial_d(S \ll_P T)$ into the source of $S \ll_P T$, and the left-hand side first includes $\partial_d(S)$ into the source of $S$ and then projects $S$ onto $S \ll_P T$ via the exterior labelling.

There is a catch with proving this equality; The exterior labelling sends $\overline{P}$ to the standard coherence, and so if $\delta_d^-(S)$ has $\overline{P}$ in its image, the equality cannot hold syntactically. We therefore further require that $d < \mathsf{lh}(P)$, which ensures this cannot happen. We now state these results in the following lemma.

**Lemma 3.4.32.** *Let $n \in \mathbb{N}$ and suppose $(S, P, T)$ is an insertion point such that $n \leq \mathsf{th}(T)$. Then:*

$$\partial_n(S) \equiv \partial_n(S \ll_P T)$$

*If we further have $n < \mathsf{lh}(P)$ then:*

$$\delta_n^\epsilon(S) \circ \kappa_{S,P,T} \equiv^{\mathsf{max}} \delta_n^\epsilon(S \ll_P T)$$

*for $\epsilon \in \{-, +\}$.*

*Proof.* See the functions insertion-bd-1 and bd-κ-comm-1 in the formalisation module Catt.Tree.Insertion.Properties. □

We now move to the second case. We again suppose we have an insertion point $(S, P, T)$ and dimension $n \in \mathbb{N}$. To perform an insertion into the boundary $\partial_n(S)$, the dimension $n$ must be high enough not to remove the branch $P$ from $S$. More specifically, we must have the inequality:

$$n > \mathsf{bh}(P)$$

which ensures that the list $P$ is still a branch of $\partial_n(S)$.

**Definition 3.4.33.** Let $S$ be a tree with a branch $P$, and let $n > \mathsf{bh}(P)$. Then there is a branch $\partial_n(P)$ of $\partial_n(S)$ given by the same list as $P$ with $\mathsf{bh}(\partial_n(P)) = \mathsf{bh}(P)$.

As $\mathsf{th}(\partial_n(T)) \geq \mathsf{bh}(P)$ when $\mathsf{th}(T) \geq \mathsf{bh}(P)$ and $n > \mathsf{bh}(P)$, we are able to insert the tree $\partial_n(T)$ into $\partial_n(S)$ along the branch $\partial_n(P)$. This is depicted in the following diagram, where $\mathsf{bh}(P) = 2$ and $n = 3$. In this diagram, the insertion $S \ll_P T$ is drawn, and dashed lines are drawn across each tree where they would be truncated by the boundary operation. Crucially, the branch is still well-formed under this line, and preforming the insertion on the truncated trees yields the truncation of the inserted tree.

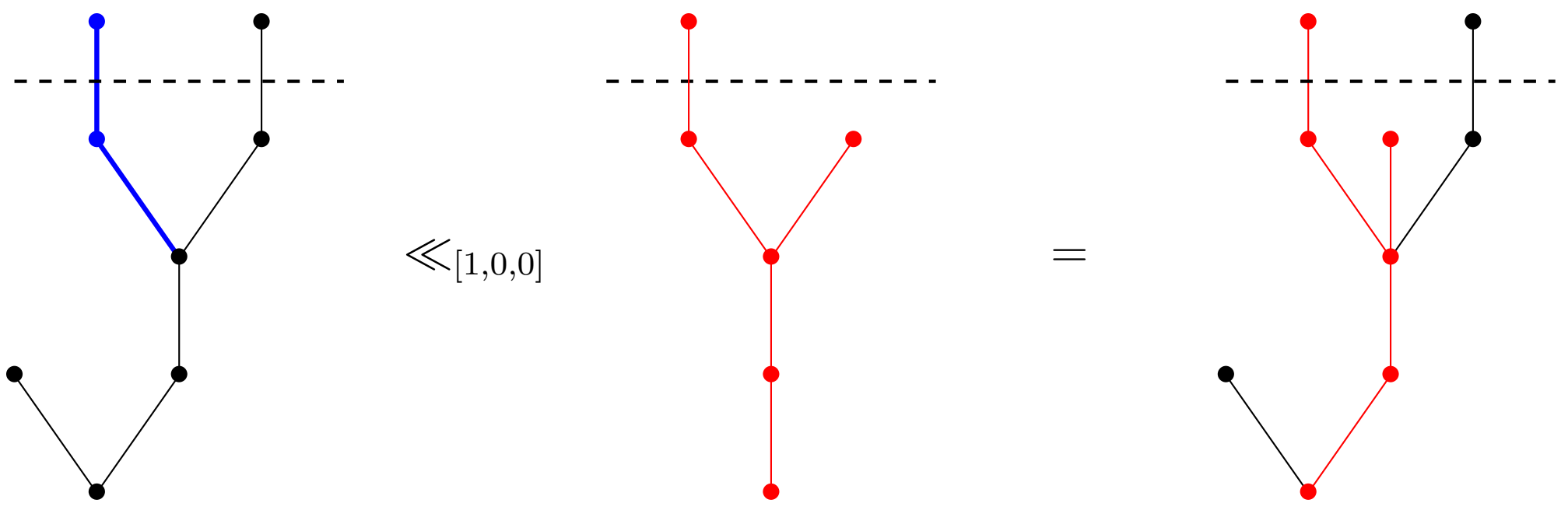

As with the previous case, we explore the interaction of the boundary inclusion labellings and the exterior labelling. We aim to give conditions under which:

$$\delta_n^-(S) \bullet \kappa_{S,P,T} \equiv \kappa_{\partial_n(S),\partial_n(P),\partial_n(T)} \bullet \delta_n^-(S \ll_P T)$$

We examine the action of each side of the equation on the path $\overline{\partial_n(P)}$. On the right-hand side, this path is sent by $\kappa$ to a standard coherence, and so on the left-hand side, $(\delta_n^-(S))(\overline{\partial_n(P)})$ must also be sent to a standard coherence by $\kappa$. If $(\delta_n^-(S))(\overline{\partial_n(P)})$ is a maximal path, which will always be the case when $n \geq \mathsf{lh}(P)$, then it will be sent to a standard coherence. Alternatively, if $n \leq \mathsf{lh}(P)$ then $\mathsf{lh}(\partial_n(P)) = n$ and if $n > \mathsf{th}(T)$ then the standard term returned by $\kappa_{S,P,T}$ will be a standard coherence. These conditions lead to the following lemma.

**Lemma 3.4.34.** *Let $n \in \mathbb{N}$ and suppose $(S, P, T)$ is an insertion point with $n > \mathsf{bh}(P)$. Then:*

$$\partial_n(S) \ll_{\partial_n(P)} \partial_n(T) \equiv \partial_n(S \ll_P T)$$

*Suppose further that one of the following holds:*

1. $n > \mathsf{th}(T)$ *and* $n \leq \mathsf{lh}(P)$
2. $n \geq \mathsf{lh}(P)$

*Then:*

$$\delta_n^\epsilon(S) \bullet \kappa_{S,P,T} \equiv^{\mathsf{max}} \kappa_{\partial_n(S),\partial_n(P),\partial_n(T)} \bullet \delta_n^\epsilon(S \ll_P T)$$

*for $\epsilon \in \{-, +\}$.*

*Proof.* See the functions insertion-bd-2 and bd-κ-comm-2 in the formalisation module Catt.Tree.Insertion.Properties. □

Both of the further conditions in Lemma 3.4.34 imply that $n > \mathsf{bh}(P)$. We have therefore seen 3 conditions that can be put on $n$, $P$, and $T$:

- $n \leq \mathsf{th}(T)$ and $n < \mathsf{lh}(P)$,
- $n > \mathsf{th}(T)$ and $n \leq \mathsf{lh}(P)$,
- $n \geq \mathsf{lh}(P)$.

One of these conditions must always hold for any $n$ and insertion point $(S, P, T)$, and hence one of Lemmas 3.4.32 and 3.4.34 can always be applied.

*Remark* 3.4.35. The further conditions in each of Lemmas 3.4.32 and 3.4.34 could be dropped in favour of weakening the syntactic equalities to definitional equalities in a theory with disc removal, as this would remove the distinction between standard terms and standard coherences. It was however more convenient to take this approach in the formalisation, and although the extra side conditions may seem arbitrary, the key result is that one of the above lemmas always holds.

**Insertion into standard constructions** Equipped with Lemmas 3.4.32 and 3.4.34, we can now prove that the standard constructions are preserved by applying an exterior labelling up to a definitional equality containing insertion and disc removal. We begin with the following lemma, whose intuition is clear from the universal property of insertion.

**Lemma 3.4.36.** *Suppose $(S, P, T)$ is an insertion point. Then $\kappa_{S,P,T} \ll_P \iota_{S,P,T} \equiv \mathsf{id}_{S \ll_P T}$.*

*Proof.* See κ-ι-prop in Catt.Tree.Insertion.Properties. □

We can then proceed to the main theorem of this section.

**Theorem 3.4.37.** *Let $\mathcal{R}$ be a tame equality rule set that has disc removal and insertion. Then for any insertion point $(S, P, T)$ and $n \in \mathbb{N}$, we have:*

$$S \ll_P T \vdash_{\mathcal{R}} \mathcal{T}^n_{\partial_n(S)}[\![\delta^\epsilon_n(S) \bullet \kappa_{S,P,T}]\!] = \mathcal{T}^n_{\partial_n(S \ll_P T)}[\![\delta^\epsilon_n(S \ll_P T)]\!]$$

$$S \ll_P T \vdash_{\mathcal{R}} \mathcal{U}^n_S[\![\kappa_{S,P,T}]\!] = \mathcal{U}^n_{S \ll_P T}$$

*for $\epsilon \in \{-,+\}$ and if $n \geq \mathrm{h}(S)$ then:*

$$S \ll_P T \vdash_{\mathcal{R}} \mathcal{C}^n_S[\![\kappa_{S,P,T}]\!] = \mathcal{C}^n_{S \ll_P T} \qquad S \ll_P T \vdash_{\mathcal{R}} \mathcal{T}^n_S[\![\kappa_{S,P,T}]\!] = \mathcal{T}^n_{S \ll_P T}$$

*Proof.* We prove all three properties by mutual induction: We begin with the equality:

$$\mathcal{T}^n_{\partial_n(S)}[\![\delta^\epsilon_n(S) \bullet \kappa_{S,P,T}]\!] = \mathcal{T}^n_{\partial_n(S \ll_P T)}[\![\delta^\epsilon_n(S \ll_P T)]\!]$$

The conditions for either Lemmas 3.4.32 and 3.4.34 must hold, and so we treat in case separately. If the conditions for Lemma 3.4.32 hold then the required equality is immediately implied by $\partial_n(S \ll_P T) \equiv \partial_n(S)$ and $\delta^\epsilon_n(S) \bullet \kappa_{S,P,T} \equiv \delta^\epsilon_n(S \ll_P T)$. If instead the conditions for Lemma 3.4.34 hold then:

$$\begin{aligned}
\mathcal{T}^n_{\partial_n(S)}[\![\delta^\epsilon_n(S) \bullet \kappa_{S,P,T}]\!] &\equiv \mathcal{T}^n_{\partial_n(S)}[\![\kappa_{\partial_n(S),\partial_n(P),\partial_n(T)} \bullet \delta^\epsilon_n(S \ll_P T)]\!] \\
&\equiv \mathcal{T}^n_{\partial_n(S)}[\![\kappa_{\partial_n(S),\partial_n(P),\partial_n(T)}]\!][\![\delta^\epsilon_n(S \ll_P T)]\!] \\
&= \mathcal{T}^n_{\partial_n(S) \ll_{\partial_n(P)} \partial_n(T)}[\![\delta^\epsilon_n(S \ll_P T)]\!] \\
&\equiv \mathcal{T}^n_{\partial_n(S \ll_P T)}[\![\delta^\epsilon_n(S \ll_P T)]\!]
\end{aligned}$$

where the definitional equality is due to the inductive hypothesis on terms.

We continue to the case for types. If $n = 0$, then both sides of the equality are $\star$. Instead, consider the $n+1$ case, where we have:

$$\mathcal{U}^{n+1}_S[\![\kappa_{S,P,T}]\!] \equiv \begin{array}{c} \mathcal{T}^n_{\partial_n(S)}[\![\delta^-_n(S)]\!][\![\kappa_{S,P,T}]\!] \\ \to_{\mathcal{U}^n_S[\![\kappa_{S,P,T}]\!]} \\ \mathcal{T}^n_{\partial_n(S)}[\![\delta^+_n(S)]\!][\![\kappa_{S,P,T}]\!] \end{array} \qquad \mathcal{U}^{n+1}_{S \ll_P T} \equiv \begin{array}{c} \mathcal{T}^n_{\partial_n(S \ll_P T)}[\![\delta^-_n(S \ll_P T)]\!] \\ \to_{\mathcal{U}^n_{S \ll_P T}} \\ \mathcal{T}^n_{\partial_n(S \ll_P T)}[\![\delta^+_n(S \ll_P T)]\!] \end{array}$$

By the inductive hypothesis on $n$, we have $\mathcal{U}^n_S[\![\kappa_{S,P,T}]\!] = \mathcal{U}^n_{S \ll_P T}$, and other necessary equalities follow from the first case we considered.

We now consider the case for standard coherences, where we must prove that:

$$\mathsf{SCoh}_{(S\,;\,\mathcal{U}^n_S)}[\kappa_{S,P,T}] = \mathsf{SCoh}_{(S \ll_P T\,;\,\mathcal{U}^n_{S \ll_P T})}[\mathsf{id}]$$

By Lemma 3.4.12, $\overline{P}[\![\kappa]\!]_{S,P,T}$ is the standard coherence $\mathcal{C}^{\mathsf{lh}(P)}_T[\![\iota_{S,P,T}]\!]$, and so the left-hand side of the above equation admits an insertion. Therefore:

$$\begin{aligned}
\mathsf{SCoh}_{(S\,;\,\mathcal{U}^n_S)}[\kappa_{S,P,T}] &= \mathsf{SCoh}_{(S \ll_P T\,;\,\mathcal{U}^n_S[\![\kappa_{S,P,T}]\!])}[\kappa_{S,P,T} \ll_P \iota_{S,P,T}] && \text{by insertion} \\
&\equiv \mathsf{SCoh}_{(S \ll_P T\,;\,\mathcal{U}^n_S[\![\kappa_{S,P,T}]\!])}[\mathsf{id}] && \text{by Lemma 3.4.36} \\
&= \mathsf{SCoh}_{(S \ll_P T\,;\,\mathcal{U}^n_{S \ll_P T})}[\mathsf{id}] && \text{by inductive hypothesis} \\
&\equiv \mathcal{C}^n_{S \ll_P T}
\end{aligned}$$

The equality for standard terms follows from the equality for standard coherences, using Theorem 3.3.26. □

**Corollary 3.4.38.** *If $\mathcal{R}$ has disc removal and insertion, then an insertion into a standard coherence is equal to the standard coherence over the inserted tree.*

*Proof.* Let $s \equiv \mathcal{C}_S^n[\![L]\!]$ be a standard coherence, and suppose $(S, P, T, \mathbf{U}, L, M)$ is an insertion redex with $\mathbf{U} \vdash s : A$ for some $A$. Then:

$$\begin{aligned}\mathcal{C}_S^n[\![L]\!] &= \mathsf{SCoh}_{(S \ll_P T\,;\,\mathcal{U}_S^n[\![\kappa_{S,P,T}]\!])}[L \ll_P M] \\ &= \mathsf{SCoh}_{(S \ll_P T\,;\,\mathcal{U}^n_{S \ll_P T})}[L \ll_P M] \\ &= \mathcal{C}^n_{S \ll_P T}[\![L \ll_P M]\!]\end{aligned}$$

and so $s$ is equal to a standard coherence over the tree $S \ll_P T$. □

**Chained insertion** We explore the situation where a term $s$ has a locally maximal argument $t$ which can be inserted, and this term $t$ admits an insertion itself. For the argument $t$ to be insertable, it must be a standard coherence, and by Corollary 3.4.38, if $t = t'$ by insertion, then $t'$ will be equal to a standard coherence over some tree $T$. For the term $t'$ to be insertable, $T$ must have sufficient trunk height. Conditions for this are given in the following lemma.

**Lemma 3.4.39.** *Let $(S, P, T)$ be an insertion point. Further, assume $S$ is not linear. Then* $\mathsf{th}(S \ll_P T) \geq \mathsf{th}(S)$.

*Proof.* See insertion-trunk-height in Catt.Tree.Insertion.Properties. □

If a tree $S$ is not linear, then any branch of $S$ has branch height greater than the trunk height of $S$, and hence any insertion into $S$ only modifies the tree above its trunk height, and so can only increase the trunk height. Therefore, if $(S, P, T)$ and $T, Q, U$ are insertion points, and $T$ is not linear, then $(S, P, T \ll_Q U)$ is also an insertion point.

Conversely, it is possible to insert the argument directly into the head term, before performing the inner insertion, looking to perform the inner insertion afterwards. For this to be possible, a branch of the inserted tree must be given. This can again be done under a non-linearity condition.

**Definition 3.4.40.** Let $(S, P, T)$ be an insertion point where $T$ is not linear. Then from a branch $Q$ of $T$ we can obtain a branch $S \ll_P Q$ of $S \ll_P T$. We first observe that $\mathsf{bh}(Q) \geq \mathsf{th}(T) \geq \mathsf{bh}(P)$. We define this branch by induction on $P$ and $Q$:

- Suppose $P = [k]$ and $Q = q :: x$. Then define:

$$S \ll_P Q = (k - 1 + q) :: x$$

- Suppose $P = k :: P_2$ with $S = [S_0, \dots, S_n]$ and $T = \Sigma(T_0)$. In this case we must have $Q = 0 :: Q_2$ where $Q_2$ is a branch of $T_0$. Then define:

$$S \ll_P Q = k :: S_k \ll_{P_2} Q_2$$

It is clear that $S \ll_P Q$ has the same branching and leaf height as $Q$.

A simple inductive proof shows that:

$$\overline{S \lll_P Q} \equiv \overline{Q}[\![\iota_{S,P,T}]\!]$$

Now given insertion points $(S, P, T)$ and $(T, Q, U)$ with $T$ non-linear we have that the triple $(S \lll_P T, S \lll_P Q, U)$ is another insertion point. Therefore, two ways of performing both insertions, which are depicted in Figure 3.8.

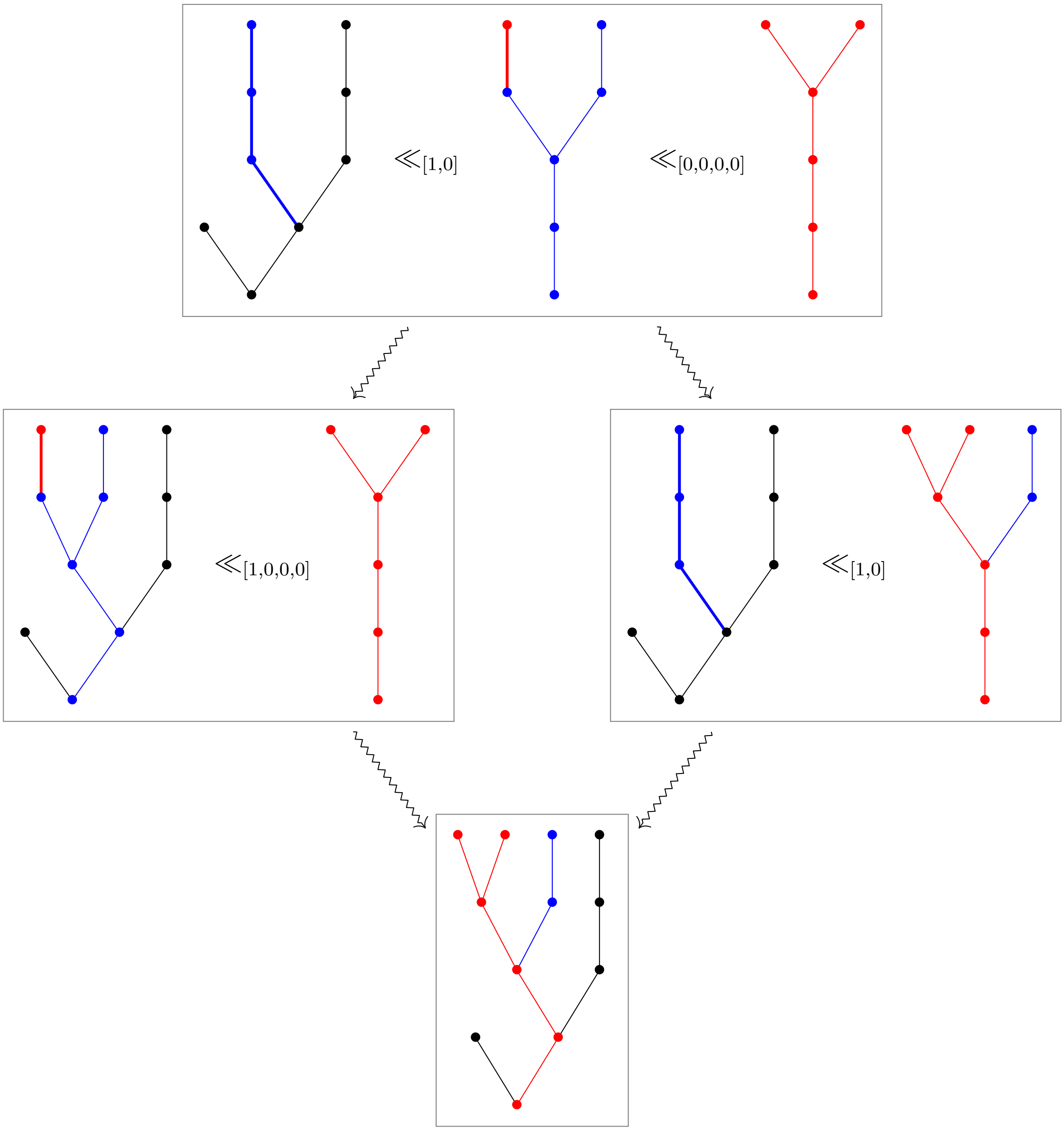

Figure 3.8: Chained insertion.

We now explore the universal property of the insertion along the branch $S \lll_P Q$. We assume that $n = \mathsf{lh}(P)$ and $m = \mathsf{lh}(Q)$ and form the following diagram:

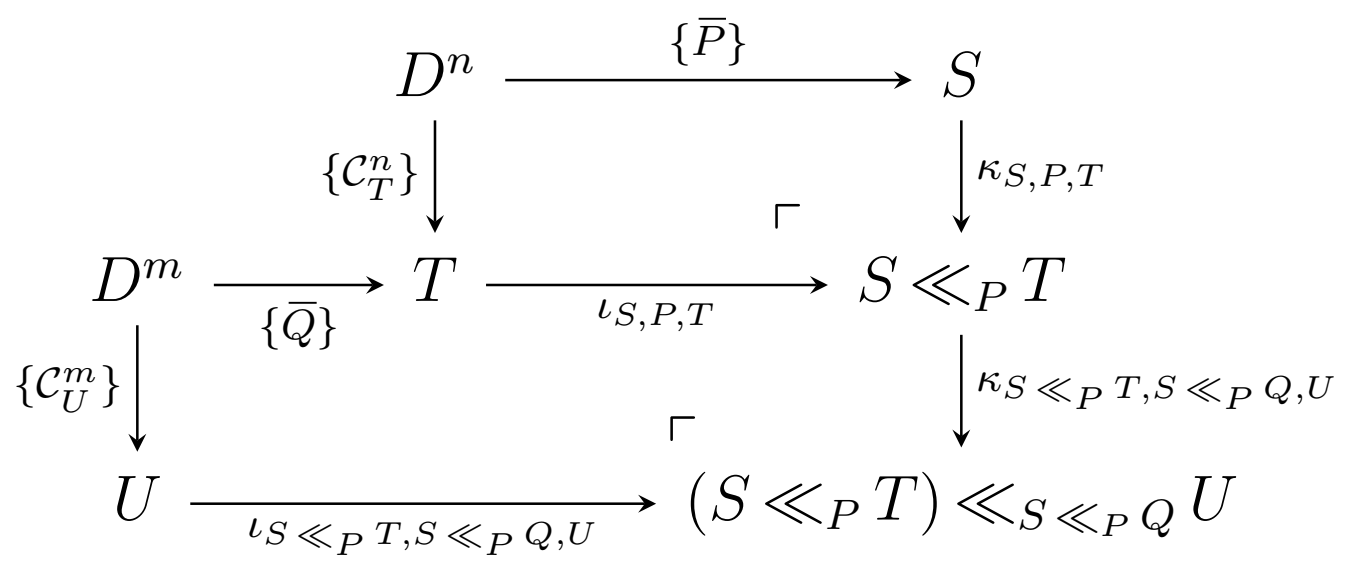

The top pushout square is given by the insertion of $T$ into $S$ along $P$. The morphism $\{\overline{Q}\} \bullet \iota_{S,P,T}$ through the middle of the diagram is then equal to $\{\overline{S \ll_P Q}\}$, allowing the bottom pushout rectangle to be formed by the insertion of $U$ into $S \ll_P T$ along $S \ll_P Q$.

We can also consider the universal property of the tree generated by first inserting $U$ into $T$, and then inserting the inserted tree into $S$, which is given by the diagram below:

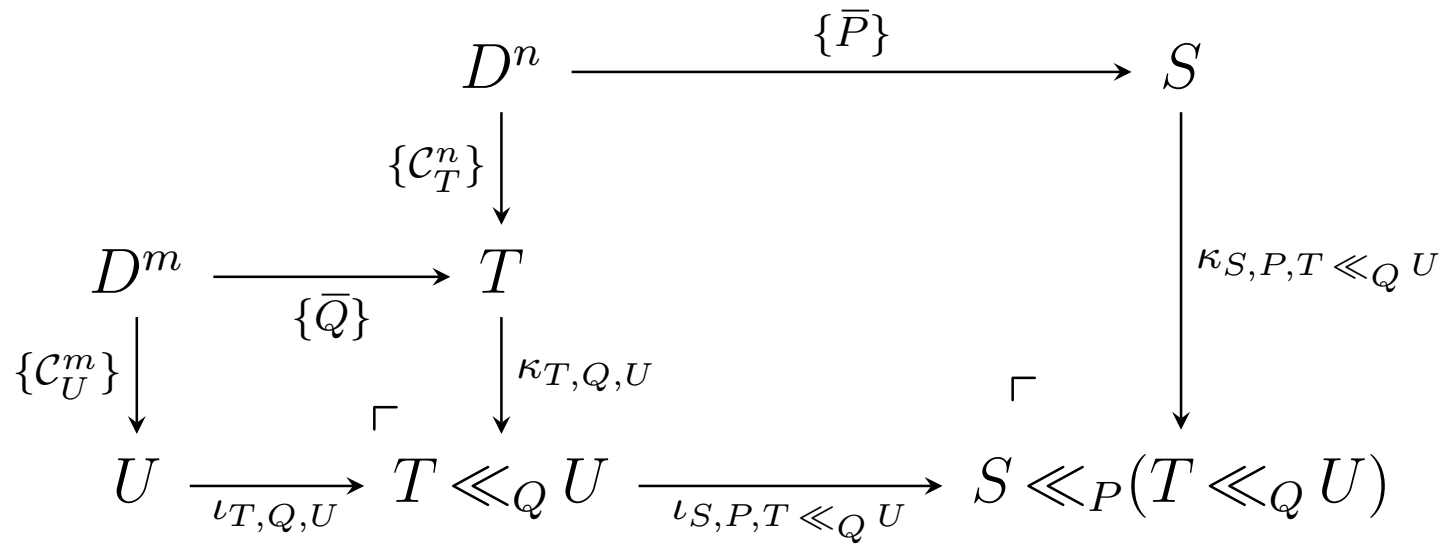

The left-hand pushout square is given by the insertion of $U$ into $T$ along $Q$. The morphism $\{\mathcal{C}_T^n\} \bullet \kappa_{T,Q,U}$ which runs vertically through the centre of the diagram is then equal to $\{\mathcal{C}^n_{T \ll_Q U}\}$ by Corollary 3.4.38, allowing for the right-hand pushout square to be formed as the insertion of $T \ll_Q U$ into $S$ along $P$. By common properties of colimits, both of these constructions then arise as colimits of the same diagram, the shared top left boundary of both constructions. The results of this section are stated in the following lemma.

**Lemma 3.4.41.** *Let $(S, P, T)$ and $(T, Q, U)$ be insertion points. Further assume $T$ is not linear. Then:*

$$\begin{aligned} S \ll_P (T \ll_Q U) &= (S \ll_P T) \ll_{S \ll_P Q} U \\ \kappa_{S,P,T \ll_Q U} &=^{\mathsf{max}} \kappa_{S,P,T} \circ \kappa_{S \ll_P T, S \ll_P Q, U} \\ L \ll_P (M \ll_Q N) &\equiv^{\mathsf{max}} (L \ll_P M) \ll_{S \ll_P Q} N \end{aligned}$$

*for any $L : S \to \mathbf{U}$, $M : T \to \mathbf{U}$, and $N : U \to \mathbf{U}$.*

*Proof.* See the functions insertion-tree-inserted-branch and label-from-inserted-branch in the formalisation module Catt.Tree.Insertion.Properties, and κ-inserted-branch in module Catt.Typing.Insertion.Equality. □

# Chapter 4

# Semistrict variants of Catt

The type theories $\text{Catt}_{\text{su}}$, a type theory for strictly unital $\infty$-categories, and $\text{Catt}_{\text{sua}}$, a type theory for strictly unital and associative $\infty$-categories, are introduced in this chapter, where we will define both theories and explore some metatheory and properties of each type theory in detail.

The results in this chapter will heavily depend on the theory developed in the previous chapters. Both type theories will be defined as instances of $\text{Catt}_{\mathcal{R}}$, which was introduced in Section 2.2, and much of the initial metatheory can be immediately derived by demonstrating that the equality rule sets that generate $\text{Catt}_{\text{su}}$ and $\text{Catt}_{\text{sua}}$ satisfy the various conditions given in Section 2.4. The theory $\text{Catt}_{\text{su}}$ is primarily generated by pruning, which was introduced in Section 3.1, and the theory $\text{Catt}_{\text{sua}}$ depends on the insertion operation, which was introduced in Section 3.4.

Section 4.2 will introduce and define the $\text{Catt}_{\text{su}}$, and Section 4.3 will do the same for $\text{Catt}_{\text{sua}}$. The main contribution of these sections is to give normalisation algorithms for their respective theories, giving a notion of computation to each theory. A normalisation algorithm is a function $\mathrm{N} : \mathsf{Term}_\Gamma \to \mathsf{Term}_\Gamma$ with the following properties:

- For any term $t : \mathsf{Term}_\Gamma$, $\Gamma \vdash \mathrm{N}(t) = t$.
- For any $s, t : \mathsf{Term}_\Gamma$ with $\Gamma \vdash s = t$, $\mathrm{N}(s) \equiv \mathrm{N}(t)$.

The term $\mathrm{N}(t)$ is called the *normal form* of $t$. Such an algorithm allows equality of two term $s$ and $t$ to be decided by taking the normal form of each term and checking if they are syntactically equal. Normalisation can be extended to types and substitutions in a natural way.

In Sections 4.2 and 4.3, the normalisation algorithm is defined by giving a reduction system on the syntax of the type theory, which we show to be terminating, meaning that there is no infinite reduction sequence and confluent, meaning that any two reduction paths converge to a common reduct. The normal form of a term can then be obtained by reducing it until there are no further reductions possible. In Section 4.1, these notions are recalled, and we demonstrate that the resulting normalisation algorithm satisfies the two properties stated above. This section also introduces a method for obtaining a reduction system from an arbitrary equality rule set $\mathcal{R}$.

Such a normalisation procedure allows a type checking algorithm to be implemented, creating an interpreter for the language. This allows us to write larger terms, and it can be automati-cally verified whether they are well-formed. In Section 4.4, we introduce our implementation of Catt, $\text{Catt}_{\text{su}}$, and $\text{Catt}_{\text{sua}}$, written in rust. This implementation supports features such as implicit arguments to terms, implicit suspension, and native support for trees and tree labellings. We will explain how the tool can be used, and use it to give larger examples of $\text{Catt}_{\text{sua}}$ terms, including proofs of Eckmann-Hilton (see Figure 1) and its higher-dimensional coherence condition, the syllepsis.

The implementation uses an approach closer to normalisation by evaluation for typechecking terms in the theory. Section 4.4 explores this algorithm and presents some perspectives on applying normalisation by evaluation to semistrict versions of Catt.

Section 4.5 provides a discussion of the models of the semistrict type theories $\text{Catt}_{\text{su}}$ and $\text{Catt}_{\text{sua}}$, demonstrating how they can be viewed as semistrict $\infty$-categories. The section proves a partial conservativity result, which allows a proof that semistrictness is a property of a weak $\infty$-category, and not additional structure. A discussion is provided on some of the challenges that must be overcome to extend this partial conservativity result.

The thesis ends with Section 4.6, which provides a review of avenues for future work in this area, including a discussion of further variants of Catt which could be defined.

## 4.1 Reduction

Reduction is a method for defining computation for a type theory. For each term, a number of reductions can be applied to it, representing the various computations that could be applied to the term. Computation can then be run on a term by repeatedly searching for positions in the term that admit a reduction, known as *redexes*, and applying this reduction, until no more redexes exist in the term. When a term admits no reductions, it is called a *normal form*.

**Definition 4.1.1.** A *reduction system* is given by a relation $s \rightsquigarrow t$ on terms. The relation $\rightsquigarrow^*$ is defined to be the reflexive transitive closure of $\rightsquigarrow$, and so $s \rightsquigarrow^* t$ exactly when there is some chain

$$s \equiv u_0 \rightsquigarrow \cdots \rightsquigarrow u_k \equiv t$$

for $k \in \mathbb{N}$ (which could be 0 with $s \equiv t$) and terms $u_i$ for $i \leq k$. Further define $\leftrightsquigarrow$ to be the reflexive symmetric transitive closure of $\rightsquigarrow$.

When a term $s$ admits no reductions, that is there is no $t$ such that $s \rightsquigarrow t$, we say it is in *normal form*.

If we have an equality rule set $\mathcal{R}$ (see Section 2.4) that generates $\text{Catt}_{\mathcal{R}}$, a reduction system can be defined on $\mathcal{R}$ modifying the rules for equality to remove the reflexivity, symmetry, and transitivity constructors and ensure that reductions do not happen "in parallel".

**Definition 4.1.2.** Let $\mathcal{R}$ be an equality rule set. Define the reduction system $\rightsquigarrow_{\mathcal{R}}$ on well-formed terms, well-formed substitutions, and well-formed types to be generated by the rules in Figure 4.1. When it is clear which equality rule set is being used, we may simply write $s \rightsquigarrow t$ instead of $s \rightsquigarrow_{\mathcal{R}} t$.

The rules for reduction are set up so that each reduction $s \rightsquigarrow_{\mathcal{R}} t$ corresponds to the application of exactly one rule from $\mathcal{R}$ at a single point in the term. Given a coherence $\mathsf{Coh}_{(\Delta\,;\,A)}[\sigma]$, we

$$\frac{(\Gamma, s, t) \in \mathcal{R}}{s \rightsquigarrow_{\mathcal{R}} t}\textsc{rule} \qquad \frac{A \rightsquigarrow_{\mathcal{R}} B}{\mathsf{Coh}_{(\Delta\,;\,A)}[\sigma] \rightsquigarrow_{\mathcal{R}} \mathsf{Coh}_{(\Delta\,;\,B)}[\sigma]}\textsc{cell}$$

$$\frac{\sigma \rightsquigarrow_{\mathcal{R}} \tau}{\mathsf{Coh}_{(\Delta\,;\,A)}[\sigma] = \mathsf{Coh}_{(\Delta\,;\,A)}[\tau]}\textsc{arg}$$

$$\frac{s \rightsquigarrow_{\mathcal{R}} s'}{s \to_A t \rightsquigarrow_{\mathcal{R}} s' \to_A t} \qquad \frac{t \rightsquigarrow_{\mathcal{R}} t'}{s \to_A t \rightsquigarrow_{\mathcal{R}} s \to_A t'} \qquad \frac{A \rightsquigarrow_{\mathcal{R}} A'}{s \to_A t \rightsquigarrow_{\mathcal{R}} s \to_{A'} t}$$

$$\frac{\sigma \rightsquigarrow_{\mathcal{R}} \tau}{\langle \sigma, s \rangle \rightsquigarrow_{\mathcal{R}} \langle \tau, s \rangle} \qquad \frac{s \rightsquigarrow_{\mathcal{R}} t}{\langle \sigma, s \rangle \rightsquigarrow_{\mathcal{R}} \langle \sigma, t \rangle}$$

Figure 4.1: Rules for $\rightsquigarrow_{\mathcal{R}}$.

call reductions generated by the CELL rule *cell reductions* and reductions generated by the ARG rule *argument reductions*. Reductions generated by RULE will be named by the rule in $\mathcal{R}$ that was used. For example a reduction generated by RULE applied with an instance of pruning will be called a pruning reduction.

We highlight that our reduction system $\rightsquigarrow_{\mathcal{R}}$ is only defined between well-formed pieces of syntax. As this reduction will be used with rule sets $\mathcal{R}$ which satisfy the preservation condition, there will be no additional burden of checking that typing is preserved while applying reductions. Therefore, we can prove that the reflexive symmetric transitive closure of reduction, $\leftrightsquigarrow_{\mathcal{R}}$, is the same relation as equality on well-formed terms, given the similarity between the rules for reduction and the rules for equality.

**Proposition 4.1.3.** *Let $\mathcal{R}$ be a rule set satisfying the preservation, support, and substitution conditions (such that the generated equality preserves typing). Letting $\leftrightsquigarrow_{\mathcal{R}}$ be the reflexive symmetric transitive closure of $\rightsquigarrow_{\mathcal{R}}$, we get:*

$$\Gamma \vdash s = t \iff s \leftrightsquigarrow_{\mathcal{R}} t$$

*for $s, t : \mathsf{Term}_\Gamma$ such that $\Gamma \vdash s : A$ and $\Gamma \vdash t : A$ for some $A : \mathsf{Type}_\Gamma$*

$$\Gamma \vdash A = B \iff A \leftrightsquigarrow_{\mathcal{R}} B$$

*for $A, B : \mathsf{Type}_\Gamma$ such that $\Gamma \vdash A$ and $\Gamma \vdash B$*

$$\Gamma \vdash \sigma = \tau \iff \sigma \leftrightsquigarrow_{\mathcal{R}} \tau$$

*for $\sigma, \tau : \Delta \to_\star \Gamma$ such that $\Gamma \vdash \sigma : \Delta$ and $\Gamma \vdash \tau : \Delta$.*

*Proof.* Each direction can be proved separately by a mutual induction on the derivation in the premise. For the right to left direction, it suffices to show that the single step reduction ($\rightsquigarrow_{\mathcal{R}}$) is contained in the equality, as equality is an equivalence relation by construction. □

Just as the preservation condition on a rule set $\mathcal{R}$ allows us to deduce that reduction preserves typing, the substitution condition can be used to prove that reduction is preserved by application of substitution.

**Proposition 4.1.4.** *Suppose $\mathcal{R}$ satisfies the substitution condition and let $\sigma : \Delta \to \Gamma$ be a well-formed substitution. Then:*

$$\begin{aligned} s \rightsquigarrow_{\mathcal{R}} t &\implies s[\![\sigma]\!] \rightsquigarrow_{\mathcal{R}} t[\![\sigma]\!] \\ A \rightsquigarrow_{\mathcal{R}} B &\implies A[\![\sigma]\!] \rightsquigarrow_{\mathcal{R}} B[\![\sigma]\!] \\ \tau \rightsquigarrow_{\mathcal{R}} \mu &\implies \tau \bullet \sigma \rightsquigarrow_{\mathcal{R}} \mu \bullet \sigma \end{aligned}$$

*for well-formed terms $s, t$, well-formed types $A, B$, and well-formed substitutions $\tau$ and $\mu$. Furthermore, if $\sigma \rightsquigarrow_{\mathcal{R}} \tau$, then:*

$$s[\![\sigma]\!] \rightsquigarrow_{\mathcal{R}}^* s[\![\tau]\!] \qquad A[\![\sigma]\!] \rightsquigarrow_{\mathcal{R}}^* A[\![\tau]\!] \qquad \mu \bullet \sigma \rightsquigarrow_{\mathcal{R}}^* \mu \bullet \tau$$

*for term $s$, type $A$, and substitution $\mu$.*

*Proof.* The first part by a simple induction on the reduction in the premise. The second holds by a mutual induction on the term $s$, type $A$, and substitution $\mu$. □

### 4.1.1 Termination

In order to obtain a normal form of each term of the theory, we perform reductions on a term until no more can be applied. This can only be done if we know that this will eventually result in a normal form, a property known as *strong termination.*

**Definition 4.1.5.** A reduction system $\rightsquigarrow$ is *strongly terminating* if there is no infinite sequence of reductions:

$$s_0 \rightsquigarrow s_1 \rightsquigarrow s_2 \rightsquigarrow \cdots$$

For such a reduction, applying reductions to a term will eventually reach a normal form.

Demonstrating the termination of the reduction systems defined in Sections 4.2 and 4.3 will be non-trivial, as each reduction adds new constructions to the term, which could themselves admit reductions. Suppose we have the following reduction due to endo-coherence removal (see Section 2.4.3):

$$\mathsf{Coh}_{(\Delta\,;\,s \to_A s)}[\sigma] \rightsquigarrow \mathsf{id}(A[\![\sigma]\!], s[\![\sigma]\!])$$

The identity term was not present in the premise of the reduction, and the term $s[\![\sigma]\!]$ is newly created by the reduction, and could itself admit any number of reductions.

To prove termination, we will exploit that although each reduction creates new subterms, these subterms are all of a lower dimension than the dimension of the term that is being reduced. In the example above, the dimension of $\mathsf{Coh}_{(\Delta\,;\,s \to_A s)}[\sigma]$ is greater than the dimension of the term $s$, and so the reduction has still made progress towards a normal form by decreasing the complexity of the term in dimension $\dim(A)$, even though it may introduce arbitrary complexity below $\dim(A)$.

To this end we define the following notion of complexity for each class of syntax, which assigns an ordinal number to each term, which we call its *syntactic complexity*. As the ordinal numbers are well-founded, we aim to prove that our reduction is terminating by proving that each single-step reduction reduces the complexity of the term. To define syntactic complexity, we will need to use ordinal numbers up to $\omega^\omega$. We will also need a construction known as the natural sum of ordinals, $\alpha \# \beta$, which is associative, commutative, and strictly monotone in both of its arguments [Lip16].

**Definition 4.1.6.** For all terms $t$, types $A$, and substitutions $\sigma$, the *syntactic complexity* $\mathsf{sc}(t)$, $\mathsf{sc}(A)$, and $\mathsf{sc}(\sigma)$ are mutually defined as follows:

- For types:
$$\mathsf{sc}(\star) = 0 \qquad \mathsf{sc}(s \to_A t) = \mathsf{sc}(s) \# \mathsf{sc}(A) \# \mathsf{sc}(t)$$
- For substitutions we have:
$$\mathsf{sc}(\langle t_0, \dots, t_n \rangle) = \mathop{\#}_{i=0}^{n} t_i$$
- For terms, we have $\mathsf{sc}(x) = 0$ for variables $x$ and for coherences we have:
$$\mathsf{sc}(\mathsf{Coh}_{(\Delta\,;\,A)}[\sigma]) = \begin{cases} \omega^{\dim(A)} \# \mathsf{sc}(\sigma) & \text{if } \mathsf{Coh}_{(\Delta\,;\,A)}[\sigma] \text{ is an identity} \\ 2\omega^{\dim(A)} \# \mathsf{sc}(\sigma) & \text{otherwise} \end{cases}$$

The syntactic complexity is given as an ordinal to leverage known results, though it should be noted that ordinals below $\omega^\omega$ can be represented by a list of natural numbers ordered reverse lexicographically. Under this interpretation the syntactic complexity effectively computes the number of coherences at each dimension. Therefore, removing a coherence of dimension $n$ reduces the complexity, even if arbitrary complexity is added at lower dimensions. Syntactic complexity also treats identities in a special way, as these play a special role in blocking reduction for the theories presented in this chapter.

The syntactic complexity does not account for the type in a coherence, as this is difficult to encode. Instead of showing that all reductions reduce syntactic complexity, we instead show that all reductions which are not cell reductions (reductions that have the rule marked CELL in their derivation) reduce syntactic complexity and deduce that a hypothetical infinite reduction sequence must only consist of cell reductions after a finite number of steps, and then appeal to an induction on dimension.

**Lemma 4.1.7.** *Let $\mathcal{R}$ be an equality set with $\mathsf{sc}(s) > \mathsf{sc}(t)$ for all $(\Gamma, s, t) \in \mathcal{R}$. Then $\rightsquigarrow_{\mathcal{R}}$ is strongly terminating.*

*Proof.* By a simple induction on reductions, we immediately have that if $s \rightsquigarrow_{\mathcal{R}} t$ then $\mathsf{sc}(s) \geq \mathsf{sc}(t)$, with the inequality strict when the reduction is not a cell reduction. We then proceed by induction on the dimension. Suppose there is an infinite reduction sequence, starting with a $k$-dimensional term:

$$s_0 \rightsquigarrow s_1 \rightsquigarrow s_2 \rightsquigarrow \cdots$$

Then by assumption, only finitely many of these reductions do not use the cell rule, as otherwise we would obtain an infinite chain of ordinals

$$\mathsf{sc}(s_0) \geq \mathsf{sc}(s_1) \geq \mathsf{sc}(s_2) \geq \cdots$$

where infinitely many of these inequalities are strict. Therefore, there is an $n$ such that:

$$s_n \rightsquigarrow s_{n+1} \rightsquigarrow \cdots$$

are all cell reductions. Each of these reductions reduces one of finitely many subterms of $s_n$, and each of these subterms has dimension less than $k$, so by inductive hypothesis, none of these subterms can be reduced infinitely often, contradicting the existence of an infinite reduction sequence. □

We can immediately prove that disc removal reduces syntactic complexity.

**Proposition 4.1.8.** *Let $s \rightsquigarrow t$ be an instance of disc removal. Then $\mathsf{sc}(s) > \mathsf{sc}(t)$.*

*Proof.* We must have $s \equiv \mathsf{Coh}_{(D^n\,;\,\mathrm{wk}(U^n))}[\{A, t\}]$ for some $n$ and $A$. Then:

$$\begin{aligned} \mathsf{sc}(s) &= \mathsf{sc}(\mathsf{Coh}_{(D^n\,;\,\mathrm{wk}(U^n))}[\{A, t\}]) \\ &= 2\omega^n \mathbin{\#} \mathsf{sc}(\{A, t\}) \\ &> \mathsf{sc}(\{A, t\}) \\ &\geq \mathsf{sc}(t) \end{aligned}$$

where the last inequality holds by a simple induction on the dimension of $A$. □

We note that as stated so far the reduction:

$$\mathsf{id}(A, s) \rightsquigarrow \mathsf{id}(A, s)$$

is a valid instance of endo-coherence removal for type $A$ and term $s$, which will break termination. We therefore let ecr′ be the equality rule set obtained by removing all triples $(\Gamma, s, t)$ from ecr where $s$ is already an identity. We justify replacing ecr by ecr' with the following lemma.

**Lemma 4.1.9.** *The following reduction holds, even when the left-hand side is an identity:*

$$\mathsf{Coh}_{(\Delta\,;\,s \to_A s)}[\sigma] \rightsquigarrow^*_{\mathit{ecr'}} \mathsf{id}(A[\![\sigma]\!], s[\![\sigma]\!])$$

*Proof.* If $\mathsf{Coh}_{(\Delta\,;\,s \to_A s)}[\sigma]$ is not an identity then it can be reduced by endo-coherence removal. Otherwise, we have $\Delta = D^n$ for some $n$, $s \equiv d_n$, $A \equiv \mathrm{wk}(U^n)$, and $\sigma \equiv \{B, t\}$ for some $B$ and $t$ and so:

$$\mathsf{id}(A[\![\sigma]\!], s[\![\sigma]\!]) \equiv \mathsf{id}(\mathrm{wk}(U^n)[\![\{B, t\}]\!], d_n[\![\{B, t\}]\!]) \equiv \mathsf{id}(B, t)$$

It follows that the reduction is trivial. □

It can then be proven that the reductions in this set reduce syntactic complexity.

**Proposition 4.1.10.** *Let $s \rightsquigarrow t$ be an instance of endo-coherence removal. If $s$ is not an identity then* $\mathsf{sc}(s) > \mathsf{sc}(t)$.

*Proof.* As $s \rightsquigarrow t$ is an instance of endo-coherence removal, we must have $s \equiv \mathsf{Coh}_{(\Delta\,;\,u \to_A u)}[\sigma]$ and $t \equiv \mathsf{id}(A[\![\sigma]\!], u[\![\sigma]\!])$. Further, $s$ is not an identity and so:

$$\begin{aligned}
\mathsf{sc}(s) &= \mathsf{sc}(\mathsf{Coh}_{(\Delta\,;\,u \to_A u)}[\sigma]) \\
&= 2\omega^{\dim(A)+1} \mathbin{\#} \mathsf{sc}(\sigma) \\
&\geq 2\omega^{\dim(A)+1} \\
&< \omega^{\dim(A)+1} \mathbin{\#} \mathsf{sc}(A[\![\sigma]\!]) \mathbin{\#} \mathsf{sc}(u[\![\sigma]\!]) && = \mathsf{sc}(\mathsf{id}(A[\![\sigma]\!], u[\![\sigma]\!])) \\
&= \mathsf{sc}(t)
\end{aligned}$$

where the last inequality holds as $\mathsf{sc}(A[\![\sigma]\!]) \mathbin{\#} \mathsf{sc}(u[\![\sigma]\!]) < \omega^{\dim(A)+1}$ due to both $A[\![\sigma]\!]$ and $u[\![\sigma]\!]$ having the same dimension as $\dim(A)$, meaning that their syntactic complexities are strictly bounded by $\omega^{\dim(A)+1}$. □

### 4.1.2 Confluence

Another crucial property of reduction systems is *confluence*. A term $s$ may have any number of redexes and could reduce to distinct terms $t$ and $u$. Confluence states that both the terms $t$ and $u$ must reduce to some common term, allowing us to apply reductions to a term in any order.

**Definition 4.1.11.** Let $\rightsquigarrow$ be a reduction system. It is *(globally) confluent* if for all terms $s$,$t$, and $u$ with $s \rightsquigarrow^* t$ and $s \rightsquigarrow^* u$, there is a term $v$ such that $t \rightsquigarrow^* v$ and $t \rightsquigarrow^* v$. This can be assembled into the following diagram:

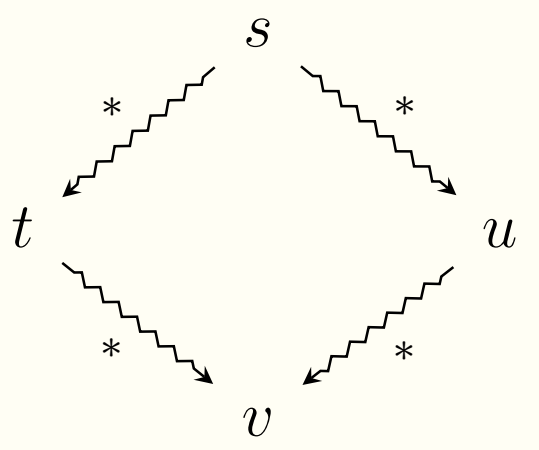

and hence is sometimes called the diamond property for $\rightsquigarrow^*$.

From global confluence, it is clear that if $s \leftrightsquigarrow_{\mathcal{R}} t$, where $\leftrightsquigarrow_{\mathcal{R}}$ is the reflexive symmetric transitive closure of $\rightsquigarrow_{\mathcal{R}}$, then there is $u$ with $s \rightsquigarrow_{\mathcal{R}}^* u$ and $t \rightsquigarrow_{\mathcal{R}}^* u$. It is sometimes simpler to show that the following weaker confluence property holds:

**Definition 4.1.12.** Let $\rightsquigarrow$ be a reduction system. It is *locally confluent* if given $s \rightsquigarrow t$ and

$s \rightsquigarrow u$ there exists a term $v$ such that:

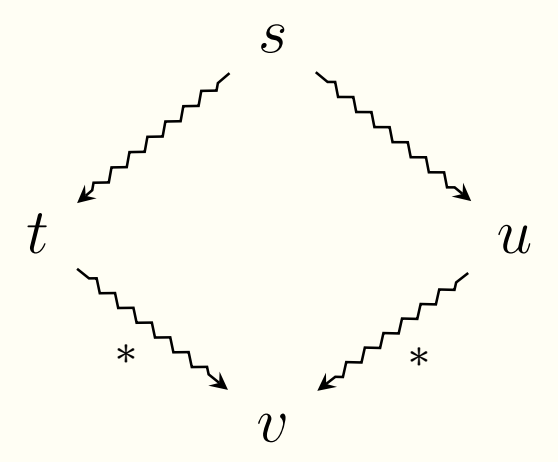

that is $t \rightsquigarrow^* v$ and $u \rightsquigarrow^* v$.

Global confluence trivially implies local confluence. If we further know that the reduction system $\rightsquigarrow$ is strongly terminating then local confluence is sufficient to show global confluence.

**Lemma 4.1.13** (Newman's lemma [NH42]). *Let $\rightsquigarrow$ be strongly terminating and locally confluent. Then $\rightsquigarrow$ is globally confluent.*

Local confluence for the reduction systems of the type theories Catt$_{\text{su}}$ and Catt$_{\text{sua}}$ will be proved using *critical pair analysis*. A critical pair is a pair of distinct reductions which apply to the same term. When analysing the critical pairs of our semistrict type theories, we will encounter terms that are structurally similar, but differ on lower-dimensional subterms up to equality. We define this precisely.

**Definition 4.1.14.** Let $\mathcal{R}$ be an equality rule set. For $n \in \mathbb{N}$, define the *bounded equality set* $\mathcal{R}_n$ as:

$$\mathcal{R}_n = \{(\Gamma, s, t) \in \mathcal{R} \mid \dim(s) = \dim(t) < n\}$$

Let the *bounded equality relation* $s =_n t$ be the equality generated by the set $\mathcal{R}_n$.

This is used to prove the following lemma, which implies that for a critical pair $t \leftsquigarrow s \rightsquigarrow u$ it is not necessary to find a common reduct of $t$ and $u$, but simply find reducts $t'$ and $u'$ of $t$ and $u$ such that $t' =_{\dim(s)} u'$.

**Lemma 4.1.15.** *Let $\mathcal{R}$ be a tame equality rule set which satisfies the preservation and support conditions, and further assume that $\rightsquigarrow_{\mathcal{R}}$ is strongly terminating. Suppose the following diagram can be formed:*

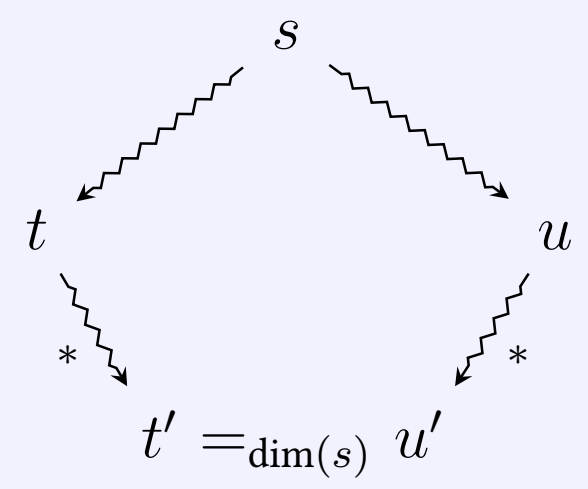

*for all critical pairs $t \leftsquigarrow_{\mathcal{R}} s \rightsquigarrow_{\mathcal{R}} u$ such that $s \rightsquigarrow_{\mathcal{R}} t$ is derived using* rule.

*Then $\rightsquigarrow_{\mathcal{R}}$ is confluent.*

*Proof.* By Lemma 4.1.13, it suffices to show local confluence. We proceed by strong induction on $n$ and $s$, proving that all critical pairs $t \leftsquigarrow_{\mathcal{R}_n} s \rightsquigarrow_{\mathcal{R}_n} u$ have a common reduct, assuming that all critical pairs $t \leftsquigarrow_{\mathcal{R}_m} s' \rightsquigarrow_{\mathcal{R}_m} u$ have a common reduct, where $s'$ is a subterm of $s$ or $m < n$. We justify this induction principle by noting that for all subterms $s'$ of $s$ we have $\dim(s') \leq \dim(s)$.

We now consider critical pair $t \leftsquigarrow_{\mathcal{R}_n} s \rightsquigarrow_{\mathcal{R}_n} u$. We first suppose that $s \rightsquigarrow_{\mathcal{R}_n} t$ is derived from RULE. Then, by definition of the set $\mathcal{R}_n$, we must have that $n > \dim(s)$. By the assumption of the lemma, there exist $t'$ and $u'$ with $t' =_{\dim(s)} u'$ and $t \rightsquigarrow^*_{\mathcal{R}} t'$ and $u \rightsquigarrow^*_{\mathcal{R}} u'$. As $n > \dim(s)$, we further have that $t \rightsquigarrow^*_{\mathcal{R}_n} t'$ and $u \rightsquigarrow^*_{\mathcal{R}_n} u'$.

By Proposition 4.1.3, $t' \leftrightsquigarrow_{\mathcal{R}_{\dim(s)}} u'$, and so as $\rightsquigarrow_{\mathcal{R}_{\dim(s)}}$ is confluent by inductive hypothesis on dimension we have $v$ such that $t' \rightsquigarrow^*_{\mathcal{R}_{\dim(s)}} v \leftsquigarrow^*_{\mathcal{R}_{\dim(s)}} u'$. The following diagram can therefore be formed, where all the reductions are $\mathcal{R}_n$ reductions (noting that $\mathcal{R}_{\dim(s)} \subseteq \mathcal{R}_n$):

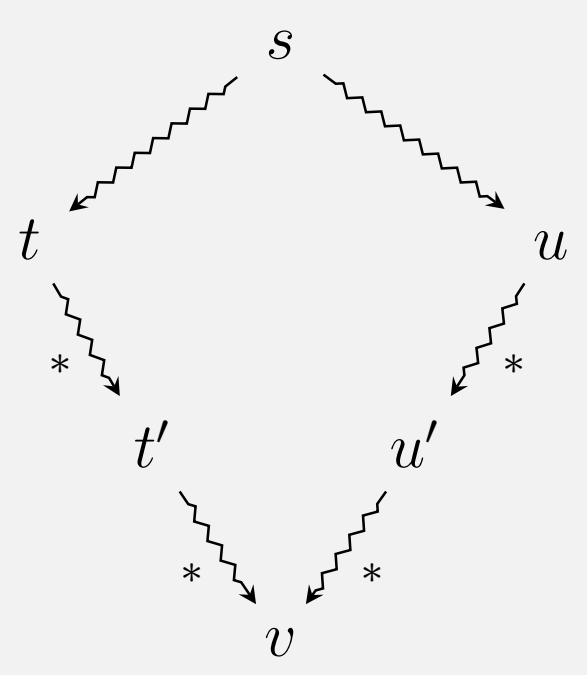

If $s \rightsquigarrow_{\mathcal{R}_n} u$ was derived from RULE, then finding a reduct can be found similarly to the first case by symmetry. We therefore consider the cases where neither $s \rightsquigarrow t$ nor $s \rightsquigarrow u$ are derived using RULE. Both reductions must be either cell or argument reductions, and so each reduces some subterm of $s$. If they reduce distinct subterms of $s$, then a common reduct $v$ can be formed by applying both reductions to $s$. Otherwise, both reductions act on the same subterm of $s$, and a common reduct can be found by applying the inductive hypothesis for subterms. □

Once termination and confluence have been proven, a normalisation function can be defined, which repeatedly applies reductions until no more can be applied.

**Lemma 4.1.16.** *Suppose that $\rightsquigarrow$ is strongly terminating and confluent. Then every term $s$ reduces to a unique normal form $\mathsf{N}(s)$. Furthermore, if $s \leftrightsquigarrow_{\mathcal{R}} t$, then $\mathsf{N}(s) \equiv \mathsf{N}(t)$.*

*Proof.* By termination, repeatedly reducing a term will reach a normal form. Suppose a term $s$ has two normal forms $t$ and $u$ such that there are reduction sequences $s \rightsquigarrow^* t$ and $s \rightsquigarrow^* u$. Then by confluence there must be a term $v$ with $t \rightsquigarrow^* v$ and $u \rightsquigarrow^* u$. However, $t$ and $u$ are normal forms and so admit no reductions, so $t \equiv v \equiv u$ as required.

Suppose $s \leftrightsquigarrow_{\mathcal{R}} t$. Then there are terms $s_i$ such that:

$$s \equiv s_0 \rightsquigarrow^* s_1 \leftsquigarrow^* s_2 \rightsquigarrow^* \cdots \leftsquigarrow^* s_k \equiv t$$

Now we must have $\mathsf{N}(s_i) \equiv \mathsf{N}(s_{i+1})$ for each $i$ as if $s_i \rightsquigarrow^* s_{i+1}$ then both $\mathsf{N}(s_i)$ and $\mathsf{N}(s_{i+1})$ are normal forms of $s_i$ and if $s_i \leftrightsquigarrow^* s_{i+1}$ then both are normal forms of $s_{i+1}$. Therefore, $\mathrm{N}(s)$ and $\mathrm{N}(t)$ are syntactically equal as required. □

**Corollary 4.1.17.** *Let $\mathcal{R}$ be tame and satisfy the preservation and support properties. Further, suppose that $\rightsquigarrow_{\mathcal{R}}$ is strongly terminating and confluent, and it is decidable whether a term admits a reduction. Then the equality $s = t$ is decidable.*

*Proof.* By Proposition 4.1.3, $s = t$ if and only if $s \leftrightsquigarrow_{\mathcal{R}} t$. By the above lemma, $s \leftrightsquigarrow_{\mathcal{R}} t$ if and only if $\mathrm{N}(s) \equiv \mathrm{N}(t)$. As syntactic equality is clearly decidable, and normal forms can be computed, equality is also decidable. □

We note that for an arbitrary rule set $\mathcal{R}$, it may not be decidable whether a specific term $s$ admits a reduction, but for the rule sets introduced in Sections 4.2 and 4.3, it will be easy to mechanically check whether any reduction applies to a term $s$.

## 4.2 Catt$_{\mathsf{su}}$

We are ready to define Catt$_{\mathsf{su}}$, the type theory for strictly unital $\infty$-categories. Catt$_{\mathsf{su}}$ is a variant of Catt$_{\mathcal{R}}$ for which the equality is built from three classes of equalities:

- Pruning: The pruning operation was introduced in Section 3.1. Pruning is the key operation in Catt$_{\mathsf{su}}$ and drives the strict unitality of the theory. The operation "prunes" identities that appear as locally maximal arguments to other terms, simplifying the overall structure of a term by removing unnecessary units.
- Endo-coherence removal: This operation was introduced in Section 2.4.3, and converts "fake identities", terms which are morally identities yet have the wrong syntactic form, into true identities. These converted identities can then be further removed from terms by pruning.
- Disc removal: Disc removal was the running example from Section 2.2, and removes unary composites from the theory. Commonly after pruning, a composite is reduced to a unary composite, for which disc removal is necessary to complete the simplification of the term.

In this section we will prove that Catt$_{\mathsf{su}}$ is a type theory satisfying many standard meta-theoretic properties by combining results from previous chapters. We also give a reduction system for Catt$_{\mathsf{su}}$ and show that this is strongly terminating and confluent.

*Example* 4.2.1. Suppose we have terms $f : x \to_\star y$, $g : y \to_\star z$, $h : x \to_\star z$, and $\alpha : f * g \to h$ in some context $\Gamma$. We can then consider the term:

$$\mathsf{Coh}_{((x:*),(y:*),(f:x\to_\star y),(z:\star),(g:y\to_\star z)\,;\,f*g\to f*g)}[\langle x, y, f, z, g\rangle] * \alpha$$

which consists of an endo-coherence composed with the variable $\alpha$. This then reduces as

follows:

$$
\begin{aligned}
&\mathsf{Coh}_{((x:*),(y:*),(f:x\to_\star y),(z:\star),(g:y\to_\star z)\,;\,f*g\to f*g)}[\langle f,g\rangle]*\alpha \\
\rightsquigarrow\ &\mathsf{id}(x\to_\star z, f*g)*\alpha && \text{by endo-coherence removal}\\
\rightsquigarrow\ &\mathsf{Coh}_{(D^2\,;\,\mathrm{wk}(U^2))}[\langle x,z,f*g,h,\alpha\rangle] && \text{by pruning}\\
\rightsquigarrow\ &\alpha && \text{by disc removal}
\end{aligned}
$$

and so uses all three reductions to fully simplify to a variable.

We define $\textsc{Catt}_{\mathsf{su}}$ by the following equality rule set.

**Definition 4.2.2.** Define the equality rule set su for $\textsc{Catt}_{\mathsf{su}}$ by:

$$\mathsf{su} = \mathsf{dr} \cup \mathsf{prune} \cup \mathsf{ecr}$$

$\textsc{Catt}_{\mathsf{su}}$ is then the variant of $\textsc{Catt}_{\mathcal{R}}$ where $\mathcal{R} = \mathsf{su}$.

When it is not specified, we will assume that the operation set $\mathcal{O}$ is given by the regular operation set Reg.

**Theorem 4.2.3.** *The rule set su is tame and satisfies the support and preservation conditions.*

*Proof.* By Propositions 2.4.10, 2.4.13, and 2.4.19, disc removal satisfies the weakening, suspension, and su-substitution conditions. Endo-coherence removal and pruning satisfy the same conditions by Propositions 2.4.37 and 3.1.17. As these conditions are closed under unions, the set su must also satisfy the weakening, suspension, and substitution conditions, and hence is tame.

We now use the proof strategy introduced in Section 2.4.2 to prove that su satisfies the support condition. Firstly, by Lemma 2.4.30 we know that $\mathsf{su}_{\mathsf{s}}$ is also tame. Disc removal then satisfies the $\mathsf{su}_{\mathsf{s}}$-support condition by Proposition 2.4.31. The same condition is satisfied by endo-coherence removal (Lemma 2.4.37(iv)) and pruning (Proposition 3.1.19) and so su satisfies the $\mathsf{su}_{\mathsf{s}}$-support condition. By Lemma 2.4.28, su satisfies the support condition.

Lastly, su satisfies the su-preservation condition as it is satisfied by disc removal (Proposition 2.4.34), endo-coherence removal (Lemma 2.4.37(v)), and pruning (Proposition 3.1.21) and is closed under unions of rule sets. □

From this theorem it can be deduced that weakening, suspension, and applications of substitution are well-formed. Furthermore, equality in $\textsc{Catt}_{\mathsf{su}}$ preserves the support of a term and preserves typing judgements. Such results are found in Section 2.4.

Before giving normalisation results for $\textsc{Catt}_{\mathsf{su}}$, we recall the Eckmann-Hilton argument (Figure 1) and give the definition of the term giving this equivalence. First let $\Delta$ be the ps-context

given by:

$$\begin{aligned}\Delta = D^2 \wedge D^2 = (x : *),&\\ &(y : *), (f : x \to y),\\ &\qquad (g : x \to y), (a : f \to g),\\ &(z : *), (h : x \to y),\\ &\qquad (j : x \to y), (b : h \to j)\end{aligned}$$

which is given by the diagram:

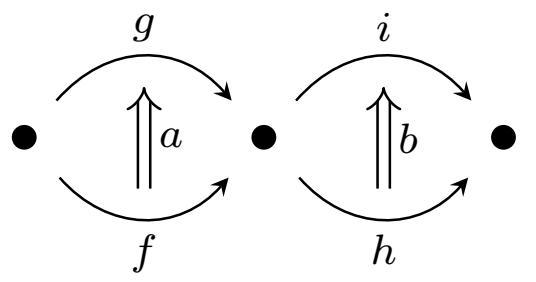

The following term can be formed, which is similar to an interchange move, and changes the order in which two whiskered terms are composed:

$$\mathsf{swap} = \mathsf{Coh}_{(\Delta\,;\,(a *_0 j) *_1 (g *_0 b) \to (f *_0 b) *_1 (a *_0 h))}[\mathsf{id}_\Delta]$$

Then given a context $\Gamma$ with terms $x : *$ and $\alpha, \beta : \mathsf{id}(x) \to \mathsf{id}(x)$, the following term, the Eckmann-Hilton term, can be formed:

$$\mathsf{EH}_{\alpha,\beta} = \mathsf{swap}[\![\langle x, x, \mathsf{id}(x), \mathsf{id}(x), \alpha, x, \mathsf{id}(x), \mathsf{id}(x), \beta\rangle]\!]$$

In $\textsc{Catt}_{\mathsf{su}}$, this term can be typed as follows:

$$\Gamma \vdash \mathsf{EH}_{\alpha,\beta} : \alpha *_1 \beta \to \beta *_1 \alpha$$

and so witnesses the Eckmann-Hilton argument.

We note that there is a clear inverse of the Eckmann Hilton term, which immediately gives rise to two morphisms $\alpha *_1 \beta \to \beta *_1 \alpha$: the original term $\mathsf{EH}_{\alpha,\beta}$ and the term $\mathsf{EH}^{-1}_{\beta,\alpha}$. These two terms manoeuvre $\alpha$ and $\beta$ round each other in opposite directions, and are not in general equivalent.

However, we can instead apply Eckmann-Hilton to terms $\phi$ and $\psi$ of type $\mathsf{id}^2(x) \to \mathsf{id}^2(x)$, which is done by suspending the Eckmann-Hilton term. By an abuse of notation we define this term to be (only giving the locally maximal arguments of the substitution):

$$\mathsf{EH}_{\phi,\psi} = \Sigma(\mathsf{swap})[\![\langle \phi, \psi\rangle]\!]$$

In this case, the extra dimension gives enough freedom to give an equivalence between the resulting two terms $\phi *_2 \psi \to \psi *_2 \phi$ which is called the *syllepsis* and has the type:

$$\mathsf{Syl}_{\phi,\psi} : \mathsf{EH}_{\phi,\psi} \to \mathsf{EH}^{-1}_{\psi,\phi}$$

To define this term, a similar approach to the approach used for Eckmann-Hilton of giving a single coherence containing a more complex type and a substitution containing multiple identity terms, and letting the $\textsc{Catt}_{\mathsf{su}}$ reduction simplify the type to the required one. We delay defining this term until Section 4.4, where the implementation presented in this section can be used to check that the resulting term is well-formed.

### 4.2.1 Normalisation for Catt$_{\mathsf{su}}$

Following Section 4.1 we aim to give a normalisation algorithm for Catt$_{\mathsf{su}}$ by exhibiting a strongly terminating and confluent reduction system.

The reduction system $\rightsquigarrow_{\mathsf{su}}$ cannot be used directly because the reduction generated from ecr is not terminating, as it allows identities to reduce to identities. Even after replacing the equality rule set ecr by ecr', the equality set obtained by removing these trivial identity to identity reductions from ecr, the generated reduction is still non-terminating. Consider the term $\mathsf{id}(t \to_A t, \mathsf{id}(A, t))$ for some term $t$ of type $A$. Then the following reduction sequence can be formed:

$$\mathsf{id}(t \to_A t, \mathsf{id}(A,t)) \rightsquigarrow \mathsf{Coh}_{(D^n\,;\,\mathrm{id}(\mathrm{wk}(U^n),d_n)\to\mathrm{id}(\mathrm{wk}(U^n),d_n))}[\{A,t\}] \rightsquigarrow \mathsf{id}(t \to_A t, \mathsf{id}(A,t))$$

where $n = \dim(A)$, the first reduction is by pruning, and the second reduction is by endo-coherence removal. We therefore choose to also restrict the pruning equality rule set to not apply when the head term is an identity, obtaining the set prune'. We can now define the reduction system for Catt$_{\mathsf{su}}$.

**Definition 4.2.4.** Define the reduction $\rightsquigarrow_{\mathsf{su}'}$ to be the reduction generated by the equality rule set $\mathsf{su}'$ where

$$\mathsf{su}' = \mathsf{dr} \cup \mathsf{prune}' \cup \mathsf{ecr}'$$

where ecr' is the endo-coherence removal set without identity to identity equalities and prune' is the pruning set restricted to the triples where the left-hand side term is not an identity.

The reduction $\rightsquigarrow_{\mathsf{su}'}$ applies equality rules from Catt$_{\mathsf{su}}$ when the redex is not an identity, effectively forcing identities to be normal forms of the theory. As applying a substitution to or suspending a non-identity term cannot result in an identity, it is clear that su' is tame. Strong termination for $\rightsquigarrow_{\mathsf{su}'}$ can now be proven using Lemma 4.1.7, by showing that all rules reduce the syntactic complexity of terms.

**Proposition 4.2.5.** *Let $s \rightsquigarrow t$ be an instance of pruning. If $s$ is not an identity then* $\mathsf{sc}(s) > \mathsf{sc}(t)$.

*Proof.* The reduction $s \rightsquigarrow t$ is an instance of pruning, and so there must be Dyck word $\mathcal{D} : \mathsf{Dyck}_0$, and peak $p : \mathsf{Peak}_{\mathcal{D}}$ such that

$$s \equiv \mathsf{Coh}_{(\lfloor \mathcal{D} \rfloor\,;\,A)}[\sigma] \qquad t \equiv \mathsf{Coh}_{(\lfloor \mathcal{D} /\!\!/ p \rfloor\,;\,A[\![\pi_p]\!])}[\sigma /\!\!/ p]$$

where $s$ is not an identity and $\lfloor p \rfloor [\![\sigma]\!]$ is. We then have $\mathsf{sc}(s) = \mathsf{sc}(\sigma)$ and $\mathsf{sc}(t) = \mathsf{sc}(\sigma /\!\!/ p)$, but $\sigma /\!\!/ p$ is simply $\sigma$ with two terms removed, one of which is known to be a coherence, and so $\mathsf{sc}(s) > \mathsf{sc}(t)$. □

**Corollary 4.2.6.** *The reduction $\rightsquigarrow_{\mathsf{su}'}$ is strongly terminating.*

*Proof.* By Lemma 4.1.7, it suffices to show that each rule of $\mathsf{su}'$ reduces syntactic complexity, which follows from the preceding proposition and Propositions 4.1.8 and 4.1.10. □

By Proposition 4.1.3, we know that the reflexive symmetric transitive closure of $\rightsquigarrow_{\mathsf{su}'}$ is the equality relation generated by su'. We therefore prove that this agrees with the equality relation from $\text{Catt}_{\mathsf{su}}$.

**Proposition 4.2.7.** *The type theories generated from su and su' are equivalent. Terms are equal or well-formed in one theory exactly when they are equal or well-formed in the other, and similar properties hold for types and substitutions.*

*Proof.* We use Lemma 2.4.2 for both directions. Since $\mathsf{su}' \subseteq \mathsf{su}$, we are only required to show that if $(\Gamma, s, t) \in \mathsf{su}$ with $\Gamma \vdash_{\mathsf{su}'} s : A$ for some $A : \mathsf{Type}_\Gamma$ then

$$\Gamma \vdash_{\mathsf{su}'} s = t$$

If $(\Gamma, s, t) \in \mathsf{su}'$, then the equality follows from the rule constructor. Otherwise, $s$ must be an identity and the rule is an instance of endo-coherence removal or pruning. Suppose $s$ reduces to $t$ by endo-coherence removal. Then $s \equiv \mathsf{id}(A, u)$ and

$$t \equiv \mathsf{id}(\mathrm{wk}(U^n)[\![\{A, u\}]\!], d_n[\![\{A, u\}]\!]) \equiv \mathsf{id}(A, u) \equiv s$$

and so the equality holds by reflexivity.

Now assume $s$ reduces by pruning to $t$. Letting $s \equiv \mathsf{id}(A, u)$ and $n = \dim(A)$, we get:

$$\begin{aligned}
t &\equiv \mathsf{Coh}_{(\lfloor \mathcal{D}^n /\!\!/ p^n \rfloor \,;\, d_n \to_{\mathrm{wk}(U^n)} d_n[\![\pi_{p^n}]\!])}[\{A, u\} /\!\!/ p] \\
&= \mathsf{id}(\mathrm{wk}(U^n)[\![\pi_{p^n}]\!][\![\{A, u\} /\!\!/ p^n]\!], d_n[\![\pi_{p^n}]\!][\![\{A, u\} /\!\!/ p^n]\!]) && \text{by endo-coherence removal} \\
&\equiv \mathsf{id}(\mathrm{wk}(U^n), d_n)[\![\pi_{p^n} \bullet \{A, u\} /\!\!/ p^n]\!] \\
&= \mathsf{id}(\mathrm{wk}(U^n), d_n)[\![\{A, u\}]\!] && \text{by Proposition 3.1.15} \\
&\equiv \mathsf{id}(\mathrm{wk}(U^n)[\![\{A, u\}]\!], d_n[\![\{A, u\}]\!]) \\
&\equiv \mathsf{id}(A, u)
\end{aligned}$$

and so the equality holds as required. □

We therefore have that two terms $s$ and $t$ are equal in $\text{Catt}_{\mathsf{su}}$ if and only if $s \leftrightsquigarrow_{\mathsf{su}'} t$. To demonstrate normalisation, it therefore remains to show that the reduction system is confluent, for which we employ the strategy introduced in Lemma 4.1.15.

**Theorem 4.2.8.** *The reduction $\rightsquigarrow_{su'}$ is confluent.*

*Proof.* By Lemma 4.1.15 it is sufficient to show that for all $t \leftsquigarrow s \rightsquigarrow u$ with $s \rightsquigarrow t$ being a reduction derived from rule, that the following diagram can be formed:

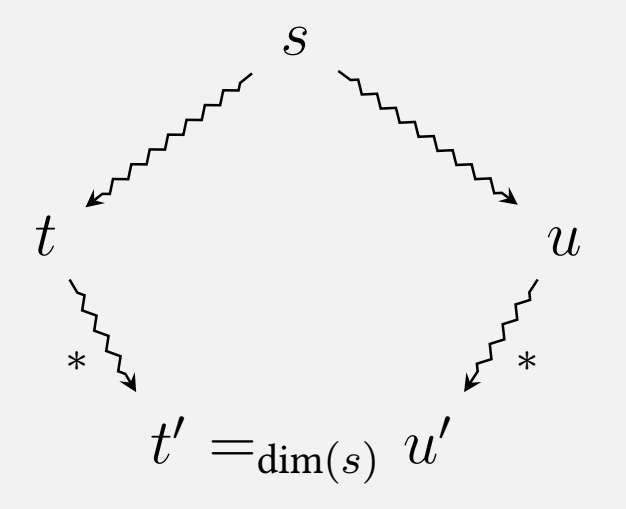

We therefore begin by case splitting on the reduction $s \rightsquigarrow t$, ignoring cases where both reductions are identical and ignoring cases which follow by symmetry of other cases.

**Disc removal:** Suppose $s \rightsquigarrow t$ is a disc removal reduction. Then $s \equiv \mathsf{Coh}_{(D^n \,;\, \mathrm{wk}(U^n))}[\{A, t\}]$. We now split on the reduction $s \rightsquigarrow u$. We immediately know that $s \rightsquigarrow u$ cannot be an endo-coherence removal reduction, as $s$ is not an endo-coherence. It also cannot be a cell reduction as $\mathrm{wk}(U^n)$ only contains variables and so is in normal form.

Let $s \rightsquigarrow u$ be an argument reduction. It must therefore be generated by a reduction on $\{A, t\}$. If it is a reduction generated by $A \rightsquigarrow A'$ then $u \rightsquigarrow t$ by endo-coherence removal and so we are done. Otherwise, it is generated by $t \rightsquigarrow t'$ and so $t$ and $u$ both reduce by disc removal to $t'$.

The only remaining case is where $s \rightsquigarrow u$ is an instance of pruning, which forces $t \equiv \mathsf{id}(B, a)$ for some $B$ and $a$. As $s$ is well-formed, we must have $n > 0$ and so $A \equiv b \to_{A'} c$. Therefore:

$$\begin{aligned} u &\equiv \mathsf{Coh}_{(\lfloor \mathcal{D}^n /\!\!/ p \rfloor \,;\, \mathrm{wk}(U^n)[\![\pi_p]\!])}[\{A, \mathsf{id}(B, a)\} /\!\!/ p] \\ &\equiv \mathsf{Coh}_{(D^{n-1} \,;\, \mathrm{wk}(U^n)[\![\{d_{n-1} \to_{\mathrm{wk}(U^{n-1})} d_{n-1}, \mathrm{id}(\mathrm{wk}(U^{n-1}), d_{n-1})\}\}]\!])}[\{A', b\}] \quad \text{by Proposition 3.1.16} \\ &\equiv \mathsf{Coh}_{(D^{n-1} \,;\, d_{n-1} \to_{\mathrm{wk}(U^{n-1})} d_{n-1})}[\{A', b\}] \\ &\equiv \mathsf{id}(A', b) \end{aligned}$$

Now as $s$ is well-formed we have $\Gamma \vdash \{A, \mathsf{id}(B, a)\} : D^n$ and so by Lemma 2.4.8, we have $\Gamma \vdash \mathsf{id}(B, a) : A$ and hence by Corollary 2.4.9 and uniqueness of typing:

$$a \to_B a = A \equiv b \to_{A'} c$$

and so $a = b$ and $B = A'$ and hence $s \equiv \mathsf{id}(A', b) = \mathsf{id}(B, a) \equiv t$. Since $\dim(a) = \dim(B) < \dim(s)$, we get $t =_{\dim(s)} u$ as required.

**Endo coherence removal:** Suppose $s \rightsquigarrow t$ is an endo-coherence removal reduction. Then:

$$s \equiv \mathsf{Coh}_{(\Delta \,;\, a \to_A a)}[\sigma] \rightsquigarrow \mathsf{id}(A[\![\sigma]\!], a[\![\sigma]\!]) \equiv t$$

with $s$ not being an identity. We now split on the reduction $s \rightsquigarrow u$. First consider when it is an argument reduction generated by $\sigma \rightsquigarrow \tau$. Then by Proposition 4.1.4, we have $t \equiv \mathsf{id}(A[\![\sigma]\!], a[\![\sigma]\!]) \rightsquigarrow^* \mathsf{id}(A[\![\tau]\!], a[\![\tau]\!])$. By endo-coherence removal, $u \rightsquigarrow \mathsf{id}(A[\![\tau]\!], a[\![\tau]\!])$, completing this case.

Now suppose the reduction $s \rightsquigarrow u$ is an instance of cell reduction. If it is generated from a reduction $A \rightsquigarrow B$ then by Proposition 4.1.4, $t \rightsquigarrow \mathsf{id}(B[\![\sigma]\!], a[\![\sigma]\!])$ and by endo-coherence removal:

$$u \equiv \mathsf{Coh}_{(\Delta \,;\, a \to_B a)}[\sigma] \rightsquigarrow \mathsf{id}(B[\![\sigma]\!], a[\![\sigma]\!])$$

We now consider when the reduction is generated by $a \to_A a \rightsquigarrow b \to_A a$, with the case where it is generated by $a \to_A a \rightsquigarrow a \to_A b$ following symmetrically. We consider the reductions sequence from $u$:

$$\begin{aligned} u &\equiv \mathsf{Coh}_{(\Delta \,;\, b \to_A a)}[\sigma] \\ &\rightsquigarrow \mathsf{Coh}_{(\Delta \,;\, b \to_A b)}[\sigma] && \text{by cell reduction} \\ &\rightsquigarrow \mathsf{id}(A[\![\sigma]\!], b[\![\sigma]\!]) && \text{by endo-coherence removal} \end{aligned}$$

Again by Proposition 4.1.4, $t \equiv \mathsf{id}(A[\![\sigma]\!], a[\![\sigma]\!]) \rightsquigarrow \mathsf{id}(A[\![\sigma]\!], b[\![\sigma]\!])$, completing the case.

Lastly, we consider when $s \rightsquigarrow u$ is a pruning reduction. We suppose $\Delta = \lfloor \mathcal{D} \rfloor$ and that the pruning is generated from peak $p : \mathcal{D}$. Then:

$$u \equiv \mathsf{Coh}_{(\lfloor \mathcal{D} /\!\!/ p \rfloor \,;\, (a \to_A a)[\![\pi_p]\!])}[\sigma /\!\!/ p]$$

Then:

$$\begin{aligned} u &\rightsquigarrow \mathsf{id}(A[\![\pi_p]\!][\![\sigma /\!\!/ p]\!], a[\![\pi_p]\!][\![\sigma /\!\!/ p]\!]) && \text{by Lemma 4.1.9} \\ &\equiv \mathsf{id}(A, a)[\![\pi_p \bullet \sigma /\!\!/ p]\!] \\ &=_{\dim(s)} \mathsf{id}(A, a)[\![\sigma]\!] \end{aligned}$$

where the last (bounded) equality is by Proposition 3.1.15 and by noting that $\dim(A) = \dim(a) < \dim(s)$.

**Pruning:** Let $s \rightsquigarrow t$ be a reduction by pruning with

$$s \equiv \mathsf{Coh}_{(\lfloor \mathcal{D} \rfloor \,;\, A)}[\sigma]$$

for some $\mathcal{D} : \mathsf{Dyck}_0$ with peak $p : \mathsf{Peak}_{\mathcal{D}}$ such that $\lfloor p \rfloor [\![\sigma]\!]$ is an identity. Then:

$$t \equiv \mathsf{Coh}_{(\lfloor \mathcal{D} /\!\!/ p \rfloor \,;\, A[\![\pi_p]\!])}[\sigma /\!\!/ p]$$

We now split on the reduction $s \rightsquigarrow u$. First suppose it is given by an argument reduction $\sigma \rightsquigarrow \tau$. Identities do not admit head reductions, meaning $\lfloor p \rfloor [\![\tau]\!]$ is still an identity. Therefore, pruning can be applied to $u$ to get:

$$u \rightsquigarrow \mathsf{Coh}_{(\lfloor \mathcal{D} /\!\!/ p \rfloor \,;\, A[\![\pi_p]\!])}[\tau /\!\!/ p]$$

Now $\sigma /\!\!/ p$ is simply $\sigma$ with two terms removed, and so $\sigma /\!\!/ p \rightsquigarrow^* \tau /\!\!/ p$, meaning $t$ reduces to the same term as $u$.

If $s \rightsquigarrow u$ is a cell reduction $A \rightsquigarrow B$, then pruning can be applied to $u$ immediately to get the term:

$$\mathsf{Coh}_{(\lfloor \mathcal{D} /\!\!/ p \rfloor \,;\, B[\![\pi_p]\!])}[\sigma /\!\!/ p]$$

but $t$ also reduces to this term by Proposition 4.1.4.

Let $s \rightsquigarrow u$ be a second pruning reduction, at a different peak $q : \mathsf{Peak}_{\mathcal{D}}$. By Proposition 3.1.22, there is a common reduct:

$$\mathsf{Coh}_{(\lfloor (\mathcal{D} /\!\!/ p) /\!\!/ q_p \rfloor \,;\, A[\![\pi_p]\!][\![\pi_{q_p}]\!])}[(\sigma /\!\!/ p) /\!\!/ q_p]$$

which both reduce to by pruning if $\lfloor q_p \rfloor$ and $\lfloor p_q \rfloor$ are identities. However:

$$\lfloor q_p \rfloor \equiv \lfloor q \rfloor [\![\pi_p]\!]$$

and $\lfloor q \rfloor$ must be an identity for $s \rightsquigarrow u$ to be a valid instance of pruning. Therefore, as identities are preserved by application of substitution, $\lfloor q_p \rfloor$ is an identity. Similarly, $\lfloor p_q \rfloor$ is an identity, and so both $t$ and $u$ reduce to the term above.

Any remaining cases follow by symmetry, completing the proof. □

### 4.2.2 Disc trivialisation

We take a brief moment to explore the theory $\textsc{Catt}_{\mathsf{su}}$ in its entirety. For this section we will further assume that we take the set of operations $\mathcal{O}$ to be the regular operations.

We begin by proving a property of terms over disc contexts, which we call *disc trivialisation*. This is the following structure theorem: in a disc context $D^n$, every term is either a variable, or the iterated identity on a variable, up to definitional equality. Restricting to those terms $t : \mathsf{Term}_{D^n}$ that are full, that is $\mathrm{Supp}(t) = \mathrm{Var}\, D^n$, then there is exactly one term (up to definitional equality) at each dimension $k \geq n$. Hence, the type theory $\textsc{Catt}_{\mathsf{su}}$ trivialises disc contexts.

This property relates the type theory $\textsc{Catt}_{\mathsf{su}}$ to the definition of strictly unital $\infty$-categories of Batanin, Cisinski, and Weber [BCW13], whose *reduced operads* enforce that there is a unique term of each dimension over a linear tree.

We now state and prove disc trivialisation, recalling the definition of an iterated canonical identity from Definition 2.2.8.

**Theorem 4.2.9** (Disc trivialisation)**.** *Suppose $D^n \vdash t : A$ in $\textsc{Catt}_{su}$. Then $t$ is equal to an iterated canonical identity on a variable, that is $t = \mathsf{id}^k(x)$ for some variable $x \in \mathrm{Var}(D^n)$ and $k \in \mathbb{N}$.*

*Proof.* Without loss of generality, we may assume that $t$ is in $\textsc{Catt}_{\mathsf{su}}$ normal form, and proceed to prove that $t$ is an iterated canonical identity. We proceed by induction on subterms of the term $t$. If $t$ is a variable then we are done. Otherwise, we assume $t$ is a coherence term $\mathsf{Coh}_{(\Delta\,;\,U)}[\sigma]$.

We now show that $\Delta$ must be a disc context by contradiction. We therefore assume that $\Delta$ is not a disc, and hence $t$ is not an identity. By induction on subterms, we must have that each term in $\sigma$ is an iterated canonical identity on a variable. No locally maximal argument can be an identity, as otherwise pruning could be performed and $t$ would not be in normal form, and so every locally maximal argument is a variable. Suppose there is some variable $x$ such that $x[\![\sigma]\!]$ is an identity, and let $x$ be a variable of maximal dimension with this property. As $x$ cannot be locally maximal, there must either be the source or target of a variable $y$, but this variable $y$ must be sent to a variable of $D^n$, which cannot have an identity as its source or target. Therefore, the substitution $\sigma$ is variable to variable.

We now let $\Gamma$ be the smallest ps-context prefix of $\Delta$ such that $\Gamma$ is not a disc. We must have:

$$\Gamma \equiv D^k, (y : A), (f : x \to_A y)$$

where $D^k \vdash_{\mathsf{ps}} x : A$. Furthermore, the last rule used in this derivation must be $\textsc{psd}$, as if it were $\textsc{pse}$ or $\textsc{pss}$ then $k = \dim(A)$ and $\Gamma \equiv D^{k+1}$, breaking the assumption that $\Gamma$ is not a disc. Therefore, $D^k \vdash_{\mathsf{ps}} g : w \to_A x$ for some variables $g$ and $x$. However, now $g[\![\sigma]\!]$, $x[\![\sigma]\!]$, and $f[\![\sigma]\!]$ are variables of $D^n$ such that $\mathsf{tgt}(g[\![\sigma]\!]) \equiv x[\![\sigma]\!] \equiv \mathsf{src}(f[\![\sigma]\!])$. No such variables exist in $D^n$ and so we reach a contradiction. We therefore deduce that $\Delta$ is a disc $D^n$ for some $n$.

Now $t \equiv \mathsf{Coh}_{(D^n\,;\,u \to_A v)}[\sigma]$ and so by induction on subterms, $u$ and $v$ are equal to iterated canonical identities. We now split on whether $t$ is a composition or equivalence. If it is a composition then $\mathrm{Supp}(u) = \partial^-_{n-1}(D^n)$ and $\mathrm{Supp}(v) = \partial^+_{n-1}(D^n)$ and therefore neither $u$ or $v$ are identities (as then $A$ would have the same support as $u$ or $v$ respectively) and so $u = d^-_{n-1}$ and $v = d^*_{n-1}$, but this makes $t$ a disc removal redex, and so $t$ is not in normal form.

We therefore assume that $t$ is an equivalence and $u$ and $v$ are full. Then $u$ and $v$ must be iterated identities on $d_n$, and must have the same dimension and so are syntactically equal. To avoid $t$ being an endo-coherence removal redex, it must be an identity $\mathsf{id}(B, s)$. Now, $s \equiv \mathsf{id}^k(x)$ for some variable $x$ (as $s$ is a subterm of $t$), and so if $k = 0$ then $\mathrm{Ty}(s) \equiv d^-_{n-1} \to d^+_{n-1}$ and if $k > 0$ then $\mathrm{Ty}(s) \equiv \mathsf{id}^{k-1}(x) \to \mathsf{id}^{k-1}(x)$. In either case, $\mathrm{Ty}(s)$ is in normal form, and so since $B$ is also a normal form and $\Gamma \vdash s : B$ (by the well-typing of $t$ and Corollary 2.4.9), we have $B \equiv \mathrm{Ty}(s)$ and so $t \equiv \mathsf{id}(s) \equiv \mathsf{id}^{k+1}(x)$ as required. □

Disc trivialisation allows us to prove the following results concerning terms and substitutions in pasting diagrams.

**Theorem 4.2.10.** *Let $\mathcal{D}$ be a Dyck word. Let $t$ be a well-formed* Catt$_{su}$ *term of $\lfloor \mathcal{D} \rfloor$. Then* $\mathrm{Supp}(t)$ *is a ps-context.*

*Proof.* Suppose, for contradiction, that we have a Dyck word $\mathcal{D}$ and a term $t$ where $\mathrm{Supp}(t)$ is not a ps-context. Assume further that $\mathcal{D}$ is minimal (in terms of length) where such a term exists.

Immediately, $\mathcal{D} \not\equiv \circleddash$, as all terms have non-empty support. We now examine the locally maximal variables of $\mathcal{D}$. There must exist some locally maximal variable $f : x \to y$ such that $f \notin \mathrm{Supp}(t)$, as otherwise $\mathrm{Supp}(t) = \mathrm{Var}(\lfloor \mathcal{D} \rfloor)$.

Now suppose that $y \notin \mathrm{Supp}(t)$. Then we let $p$ be the peak corresponding to $f$ and consider the term:

$$t[\![\pi_p]\!] : \mathsf{Term}_{\lfloor \mathcal{D} /\!\!/ p \rfloor}$$

Then $\mathrm{Supp}(t[\![\pi_p]\!]) = \mathrm{Supp}(t)$, which contradicts the minimality of $\mathcal{D}$. By a similar argument, $x$ must also be in $\mathrm{Supp}(t)$. It is also the case that if such a variable $f : x \to y$ with $f \notin \mathrm{Supp}(t)$ and $\{x, y\} \subseteq \mathrm{Supp}(t)$ exists, then $\mathrm{Supp}(t)$ cannot be a ps-context, by an argument involving the linear order on ps-contexts introduced by Finster and Mimram [FM17].

Now suppose $\mathcal{D}$ has a peak $p$ that is not $f$. Then $f[\![\pi_p]\!] : x[\![\pi_p]\!] \to y[\![\pi_p]\!]$ is a locally maximal variable of $\lfloor \mathcal{D} /\!\!/ p \rfloor$ with $f[\![\pi_p]\!] \notin \mathrm{Supp}(t[\![\pi_p]\!])$ and $\{x[\![\pi_p]\!], y[\![\pi_p]\!]\} \subseteq \mathrm{Supp}(t[\![\pi_p]\!])$. Hence, $\mathrm{Supp}(t[\![\pi_p]\!])$ is not a ps-context, again breaking the minimality of $\mathcal{D}$.

Therefore, $\mathcal{D}$ has one peak, and so $\lfloor \mathcal{D} \rfloor \equiv D^n$ for some $n$. Now by Theorem 4.2.9, $t$ is Catt$_{su}$ equal to a variable $z$ or an iterated identity on a variable $z$. Since Catt$_{su}$ preserves support, we must have $\mathrm{Supp}(t) = \mathrm{Supp}(z)$, but $\mathrm{Supp}(z)$ is a disc and so is a ps-context.

Hence, no such term $t$ existed. □

Since any Catt term is also a Catt$_{su}$ term, we get the following corollary.

**Corollary 4.2.11.** *If $\Gamma \vdash t : A$ in* Catt*, and $\Gamma$ is a ps-context, then* $\mathrm{Supp}(t)$ *is a ps-context.*

## 4.3 $\textsc{Catt}_{\mathsf{sua}}$

We now move on to defining $\textsc{Catt}_{\mathsf{sua}}$, the type theory for strictly unital and associative $\infty$-categories. $\textsc{Catt}_{\mathsf{sua}}$ extends $\textsc{Catt}_{\mathsf{su}}$ by replacing the pruning equality with the more general insertion equality, which was introduced in Section 3.4. Under certain conditions, insertion can merge more complex terms into a single coherence. As an example, the term $(f * g) * h$, which is a composite which has a composite as one of its arguments, is reduced by insertion to the ternary composite $f * g * h$, reducing the height of the term.

As we did for $\textsc{Catt}_{\mathsf{su}}$, we will prove in this section that $\textsc{Catt}_{\mathsf{sua}}$ satisfies standard meta-theoretic properties, and provide a reduction system for it which is strongly terminating and confluent.

*Example* 4.3.1. We consider the associator term, and its reductions in $\textsc{Catt}_{\mathsf{sua}}$. The associator witnesses the associativity law in a weak $\infty$-category. Letting $\Delta$ be the following ps-context:

$$\begin{aligned}\Delta = \lfloor[[\,],[\,],[\,]]\rfloor = (w : *)&\\ &(x : *)\,(f : w \to x)\\ &(y : *)\,(g : x \to y)\\ &(z : *)\,(h : y \to z)\end{aligned}$$

we can define the associator as:

$$\alpha = \mathsf{Coh}_{(\Delta\,;\,(f*g)*h \to f*(g*h))}[\mathsf{id}_\Delta]$$

This then admits the following reduction sequence in $\textsc{Catt}_{\mathsf{sua}}$:

$$\begin{aligned}\alpha &\rightsquigarrow \mathsf{Coh}_{(\Delta\,;\,f*g*h \to f*(g*h))}[\mathsf{id}_\Delta] && \text{by insertion}\\ &\rightsquigarrow \mathsf{Coh}_{(\Delta\,;\,f*g*h \to f*g*h)}[\mathsf{id}_\Delta] && \text{by insertion}\\ &\rightsquigarrow \mathsf{id}(f*g*h) && \text{by endo-coherence removal}\end{aligned}$$

We formally define $\textsc{Catt}_{\mathsf{sua}}$ as the version of $\textsc{Catt}_{\mathcal{R}}$ generated by the rule set sua, which we define below:

**Definition 4.3.2.** We define the equality rule set sua for $\textsc{Catt}_{\mathsf{sua}}$ by:

$$\mathsf{sua} = \mathsf{dr} \cup \mathsf{ecr} \cup \mathsf{insert}$$

$\textsc{Catt}_{\mathsf{sua}}$ is then the variant of $\textsc{Catt}_{\mathcal{R}}$ where $\mathcal{R} = \mathsf{sua}$.

As before, when we do not specify an operation set, it should be assumed that the regular operation set is used. When we use the groupoidal operation set, we refer to the resulting theory as *groupoidal* $\textsc{Catt}_{\mathsf{sua}}$.

**Theorem 4.3.3.** *The rule set sua is tame and satisfies the support condition. If $\mathcal{O}$ supports insertion, then sua also satisfies the preservation condition.*

*Proof.* By Propositions 2.4.10, 2.4.13, 2.4.19, 2.4.37, and 3.4.17, each of the disc removal, endo-coherence removal, and insertion sets satisfy the weakening, suspension, and sua-substitution conditions. It follows that sua satisfies the weakening, suspension, and substitution conditions. Hence, sua is tame.

To prove that the support condition holds for sua, we use the strategy introduced in Section 2.4.2 and instead show that sua satisfies the $\mathsf{sua}_\mathsf{S}$-support condition. By Lemma 2.4.30, the equality rule set $\mathsf{sua}_\mathsf{S}$, the restriction of sua to support preserving equalities, is also tame. As it trivially satisfies the support condition, we have by Propositions 2.4.31 and 3.4.19 and Lemma 2.4.37(iv) that disc removal, endo-coherence removal, and insertion satisfy the $\mathsf{sua}_\mathsf{S}$-support condition. Therefore, sua satisfies the $\mathsf{sua}_\mathsf{S}$-support condition and so by Lemma 2.4.28 sua satisfies the support condition.

The sua-preservation condition is satisfied by disc removal (by Proposition 2.4.34) and endo-coherence removal (by Lemma 2.4.37(v)). If $\mathcal{O}$ supports insertion, then insertion also satisfies the sua-preservation condition by Proposition 3.4.21. Therefore, sua satisfies the preservation condition, completing the proof. ☐

While the groupoidal operation set trivially supports insertion, we have not yet proven that the regular operation set, Reg, supports insertion. This is done now using Theorem 4.3.3.

**Proposition 4.3.4.** *The regular operation set, Reg, supports insertion.*

*Proof.* Using that the regular operation set is equal to the standard operation set, we instead prove that the standard operation set supports insertion. For this it will be sufficient to prove that for an insertion point $(S, P, T)$, dimension $n \in \mathbb{N}$ and $\epsilon \in \{-, +\}$ that:

$$\partial_n^\epsilon(S)[\![\kappa_{S,P,T}]\!] = \partial_n^\epsilon(S \ll_P T)$$

Then:

$$\begin{aligned}
\partial_n^\epsilon(S)[\![\kappa_{S,P,T}]\!] &= \mathsf{Supp}(\mathcal{T}^n_{\partial_n(S)}[\![\delta_n^\epsilon(S)]\!])[\![\kappa_{S,P,T}]\!] && \text{by Lemma 3.3.22}\\
&= \mathsf{Supp}(\mathcal{T}^n_{\partial_n(S)}[\![\delta_n^\epsilon(S) \bullet \kappa_{S,P,T}]\!])\\
&= \mathsf{Supp}(\mathcal{T}^n_{\partial_n(S \ll_P T)}[\![\delta_n^\epsilon(S \ll_P T)]\!]) && \text{by (*)}\\
&= \partial_n^\epsilon(S \ll_P T) && \text{by Lemma 3.3.22}
\end{aligned}$$

where the equality $(*)$ holds as sua satisfies the support condition by Theorem 4.3.3 and:

$$S \ll_P T \vdash_{\mathsf{sua}} \mathcal{T}^n_{\partial_n(S)}[\![\delta_n^\epsilon(S) \bullet \kappa_{S,P,T}]\!] = \mathcal{T}^n_{\partial_n(S \ll_P T)}[\![\delta_n^\epsilon(S \ll_P T)]\!]$$

by Theorem 3.4.37. ☐

### 4.3.1 Reduction for $\textsc{Catt}_{\mathsf{sua}}$

Using the results of Section 4.1, we give a normalisation algorithm for $\textsc{Catt}_{\mathsf{sua}}$ by defining a reduction system which generates the equality relation and proving that this reduction system is strongly terminating and confluent.

As with $\textsc{Catt}_{\mathsf{su}}$, we cannot directly use the reduction $\rightsquigarrow_{\mathsf{sua}}$ directly, as we have seen already that the reduction $\rightsquigarrow_{\mathsf{ecr}}$ alone is non-terminating. Similarly to pruning, allowing insertions into identity terms also creates non-terminating loops of reductions when combined with endo-coherence removal, as was explained in Section 4.2.1. We therefore restrict our reduction so that no head-reductions can be applied to identity terms.

Although these restrictions are sufficient to ensure termination, we choose to further restrict the set of insertion reductions, in order to streamline the proof of confluence. Firstly, we only allow insertions of a locally maximal argument when that argument is either an identity or a standard composition. The motivation for this restriction is that identities and standard compositions are the only standard coherences that are in normal form. Moreover, not allowing the insertion of endo-coherences avoids a difficult insertion/argument endo-coherence removal confluence case.

We also disallow insertions into a unary composite and insertions of a unary composite, as we have already seen in Section 3.4.3 that discs act as a left and right unit for insertion, and so these two insertion reductions are subsumed by disc removal. Further, disallowing the insertion of discs removes another case where an insertable standard coherence is not in normal form. We now define the resulting reduction system.

**Definition 4.3.5.** Define the reduction $\rightsquigarrow_{\mathsf{sua}'}$ to be the reduction generated by the equality rule set $\mathsf{sua}'$ where:

$$\mathsf{sua}' = \mathsf{dr} \cup \mathsf{ecr}' \cup \mathsf{insert}'$$

where $\mathsf{ecr}'$ is the endo-coherence removal set without the identity-to-identity reductions, and $\mathsf{insert}'$ is the insertion rule set restricted to insertion redexes $(S, P, T, \Gamma, L, M)$ and types $A$ such that $\mathsf{SCoh}_{(S\,;\,A)}[L]$ is not an identity or a unary composite, and $L(\overline{P}) \equiv \mathcal{C}_T^{\mathsf{lh}(P)}[\![M]\!]$ is an identity or a standard composite which is not a unary composite.

It can be determined by a simple observation that $\mathsf{sua}'$ is tame, as suspension and the application of substitution cannot transform a term into an identity or unary composite where it wasn't before. We further justify the restrictions made to insertion by showing that many insertion reductions can still be performed, starting with the following technical lemma.

**Lemma 4.3.6.** *If $P$ is a branch of $S$, and $L, L' : S \to \Gamma$ are labellings differing only on $\overline{P}$, then the following holds for insertion redex $(S, P, T, \Gamma, L, M)$:*

$$L \ll_P M \equiv L' \ll_P M$$

*Proof.* By inspection of the definition, $L \ll_P M$ does not use the term $L(\overline{P})$. □

We now show that many insertion reductions can still be simulated up to bounded equality.

**Lemma 4.3.7.** *Let $(S, P, T, \Gamma, L, M)$ be an insertion redex. Further suppose that $a \equiv \mathsf{SCoh}_{(S\,;\,A)}[L]$ is not an identity or disc. Then there exists a term $s$ with:*

$$a \rightsquigarrow_{sua'}^{*} s =_{\dim(a)} \mathsf{SCoh}_{(S \ll_P T\,;\,A[\![\kappa_{S,P,T}]\!])}[L \ll_P M]$$

*even when $L(\overline{P})$ is a unary composite or is not a standard composite or identity.*

*Proof.* We proceed by induction on $\mathsf{lh}(P) - \mathrm{h}(T)$. If $\mathsf{lh}(P) - \mathrm{h}(T) = 0$ then $\mathcal{C}_T^{\mathsf{lh}(P)}$ is a composite. The only case for which insertion cannot be performed is when $\mathcal{C}_T^{\mathsf{lh}(P)}$ is a unary composite, such that $T = D^{\mathsf{lh}(P)}$. Now by Lemma 3.4.23, $S \ll_P T \equiv S$, $L \ll_P M \equiv^{\max} L$ and $\kappa_{S,P,T} = \mathsf{id}_S$ and so

$$a =_{\dim(a)} \mathsf{SCoh}_{(S \ll_P T\,;\,A[\![\kappa_{S,P,T}]\!])}[L \ll_P M]$$

We now assume that $\mathsf{lh}(P) > \dim(T)$. We may also assume without loss of generality that $\mathcal{C}_T^{\mathsf{lh}(P)}$ is not an identity, as otherwise it would be immediately insertable. This allows us to perform endo-coherence removal to get:

$$\mathcal{C}_T^{\mathsf{lh}(P)} \rightsquigarrow \mathsf{id}(\mathcal{U}_T^{\mathsf{lh}(P)-1}, \mathcal{T}_T^{\mathsf{lh}(P)-1})[\![M]\!]$$

Now suppose $b \equiv \mathsf{Coh}_{(S\,;\,A)}[L']$ where $L'$ is the result of applying the above reduction to the term of $L$ corresponding to $\overline{P}$. Since $L'(\overline{P})$ is now an identity it can be inserted to get $b \rightsquigarrow c$ where:

$$\begin{aligned} c &\equiv \mathsf{SCoh}_{(S /\!\!/ P\,;\,A[\![\pi_P]\!])}[L' \ll_P (\{\mathcal{T}_T^{\mathsf{lh}(P)-1}\} \bullet M)] \\ &\equiv \mathsf{SCoh}_{(S /\!\!/ P\,;\,A[\![\pi_P]\!])}[L' \ll_P (\{\mathcal{C}_T^{\mathsf{lh}(P)-1}\} \bullet M)] \end{aligned}$$

where $\mathcal{T}_T^{\mathsf{lh}(P-1)} \equiv \mathcal{C}_T^{\mathsf{lh}(P-1)}$ as if $\mathcal{T}_T^{\mathsf{lh}(P)-1}$ was a variable then $\mathcal{C}_T^{\mathsf{lh}(P)}$ would be an identity.

We now wish to show that $2 + \mathsf{bh}(P) \leq \mathsf{lh}(P)$ so that $P'$ exists as a branch of $S /\!\!/ P$. Since we always have $1 + \mathsf{bh}(P) \leq \mathsf{lh}(P)$, we consider the case where $1 + \mathsf{bh}(P) = \mathsf{lh}(P)$. We know that $\mathsf{bh}(P) \leq \mathrm{h}(T) \leq \mathsf{lh}(P)$ and so one of these inequalities must be an equality. If $\mathrm{h}(T) = \mathsf{lh}(P)$ then $\mathcal{C}_T^{\mathsf{lh}(P)}$ is a standard composite. If $\mathrm{h}(T) = \mathsf{bh}(P)$ then $\mathsf{th}(T) = \mathrm{h}(T)$ and so $T$ is linear. However, this makes $\mathcal{C}_T^{\mathsf{lh}(P)}$ an identity. Either case is a contradiction and so $2 + \mathsf{bh}(P) \leq \mathsf{lh}(P)$ and so $P'$ is a branch of $S /\!\!/ P$.

By Lemmas 3.4.12 and 3.4.26, we now have:

$$\begin{aligned} &\overline{P'}[\![L' \ll_P (\{\mathcal{C}_T^{\mathsf{lh}(P)-1}\} \bullet M)]\!] \\ \equiv\; &d_{\mathsf{lh}(P)-1}[\![\iota_{S,P,D^{\mathsf{lh}(P)-1}} \bullet (L' \ll_P (\{\mathcal{C}_T^{\mathsf{lh}(P)-1}\} \bullet M))]\!] \\ \equiv\; &d_{\mathsf{lh}(P)-1}[\![\{\mathcal{C}_T^{\mathsf{lh}(P)-1}\} \bullet M]\!] \\ \equiv\; &\mathcal{C}_T^{\mathsf{lh}(P)-1}[\![M]\!] \end{aligned}$$

As $\mathsf{lh}(P') - \dim(T) = \mathsf{lh}(P) - \dim(T) - 1$ we can use the induction hypothesis to get that $c \rightsquigarrow d$ and:

$$\begin{aligned} d =_{\dim(a)} \mathsf{SCoh}&_{((S /\!\!/ P) \ll_{P'} T\,;\,A[\![\pi_P \bullet \kappa_{S /\!\!/ P,P',T}]\!])}[ \\ &(L' \ll_P (\{\mathcal{C}_T^{\mathsf{lh}(P)-1}\} \bullet M)) \ll_{P'} M] \end{aligned}$$

By Lemmas 3.4.26 and 4.3.6,

$$d =_{\dim(a)} \mathsf{SCoh}_{(S \ll_P T\,;\,A[\![\kappa_{S,P,T}]\!])}[L \ll_P M]$$

which completes the proof as $a \rightsquigarrow^* d$. □

We further show that insertions into discs can be simulated by disc removal.

**Lemma 4.3.8.** *Let $(D^n, P, T, \Gamma, L, M)$ be an insertion redex and let $a \equiv \mathcal{C}^n_{D^n}[\![L]\!]$. Then:*

$$a \rightsquigarrow_{\mathit{sua}'} s =_n \mathsf{SCoh}_{(D^n \ll_P T\,;\,\mathcal{U}^n_{D^n}[\![\kappa]\!])}[L \ll_P M]$$

*Proof.* We have the equality:

$$\begin{aligned}
\mathsf{SCoh}_{(D^n \ll_P T\,;\,\mathcal{U}^n_{D^n}[\![\kappa]\!])}[L \ll_P M] &\equiv \mathsf{SCoh}_{(T\,;\,\mathcal{U}^n_{D^n}[\![\kappa_{D^n,P,T}]\!])}[M] && \text{Lemma 3.4.22}\\
&=_n \mathsf{SCoh}_{(T\,;\,\mathcal{U}^n_T)}[M] && \text{by Theorem 3.4.37}\\
&\equiv \mathcal{C}^n_T[\![M]\!]\\
&\equiv L(\overline{P})
\end{aligned}$$

Therefore, the reduction $a \rightsquigarrow s \equiv L(\overline{P})$ is given by disc removal. □

Using these lemmas, we now show that the type theories Catt$_{\mathsf{sua}}$ and Catt$_{\mathsf{sua}'}$ are equivalent.

**Proposition 4.3.9.** *The type theories generated by sua and sua' are equivalent. Terms, types, and substitutions are equal or well-formed in one theory exactly when they are equal or well-formed in the other.*

*Proof.* Both directions proceed by Lemma 2.4.2. Since $\mathsf{sua}' \subseteq \mathsf{sua}$, it suffices to show that if $(\Gamma, s, t) \in \mathsf{sua}$ with $\Gamma \vdash_{\mathsf{sua}'} s : A$ for some type $A$ then:

$$\Gamma \vdash_{\mathsf{sua}'} s = t$$

If $(\Gamma, s, t) \in \mathsf{sua}'$, then there is nothing to do. If it is in $\mathsf{ecr}'$, then the argument is the same as in the proof of Proposition 4.2.7. We therefore assume $(\Gamma, s, t) \in \mathsf{insert}$, and so there must be some insertion redex $(S, P, T, \Gamma, L, M)$ such that $s \equiv \lfloor \mathsf{SCoh}_{(S\,;\,B)}[L] \rfloor$ and

$$t \equiv \lfloor \mathsf{SCoh}_{(S \ll_P T\,;\,B[\![\kappa_{S,P,T}]\!])}[L \ll_P M] \rfloor$$

By an induction on dimension, we assume that the theories generated by sua and $\mathsf{sua}'$ are already equivalent for terms of dimension less than $\dim(s)$. We begin a case analysis of such reductions than are not in insert. If $s$ is an identity, then $B \equiv b \to b$ for some term $b$ and so $t$ is an endo-coherence. If $t$ is already an identity, then $s \equiv t$. Otherwise:

$$\begin{aligned}
\Gamma \vdash_{\mathsf{sua}'} t &= \mathsf{id}(b[\![\kappa_{S,P,T}]\!])[\![L \ll_P M]\!]\\
&\equiv \mathsf{id}(b)[\![\kappa_{S,P,T} \bullet (L \ll_P M)]\!]\\
&= \mathsf{id}(b)[\![L]\!]\\
&\equiv s
\end{aligned}$$

where the first equality is by endo-coherence removal, and the second equality is by Lemma 3.4.13, appealing to the induction on dimension.

If $s$ is a unary composite we apply Lemma 4.3.8 and use the inductive hypothesis on dimension. Otherwise, we are done by Lemma 4.3.7 and the inductive hypothesis on dimension. □

Having shown that the reflexive symmetric transitive closure of the reduction $\rightsquigarrow_{\mathsf{sua}'}$ agrees with the equality of $\textsc{Catt}_{\mathsf{sua}}$, we move on to showing that this reduction is strongly terminating. To do this we appeal to Lemma 4.1.7, and show that all reductions reduce the syntactic complexity of the terms involved.

**Lemma 4.3.10.** *The following inequality holds for any insertion redex* $(S, P, T, \Gamma, L, M)$*:*

$$\mathsf{sc}(L \ll_P M) < \mathsf{sc}(L)$$

*Proof.* We extend the notion of syntactic complexity to labellings in the obvious way. We begin by noting that:

$$\begin{aligned}
\mathsf{sc}(L) &= \left( \mathop{\#}_{p \neq \overline{P}} \mathsf{sc}(L(p)) \right) \# \, \mathsf{sc}(L(\overline{P})) \\
&= \left( \mathop{\#}_{p \neq \overline{P}} \mathsf{sc}(L(p)) \right) \# \, \mathsf{sc}(\mathcal{C}_T^{\mathsf{lh}(P)}[\![M]\!]) \\
&> \left( \mathop{\#}_{p \neq \overline{P}} \mathsf{sc}(L(p)) \right) \# \, \mathsf{sc}(M)
\end{aligned}$$

Further, we show that for all labels $L$ and $M$ with appropriate conditions that:

$$\mathsf{sc}(L \ll_P M) \leq \mathop{\#}_{p \neq \overline{P}} \mathsf{sc}(L(p)) \# \, \mathsf{sc}(M)$$

which we do by induction on $P$. If $P = [k]$ then it is clear that $L \ll_P M$ contains all the terms of $M$ and some terms of $L$, and crucially not $L(\overline{P})$. If instead $P = k :: P_2$ then by induction hypothesis we get that:

$$\mathsf{sc}(L_k \ll_{P_2} M_0) \leq \mathop{\#}_{p \neq \overline{P_2}} \mathsf{sc}(L_k(p)) \# \, \mathsf{sc}(M_1)$$

It is then clear again that $L \ll_P M$ contains terms from $M$ and terms of $L$ which are not $L(\overline{P})$, and so the inequality holds. □

We can now show that insertion reductions reduce syntactic complexity.

**Proposition 4.3.11.** *Let* $s \rightsquigarrow t$ *be an instance of insertion. If* $s$ *is not an identity then* $\mathsf{sc}(s) > \mathsf{sc}(t)$.

*Proof.* Let $(S, P, T, \Gamma, L, M)$ be an insertion redex so that:

$$\mathsf{SCoh}_{(S\,;\,A)}[L] \rightsquigarrow \mathsf{SCoh}_{(S \ll_P T\,;\,A[\![\kappa]\!])}[L \ll_P M]$$

by insertion. By assumption $\mathsf{Coh}_{(S\,;\,A)}[L]$ is not an identity. Then:

$$\begin{aligned}
\mathsf{sc}(t) &= \mathsf{sc}(\mathsf{SCoh}_{(S \ll_P T\,;\,A[\![\kappa]\!])}[L \ll_P M]) \\
&\leq 2\omega^{\dim(A)} \mathbin{\#} \mathsf{sc}(L \ll_P M) \\
&< 2\omega^{\dim(A)} \mathbin{\#} \mathsf{sc}(L) && \text{by Lemma 4.3.10} \\
&\leq \mathsf{SCoh}_{(S\,;\,A)}[L] \\
&= \mathsf{sc}(s)
\end{aligned}$$

and so $\mathsf{sc}(s) > \mathsf{sc}(t)$, completing the proof. □

**Corollary 4.3.12.** *The reduction system $\rightsquigarrow_{\mathcal{R}}$ is strongly terminating.*

*Proof.* By Lemma 4.1.7, it suffices to show that each rule of $\mathsf{sua}'$ reduces syntactic complexity, which follows from Propositions 4.1.8, 4.1.10, and 4.3.11. □

## 4.3.2 Confluence of $\textsc{Catt}_{\mathsf{sua}}$

In this section, we prove the following theorem:

**Theorem 4.3.13.** *The reduction $\rightsquigarrow_{\mathsf{sua}'}$ is confluent.*

The confluence proof for $\textsc{Catt}_{\mathsf{sua}}$ is significantly more complex than the corresponding proof for $\textsc{Catt}_{\mathsf{su}}$. The primary difficulty with $\textsc{Catt}_{\mathsf{sua}}$ is that a term can have an insertion redex where the term to be inserted admits a head reduction. In particular, consider the case where $a \equiv \mathsf{SCoh}_{(S\,;\,A)}[L] \rightsquigarrow b$ is an instance of insertion along some branch $P$, and $a \rightsquigarrow c$ is an insertion on the argument $L(\overline{P})$. The difficulty of this critical pair is that $L(\overline{P})$ need not be in head normal form, and furthermore, the reduction $a \rightsquigarrow c$ can make the original insertion invalid. This does not occur in the predecessor theory $\textsc{Catt}_{\mathsf{su}}$, where only identities can be pruned, and all reducts of identities are again identities.

We will prove this theorem using Lemma 4.1.15. It is therefore sufficient to show that whenever $b \leftsquigarrow a \rightsquigarrow c$, with $a \rightsquigarrow b$ being a reduction derived from $\textsc{rule}$, that the following diagram can be formed:

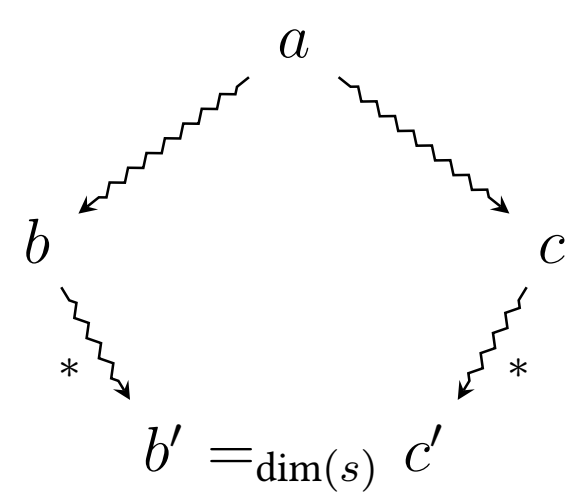

We split by cases on the reduction $a \rightsquigarrow b$, ignoring cases where both reductions are identical and ignoring cases which follow by symmetry of other cases. Any cases which do not mention insertion will follow from an identical argument to the one given in Theorem 4.2.8, and so we omit these here. We can therefore assume without loss of generality that $a \rightsquigarrow b$ is an insertion along redex $(S, P, T, \Gamma, L, M)$ such that $a$ is not an identity or unary composite and $\mathcal{C}_T^{\mathsf{lh}(P)}$ is an identity or a standard composite which is not unary. We now split on the reduction $a \rightsquigarrow c$.

**Insertion on the inserted argument $L(\overline{P})$** Suppose $\mathcal{C}_T^{\mathsf{lh}(P)}[\![M]\!]$ admits an insertion along redex $(T, Q, U, \Gamma, M, N)$. Then:

$$\mathcal{C}_T^{\mathsf{lh}(P)}[\![M]\!] \rightsquigarrow \mathsf{SCoh}_{(T \ll_Q U \,;\, \mathcal{U}_T^{\mathsf{lh}(P)}[\![\kappa_{T,Q,U}]\!])}[M \ll_Q N]$$

We then have $c \equiv \mathsf{SCoh}_{(S\,;\,A)}[L']$ where $L'$ is $L$ with the reduction above applied. We can conclude that $\mathcal{C}_T^{\mathsf{lh}(P)}$ must be a composite (i.e. not an identity) as otherwise the second insertion would not be possible. Similarly, $T$ cannot be linear as otherwise $\mathcal{C}_T^{\mathsf{lh}(P)}$ would be a unary composite.

We now need the following lemmas, the second of which is a directed version of Theorem 3.4.37 with more conditions.

**Lemma 4.3.14.** *For all $n$ and $S$, $\mathcal{C}_S^n \rightsquigarrow^* \mathcal{T}_S^n$.*

*Proof.* The only case in which $\mathcal{C}_S^n \neq \mathcal{T}_S^n$ is when $S = D^n$, in which case a single disc removal gives the required reduction. ☐

**Lemma 4.3.15.** *Let $(S, P, T)$ be an insertion point. Then if $S$ is not linear or $n \leq \mathrm{h}(S)$, $\mathcal{U}_S^n[\![\kappa_{S,P,T}]\!] \rightsquigarrow^* \mathcal{U}_{S \ll_P T}^n$ and if $\mathrm{h}(S) \leq n$ and $S$ is not linear or $\mathrm{h}(S) = n$ then $\mathcal{T}_S^n[\![\kappa_{S,P,T}]\!] \rightsquigarrow^* \mathcal{T}_{S \ll_P T}^n$.*

*Proof.* We proceed by induction on $n$, starting with the statement for types. If $n = 0$ then both standard types are $\star$, so we are done. Otherwise, we have:

$$\mathcal{U}_S^{n+1}[\![\kappa_{S,P,T}]\!] \equiv \mathcal{T}_{\partial_n(S)}^n[\![\delta_n^-(S)]\!][\![\kappa_{S,P,T}]\!] \to_{\mathcal{U}_S^n[\![\kappa_{S,P,T}]\!]} \mathcal{T}_{\partial_n(S)}^n[\![\delta_n^+(S)]\!][\![\kappa_{S,P,T}]\!] \qquad \mathcal{U}_{S \ll_P T}^{n+1} \equiv \mathcal{T}_{\partial_n(S \ll_P T)}^n[\![\delta_n^-(S \ll_P T)]\!] \to_{\mathcal{U}_{S \ll_P T}^n} \mathcal{T}_{\partial_n(S \ll_P T)}^n[\![\delta_n^+(S \ll_P T)]\!]$$

By inductive hypothesis: $\mathcal{U}_S^n[\![\kappa_{S,P,T}]\!] \rightsquigarrow^* \mathcal{U}_{S \ll_P T}^n$, and so we need to show that:

$$\mathcal{T}_{\partial_n(S)}^n[\![\delta_n^\epsilon(S) \bullet \kappa_{S,P,T}]\!] \rightsquigarrow^* \mathcal{U}_{\partial_n(S \ll_P T)}^n[\![\delta_n^\epsilon(S \ll_P T)]\!]$$

We now note that either the conditions for Lemma 3.4.32 or Lemma 3.4.34 must hold. If conditions for Lemma 3.4.32 hold then (as everything is well-formed in Catt) we get that the required reduction is trivial. Therefore, we focus on the second case. Here we get from Lemma 3.4.34 that:

$$\mathcal{T}_{\partial_n(S)}^n[\![\delta_n^\epsilon(S) \bullet \kappa_{S,P,T}]\!] \equiv \mathcal{T}_{\partial_n(S)}^n[\![\kappa_{\partial_n(S),\partial_n(P),\partial_n(T)} \bullet \delta_n^\epsilon(S \ll_P T)]\!]$$

Then we can apply the inductive hypothesis for terms as if $n \leq \dim(S)$ then $\mathrm{h}(\partial_n(S)) = n$ and otherwise $\partial_n(S) = S$ is not linear, and so we get the required reduction.

Now we move on to the case for terms. If $\mathcal{T}_S^n$ is a variable, then we must have that $S$ is linear and so $S = D^n$. We must also have in this case that $\mathcal{T}_S^n = \overline{P}$. Then by Lemma 3.4.12, $\mathcal{T}_S^n[\![\kappa_{S,P,T}]\!] \equiv \mathcal{C}_T^n[\![\iota_{S,P,T}]\!]$ and then by Lemmas 3.4.22 and 4.3.14 this reduces to $\mathcal{T}^n_{S \ll_P T}$ as required. If $\mathcal{T}_S^n$ is not a variable, then $\mathcal{T}_S^n \equiv \mathcal{C}_S^n$, and $\mathcal{C}_S^n$ cannot be an identity (as either $S$ is non-linear or $n = \dim(S)$). By Lemma 3.4.12 and other assumptions we get that $\mathcal{C}_S^n[\![\kappa_{S,P,T}]\!]$ admits an insertion along branching point $P$ and so:

$$\begin{aligned}
\mathcal{T}_S^n[\![\kappa_{S,P,T}]\!] &\equiv \mathcal{C}_S^n[\![\kappa_{S,P,T}]\!] \\
&\rightsquigarrow \mathsf{SCoh}_{(S \ll_P T\,;\,\mathcal{U}_S^n[\![\kappa_{S,P,T}]\!])}[\kappa_{S,P,T} \ll_P \iota_{S,P,T}] \\
&\equiv \mathsf{SCoh}_{(S \ll_P T\,;\,\mathcal{U}_S^n[\![\kappa_{S,P,T}]\!])}[\mathsf{id}] \\
&\rightsquigarrow^* \mathsf{SCoh}_{(S \ll_P T\,;\,\mathcal{U}^n_{S \ll_P T})}[\mathsf{id}] \\
&\equiv \mathcal{C}^n_{S \ll_P T} \\
&\rightsquigarrow^* \mathcal{T}^n_{S \ll_P T}
\end{aligned}$$

With the second equivalence coming from Lemma 3.4.36, the second reduction coming from inductive hypothesis (which is well-founded as the proof for types only uses the proof for terms on strictly lower values of $n$), and the last reduction coming from Lemma 4.3.14. □

By this lemma (as $T$ is not linear), we have

$$\mathcal{U}_T^{\mathsf{lh}(P)}[\![\kappa_{T,Q,U}]\!] \rightsquigarrow^* \mathcal{U}^{\mathsf{lh}(P)}_{T \ll_P Q}$$

and so $\mathcal{C}_T^{\mathsf{lh}(P)}[\![M]\!] \rightsquigarrow^* \mathcal{C}^{\mathsf{lh}(P)}_{T \ll_Q U}[\![M \ll_Q N]\!]$. Let $c'$ be the term obtained by applying this further reduction to the appropriate argument. Now by Lemma 3.4.39, we have that $\mathsf{th}(T \ll_Q U) \geq \mathsf{th}(T)$ and so by Lemma 4.3.7, there is $c' \rightsquigarrow^* c''$ with:

$$c'' =_{\dim(a)} \mathsf{SCoh}_{(S \ll_P (T \ll_Q U)\,;\,A[\![\kappa_{S,P,T \ll_Q U}]\!])}[L \ll_P (M \ll_Q N)]$$

We now examine how $b$ reduces. As $T$ is not linear, there is a branch $S \ll_P Q$ of $S \ll_P T$ and we get the following by Lemma 3.4.13:

$$\overline{S \ll_P Q}[\![L \ll_P M]\!] \equiv \overline{Q}[\![\iota_{S,P,T} \bullet (L \ll_P M)]\!] \equiv \overline{Q}[\![M]\!] \equiv \mathcal{C}_U^{\mathsf{lh}(Q)}[\![N]\!]$$

Since $\mathsf{th}(U) \geq \mathsf{bh}(Q) = \mathsf{bh}(S \ll_P Q)$ we can reduce $b$ to $b'$ by insertion as follows:

$$b' \equiv \mathsf{SCoh}_{((S \ll_P T) \ll_{S \ll_P Q} U\,;\,A[\![\kappa_{S,P,T} \bullet \kappa_{S \ll_P T, S \ll_P Q, U}]\!])}[(L \ll_P M) \ll_{S \ll_P Q} N]$$

and then by Lemma 3.4.41 we get $b' =_{\dim(a)} c''$ as required.

**Argument reduction on the inserted argument $L(\overline{P})$** Suppose $M \rightsquigarrow M'$, and $L'$ is $L$ but with the argument for $\overline{P}$ replaced by $\mathcal{C}_T^{\mathsf{lh}(P)}[\![M']\!]$, such that $L \rightsquigarrow L'$ and $a \rightsquigarrow c \equiv \mathsf{Coh}_{(S\,;\,A)}[L']$. Then $c$ admits an insertion and reduces as follows:

$$c \rightsquigarrow c' \equiv \mathsf{Coh}_{(S \ll_P T\,;\,A[\![\kappa_{S,P,T}]\!])}[L' \ll_P M']$$

Since each term in $L \ll_P M$ is a term of $L$ or a term of $M$, we can simply apply the same reductions from $L \rightsquigarrow L'$ and $L \rightsquigarrow M'$ to get $L \ll_P M \rightsquigarrow^* L' \ll_P M'$. Therefore, $b \rightsquigarrow^* c'$.

**Other reduction on the inserted argument $L(\overline{P})$** The argument $L(\overline{P})$ is either a standard composite which is not unary or an identity. Therefore, the type contained in the coherence is in normal form and hence a cell reduction cannot be applied. Further, disc removal cannot be applied, as $L(\overline{P})$ is not a unary composite, and endo-coherence removal cannot be applied as if $L(\overline{P})$ is an endo-coherence then it is an identity. Hence, there are no other reductions that can be applied to the inserted argument and so this case is vacuous.

**Reduction of non-inserted argument** Suppose $L \rightsquigarrow L'$ along an argument which is not $\overline{P}$ and $c \equiv \mathsf{Coh}_{(S\,;\,A)}[L']$. Then as $L'(\overline{P}) \equiv \mathcal{C}_T^{\mathsf{lh}(P)}$, an insertion can still be performed on $c$ to get:

$$c \rightsquigarrow c' \equiv \mathsf{SCoh}_{(S \ll_P T\,;\,A[\![\kappa_{S,P,T}]\!])}[L' \ll_P M]$$

Since the terms of $L \ll_P M$ are a subset of the terms of $L$ and $M$, we get $L \ll_P M \rightsquigarrow^* L' \ll_P M$ and so $b \rightsquigarrow^* c'$.

**Disc removal** By assumption, insertion cannot be applied to unary composites, and so this case is vacuous.

**Endo-coherence removal** Suppose $A \equiv s \to_B s$ and $a \rightsquigarrow c$ by endo-coherence removal. In this case $c \equiv \mathsf{id}(A, s)[\![L]\!]$ and

$$b \equiv \mathsf{Coh}_{(S \ll_P T\,;\,(s \to_B s)[\![\kappa_{S,P,T}]\!])}[L \ll_P M]$$

which reduces by endo-coherence removal to:

$$b' \equiv \mathsf{id}(A, s)[\![\kappa_{S,P,T} \bullet (L \ll_P M)]\!]$$

By Lemma 3.4.13, we have that $\kappa_{S,P,T} \circ (L \ll_P M) =_{\dim(S)} L$ and so $b' =_{\dim(S)} c$ and since $\dim(S) \leq \dim(a)$, we get $b' =_{\dim(a)} c$ as required.

**Cell reduction** If $A \rightsquigarrow B$ and $c \equiv \mathsf{SCoh}_{(S\,;\,B)}[L]$ from cell reduction, then if $c$ is not an identity or disc it admits an insertion to reduce to:

$$c' \equiv \mathsf{SCoh}_{(S \ll_P T\,;\,B[\![\kappa_{S,P,T}]\!])}[L \ll_P M]$$

As reduction is compatible with substitution, $b$ also reduces to $c'$. If instead $c$ was an identity then

$$\begin{aligned} b &\equiv \mathsf{SCoh}_{(D^n \ll_P T\,;\,A[\![\kappa_{S,P,T}]\!])}[L \ll_P M] \\ &\rightsquigarrow \mathsf{SCoh}_{(D^n \ll_P T\,;\,\mathcal{U}_{D^n}^{n+1}[\![\kappa_{S,P,T}]\!])}[(L \ll_P M)] \\ &\rightsquigarrow^* \mathsf{id}(d_n)[\![\kappa_{S,P,T} \bullet L \ll_P M]\!] \\ &=_{n+1} \mathsf{id}(d_n)[\![L]\!] \\ &\equiv c \end{aligned}$$

Where the second reduction is due to Lemma 4.1.9 and the equality is due to Lemma 3.4.13. If $c$ is a disc then Lemma 4.3.8 can be applied to get that $c$ reduces to a term $c''$ with $c'' =_{n+1} c'$ and $b \rightsquigarrow c'$, completing this case.

**Insertion** Suppose $a \rightsquigarrow c$ is also an insertion, along a branch $Q$ of $S$. We now split on whether $\overline{P} = \overline{Q}$. First suppose $\overline{P} = \overline{Q}$; then by Lemma 3.4.27, we have $b =_{\dim(a)} c$. Suppose now that $\overline{P} \neq \overline{Q}$, and that $L(\overline{Q}) \equiv \mathcal{C}_U^{\mathsf{lh}(Q)}[\![N]\!]$, such that:

$$c \equiv \mathsf{SCoh}_{(S \ll_Q U\,;\,A[\![\kappa_{S,Q,U}]\!])}[L \ll_Q N]$$

We now consider the case where $b$ is an identity. As $P$ and $Q$ are distinct branches of $S$, we must have that $S$ itself is not linear. Therefore, the insertion along $P$ must be an insertion of an identity. Further, for $b$ to have the correct type for an identity, we must have that $A[\![\pi_P]\!] \equiv \mathsf{SPath}(\overline{Q}) \to \mathsf{SPath}(\overline{Q})$. The only path sent to $\overline{Q}$ by $\pi_P$ is $\overline{Q}$ itself, and so $A \equiv \mathsf{SPath}(\overline{Q}) \to \mathsf{SPath}(\overline{Q})$. Now, by Lemma 3.4.12:

$$\begin{aligned} c &\equiv \mathsf{SCoh}_{(S \ll_Q U\,;\,\mathcal{C}_U^{\mathsf{lh}(Q)}[\![\iota]\!] \to \mathcal{C}_U^{\mathsf{lh}(Q)}[\![\iota]\!])}[L \ll_Q N] \\ &\rightsquigarrow \mathsf{id}(\mathcal{C}_U^{\mathsf{lh}(Q)}[\![\iota]\!])[\![L \ll_Q N]\!] && \text{by endo-coherence removal} \\ &\equiv \mathsf{id}(\mathcal{C}_U^{\mathsf{lh}(Q)})[\![N]\!] && \text{by Lemma 3.4.13} \end{aligned}$$

Then, $L \ll_P M$ sends $\overline{Q}$ to $L(\overline{Q}) \equiv \mathcal{C}_U^{\mathsf{lh}(Q)}[\![N]\!]$, and so $b \equiv \mathsf{id}(\mathcal{C}_U^{\mathsf{lh}(Q)})[\![N]\!]$.

The case where $c$ is an identity is symmetric, so we now consider when neither $b$ or $c$ are identities. We now observe that $b$ and $c$ further reduce as follows:

$$\begin{aligned} b &\rightsquigarrow^* b' =_{\dim(a)} \mathsf{SCoh}_{((S \ll_P T) \ll_{Q \ll_P T} U\,;\,A[\![\kappa_{S,P,T} \bullet \kappa_{S \ll_P T, Q \ll_P T, U}]\!])}[(L \ll_P M) \ll_{Q \ll_P T} N] \\ c &\rightsquigarrow^* c' =_{\dim(a)} \mathsf{SCoh}_{((S \ll_Q U) \ll_{P \ll_Q U} T\,;\,A[\![\kappa_{S,Q,U} \bullet \kappa_{S \ll_Q U, P \ll_Q U, T}]\!])}[(L \ll_Q N) \ll_{P \ll_Q U} M] \end{aligned}$$

We show that the first reduction is valid with the validity of the second holding by symmetry. If $b$ is a unary composite then we apply Lemma 4.3.8 to obtain a suitable $b'$: Otherwise, we obtain the reduction via insertion, noting that:

$$\begin{aligned} \overline{Q \ll_P T}[\![L \ll_P M]\!] &\equiv \overline{Q}[\![\kappa]\!][\![L \ll_P M]\!] \\ &\equiv L(Q) \\ &\equiv \mathcal{C}_U^{\mathsf{lh}(Q)}[\![N]\!] \\ &\equiv \mathcal{C}_U^{\mathsf{lh}(Q \ll_P T)}[\![N]\!] \end{aligned}$$

as required for the insertion, with the third equality coming from Lemma 3.4.13. Lastly, the trunk height condition is satisfied as $\mathsf{bh}(Q) = \mathsf{bh}(Q \ll_P T)$.

Therefore, both reductions are valid. We now need the following lemma to complete the proof:

**Lemma 4.3.16.** *Let $(S, P, T, \Gamma, L, M)$ be an insertion redex. Then:*

$$L \ll_P M =_{\mathsf{bh}(P)+1} L \ll'_P M$$

*Proof.* By Proposition 3.4.30, the two labellings are equal. By inspection of the definition, the maximum dimension of terms that differ is $\dim(\mathsf{bh}(P))$. ☐

By the above and Lemma 3.4.31, $b' =_{\dim(a)} c'$. This completes all cases of Theorem 4.3.13.

## 4.4 Towards normalisation by evaluation

In this section, the Rust implementation of Catt, $\text{Catt}_{\text{su}}$, and $\text{Catt}_{\text{sua}}$, which can be found at [Ric24b], is introduced. This implementation takes the form of an interpreter, allowing terms of Catt to be written in a convenient syntax which can be mechanically checked. The implementation aids the user in writing Catt terms by automatically constructing standard composites, allowing terms to be bound to top level syntax, implicitly suspending terms, automatically filling arguments which are not locally maximal, and providing informative error messages to the user when typechecking fails.

We highlight three points of our implementation:

- The typechecker uses *bidirectional typing* [DK21] to mix "inference" and "checking" rules. Although types for Catt can always be inferred, we find ourselves in the unusual situation where in some cases the context a term lives in can be inferred, and in some cases it must be provided. We expand on this type system in Section 4.4.3.
- Tree contexts (see Section 3.2) are given an explicit representation in the tool. The syntax in the theory is then split into syntax over a tree context and syntax over an arbitrary context. Syntax over a tree context can then use paths instead of de Bruijn levels to reference positions in the context, and substitutions from tree contexts can be given by labellings. We explore this syntax in Section 4.4.1.
- During typechecking, the equality between types must be checked, which is done by syntactically comparing the normal form of each type. In this implementation, an approach inspired by *normalisation by evaluation* is taken, as opposed to the reduction based approaches used in the previous sections.

Normalisation by evaluation (NbE) (see [Abe13] for an introduction), can be viewed as a method of evaluating terms with "unknowns". Equivalently, NbE defines a semantic model of the theory, and interprets each constructor of the type theory in these semantics. When equipped with a method for transforming elements of this model back to terms of the type theory (referred to as *quoting*), the normal form of a term can be calculated directly by recursion on its structure. Compared to the reduction based approach taken in the previous sections, which simplifies the term via a series of locally applied reduction rules, NbE takes a more global approach, deconstructing the original term and using it to synthesise a normal form.

The form of NbE implemented in the tool is largely inspired by the paper "Implementing a modal dependent type theory" [GSB19], although we note that the form of the theory Catt is vastly different to the modal type theory they present; Catt does not have lambda abstraction or application in the usual sense, which makes adapting NbE techniques from the literature difficult. Nevertheless, the overall form of the evaluation is similar.

A high-level overview of the implementation is given in Figure 4.2. We pause to explain the purpose of each component:

- The *raw syntax* is the syntax that the user of the tool interacts with. We maintain no invariants over the well-formedness of the raw syntax, and it allows the user to omit arbitrary arguments. The primary purpose of the raw syntax is to be the target of parsing, and conversely to facilitate the pretty-printing of terms. We also specify a command language around this raw syntax which is used to interact with the tool.

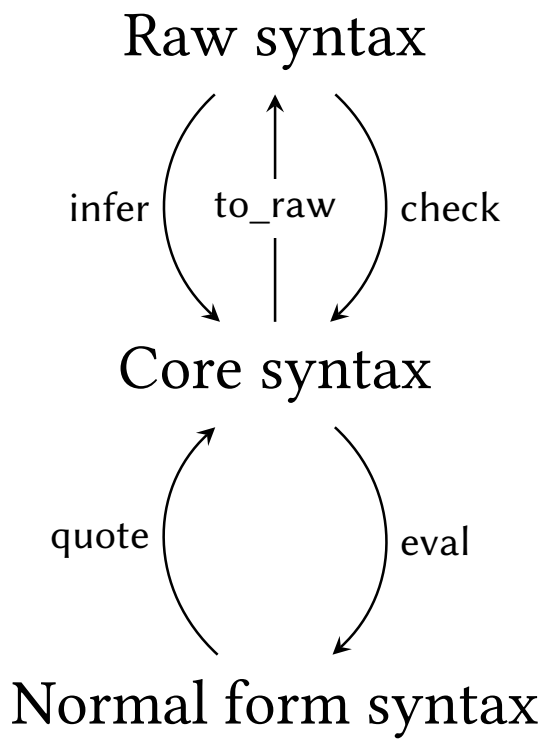

Figure 4.2: Implementation overview.

- The *core syntax* is the result of the typechecking procedure. Syntax of this form is known to be well-formed, and all implicit arguments have been filled in at this point. The terms of this syntax resemble the structured terms of Section 3.3, with various common operations of Catt being defined as constructors. Contrary to previous representations of Catt in this thesis, the application of substitution is treated as a term former, instead of an operation.
- The *normal form syntax* represents the normal forms of each of the type theories $\text{Catt}_{\mathsf{sua}}$, $\text{Catt}_{\mathsf{su}}$, and Catt itself. This syntax is also always assumed to be well-formed, and is the closest to original syntax of Catt.
- The eval and quote functions convert syntax between core syntax and normal form syntax. For each constructor in the core syntax, evaluation computes the result of the corresponding operation, quotienting by the rules of $\text{Catt}_{\mathsf{su}}$ or $\text{Catt}_{\mathsf{sua}}$ when applicable. We note that despite Catt itself having no computation, evaluation must still process operations such as suspension and substitution application. Quotation converts normal form syntax back to core syntax, and in our case is a trivial inclusion.
- The infer and check functions perform typechecking while converting raw syntax into core syntax. Both functions are mutually dependent on each other, and also may need to convert types to normal form syntax to check equality. The to_raw functions "forget" that a piece of core syntax is well-formed, returning a piece of raw syntax, and can optionally remove all non-locally maximal arguments from terms.

In the following subsections, we expand on these points, fully defining each class of syntax, and describing the typechecking and evaluation procedures.

### 4.4.1 Syntax

Before defining each of the syntactic classes in the tool, we introduce some common notation that will be used in the definitions below:

- The letter $v$ will be used to represent *names* in the syntax: strings that represent a valid identifier.
- A $\mathsf{Maybe}(x)$ is either of the form $\mathsf{Some}(x)$ or $\mathsf{None}$.
- The notation $\mathsf{Tree}(x)$ represents a tree structure which is given by a list of $x$'s which we

call the *elements* and a list of trees, which we call the *branches*, whose length is one less than the list of elements. These resemble labellings from Section 3.2.2, but will allow trees to be labelled with arbitrary objects.

We begin our study of the syntax with the raw syntax, which is defined by the following grammar:

| | | | |
|---|---|---|---|
| (Terms) | $s, t$ | $::=$ | $v \mid \mathsf{coh}[T : A] \mid _ \mid \mathsf{id} \mid \mathsf{comp} \mid \mathsf{inc}_n^m(s) \mid s[\![\sigma]\!] \mid \Sigma(s)$ |
| (Types) | $A$ | $::=$ | $\star \mid s \to_{\mathsf{Maybe}(A)} t \mid _ \mid A[\![\sigma]\!] \mid \Sigma(A)$ |
| (Arguments) | $\sigma$ | $::=$ | $(\mathsf{Tree}(\mathsf{Maybe}(s)), \mathsf{Maybe}(A)) \mid (\mathsf{Maybe}(A, )s_0, \dots, s_n)$ |
| (Contexts) | $\Gamma$ | $::=$ | $T \mid (v_0, A_0), \dots, (v_n : A_n)$ |
| (Tree Contexts) | $T$ | $::=$ | $\mathsf{Tree}(\mathsf{Maybe}(v))$ |

The primary purpose of the raw syntax is to accurately represent the written plaintext syntax. For most cases, each constructor is written in plaintext exactly how it is written above, apart from a few cases:

- The application of substitution $s[\![\sigma]\!]$ and $A[\![\sigma]\!]$ is simply written $s\ \sigma$ and $A\ \sigma$ respectively.
- The constructor $\mathsf{inc}_n^m$ is not parsed and is used as an internal operation for defining the external labelling (see Section 3.4). It is displayed as `inc<n-m>`.
- The suspension can be given by the characters $\Sigma$ or $S$, to avoid the user being forced to type Unicode characters.
- The type $s \to_{\mathsf{None}} t$ is written simply as $s \to t$, and the type $s \to_{\mathsf{Some}(A)} t$ is written as $A \mid s \to t$, where the symbol $\to$ can be replaced by `->` in either case.
- For the construction Maybe, $\mathsf{Some}(s)$ is printed the same as $s$, and None is printed as the empty string.
- We provide two ways to write trees:
  - The curly bracket notation from Section 3.2 can be used. The string:

    $$s_0\{T_0\}s_1 \cdots \{T_n\}s_{n+1}$$

    is parsed as a tree with elements given by (the parse of) $s_0$ to $s_{n+1}$ and branches given by the parse of $T_0$ to $T_n$.
  - We provide a notation for specifying the locally maximal arguments of a tree. We parse the string:

    $$[a_1, a_2, \dots, a_n]$$

    As a tree that has None as each of elements branches by given by each of the $a_i$, where if $a_i$ does not recursively parse as a tree, it is parsed as an element and wrapped in a singleton tree.

  To compare these two notations, the two trees below are equal:

  $$\{f\}\{\{a\}\{b\}\} = [f, [a, b]]$$

  When using the full (curly bracket) notation to specify a labelling, it must be wrapped in angle brackets to avoid parse ambiguity.

We highlight the use of the extended substitution introduced in Section 2.1 in the raw syntax. This allows the tool to perform "implicit suspension", the automatic suspension of a term, by reducing it to a problem of type inference. These extended substitutions are converted to regular substitutions by the evaluation function introduced in Section 4.4.2, which applies the appropriate number of suspensions to the head term. An example of this is given in Section 4.4.4.

The syntax includes two new base constructors for terms alongside the coh constructor: comp and id. These allow easy access to common terms that are prevalent in most CATT terms; the id constructor produces an identity term and the comp constructor produces a standard composite (a term of the form $\mathcal{C}_T^{\dim(T)}$, see Section 3.3.2). The dimension of the identity term and the tree $T$ that the comp term composes over can be inferred during typechecking (Section 4.4.3), simplifying the syntax of many common terms. The evaluator (Section 4.4.2), converts coherence terms of the appropriate form to these constructors and leverages this to identify regexes, noting that composites and identities are the normal forms of standard coherences (see Theorem 3.3.27) in $\text{CATT}_{\mathsf{su}}$ and $\text{CATT}_{\mathsf{sua}}$.

We also provide a command language on top of the raw syntax for CATT, which allows the user to perform various operations on terms, such as binding them to a top-level name, or normalising them. These commands are given by the following syntax:

$$\begin{array}{llll} & \mathsf{def}\ v = s & \mid \mathsf{def}\ v\ \Gamma = s & \mid \mathsf{def}\ v\ \Gamma : A = s \\ \mid & \mathsf{normalise}\ s\ \mathsf{in}\ \Gamma & \mid \mathsf{assert}\ s = t\ \mathsf{in}\ \Gamma & \mid \mathsf{size}\ s\ \mathsf{in}\ \Gamma \\ \mid & \mathsf{import}\ \mathtt{filename} & & \end{array}$$

The first three commands define the name $v$ to be given by the term $s$, where the context $\Gamma$ and type $A$ can optionally be given, determining whether the term $s$ will be inferred or checked. The next three commands take a context $\Gamma$ and respectively calculate the normal form of $s$ in $\Gamma$, assert that $s$ and $t$ are equal in $\Gamma$, or count the number of coherence constructors in $s$. The last command parses the file `filename` and runs the commands it contains.

In the implementation, each piece of syntax is paired with a piece of span information, which specifies where in the source file it originated. This is done by making the raw syntax generic over a type $S$ of spans. When obtaining the raw syntax from parsing, this $S$ is given by a range $n < m$ specifying the start and end indices of where the syntax appeared in the text. When the syntax is obtained from the to_raw functions, $S$ is given by the unit type.

The span information allows more informative error messages to be given, which are formatted using the ariadne crate. An example error message is given below:

```
Error: Given term "x" does not match inferred term "y"
    ╭─[examples/test.catt:19:34]
    │
 19 │ def error_test x{f}y = comp ⟨x{f}x⟩
    │                                  ┬
    │                                  ╰── Given term
────╯
```

In the above example, the span attached to the erroneous term $x$ is used to identify the line of the code which should be displayed and determine where the "Given term" label should point to.

We now move on to the core syntax of the tool. The overall form of the core syntax is similar to that of the raw syntax with the following differences:

- Span information is no longer included in the core syntax.
- Terms that were optional in the raw syntax (as specified using the Maybe construction) are no longer optional. The hole constructor is also removed.
- The syntax is generic over a type $P$ of positions. In practice, the type $P$ is either the type of de Bruijn levels, for terms over a standard context, or paths, for terms over a tree context. Both of these types implement the Position trait, allowing us to call the required functions that depend on the type of positions.
- The constructor for variables $v$ in the raw syntax is replaced by two separate constructions, a constructor $\mathsf{var}_p$ for local variables, where $p : P$, and $\mathsf{top_lvl}(v, s)$ for a top-level symbol $v$ which is bound to term $s$.
- Application of substitutions and labellings are now separate constructors. This is primarily due to a limitation of the Rust type system which doesn't allow an App constructor to existentially quantify over the type of positions in the input term.
- The comp constructor now records the tree it is composing over, and will be written $\mathsf{comp}_T$. Similarly, the id constructor becomes $\mathsf{id}_n$, with $n$ recording the dimension of the term the identity is applied to.

The core syntax is defined by the following grammar. Following [GSB19], a different colour is used for each of the three categories of syntax.

$$\begin{array}{llcl}
\text{(Terms)} & s, t & ::= & \mathsf{var}_p \mid \mathsf{top_lvl}(v, s) \mid \mathsf{coh}[T : A] \mid \\
& & & \mathsf{id}_n \mid \mathsf{comp}_T \mid \mathsf{inc}_n^m(s) \mid s[\![\sigma]\!] \mid s[\![L]\!] \mid \Sigma(s) \\
\text{(Types)} & A & ::= & \star \mid s \to_A t \mid A[\![\sigma]\!] \mid A[\![L]\!] \mid \Sigma(A) \\
\text{(Substitutions)} & \sigma & ::= & (A, s_0, \dots, s_n) \\
\text{(Labellings)} & L & ::= & (\mathsf{Tree}(s), A) \\
\text{(Contexts)} & \Gamma & ::= & (v_0, A_0), \dots, (v_n : A_n) \\
\text{(Tree Contexts)} & T & ::= & \mathsf{Tree}(\mathsf{Maybe}(v))
\end{array}$$

The above classes of syntax are parameterised by the type of positions $P$. We enforce that for the application of substitution $s[\![\sigma]\!]$ that the term $s$ is a term with $P = \mathbb{N}$, the type of de Bruijn levels, and for $s[\![L]\!]$, the application of a labelling, that $s$ is a term with $P = \mathsf{Path}$. The type of paths is given by a non-empty list of natural numbers, as in Section 3.2.

The type of positions must satisfy the Position trait, which provides:

- An associated Ctx type, given by contexts for $\mathbb{N}$ and tree contexts for Path.
- An associated container type Container which can be indexed by $P$. This is given by Vec (Rust's dynamically sized array type) for levels and Tree for paths. This container type defines the form of substitutions and labellings respectively.
- A function to_name, which gives a canonical name to each $p : P$, used when converting back to a raw term.

In the core syntax, all the variable names have been replaced by an index into the context, with all the original variable names being moved to the context. The to_raw function then takes a piece of syntax and a context and simply undo this, mapping a $\mathsf{var}_p$ to $v$, where $p$ maps to $v$ in the supplied context. The context argument of to_raw is optional as the context is not always available, for instance with the term $s[\![\sigma]\!]$, we do not know the context that $s$ should live in. When the context is not available, or no name is known for $p$ in the context, $\mathsf{var}_p$ is mapped to $\mathsf{to_name}(p)$.

For the remainder of the syntax, to_raw removes extra information that was obtained during typechecking, such as the tree associated to a comp, the dimension of an id, or the term associated to a top-level binding. We also accept a boolean parameter declaring whether implicit arguments should be kept; if this is true then the Some constructor of the Maybe type is used whenever possible, otherwise only locally maximal arguments are kept, replacing the rest with the None constructor.

Lastly, we introduce the normal form syntax. This is the simplest of the three categories of syntax and is closest to the base syntax of Catt. It is given by the following grammar:

$$\begin{array}{llcl}
\text{(Head)} & H & ::= & \mathsf{coh}[T : A] \mid \mathsf{id}_n \mid \mathsf{comp}_T \\
\text{(Terms)} & s, t & ::= & \mathsf{var}_p \mid H[\![L]\!] \\
\text{(Labellings)} & L & ::= & \mathsf{Tree}(s) \\
\text{(Types)} & A & ::= & [(s_0, t_0), \dots, (s_n, t_n)] \\
\text{(Tree Contexts)} & T & ::= & \mathsf{Tree}(\mathsf{Maybe}(v))
\end{array}$$

As with the core syntax, this syntax is parameterised by a type of positions $P$. In the normal form syntax, application of substitution is removed, and labellings can no longer be applied to an arbitrary term, but only to a head term, which is a single coherence, composite, or identity. The composite and identity constructions are prioritised as head terms due to the role they play in $\textsc{Catt}_{\mathsf{sua}}$, being the only insertable arguments.

Many of the extra operations such as suspension are not present in the normal form syntax. In particular, the syntax for types is far simpler, allowing them to be represented by a vector of pairs of terms. A type $[(s_0, t_0), \dots, (s_n, t_n)]$ represents the type $s_0 \to t_0$, with the tail of the list giving the lower-dimensional part of the type. The type $\star$ is represented by the empty list.

### 4.4.2 Evaluation

We now describe the core technical part of the tool, the evaluation of a piece of core syntax to its normal form, which is crucial in checking the equality of types. Various pieces of reduction can be configured in the evaluation process:

- Disc removal can be turned off or on.
- Endo-coherence removal can be enabled or disabled.
- Insertion can be set to never happen, only allow insertion of identities, or allow insertion of identities and standard composites.

Following the NbE style we have already introduced, we define *(semantic) environments*, which are required to evaluate a term to a normal form.

**Definition 4.4.1.** For a type of positions $P$, a *$P$-environment* takes the form of a $P$-Container of normal form terms, along with a normal form type. More concretely this is given by a tree of terms when $P = \mathsf{Path}$ and a vector of terms when $P = \mathbb{N}$.

For an environment $\rho$, we write $\mathsf{Ty}(\rho)$ for the type associated with $\rho$ and $\rho(p)$ for the $p^{\text{th}}$ element of the container, where $p : P$. This mirrors the syntax used for a term-labelling in Section 3.2, as environments are simply an abstraction over labellings and substitutions. Due to this similarity, the *restriction* of an environment can be defined, similarly to its definition for substitutions. A Path-environment can be repeatedly *unrestricted* until the contained type is $\star$, returning a labelling.

**Definition 4.4.2.** We define the restricted environment $\uparrow\rho$ for each $P$-environment $\rho$. For $P = \mathbb{N}$, we let $\mathsf{Ty}(\uparrow\rho) = (\rho(0), \rho(1)) :: \mathsf{Ty}(\rho)$ and $\uparrow\rho(n) = \rho(n+2)$. For $P = \mathsf{Path}$, we let $\mathsf{Ty}(\uparrow\rho) = (\rho([0]), \rho([1])) :: \mathsf{Ty}(\rho)$ and $\uparrow\rho(p) = \rho(0 :: p)$.

If $\rho$ is a Path-environment, then there is a labelling $\downarrow\rho$, obtained by popping pairs of terms $(s, t)$ from $\mathsf{Ty}(\rho)$ and pushing them to the bottom of the tree contained in $\rho$ (applying the map $T \mapsto s\{T\}t$) until $\mathsf{Ty}(\rho) = [\,]$. Lastly, for the same $\rho$, its inclusion $\rho_n^m$ can be defined letting $\mathsf{Ty}(\rho_n^m) = \mathsf{Ty}(\rho)$ and if the tree part of $\rho$ is given by:

$$s_0\{T_0\}\cdots s_n\{T_n\}\cdots\{T_{m-1}\}s_m\cdots\{T_k\}s_{k+1}$$

then the tree part of $\rho_n^m$ is given by $s_n\{T_n\}\cdots\{T_{m-1}\}s_m$.

There are also identity environments, that play the role of the identity substitution and identity labelling.

**Definition 4.4.3.** From a (core) context $\Gamma$, an $\mathbb{N}$-environment $\mathsf{id}_\Gamma$ can be formed with type $[\,]$ and $\mathsf{id}_\Gamma(i) = \mathsf{var}_i$ for each $i < \mathsf{len}(\Gamma)$. For each tree $T$, a Path-environment $\mathsf{id}_T$ can be formed again with type $[\,]$ such that $\mathsf{id}_T(p) = \mathsf{var}_p$ for each path $p$ of $T$.

We can now define the functions $\mathsf{eval}_\rho$, which takes a piece of syntax and evaluates it in the environments $\rho$ to a normal form, and quote which includes normal forms back into core syntax. Intuitively, $\mathsf{eval}_\rho(s)$ computes the normal form of $s[\![\rho]\!]$.

We define $\mathsf{eval}_\rho(s)$, $\mathsf{eval}_\rho(A)$, $\mathsf{eval}_\rho(\sigma)$, $\mathsf{eval}_\rho(L)$, to be respectively a normal form term, a normal form type, an $\mathbb{N}$-environment, and a Path-environment. We proceed by case analysis, giving the easier cases below.

$$\mathsf{eval}_\rho(\mathsf{var}_p) = \rho(p) \qquad \mathsf{eval}_\rho(\mathsf{top_lvl}(v, s)) = \mathsf{eval}_\rho(s) \qquad \mathsf{eval}_\rho(\mathsf{id}_n) = \mathsf{id}_n[\![\downarrow\rho]\!]$$

$$\mathsf{eval}_\rho(\mathsf{inc}_n^m(s)) = \mathsf{eval}_{\rho_n^m}(s) \qquad \mathsf{eval}_\rho(s[\![\sigma]\!]) = \mathsf{eval}_{\mathsf{eval}_\rho(\sigma)}(s) \qquad \mathsf{eval}_\rho(s[\![L]\!]) = \mathsf{eval}_{\mathsf{eval}_\rho(L)}(s)$$

$$\mathsf{eval}_\rho(\Sigma(s)) = \mathsf{eval}_{\uparrow\rho}(s)$$

$$\mathsf{eval}_\rho(\star) = \mathsf{Ty}(\rho) \qquad \mathsf{eval}_\rho(s \to_A t) = (\mathsf{eval}_\rho(s), \mathsf{eval}_\rho(t)) :: \mathsf{eval}_\rho(A)$$

$$\mathsf{eval}_\rho(A[\![\sigma]\!]) = \mathsf{eval}_{\mathsf{eval}_\rho(\sigma)}(A) \qquad \mathsf{eval}_\rho(A[\![L]\!]) = \mathsf{eval}_{\mathsf{eval}_\rho(L)}(A) \qquad \mathsf{eval}_\rho(\Sigma(A)) = \mathsf{eval}_{\uparrow\rho}(A)$$

The environments $\mathsf{eval}_\rho(\sigma)$ and $\mathsf{eval}_\rho(L)$ are then obtained by evaluating the type and all the terms in $\sigma$ and $L$ respectively. We note that suspension is evaluated by restricting the environment, and does not require a full traversal of the term, demonstrating the further utility of the extended substitution introduced in Section 2.1.

This leaves the cases for the coh and comp terms. For these cases, we need definitions of the standard type of dimension $n$ over a tree $S$ and the exterior labelling for an insertion point $(S, P, T)$, which we define as the core syntax $\mathcal{U}_T^n$ and $\kappa_{S,P,T}$. We omit the definitions of these, as they are similar to those given in Section 3.4, using the inc and $\Sigma$ constructors in place of the Inc constructor of structured terms. For (labelled) trees $S$ and $T$ such that $(S, P, T)$ is an insertion point, we also define the inserted (labelled) tree $S \ll_P T$ identically to the inserted labelling.

We now proceed with the case for $\mathsf{coh}[S : A]$, assuming we are evaluating it in environment $\rho$. We begin by letting $d = \mathsf{dim}(\mathsf{Ty}(\rho))$ and obtaining the labelling $L = \downarrow \rho$. The number $d$ represents the number of times the term must be suspended, and so $S$ and $A$ are each suspended $d$ times, where the type $A$ is suspended by applying a $\Sigma$ constructor, and $S$ is suspended by replacing it with the tree $\mathsf{None}\{S\}\mathsf{None}$.

We now search for insertion redexes in the labelling $L$, splitting on the type of insertion that is enabled:

- If no insertion is enabled, this phase is skipped.
- If insertion of identities is enabled, and there is a locally maximal argument given by branch $P$ (where we take $P$ to be the branch of minimal branching height) that is of the form $\mathsf{id}_n[\![M]\!]$, we return the insertion redex $(S, P, D^n, _, L, M)$, where the target of $L$ and $M$ is unspecified.
- If full insertion is enabled, and there is a locally maximal argument given by branch $P$ that is of the form $\mathsf{comp}_T[\![M]\!]$, where $\mathsf{bh}(P) > \mathsf{lh}(T)$, we return the insertion redex $(S, P, T, _, L, M)$.

If an insertion redex $(S, P, T, _, L, M)$ is found, then $S$ is replaced by $S \ll_P T$, $L$ is replaced by $L \ll_P M$, and $A$ is replaced by $A[\![\kappa_{S,P,T}]\!]$. This step is then repeated until no insertion redexes are found.

*Remark* 4.4.4. At this critical step, the evaluation proceeds in a fashion closer to reduction than NbE, with insertions repeatedly applied by searching for redexes and applying reductions to the head term. This seems unavoidable; even if one could define a parallel insertion which inserted all insertable arguments at once, it is not clear how to deal with locally maximal arguments that are iterated identities. Despite this, we still claim that the overall structure of the evaluation follows an NbE style, especially regarding the treatment of suspension and application of substitutions and labellings.

We next obtain the type $B = \mathsf{eval}_{\mathsf{id}_S}(A)$, and split into cases:

- If endo-coherence removal is enabled, and $B$ is of the form $(s, s) :: B'$, then we let $t \to_C t = \mathsf{quote}(B)$, interpret $L$ as an environment by letting $\mathsf{Ty}(L) = \star$ and let:

  $$\mathsf{eval}_\rho(\mathsf{coh}[S : A]) = \mathsf{id}_{\mathsf{dim}(B')}[\![\{\mathsf{eval}_L(C), \mathsf{eval}_L(t)\}]\!]$$

  where the labelling $\{_, _\}$ from a disc can be trivially constructed by deconstructing the type.

- Suppose endo-coherence removal is disabled, $S$ is a disc $D^n$, and $B$ is of the form $(\mathsf{var}_{p^n}, \mathsf{var}_{p^n}) :: B'$, where we recall the path $p^n$ is the unique locally maximal variable of $D^n$, then we let:
$$\mathsf{eval}_\rho(\mathsf{coh}[S : A]) = \mathsf{id}_n[\![L]\!]$$

- If disc-removal is enabled, $S = D^n$, and $B$ is equal to the standard type of dimension $n$, then:
$$\mathsf{eval}_\rho(\mathsf{coh}[S : A]) = L(p^n)$$

- If none of the above cases hold, and $B$ is equal to the standard type of dimension $\dim(S)$, then:
$$\mathsf{eval}_\rho(\mathsf{coh}[S : A]) = \mathsf{comp}_S[\![L]\!]$$

- If none of the above cases hold, then:
$$\mathsf{eval}_\rho(\mathsf{coh}[S : A]) = \mathsf{coh}[S : B][\![L]\!]$$

The $\mathsf{comp}_T$ case is treated in much the same way, removing any step involving $A$ and instead setting $B = \mathsf{eval}_{\mathsf{id}_T}(\mathcal{U}_T^n)$, where $n$ is given by the dimension of $T$ before any insertion was performed. This completes all cases for the evaluation function.

In contrast, the quote function is defined completely trivially by recursion, converting head terms and normal form terms to core terms, normal form labellings to core labellings, and converting normal form types to an iterated arrow type in the obvious way. We note that this is unusual for NbE, where the quote function is often mutually defined with evaluation, and performs a significant portion of the work of converting terms to normal form.

### 4.4.3 Typechecking

Now that the three classes of syntax and the evaluation function have been introduced, the bidirectional typechecking algorithm in the tool can be described. Bidirectional typing allows us to mix typing rules which "check" a term, and typing rules which "infer" the type for a term. In the implementation, this will determine which pieces of data are inputs to a procedure, and which pieces of data are outputs.

By Lemma 2.2.7, all Catt terms $s$ have a unique type, which is given by the canonical type $\mathrm{Ty}(s)$. However, for certain terms, such as the coherence term $\mathsf{coh}[T : A]$, we will be able to further infer the context that a term lives in, which in this case is the tree context $T$. In this case the pair of the inferred context and type is known as a *principal typing* [Jim96], which is not to be confused with a *principal type* of a term in a fixed context.

Due to our unique case where all types are inferable, but the context in a judgement may or may not be inferable, we refer to judgements where the context is an input as *checking* judgements and judgements where the context is output as *inferring* judgements.

*Remark* 4.4.5. We justify this choice of terminology by noting the similarity of the judgements $\Gamma \vdash s : A$ and $\cdot \vdash \Pi_\Gamma\, s : \Gamma \to A$ in a type theory with (dependent) function types, where inferring the type of the second judgement would infer the context of the first. Of course, Catt does not have function types, yet the intuition can still apply.

The typing system will be defined with respect to a *Signature* $\Psi$, which contains a mapping from names to triples $(\mathbf{U}, s, A)$ where $s$ is a term of type $A$ in (tree) context $\mathbf{U}$. In the implementation, the signature also stores all relevant settings for the tool: which reductions are active, the operation set $\mathcal{O}$ (which can only be configured to the groupoidal or regular operation sets), and whether implicit variables should be kept in the to_raw functions. We write:

$$\Psi(v) = (\mathbf{U}, s, A)$$

if the signature $\Psi$ maps $v$ to the triple above.

We further define the notation $\mathbf{U}(i) = (v : A)$ to mean that at the $i^{\text{th}}$ index of $\mathbf{U}$ (with $\mathbf{U}$ being a tree or a context), contains a variable name $v$, which is given type $A$ by $\mathbf{U}$.

Lastly we define two conversion functions: from_sub and flatten. The first is a (partial) function which takes a tree $T$ and a substitution $\sigma$ and creates a labelling $\text{from_sub}_T(\sigma)$ by letting the locally maximal arguments be given by the terms of $\sigma$, if $\sigma$ contains the correct number of terms. The function flatten acts on the Maybe construction applied to a term or type. It takes $\text{Some}(s)$ and $\text{Some}(A)$ to $s$ and $A$ respectively, and None to _, the hole constructor for terms and types.

Our bidirectional typing system will be based on the following judgements, letting $\mathbf{U}$ refer to either a context or tree context:

| | |
|---|---|
| $s \rightsquigarrow \mathbf{U} \vdash t : A$ | Convert $s$ to $t$ inferring its type $A$ in inferred (tree) context $\mathbf{U}$ |
| $\mathbf{U} \vdash s \rightsquigarrow t : A$ | Given $\mathbf{U}$, convert $s$ to $t$ checking it has some type $A$ in $\mathbf{U}$ |
| $\mathbf{U} \vdash s = t \rightsquigarrow ()$ | In $\mathbf{U}$, check $s$ has normal form $t$ |
| $\mathbf{U} \vdash A \rightsquigarrow B = C$ | In $\mathbf{U}$, convert $A$ to $B$, inferring its normal form $C$ |
| $\mathbf{U} \vdash A = C \rightsquigarrow ()$ | In $\mathbf{U}$, check $A$ has normal form $C$ |
| $\Gamma \vdash \ \rightsquigarrow \mathbf{U}$ | Check $\Gamma$, producing (tree) context $\mathbf{U}$ |
| $\mathbf{U} \vdash \sigma : \Gamma \rightsquigarrow \tau$ | Check $\sigma$ is a substitution from $\Gamma$ to $\mathbf{U}$, producing $\tau$ |
| $\mathbf{U} \vdash L : T \rightsquigarrow M : A$ | Check labelling $L$ in $\mathbf{U}$, producing $M$ with type $A$ |

for each judgement, the syntax to the left of $\rightsquigarrow$ are the inputs to the judgements, and the syntax to the right are the outputs.

The typing rules for all judgements of this system are given in Figure 4.3. In this figure, $D^n$ always refers to the linear tree of height $n$, rather than the disc context, $\emptyset$ refers to the empty context, and $[\,]$ refers to the singleton tree. In the final rules, $i$ should be treated as if it is universally quantified. We pause to highlight some of these rules:

- In the rule for coherences, marked $\alpha$, the support conditions are checked. This is done using the normal form syntax for the type, due to the simplicity of this syntax. The variable sets of a term can easily be collected by recursion, and in the implementation are stored in a hash set, using Rust's HashSet type.
- The rule for composites, marked $\beta$, is crucially a checking rule as there is no way to infer the tree $T$ for the term $\text{comp}_T$.
- For the rule for the application of labellings, marked $\delta$, the premise for the typing of the term is given by a checking judgement instead of an inferring judgement, as the tree $T$ can be inferred form the labelling. This is in contrast to the corresponding rule for application of substitutions, where the context must be inferred from the inner term before the substitution can be checked. Combined with the point above, this allows a labelling applied to a comp term to be checked.

  As an extra convenience feature, the tree context $T$ generated from the labelling has a name assigned to each position given by the letter p followed by its path. For example, the first variable of the tree is given by `p0`, allowing the user to write an expression such as `p0{x{f}y}`.

- The rule marked $\gamma$ allows a substitution to be applied to a term over a tree context, by converting the substitution to a labelling. This is mainly a convenience feature, as given a term $s$ where it can be inferred that the context of $s$ is a tree $T$, it can be easier to give the locally maximal arguments for $s$ as a list rather than describing the labelling.

- Lastly, we explain each component of the rule for the typing of a substitution, marked $\varepsilon$. We note that the first type in any Catt context, which in the rule is given by the type $A_0$, is always $\star$. Therefore, the type of the first term in a substitution $\sigma$ should be equal to $\star[\![\sigma]\!] \equiv \text{Ty}(\sigma)$. In the rule, the type of the first term is given by $B_0$, explaining its presence as the type of the substitution that gets evaluated to $\rho$. We further note that $\text{Ty}(\rho)$ is simply the evaluation of $B_0$, which is why $X$ is checked against it.

  Due to the choice to use de Bruijn levels instead of indices, weakening a term is the identity, and so $s[\![\sigma]\!] \equiv s[\![\langle\sigma, t\rangle]\!]$ for any $t$. Therefore, by inspecting the typing rules for substitutions in Catt, it can be proven that to type $\Gamma \vdash \sigma : \Delta$, it is sufficient to show that $\Gamma \vdash x[\![\sigma]\!] : A[\![\sigma]\!]$ for all $(x : A) \in \Delta$. Observing the rule $\varepsilon$, this translates to proving that $A_i[\![(B_0, t_0, \ldots, t_n)]\!] = B_i$ recalling that $B_0$ is the core syntax version of the type of the substitution. These equations can be shown by proving that the evaluation of each side is the same, but the evaluation of the left-hand side is given by $\text{eval}_\rho(A_i)$ for each $i$, and so for efficiency we factor out the calculation of $\rho$.

The typing rules in Figure 4.3 can easily be translated into an algorithm for mechanically checking each of these typing judgements. In some cases, some equalities of normal forms are left implicit, such as in the final rule concerning the typing of a non-singleton labelling, and must be made explicit in the final algorithm.

Many of the choices for the form of these rules was made to improve the quality of error messages. Each of these rules can fail for a variety of reasons, at which point an error is created by converting the relevant syntax back to raw syntax using the to_raw functions so that it can be displayed to the user. The use of Rust's Result type, which allows each of these functions to return either the well-formed core syntax or an appropriate error message, is essential, and benefits greatly from the question mark syntax in Rust, which allows errors to easily be propagated through the code.

We end this section by describing the function of each of the commands introduced in Section 4.4.1. Each of these commands is run with a mutable reference to a signature $\Psi$. The commands use this signature for typechecking, and may modify the signature.

The three def commands are used to add a new binding to the signature $\Psi$. For the first command, which omits the context, the term $s$ must be inferred, producing a core syntax context, term, and type, which is inserted into the signature with key $v$ and printed to the user. The second command is given a raw context and so first checks this raw context to produce a

$$\frac{\Psi(v) = (\mathbf{U}, t, A)}{v \rightsquigarrow \mathbf{U} \vdash \mathsf{top_lvl}(v, t) : A} \qquad \frac{T \vdash A \rightsquigarrow B = C \qquad (T, \mathsf{src}(C), \mathsf{tgt}(C)) \in \mathcal{O}}{\mathsf{coh}[T : A] \rightsquigarrow T \vdash \mathsf{coh}[T : B] : B}\,\alpha$$

$$\frac{}{\mathsf{id} \rightsquigarrow D^1 \vdash \mathsf{id}_0 : \mathsf{var}_{[0]} \to_\star \mathsf{var}_{[0]}} \qquad \frac{s \rightsquigarrow \mathbf{U} \vdash t : A}{\Sigma(s) \rightsquigarrow \Sigma(\mathbf{U}) \vdash \Sigma(t) : \Sigma(A)}$$

$$\frac{s \rightsquigarrow T \vdash t : A}{T \vdash s \rightsquigarrow t : A} \qquad \frac{\mathbf{U}(i) = (v : A)}{\mathbf{U} \vdash v \rightsquigarrow \mathsf{var}_i : A} \qquad \frac{}{D^n \vdash \mathsf{id} \rightsquigarrow \mathsf{id}_n : \mathcal{U}^{n+1}_{D^n}}$$

$$\frac{}{T \vdash \mathsf{comp} \rightsquigarrow \mathsf{comp}_T : \mathcal{U}^n_T}\,\beta \qquad \frac{s \rightsquigarrow \Gamma : t : A \qquad \mathbf{U} \vdash \sigma : \Gamma \rightsquigarrow \tau \qquad \mathbf{U} \vdash \mathsf{Ty}(\sigma) = B \rightsquigarrow ()}{\mathbf{U} \vdash s[\![\sigma]\!] \rightsquigarrow t[\![\tau]\!] : A[\![\tau]\!]}$$

$$\frac{T \vdash s \rightsquigarrow t : A \qquad \mathbf{U} \vdash \mathsf{from_sub}_T(\sigma) : T \rightsquigarrow M : B \qquad \mathbf{U} \vdash \mathsf{Ty}(\sigma) = B \rightsquigarrow ()}{\mathbf{U} \vdash s[\![\sigma]\!] \rightsquigarrow t[\![M]\!] : A[\![M]\!]}\,\gamma$$

$$\frac{T : s \rightsquigarrow t : A \qquad \mathbf{U} \vdash L : T \rightsquigarrow M : B \qquad \mathbf{U} \vdash \mathsf{Ty}(L) = B \rightsquigarrow ()}{\mathbf{U} \vdash s[\![L]\!] \rightsquigarrow t[\![M]\!] : A[\![M]\!]}\,\delta$$

$$\frac{}{\mathbf{U} \vdash _ = t \rightsquigarrow ()} \qquad \frac{\mathbf{U} \vdash s \rightsquigarrow t : A}{\mathbf{U} \vdash s = \mathsf{eval}_{\mathsf{id}_{\mathbf{U}}}(t) \rightsquigarrow ()}$$

$$\frac{}{\mathbf{U} \vdash \star \rightsquigarrow \star = [\,]} \qquad \frac{\mathbf{U} \vdash s \rightsquigarrow s' : A \qquad \mathbf{U} \vdash t \rightsquigarrow t' : B \qquad \mathsf{eval}_{\mathsf{id}_{\mathbf{U}}} A = \mathsf{eval}_{\mathsf{id}_{\mathbf{U}}} B}{\mathbf{U} \vdash s \to t \rightsquigarrow s' \to_A t' = (\mathsf{eval}_{\mathsf{id}_{\mathbf{U}}} s', \mathsf{eval}_{\mathsf{id}_{\mathbf{U}}} t') :: \mathsf{eval}_{\mathsf{id}_{\mathbf{U}}} A}$$

$$\frac{\mathbf{U} \vdash s \rightsquigarrow s' : B \qquad \mathbf{U} \vdash t \rightsquigarrow t' : C \qquad \mathbf{U} \vdash A \rightsquigarrow A' = A'' \qquad A'' = \mathsf{eval}_{\mathsf{id}_{\mathbf{U}}} B = \mathsf{eval}_{\mathsf{id}_{\mathbf{U}}} C}{\mathbf{U} \vdash s \to_A t \rightsquigarrow s' \to_{A'} t' = (\mathsf{eval}_{\mathsf{id}_{\mathbf{U}}} s', \mathsf{eval}_{\mathsf{id}_{\mathbf{U}}} t') :: A''}$$

$$\frac{\mathbf{U} \vdash A \rightsquigarrow B = C}{\mathbf{U} \vdash A = C \rightsquigarrow ()} \qquad \frac{\mathbf{U} \vdash s = s' \rightsquigarrow () \qquad \mathbf{U} \vdash t = t' \rightsquigarrow ()}{\mathbf{U} \vdash s \to t = s' \to_A t' \rightsquigarrow ()}$$

$$\frac{\mathbf{U} \vdash s = s' \rightsquigarrow () \qquad \mathbf{U} \vdash t = t' \rightsquigarrow () \qquad \mathbf{U} \vdash A = A' \rightsquigarrow ()}{\mathbf{U} \vdash s \to_A t = s' \to_{A'} t' \rightsquigarrow ()} \qquad \frac{}{\mathbf{U} \vdash _ = C \rightsquigarrow ()}$$

$$\frac{}{T \vdash \ \rightsquigarrow T} \qquad \frac{}{\emptyset \vdash \ \rightsquigarrow \emptyset} \qquad \frac{\Gamma \vdash \ \rightsquigarrow \Delta \qquad \Delta \vdash A \rightsquigarrow B = C}{\Gamma, (v : A) \vdash \ \rightsquigarrow \Delta, (v : B)}$$

$$\frac{\begin{array}{c}\mathbf{U} \vdash s_i \rightsquigarrow t_i : B_i \qquad \rho := \mathsf{eval}_{\mathsf{id}_{\mathbf{U}}}((B_0, t_0, \dots, t_n)) \\ \mathsf{eval}_{\mathsf{id}_{\mathbf{U}}}(B_i) = \mathsf{eval}_\rho(A_i) \qquad \mathbf{U} \vdash \mathsf{flatten}(X) = \mathsf{Ty}(\rho) \rightsquigarrow ()\end{array}}{\mathbf{U} \vdash (X, s_0, \dots, s_n) : (v_0 : A_0), \dots, (v_n : A_n) \rightsquigarrow (B_0, t_0, \dots, t_n)}\,\varepsilon$$

$$\frac{\mathbf{U} \vdash \mathsf{flatten}(x) \rightsquigarrow A}{\mathbf{U} \vdash x : [\,] \rightsquigarrow t : \mathsf{eval}_{\mathsf{id}_{\mathbf{U}}}(A)} \qquad \frac{\mathbf{U} \vdash L_i \rightsquigarrow M_i : (s_i, s_{i+1}) :: A \qquad \mathbf{U} \vdash \mathsf{flatten}(x_i) = s_i \rightsquigarrow ()}{\mathbf{U} \vdash x_0\{L_0\} \cdots \{L_n\} x_{n+1} \rightsquigarrow s_0\{M_0\} \cdots \{M_n\} s_{n+1} : A}$$

Figure 4.3: Bidirectional typing rules.

core (possibly tree) context $\mathbf{U}$, before checking the term $s$ in this context. Checking the term then produces a core syntax term and type, which are inserted into the signature along with the context $\mathbf{U}$. The last def command proceeds as before, checking the context to get a context $\mathbf{U}$ and then checking the term in $\mathbf{U}$, producing a core term $t$ and type $B$. The supplied type $A$ is then checked against $\mathsf{eval}_{\mathsf{id}_{\mathbf{U}}}(B)$. If this check succeeds, the key-value pair $(v, (\mathbf{U}, t, B))$ is added to the signature $\Psi$, identically to the previous case.

The normalise command is used to print the normal form of a term $s$. As with the final two def cases, we begin by checking the context, and checking the term $s$ in the resulting core context to get term $t$ of type $A$. Both $t$ and $A$ are then evaluated to normal form, quoted, and converted back to raw syntax, before being pretty-printed to the user. The size command calculates a primitive estimate of the complexity of a term (which we note is not the same as the syntactic complexity given in Section 4.1.1) by counting the number of constructors in the normal form. To run this command, the term $s$ is checked as before, and converted to a normal form term $t$. Then $\mathsf{size}(t)$ is then calculated by induction by the rules given in Figure 4.4 and this size is printed to the user. The assert command checks both input terms $s$ and $t$, and evaluates the resulting core syntax terms to normal form to check that they are equal. None of the normalise, size, or assert commands modify the signature $\Psi$.

$$\mathsf{size}(\mathsf{coh}[T : A]) = 1 + \mathsf{size}(A) \qquad \mathsf{size}(\mathsf{id}_n) = \mathsf{size}(\mathsf{comp}_T) = 1 \qquad \mathsf{size}(\mathsf{var}_p) = 0$$

$$\mathsf{size}(H⟦L⟧) = \mathsf{size}(H) + \mathsf{size}(L) \qquad \mathsf{size}(L) = \sum_{p:\mathsf{Path}_T} \mathsf{size}(L(p))$$

$$\mathsf{size}([(s_0, t_0), \dots, (s_n, t_n)]) = \sum_{i=0}^{n} (\mathsf{size}(s_i) + \mathsf{size}(t_i))$$

Figure 4.4: Size of normal form syntax.

Finally, the import command reads the contents of the supplied file, parses it as a list of commands, and runs each of these commands with the same signature. The tool has a command line interface, which allows files to be loaded at startup, as well as providing a REPL (read-eval-print loop) which parses one command at a time.

### 4.4.4 Examples

We now demonstrate the use of the tool with some examples. All the examples below can be found in the `/examples` directory of the implementation code base [Ric24b].

We begin defining some standard operations that can be found in a monoidal category or bicategory, which can be found in the file `/examples/monoidal.catt`. We start by defining 1-composition as a coherence:

```
def comp1coh [f,g] = coh [ x{}{}z : x -> z ] (f,g)
```

This example demonstrates the two ways of giving a tree context: in the def command we give the context using the square bracket notation, which only labels the maximal elements, and in the coherence it is given by the full labelling, as we require access to the variables $x$ and $z$ (we note that all other variables of the context have been omitted). This example further demonstrates that a substitution can be applied to a term over a tree context, where we have only specified the locally maximal arguments.

This composite can of course also be given using the comp construction.

```
def comp1 [f,g] = comp
assert comp1coh(f,g) = comp1(f,g) in [f,g]
```

The tree for comp is inferred from the labelling `[f,g]`. The assert statement ensures that these two ways of giving the 1-composition are equal in the theory. The assert passes even with no reduction enabled, demonstrating the value of evaluation in the fully weak case. The horizontal and vertical composites of 2-cells can be given similarly:

```
def horiz [[a],[b]] = comp
def vert [[a,b]] = comp
```

As the vertical composite is the suspension of 1-composition, it can also be given using implicit suspension:

```
def vertsusp [[a,b]] = comp1[a,b]
assert vert(a,b) = vertsusp(a,b) in [[a,b]]
```

In this case, the labelling applied to `comp1` is a tree of height 1 where the locally maximal arguments are given by 2-dimensional terms. Type inference then deduces that the type component of this labelling should be 1-dimensional, and hence evaluation causes the head term `comp1` to be suspended, making it equal to the composite `vert`, as demonstrated by the assertion.

The unitors and associator are then given by the following coherences, using the id builtin for the unitors:

```
def unitor_l = coh [ x{f}y : comp1(id(x),f) -> f ]
def unitor_r = coh [ x{f}y : comp1(f, id(y)) -> f ]
def assoc    = coh [ {f}{g}{h} : comp1(comp1(f,g),h) -> comp1(f,comp1(g,h)) ]
```

which allows definitions to be given for terms which witness the triangle and pentagon equations of monoidal categories:

```
def triangle = coh [ x{f}y{g}z
                   : vert(assoc(f,id(y),g), horiz(id(f),unitor_l(g)))
                     ->
                     horiz(unitor_r(f),id(g))
                   ]

def pentagon = coh [ v{f}w{g}x{h}y{i}z
                   : vert(assoc(comp1(f,g),h,i),assoc(f,g,comp1(h,i)))
                     ->
                     comp [
                       horiz(assoc(f,g,h),id(i)),
                       assoc(f,comp1(g,h),i),
                       horiz(id(f),assoc(g,h,i))
                     ]
                   ]
```

We note the direct use of the comp constructor to easily provide a ternary composite without needing to give a new top-level definition. Using the normalise command, it can be shown that the triangle reduces to the identity with $\textsc{Catt}_{su}$ normalisation enabled, and the pentagon reduces to the identity with $\textsc{Catt}_{sua}$ normalisation enabled.

In the files `/examples/eh.catt` and `/examples/eh-cyll.catt`, we give two Catt proofs of the Eckmann-Hilton argument (see Proposition 1.1.5). In $\textsc{Catt}_{su}$, these both normalise to the following vastly smaller term:

```
def swap = coh [ x{f{a}g}y{h{b}k}z
               : comp[comp [[a],h], comp[g,[b]]]
                 ->
                 comp[comp [f,[b]], comp[[a],k]]
               ]
```

The size command demonstrates that the Catt Eckmann-Hilton proof in `/examples/eh.catt` has size 1807 whereas its $\textsc{Catt}_{su}$ normalisation has a size of only 19. Due to the simplicity of Eckmann-Hilton in $\textsc{Catt}_{su}$, we are able to give $\textsc{Catt}_{su}$ and $\textsc{Catt}_{sua}$ proofs of the syllepsis (see Section 4.2) in `/examples/syllepsis-su.catt` and `/examples/syllepsis.catt` respectively. It can be verified that in $\textsc{Catt}_{sua}$, the $\textsc{Catt}_{su}$ proof of syllepsis, which has size 2745, reduces to the $\textsc{Catt}_{sua}$ proof, which has size 1785.

### 4.4.5 Further work

We end the discussion of this implementation with some options for improving the tool. Each of these suggestions could make the tool easier to use and interact with, which in turn extends what can be achieved with it.

Currently, the tool completely relies on the bidirectional typing rules to perform all of its type inference. While this is effective in some scenarios, for example labellings and implicit suspension, it is lacking in others, such as the lack of implicit arguments in substitutions.

One could try to implement such features by adding meta-variables and a unification procedure to the typechecker. Contrary to the situation for the fully weak Catt, unification for $\textsc{Catt}_{su}$ and $\textsc{Catt}_{sua}$ is non-trivial. Suppose we wished to unify the following two terms:

$$f *_0 g = h *_0 i$$

where $f$,$g$,$h$, and $i$ may contain meta-variables. In Catt, this problem could be reduced to the unification problems $f = h$ and $g = i$. In $\textsc{Catt}_{su}$ however, this cannot be done, as a potential solution is $f = h *_0 i$ and $g = \mathsf{id}$. It is likely that any unification that can be implemented for $\textsc{Catt}_{su}$ (and $\textsc{Catt}_{sua}$) is quite limited, but an investigation into the limits of unification in these settings could be valuable.

Even without a powerful unification algorithm, there are still instances where an argument could be inferred by the tool. One such example is the Eckmann-Hilton term presented in the previous section. This term is defined in the context:

$$(x : \star)\ (\alpha : \mathsf{id}(x) \to \mathsf{id}(x))\ (\beta : \mathsf{id}(x) \to \mathsf{id}(x))$$

Here, the $x$ should be inferable as it is the 0-source of $\alpha$. The tool currently has no way to deduce this.

Separately, improvements could be made to the treatment of unfolding of top-level definitions in the tool. Whenever a term is evaluated by the tool, any top-level definition is unfolded to its normal form. This is not always desirable, as it means that error messages frequently contain fully expanded terms, increasing the length and readability of terms in addition to losing the information associated with the name given to the definition.

Conversely, the full unfolding of evaluation often means that we avoid evaluating terms before displaying them to the user, even when a (partial) evaluation would simplify the term. A notable example is that when giving a new definition, its type is not simplified before being displayed, often resulting in terms such as `p0{x{f}y}`.

A better approach would likely add top-level definitions to the normal form syntax as a head term, allowing their unfolding to be optional. One potential approach for efficient unfolding is given by Kovács [Kov24].

Finally, the accessibility of the tool could be improved with proper editor integration, for example by implementing the language server protocol (see `https://microsoft.github.io/language-server-protocol/`), which would allow errors to be displayed directly in the editor, among other code refactoring features.

## 4.5 Models

Despite claiming that the type theories $\textsc{Catt}_{\mathsf{su}}$ and $\textsc{Catt}_{\mathsf{sua}}$ model semistrict $\infty$-categories, we are yet to discuss their models. In this section we recall the definition of a model for these theories, and discuss some properties of these models.

The definitions of *globular structure* and *globular sum* were given in Chapter 1. Any variant of $\textsc{Catt}_{\mathcal{R}}$ can be equipped with a globular structure by choosing the disc objects to be the disc contexts and letting the source and target maps be given by the inclusions $\lfloor \delta_n^{\epsilon}(D^{n+1}) \rfloor$ for $\epsilon \in \{-,+\}$. We then define the category of models of $\textsc{Catt}_{\mathsf{su}}$ and $\textsc{Catt}_{\mathsf{sua}}$.

**Definition 4.5.1.** Recall that for any tame variant of $\textsc{Catt}_{\mathcal{R}}$, the category $\mathsf{Catt}_{\mathcal{R}}^{\mathsf{ps}}$ is defined to be the restriction of the syntactic category $\mathsf{Catt}_{\mathcal{R}}$ to the ps-contexts. We define the category of models to be the full subcategory of the presheaf category on $\mathsf{Catt}_{\mathcal{R}}^{\mathsf{ps}}$ consisting of functors:

$$F : (\mathsf{Catt}_{\mathcal{R}}^{\mathsf{ps}})^{\mathsf{op}} \to \mathbf{Set}$$

such that $F^{\mathsf{op}}$ preserves globular sums.

Each element of the category of models has the structure of a weak $\infty$-category. For a model $F : (\mathsf{Catt}_{\mathcal{R}}^{\mathsf{ps}})^{\mathsf{op}} \to \mathbf{Set}$, the set of $n$-cells is given by $F(D^n)$, with source and target maps given by the functions:

$$F(\lfloor \delta_{n-1}^{-}(D^n) \rfloor), F(\lfloor \delta_{n-1}^{+}(D^n) \rfloor) : F(D^n) \to F(D^{n-1})$$

for which the globularity equations follow from the globularity of the inclusion maps. For each term over a ps-context in $\textsc{Catt}_{\mathcal{R}}$, an operation on each of the models can be derived. We consider the action of the 1-composition term, given by $\mathcal{C}^1_{[[\,],[\,]]}$. For the model $F$, this induces an operation:

$$F(\{\lfloor \mathcal{C}^1_{[[\,],[\,]]} \rfloor\}) : F(D^1 \vee D^1) \to F(D^1)$$

Due to the preservation of globular sums, we have $F(D^1 \vee D^1) = F(D^1) \amalg_{F(D^0)} F(D^1)$, which is exactly the set of composable 1-cells, which the function above sends to their composition. Similarly, the identity $\mathsf{id}(d_0)$ induces a map $F(D^0) \to F(D^1)$, giving the identity on each 0-cell.

These operations can be combined, to get a compound operation of the following form:

$$F(D^1) = F(D^1) \amalg_{F(D^0)} F(D^0) \xrightarrow{\mathsf{id} \amalg F(\mathsf{id}(d_0))} F(D^1) \amalg_{F(D^0)} F(D^1) \xrightarrow{F(\{\mathcal{C}^1_{[[\,],[\,]]}\})} F(D^1)$$

By the functoriality of $F$ (and preservation of globular sums), this composite should be equal to:

$$F(\{\mathcal{C}^1_{[[\,],[\,]]}\} \bullet \langle d_1, \mathsf{id}_{d_0^+} \rangle) : F(D^1) \to F(D^1)$$

Therefore, if $F$ is further a Catt$_{\mathsf{su}}$ model, then this operation must equal $F(\mathsf{id}) = \mathsf{id}$, enforcing the semistrict properties of Catt$_{\mathsf{su}}$ onto the model.

Throughout the thesis, contexts in Catt$_{\mathcal{R}}$ have been viewed as semistrict $\infty$-categories themselves. This viewpoint can be made precise by the Yoneda embedding, as for each context $\Gamma$ of Catt$_{\mathcal{R}}$, we obtain the presheaf:

$$Y(\Gamma) : \mathsf{Catt}_{\mathcal{R}}^{\mathsf{op}} \to \mathbf{Set}$$

which sends $\Delta$ to $\mathrm{Hom}(\Delta, \Gamma)$, the substitutions from $\Delta$ to $\Gamma$. This map preserves all limits, so in particular its opposite preserves the globular sums, meaning it can be restricted to a model of Catt$_{\mathcal{R}}$. Furthermore, the $n$-cells are given by substitutions $D^n \to \Gamma$, which are precisely the $n$-dimensional terms of $\Gamma$ up to definitional equality.

Since every Catt term is also a Catt$_{\mathcal{R}}$ term, there is an evident functor:

$$K_{\mathcal{R}} : \mathsf{Catt} \to \mathsf{Catt}_{\mathcal{R}}$$

which sends each context and substitution to its equivalence class in Catt$_{\mathcal{R}}$. This functor can be restricted to the functor:

$$K_{\mathcal{R}}^{\mathsf{ps}} : \mathsf{Catt}^{\mathsf{ps}} \to \mathsf{Catt}_{\mathcal{R}}^{\mathsf{ps}}$$

which is the identity on objects. We now prove that this functor preserves globular sums. By [BFM24, Lemma 64], the functor **FinGlob** $\to$ Catt from the category of finite globular sets preserves globular sums, and so it suffices to show that the functor **FinGlob** $\to$ Catt$_{\mathcal{R}}$ preserves globular sum. By [BFM24, Lemmas 25 and 29], it suffices to show that this functor preserves the initial object and preserves pushouts along the inclusion maps $S^n \to D^n$. The empty context is clearly the initial object, and this is preserved by the above functor. For the second property it suffices to show that:

$$\begin{array}{ccc} S^n & \xrightarrow{\{A\}} & \Gamma \\ {\scriptstyle \{\mathrm{wk}(U^n)\}} \downarrow & & \downarrow \\ D^n & \xrightarrow[\{A,x\}]{} & \Gamma, (x : A) \end{array}$$

is a pushout for each $\Gamma \vdash A$ in Catt$_{\mathcal{R}}$. Suppose there is context $\Delta$ with substitutions $\sigma : \Gamma \to \Delta$ and $\{B, t\} : D^n \to \Delta$ such that:

$$\{B\} \equiv \{\mathrm{wk}(U^n)\} \bullet \{B, t\} = \{A\} \bullet \sigma \equiv \{A[\![\sigma]\!]\}$$

Then the universal map is given by $\langle \sigma, t \rangle$, with this map being well-formed as $\Delta \vdash t : B$ and $B = A[\![\sigma]\!]$. The uniqueness of this universal map is clear. Hence, the square above is cocartesian. From this we get the following proposition.

**Proposition 4.5.2.** *The functors $K_{\mathcal{R}}$ and $K_{\mathcal{R}}^{\mathsf{ps}}$ preserve globular sums.*

*Proof.* As the maps **FinGlob** $\to$ Catt and **FinGlob** $\to$ $\text{Catt}_{\mathcal{R}}$ preserve globular sums, the globular sums in both Catt and $\text{Catt}_{\mathcal{R}}$ are given exactly by the ps-contexts. The two functors $K_{\mathcal{R}}$ and $K_{\mathcal{R}}^{\mathsf{ps}}$ are the identity on ps-contexts, and hence preserve globular sums. □

Due to this proposition, any model of $\text{Catt}_{\mathcal{R}}$ can be also seen as a model of Catt, by precomposing with the functor $K_{\mathcal{R}}^{\mathsf{ps}}$. This is to be expected, as intuitively every semistrict $\infty$-category should also be a weak $\infty$-category, where certain operations are given by identities.

### 4.5.1 Rehydration for pasting diagrams

We have shown a way in which every model of $\text{Catt}_{\mathcal{R}}$ can be viewed as a model of Catt. In this section we prove that this mapping from $\text{Catt}_{\mathcal{R}}$ models to Catt models is injective. This implies that being semistrict is a *property* of the model, a particular Catt model can only arise from a unique $\text{Catt}_{\mathcal{R}}$ model, if such a $\text{Catt}_{\mathcal{R}}$ model exists.

We prove this result by demonstrating a partial conservativity result for $\text{Catt}_{\mathcal{R}}$, which we call *rehydration for pasting contexts.* Rehydration refers to the process of taking a term in the semistrict theory, and inserting the necessary coherence morphisms into the term such that it can be typed in Catt. We discuss the difficulties involved with rehydrating an arbitrary term in Section 4.5.2, but for now we are only concerned with the simpler case of rehydrating a term $t : \mathsf{Term}_\Gamma$ where $\Gamma$ is a ps-context. We work towards the following theorem:

**Theorem 4.5.3.** *Let $\mathcal{R}$ be a tame equality rule set that satisfies the support condition and has pruning, disc removal, and endo-coherence removal. Then for any ps-context $\Delta$ and term $t : \mathsf{Term}_\Delta$, there is a Catt term $s : \mathsf{Term}_\Delta$ such that $\Delta \vdash s = t$ in $\text{Catt}_{\mathcal{R}}$.*

We begin with an example for $\text{Catt}_{\mathsf{su}}$. Take the pasting context given by the following diagram:

$$\Delta = \quad w \xrightarrow{f} x \xrightarrow{g} y \xrightarrow{h} z$$

The associator $\alpha$ is a $\text{Catt}_{\mathsf{su}}$ normal form term over $\Delta$, and we can further define the term:

$$\eta : \mathsf{id}((f * g) * h) \to \alpha_{f,g,h} * \alpha_{f,g,h}^{-1}$$

as a single coherence over $\Delta$. This term is also a $\text{Catt}_{\mathsf{su}}$ normal form. Finally the term:

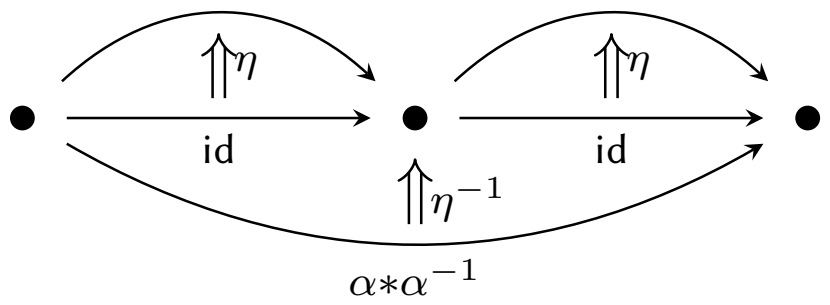

is a $\text{Catt}_{\mathsf{su}}$ normal form term over a pasting context, which is not well-formed in Catt. Such a term can be rehydrated by inserting the equivalence $\mathsf{id} \cong \mathsf{id} * \mathsf{id}$ into the centre of the term.

Performing a similar construction with the interchanger instead of the associator creates a $\text{CATT}_{\text{sua}}$ normal form term over a pasting context which is not a CATT term.

We now proceed with the proof of Theorem 4.5.3. We introduce three operations, which are mutually defined on terms of $\text{CATT}_{\mathcal{R}}$ over pasting contexts.

- The *rehydration* $R(t)$ of a term $t$ recursively rehydrates all subterms of $t$, and then pads the resulting term. For any $\text{CATT}_{\mathcal{R}}$ term $t$, the rehydration is a CATT term over the same context. For any term $t$, we call $R(N(t))$ its *rehydrated normal form*, where $N$ is the function taking any term to its normal form. We similarly define the rehydration $R(A)$ of a type $A$ over a pasting context and $R(\sigma)$ of a substitution $\sigma$ whose domain and codomain are pasting contexts.
- The *padding* $P(t)$ of a CATT term $t$, which composes the term with coherences to ensure that its boundaries are in rehydrated normal form.
- The normaliser $\phi(t)$, a coherence term from $t$ to its rehydrated normal form $R(N(t))$ for any CATT term $t$.

We give formal definitions for each of these, which we define mutually with proofs of the following statements, where we assume $\Delta$ and $\Gamma$ are pasting contexts:

(1) Suppose $\Delta \vdash_{\mathcal{R}} t : A$. Then $\Delta \vdash R(t) : R(N(A))$. Similarly, if $\Delta \vdash_{\mathcal{R}} A$ or $\Delta \vdash_{\mathcal{R}} \sigma : \Gamma$, then $\Delta \vdash R(A)$ and $\Delta \vdash R(\sigma) : \Gamma$.

(2) For a $\text{CATT}_{\mathcal{R}}$ well-formed term $t$, type $A$, and substitution $\sigma$, we have $\Delta \vdash t = R(t)$, $\Delta \vdash A = R(A)$, and $\Delta \vdash \sigma = R(\sigma)$ in $\text{CATT}_{\mathcal{R}}$.

(3) Suppose $\Delta \vdash t : A$ for a CATT term $t$, then $P_k(t)$ is well-formed for $k \leq \text{dim}(t)$ and $\Delta \vdash P(t) : R(N(A))$.

(4) Suppose $t$ is a well-formed CATT term. Then for each $k \leq \text{dim}(t)$, $P_k(t) = t$.

(5) If $\Delta \vdash t : R(N(A))$ in CATT, then $\Delta \vdash \phi(t) : t \to_A R(N(t))$.

(6) Let $t$ be a well-formed CATT term over a pasting context. Then $\phi(t) = \text{id}(t)$.

Each of these definitions and proofs are given by an induction on dimension and subterms, ensuring that they are well-founded.

We begin with the definition of the rehydrated term, type, and substitution.

**Definition 4.5.4.** Let $\Delta$ and $\Gamma$ be a pasting context. For a term $t$ or type $A$ over $\Delta$, or a substitution $\sigma : \Gamma \to \Delta$, we define the rehydrations:

$$R(t) : \mathsf{Term}_\Delta \qquad R(A) : \mathsf{Type}_\Delta \qquad R(\sigma) : \Gamma \to \Delta$$

by mutual recursion. For a variable $x$, we let $R(x) = x$, and for a coherence term we define:

$$R(\mathsf{Coh}_{(\Gamma\,;\,A)}[\sigma]) = P(\mathsf{Coh}_{(\Gamma\,;\,R(A))}[R(\sigma)])$$

For types and substitutions, we recursively apply the rehydration to all subterms.

To define the padding, we need the composites over certain trees $T_k^n$ for $k < n$ which are defined by:

$$T_0^n = [[\,], D^{n-1}, [\,]] \qquad T_{k+1}^{n+1} = \Sigma(T_k^n)$$

As an example $T_1^3$ produces the following context:

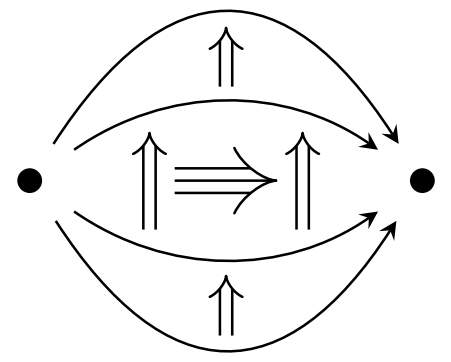

The composite over this context allows us to "fix" the 1-dimensional boundary of a 3-dimensional term.

**Definition 4.5.5.** Let $t$ be an $n$-dimensional term of a pasting diagram $\Delta$. Define its padding $P(t)$ to be equal to $P_n(t)$ where:

$$P_0(t) = t \qquad P_{k+1}(t) = \mathcal{C}^n_{T^n_k}[\![\langle \phi(\mathsf{src}_k(P_k(t)))^{-1}, P_k(t), \phi(\mathsf{tgt}_k(P_k(t)))\rangle]\!]$$

where $\mathsf{src}_k$ and $\mathsf{tgt}_k$ give the $k$ dimensional source and target of a term.

Consider the term $\alpha : f \to_{x \to_\star y} g$. As an example, we build the following sequence of paddings:

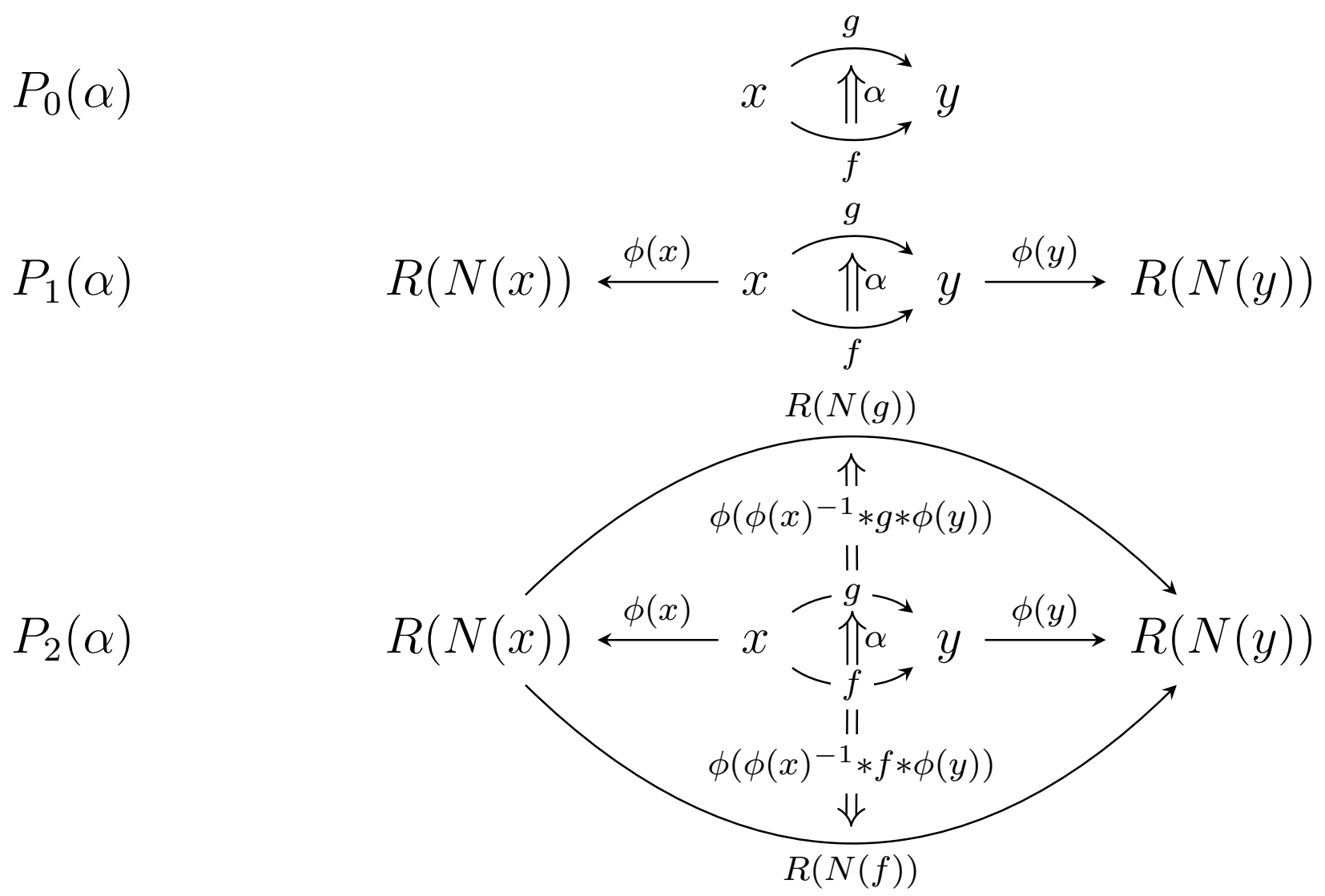

We lastly define the normaliser coherences. As these are each built from a coherence constructor with the rule for equivalences, they can all be inverted.

**Definition 4.5.6.** Let $t$ be a term of a pasting diagram $\Delta$. By Corollary 4.2.11, $\mathsf{Supp}(t)$ is a pasting diagram, and we let $i_t$ be the inclusion $\mathsf{Supp}(t) \to \Delta$. Then we define the normaliser $\phi(t)$:

$$\phi(t) = \mathsf{Coh}_{(\mathsf{Supp}(t)\,;\, t \to R(N(t)))}[i_t]$$

By assumption, $R(N(t)) = N(t) = t$ and so $\mathsf{Supp}(R(N(t))) = \mathsf{Supp}(t)$, making the above term well-formed.

We now prove the required properties, starting with statement (1). The statements for types and substitutions follow by a simple induction using the case for terms, as if $A = B$ then $R(N(A)) = R(N(B))$ (as $N(A) = N(B)$). The case for a variable is also trivial, so assume that:

$$\Delta \vdash_{\mathcal{R}} \mathsf{Coh}_{(\Gamma\,;\,B)}[\sigma] : A$$

Then it follows from induction on subterms that $\Gamma \vdash R(B)$ and $\Delta \vdash R(\sigma) : \Gamma$, and so:

$$\Delta \vdash \mathsf{Coh}_{(\Gamma\,;\,R(B))}[R(\sigma)] : R(B)[\![R(\sigma)]\!]$$

Then by induction on statement (3), we get:

$$\Delta \vdash P(\mathsf{Coh}_{(\Gamma\,;\,R(B))}[R(\sigma)]) : R(N(R(B)[\![R(\sigma)]\!]))$$

By induction on statement (2), we have $R(B)[\![R(\sigma)]\!] = B[\![\sigma]\!]$. By inspection of the original typing derivation, we have $B[\![\sigma]\!] = A$, and so $R(N(R(B)[\![R(\sigma)]\!])) \equiv R(N(A))$, as required.

Now consider statement (2). The cases for types and substitutions follow by an easy induction from the result for terms. Since the case for variables is trivial, we restrict to the cases for the coherence terms, where we must prove that:

$$\Gamma \vdash_{\mathcal{R}} \mathsf{Coh}_{(\Delta\,;\,A)}[\sigma] = P(\mathsf{Coh}_{(\Delta\,;\,R(A))}[R(\sigma)])$$

By (1), $\mathsf{Coh}_{(\Delta\,;\,R(A))}[R(\sigma)]$ is a well-formed Catt term, and so by (4) and induction on subterms we have:

$$P(\mathsf{Coh}_{(\Delta\,;\,R(A))}[R(\sigma)]) = \mathsf{Coh}_{(\Delta\,;\,R(A))}[R(\sigma)] = \mathsf{Coh}_{(\Delta\,;\,A)}[\sigma]$$

For statement (3), we let $\Delta \vdash t : A$ and prove for each $k$ that $P_k(t)$ is well-formed and that $\mathsf{src}_m(P_k(t)) \equiv R(N(\mathsf{src}_m(t)))$ and $\mathsf{tgt}_m(P_k(t)) \equiv R(N(\mathsf{tgt}_m(t)))$ for $m \leq k$. We proceed by induction on $k$. The case for $k = 0$ is trivial, so we must prove that $P_{k+1}(t)$ is well-formed, which is the term:

$$\mathcal{C}^n_{T^n_k}[\![\langle \phi(\mathsf{src}_k(P_k(t)))^{-1}, P_k(t), \phi(\mathsf{tgt}_k(P_k(t)))\rangle]\!]$$

By (5), noting that the inductive hypothesis on $k$ implies that the types of $\mathsf{src}_k(P_k(t))$ and $\mathsf{tgt}_k(P_k(t))$ are in rehydrated normal form, we have that the normalisers are well-typed. Therefore, $P_{k+1}(t)$ is well-formed by the previous fact and the inductive hypothesis on $k$. By simple calculation it follows that:

$$\mathsf{src}_m(P_k(t)) \equiv \mathsf{src}_m(P_m(t)) \equiv \mathsf{src}(\phi(\mathsf{src}_m(t))^{-1}) \equiv R(N(\mathsf{src}_m(t)))$$

with a similar equation holding for the target. It then follows that $\Delta \vdash P(t) : R(N(A))$.

Statement (4) holds by a simple induction on $k$, using statement (6) to reduce each normaliser to an identity, and then using pruning and disc removal to get the equality:

$$\mathcal{C}^n_{T^n_k}[\![\langle \mathsf{id}(\mathsf{src}_k(P_k(t))), P_k(t), \mathsf{id}(\mathsf{tgt}_k(P_k(t)))\rangle]\!] = P_k(t)$$

which along with the inductive hypothesis on $k$ is sufficient.

For statement (5), we assume $\Delta \vdash t : R(N(A))$. Then, by (1) and the preservation rule, we have $\Delta \vdash \Delta \vdash R(N(t)) : R(N(R(N(A)))) \equiv R(N(A))$, where the equality follows from (2) and the idempotency of the normal form functor. The typing for the normaliser then trivially follows, as $t$ and $R(N(t))$ are full in Supp($t$).

For statement (6), we apply statement (1) to get that $t = N(t) = R(N(t))$. Therefore:

$$\begin{aligned}
\phi(t) &\equiv \mathsf{Coh}_{(\mathrm{Supp}(t)\,;\,t\to R(N(t)))}[i_t] \\
&= \mathsf{Coh}_{(\mathrm{Supp}(t)\,;\,t\to t)}[i_t] \\
&= \mathsf{id}(t)[\![i_t]\!] && \text{by endo-coherence removal} \\
&\equiv \mathsf{id}(t)
\end{aligned}$$

This completes all parts of the definitions and proofs. Then for any well-formed $\text{Catt}_{\mathcal{R}}$ term $t$, $R(N(t))$ is a well-formed Catt term with $R(N(t)) = t$ in $\text{Catt}_{\mathcal{R}}$ completing the proof of Theorem 4.5.3. Moreover, if $t = t'$ then $R(N(t)) \equiv R(N(t'))$, and so the rehydrated of $\text{Catt}_{\mathcal{R}}$ terms over pasting contexts can be chosen to respect $\text{Catt}_{\mathcal{R}}$ equality. From this we get the following corollary.

**Corollary 4.5.7.** *Semistrictness is a property. Let $\mathcal{R}$ is a tame equality rule set satisfying the support and preservation conditions in addition to having pruning, disc removal, and endo-coherence removal. If $F$ and $G$ are $\text{Catt}_{\mathcal{R}}$ models such that:*

$$F \circ K^{\mathsf{ps}}_{\mathcal{R}} = G \circ K^{\mathsf{ps}}_{\mathcal{R}}$$

*then $F = G$.*

*Proof.* Since $K^{\mathsf{ps}}_{\mathcal{R}}$ is the identity on objects, it follows that $F$ and $G$ must be equal on objects. Now let $\Gamma$ and $\Delta$ be pasting diagrams, and let $\Gamma \vdash_{\mathcal{R}} \sigma : \Delta$. Then by Theorem 4.5.3 we have, $\Gamma \vdash R(\sigma) : \Delta$ and so:

$$F(K^{\mathsf{ps}}_{\mathcal{R}}(R(\sigma))) = G(K^{\mathsf{ps}}_{\mathcal{R}}(R(\sigma)))$$

but $K^{\mathsf{ps}}_{\mathcal{R}}$ is simply an inclusion, so $F(R(\sigma)) = G(R(\sigma))$ and since $R(\sigma) = \sigma$ in $\text{Catt}_{\mathcal{R}}$, we have $F(\sigma) = G(\sigma)$. The substitution $\sigma$ was arbitrary, so $F = G$ as required. □

The above result holds in particular for the equality rule sets su and sua, meaning that a model of Catt can be a model of $\text{Catt}_{\mathsf{su}}$ or $\text{Catt}_{\mathsf{sua}}$ in at most one way.

### 4.5.2 Towards generalised rehydration

The rehydration result of the previous section can be viewed as a partial conservativity result, stating that in a pasting context, $\text{Catt}_{\mathsf{su}}$ and $\text{Catt}_{\mathsf{sua}}$ have the same expressive power as Catt. The original motivation of semistrictness was to strictify parts of the theory without losing the expressiveness of the fully weak setting. We would therefore hope that the rehydration results of Section 4.5.1 extend to arbitrary contexts.

Such a result would be a powerful tool for constructing terms in a weak setting; a term could be constructed by constructing it in the semistrict setting, before applying rehydration to the resulting term to get term in the fully weak setting. Such a technique would allow a Catt proof of Eckmann-Hilton to be constructed mechanically from the vastly simpler $\text{Catt}_{\mathsf{su}}$ Eckmann-Hilton proof, or even give a proof of the Syllepsis in Catt, for which no proof has been given as of writing.

By observing the proof of Theorem 4.5.3, we see that the main part that would need replacing for a general rehydration result is the construction of the normalisers, as we can no longer rely on the source and target term of our normaliser living over a pasting diagram that allows the construction of a single coherence. A natural way to proceed is to attempt to build a normaliser $\phi(t) : t \to R(N(t))$ by recursion on the reduction sequence $t \rightsquigarrow^* N(t)$. We consider a context with $x : *$ and a scalar $\alpha : \mathsf{id}(x) \to \mathsf{id}(x)$, and consider the reduction by pruning:

$$\alpha *_0 \mathsf{id}(x) \rightsquigarrow (\alpha)$$

where $(\alpha)$ is the unary composite on $\alpha$. We immediately encounter two problems:

- For each individual reduction, the source and target of the reduction may not have the same type. In the example above, the source has type $\mathsf{id}(x) * \mathsf{id}(x) \to \mathsf{id}(x) * \mathsf{id}(x)$, but the target has type $\mathsf{id}(x) \to \mathsf{id}(x)$. A normaliser between these two terms can therefore not be directly constructed.
- If the source term is padded such that it has the same type as the target term, we can run into a separate problem. Consider the reduction given above again. The following normaliser can be formed:

  $$\mathsf{Coh}_{(D^2\,;\,\rho^{-1}_{d_1^-} *_1 (d_2 *_0 \mathsf{id}(d_0^+)) *_1 \rho_{d_1^+} \to (d_2))}[\langle\{\alpha\}\rangle]$$

  which has source given by the padded term:

  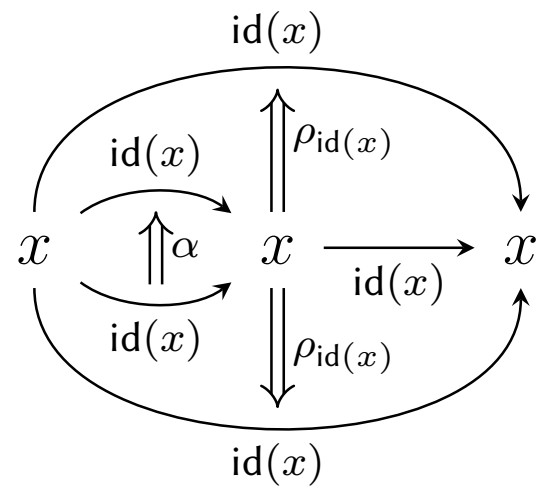
  

  However this term is padded by the right unitor on each side, which is not the canonical normaliser from $\mathsf{id}(x) * \mathsf{id}(x)$ to $\mathsf{id}(x)$, the unbiased unitor.

The reduction above was not only chosen to demonstrate both of these problems, but was chosen as it is the problematic reduction that is encountered if one tries to rehydrate the Eckmann-Hilton term from $\textsc{Catt}_{\mathsf{su}}$. To give a proof of Eckmann-Hilton, one reaches a critical point where a left unitor and right unitor on the identity must be cancelled out, highlighting the second of the two problems.

To solve the second problem one could attempt to prove that for any two reductions paths from $t$ to $N(t)$, that there is a higher cell between the normalisers generated from each reduction path, critically relying on the confluence proof for the theory to modularise the problem into finding fillers for each confluence diamond. Such an approach seems infeasible for the following reasons: To find fillers for a confluence diamond, we presumably must already know the form of all rehydrations in the dimension below, which themselves could depend on filling confluence diamonds of the dimension below. This seems to necessitate rehydrating on a dimension by dimension basis, making the full rehydration problem infeasible. It is also likely that at some point it would be necessary to show that two different fillers of a confluence diamond have a higher cell between them, leading to some form of $\infty$-groupoid flavoured confluence problem. Such a problem also seems infeasible with the tools currently available to us.

An alternative approach could be to show that the "space" of all rehydrations is contractible. This can be made precise in the following way. Let $t$ be a $\textsc{Catt}_{\mathcal{R}}$ term. Then consider the globular set whose 0-cells are $\textsc{Catt}$ terms $s$ which are equal to $t$ in $\textsc{Catt}_{\mathcal{R}}$, 1-cells are given by $\textsc{Catt}$ terms $f : s \to s'$ which are equal to $\mathsf{id}(t)$ in $\textsc{Catt}_{\mathcal{R}}$, in general $n$-cells given by $\textsc{Catt}$ terms that are equal to $\mathsf{id}^n(t)$. The contractability of such a globular set is exactly the property needed for rehydration, as it gives the existence of a 0-cell $s$ which gives the rehydration, and witnesses the essential uniqueness of this rehydration.

Such a contractability proof can be given when the term $t$ is a term of a pasting diagram, as any higher cells can be given by a simple coherence. This allows us to fix the padding in the example above, observing that the right unitor is equivalent to the unbiased unitor. It is however unclear how such a contractability proof could be extended to arbitrary contexts.

We now turn our attention to the first problem presented above. One method for tackling this problem is to give normalisers as a *cylindrical equivalence* instead of a regular equivalence. A cylindrical equivalence can be viewed as the canonical notion of equivalence between two objects of different types. We introduce the first few dimensions of cylinder terms. A 0-cylinder is simply a 1-dimensional term. A 1-cylinder from a cylinder $f : w \to x$ to a cylinder $g : y \to z$ can be defined by the square:

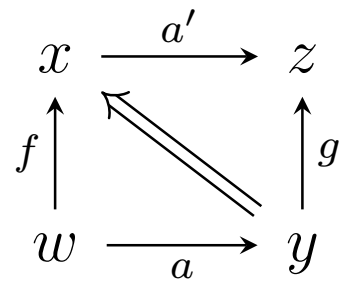

where the central arrow has type $a * g \to f * a'$. If such a cylinder was invertible, which is the case when $a$, $b$, and the two-dimensional cell are invertible, then it would be a cylindrical equivalence and would witness the equivalence of $f$ and $g$. Suppose two 1-cylinders $\alpha : f \to g$ and $\beta : g \to h$ as below:

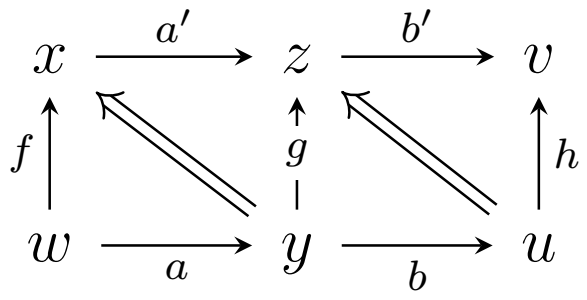

Then a composite cylinder $f \to h$ could be formed by letting the front "face" be given by $a * b$, the back "face" be given by $a' * b'$ and the filler given by a combination of associators and whiskerings of the two fillers in the diagram. A 2-cylinder could be given by the following diagram:

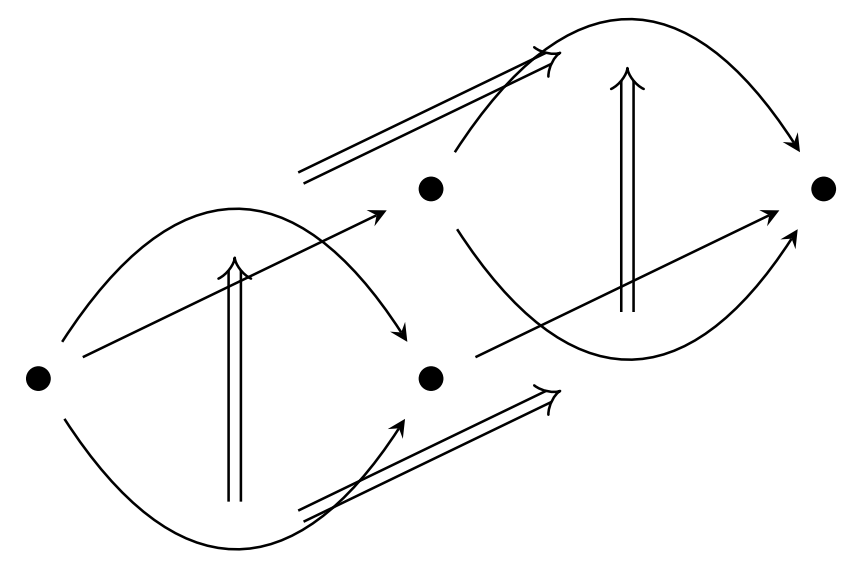

where the top and bottom faces of this diagram are 1-cylinders, and the whole digram should be filled by a 3-dimensional term with appropriate source and target. The shape of this diagram gives the name to this construction.

When using cylinders to represent the normalisers in a rehydration process, the inductive step for coherences would require a cylinder to be generated from a cylindrical version of the substitution attached to the coherence. We have seen that this can be done when the coherence is given by 1-composition, but achieving full rehydration would involve giving cylindrical versions of every operation in Catt. No such proof has been given for any variety of globular weak $\infty$-categories.

We offer an alternative solution which avoids defining cylinder composition, which we call *rehydration by dimension*. From an equality rule set $\mathcal{R}$, we can form the rule sets $\mathcal{R}_n$ which consists of the rules in $(\Gamma, s, t) \in \mathcal{R}$ such that $\dim(s) = \dim(t) \leq n$. Rehydration by dimension attempts to rehydrate an $n$-dimensional term $t$ by constructing terms $t_n, \dots, t_0$ such that $t_i$ is a term which is well-formed in $\textsc{Catt}_{\mathcal{R}_i}$, creating a rehydration sequence:

$$\textsc{Catt}_{\mathcal{R}_n} \to \textsc{Catt}_{\mathcal{R}_{n-1}} \to \cdots \to \textsc{Catt}_{\mathcal{R}_1} \to \textsc{Catt}_{\mathcal{R}_0}$$

The term $t_n$ is given immediately by $t$, and $t_0$ is then a term of $\textsc{Catt}_{\mathcal{R}_0} = \textsc{Catt}$, giving the rehydration of $t$. The key insight of this method is that when generating the normaliser for a particular $k$-dimensional generating rule $s \rightsquigarrow t$, we know by the preservation property that the types of $s$ and $t$ are equal, and so are further equal in $\textsc{Catt}_{\mathcal{R}_{k-1}}$. By factoring through these partial rehydrations, the normaliser of a dimension $k$ generating rule only has to be valid in $\textsc{Catt}_{\mathcal{R}_{k-1}}$, meaning that the normalisers can again be given by regular equivalences.

Unfortunately, this method does not avoid the need to define new classes of operations in Catt, as we could be required to prove that arbitrary Catt operations are natural in their lower-dimensional arguments. Consider terms $f : x \to y$ and $g : y \to z$ and suppose the $\textsc{Catt}_{\mathcal{R}_1}$ normal form of $y$ is $y'$ with normaliser $\phi(y)$. Then, during a rehydration proof to $\textsc{Catt}_{\mathcal{R}_0}$, it may be required to give a normaliser from $f * g$ to $(f * \phi(y)) * (\phi(y)^{-1} * g)$, effectively requiring us to prove that 1-composition is natural in its central 0-cell. Similarly to the case with cylinders, in this case for 1-composition, such a normaliser can easily be given, but we possess no way of creating such naturality arguments on arbitrary coherences.

The proofs of Eckmann-Hilton given in Section 4.4.4 give an example of the result of each of these methods, with the proof in `/examples/eh.catt` proceeding by "rehydration by dimension", and the proof in `/examples/eh-cyll.catt` using cylinders. In both proofs, the only example of the second problem we encounter is proving that the left and right unitors on the identity are equivalent to the unbiased unitor. For the cylinder proof, the composition of 1-cylinders is used and is given by the term `cyl_comp`, which is then implicitly suspended by the tool. The rehydration by dimension proof needs a naturality move like the one described above, which is given by the term `compat_move`.

## 4.6 Future ideas

In this final section, we collect together some ideas for the continuation of this work, including ideas for different semistrict theories based on $\textsc{Catt}_{\mathcal{R}}$, and modifications to the existing theories. Some ideas for future avenues of research have already been discussed, such as the potential improvements to the implementation discussed in Section 4.4.5, and the discussion of full rehydration given in Section 4.5.2, which we will not repeat here.

**Further results for $\text{Catt}_{\text{sua}}$** The metatheory of $\text{Catt}_{\text{sua}}$ is more complicated than the corresponding metatheory of $\text{Catt}_{\text{su}}$, though at first glance the relative increase in power does not match this complexity. The jump from Catt to $\text{Catt}_{\text{su}}$ vastly simplified the proof of Eckmann-Hilton, allowed the syllepsis to be proven, and lead to results such as disc trivialisation. In contrast, $\text{Catt}_{\text{sua}}$ provides no further simplification to Eckmann-Hilton and only slightly simplifies the syllepsis, removing some associators from the proof.

One potential utility of $\text{Catt}_{\text{sua}}$ could be simplifying the composites of cylinders, as briefly introduced in Section 4.5.2. Consider the following diagram from that section which contains two composable 1-cylinders.

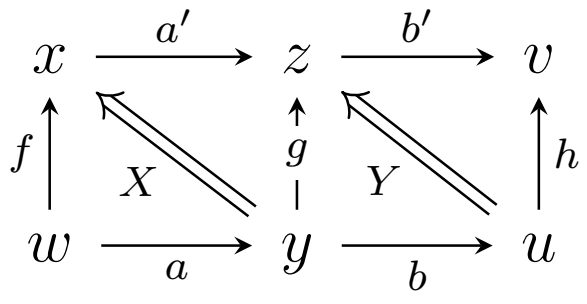

In Catt, the 1-composite of these cylinders is a term $(a * b) * h \to f * (a' * b')$ given by:

$$\alpha_{a,b,h} *_1 (a *_0 Y) *_1 \alpha^{-1}_{a,g,b'} *_1 (X *_0 b') * \alpha_{f,a',b'}$$

where each $\alpha$ term is an associator. This would of course simplify in $\text{Catt}_{\text{sua}}$ to $(a *_0 Y) *_1 (X *_0 b')$. Such a simplification could make it simpler to define higher cylinder coherences, such as associator for 1-cylinders, which would be trivial in $\text{Catt}_{\text{sua}}$, but far more involved in Catt.

Further future work for $\text{Catt}_{\text{sua}}$ could involve the search for an analogue of disc trivialisation for $\text{Catt}_{\text{sua}}$. We would expect there to be a more general class of contexts that are trivialised by $\text{Catt}_{\text{sua}}$ but are not trivialised. The contexts present in the cylinder contexts presented above could form a starting point for such a study.

A separate avenue for further study is to explore the links between $\text{Catt}_{\text{sua}}$ and more graphical presentations of semistrict $\infty$-categories. String diagrams are a common graphical method for working with monoidal categories and bicategories [Sel11], and their higher-dimensional counterparts, such as those implemented in the tool homotopy.io, can be viewed as strictly associative and unital finitely presented $\infty$-categories, much like contexts of $\text{Catt}_{\text{sua}}$. Translation results in either direction between these two settings, while highly non-trivial due to the contrast in the way each system approaches composition, would be valuable.

**Generalised insertion** The conditions given for insertion in Section 3.4 were not the most general conditions possible. In this section, we stated that to perform insertion we required an insertion redex $(S, P, T, \mathbf{U}, L, M)$, and one of the conditions of this insertion redex was that:

$$L(\overline{P}) \equiv \mathcal{C}_T^{\mathsf{lh}(P)}[\![M]\!]$$

It turns out that it is sufficient to give the weaker condition that the locally maximal argument is a coherence where the type contained in the coherence is sufficiently suspended:

$$L(\overline{P}) \equiv \mathsf{Coh}_{(T\,;\,\Sigma^{\mathsf{bh}(P)}(A))}[M]$$

As $\mathcal{C}^{n+1}_{\Sigma(T)} \equiv \Sigma(\mathcal{C}^n_T)$, and the original condition required that $\mathsf{th}(T) \geq \mathsf{bh}(P)$, this alternative condition is a strict generalisation of the previous condition. Under the new condition, the exterior labelling must be modified. It firstly must take the type $A$ as an argument. The case for $P = [k]$ is then modified such that $\kappa_{S,[k],T,A}$ (noting the extra type subscript) is given by:

$$\begin{array}{ccccc}
[S_0, \dots, S_{k-1}] & \mathbin{+\!\!+} & T & \mathbin{+\!\!+} & [S_{k+1}, \dots, S_n] \\
\uparrow \mathsf{id} & & \uparrow \{A, \mathsf{Coh}_{(T\,;\,A)}[\mathsf{id}_T]\} & & \uparrow \mathsf{id} \\
[S_0, \dots, S_{k-1}] & \vee & \Sigma S_k & \vee & [S_{k+1}, \dots, S_n]
\end{array}$$

when $S = [S_0, \dots, S_n]$. The inductive step of the exterior labelling then relies on the type $A$ being sufficiently suspended to proceed, just as the original version depends on the trunk height of $T$ being sufficient to proceed (we note that the trunk height condition is still needed in this generalisation). For the necessary typing judgements to be satisfied, we must have $\mathrm{src}_0(A) \equiv \mathrm{fst}(\lfloor T \rfloor)$ and $\mathrm{tgt}_0(A) \equiv \mathrm{snd}(\lfloor T \rfloor)$, but no other extra condition is necessary.

In some ways, this definition of insertion is more natural than the definition given earlier. We no longer rely on the syntactic condition of the locally maximal argument being a standard coherence, only relying on the far weaker suspendability property. In the proof for confluence of Catt$_{\mathsf{sua}}$, a large focus was cases where a reduction modified a standard coherence into a term which was no longer a standard coherence. Cases like these do not happen with generalised insertion, as reductions do not break the suspendability property. More generally, a confluence proof for generalised insertion does not require any proof about the interaction of insertion with boundary inclusion maps and standard coherences (given in Section 3.4.3 for the original definition).

Unfortunately, this generalised form of insertion cannot be directly used in Catt$_{\mathsf{sua}}$ without breaking confluence. Let $\Gamma$ be the following context given by the following diagram:

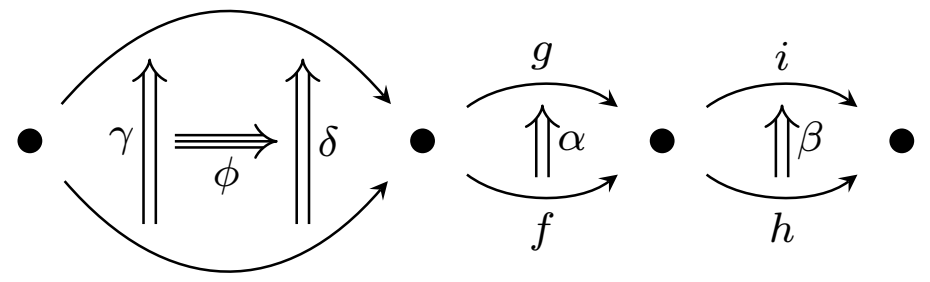

and consider the terms:

$$\begin{aligned}
I &= (\alpha *_0 h) *_1 (g *_0 \beta) \\
E &= \mathsf{Coh}_{(\mathrm{Supp}(I)\,;\,I \to I)}[\mathsf{id}] \\
X &= \phi *_0 E
\end{aligned}$$

We now have the following critical pair: $X$ can reduce by inserting the locally maximal argument $E$, as the branch has branching height $0$ making the suspendability condition vacuous, but $E$ also reduces by endo-coherence removal. By performing the generalised insertion we obtain the coherence:

$$\mathsf{Coh}_{(\Gamma\,;\,\gamma *_0 I \to \delta *_0 I)}[\mathsf{id}]$$

Let $W(x, y, z)$ refer to the standard composite over the diagram:

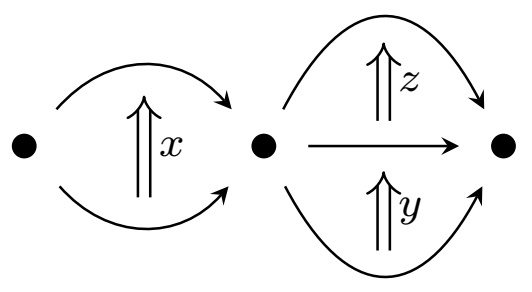

Then the coherence term above admits further cell reductions which convert the composites $\gamma *_0 I$ and $\delta *_0 I$ to $W(\gamma, (\alpha *_0 h), (g *_0 \beta))$ and $W(\delta, (\alpha *_0 h), (g *_0 \beta))$. The resulting term reduces no further. If the endo-coherence removal is performed, then $E$ reduces to $\mathsf{id}(I)$, which can be pruned from the original composite. After further reductions, we obtain a coherence over the context $\Delta$ given by the following diagram:

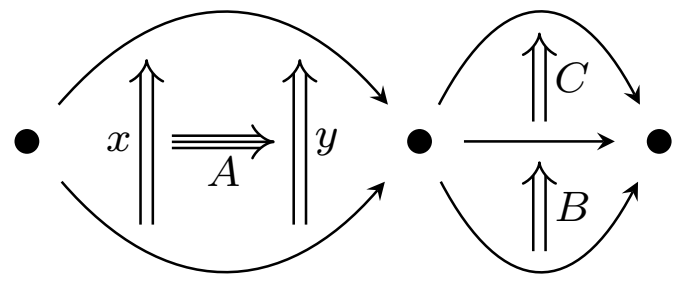

In particular, the result of these reductions is the following coherence:

$$\mathsf{Coh}_{(\Delta\,;\,W(x,B,C) \to W(y,B,C))}[\langle \phi, (\alpha *_0 h), (g *_0 \beta) \rangle]$$

which admits no further reductions, hence breaking confluence. It is even unclear which of these reduction paths is the more canonical for such a system, the first moves the complexity of $I$ to the type in the coherence, whereas the second keeps the complexity of $I$ in the arguments of the coherence. Conjecturally, one could consider generalisations to endo-coherence removal which could factor out the common structure of $W(\gamma, (\alpha *_0 h), (g *_0 \beta))$ and $W(\delta, (\alpha *_0 h), (g *_0 \beta))$, reducing the result of the first reduction path to the result of the second reduction path, though we have not explored any such definition.

**A further strictification to $\textsc{Catt}_{\mathsf{sua}}$** Douglas and Henriques give an explicit representation a Gray category [DH16, Definition 2.8], which can be used as a direct point of comparison to $\textsc{Catt}_{\mathsf{sua}}$, as Gray categories are semistrict 3-categories with strict unitors and associators. The weak structure in their presentation of Gray categories is given by an invertible 3-cell they call *switch*, which has the same form as the Catt term which we called swap in Section 4.2.

In their paper, all of the equalities between 2-cells are generated by a set of axioms [S2-4] to [S2-15]. Each of these equalities is contained in the definitional equality of $\textsc{Catt}_{\mathsf{sua}}$, with the exception of [S2-9] and [S2-10], which witness a compatibility between whiskering and vertical composition. We consider the axiom [S2-9], as [S2-10] can be treated symmetrically. Let $\Delta$ be the context given by diagram:

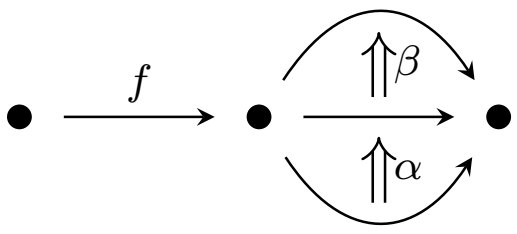

and consider the following terms of $\Delta$:

$$(f *_0 \alpha) *_1 (f *_0 \beta) \qquad f *_0 (\alpha *_1 \beta)$$

while the second term reduces to the standard composite over $\Delta$, the first does not reduce, as no insertion can be performed due to the condition on trunk height, and hence these two terms are not equal in $\textsc{Catt}_{\mathsf{sua}}$, unlike in Gray categories. Although it could be argued that these axioms reside in the interchange family of laws for $\infty$-categories, one could attempt to define a stricter version of $\textsc{Catt}_{\mathsf{sua}}$ which incorporates these equalities, with the aim of proving that 3-truncated models of this stricter type theory are equivalent to Gray categories.

**Strict interchange** In contrast to the reductions in this thesis which strictify units, one could instead consider reductions that strictify all composition, making the associativity and interchange laws strict, leaving only units weak. Such a form of semistrictness is often called *Simpson semistrictness*, due to a conjecture of Simpson [Sim98] that leaving units weak is sufficient to retain the full expressiveness of weak $\infty$-categories.

To achieve this, one could try an approach similar to insertion of merging arguments of a term into the head coherence, when all the involved terms are standard coherences. To be able to strictify terms such as the swap term given in Section 4.2, the trunk height condition of insertion must be dropped. This immediately leads to composites over contexts which are not pasting diagrams: Consider the context generated by the diagram:

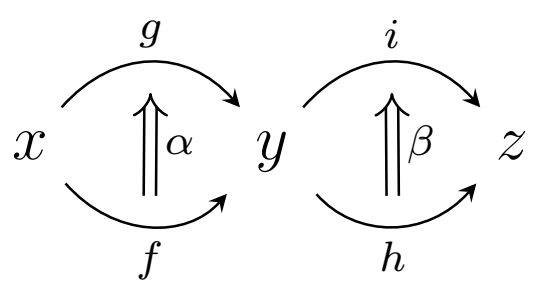

and then consider the following composite in this context:

$$\alpha *_0 ((\beta *_0 \mathsf{id}(\mathsf{id}(z))) *_1 \rho_i)$$

where $\rho_i$ is the right unitor on $i$. Allowing a more general form of merging would lead to this term becoming a composite of the following form:

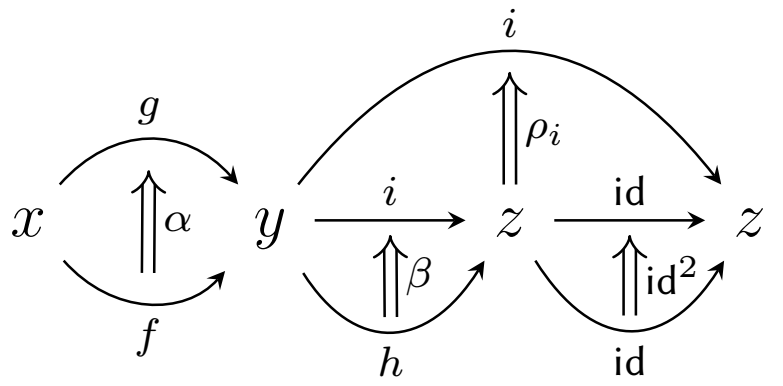

Although this diagram is not a pasting diagram, as it is not a globular set, we would still expect it to fulfil a similar contractability property to the one pasting diagrams do. One may therefore be lead to believe that strict interchange could be achieved in a type theory similar to Catt by allowing a more general class of pasting diagrams. This, however, does not work. We consider the following counterexample due to Forest [For22]: let $\Gamma$ be the context generated by the following diagram.

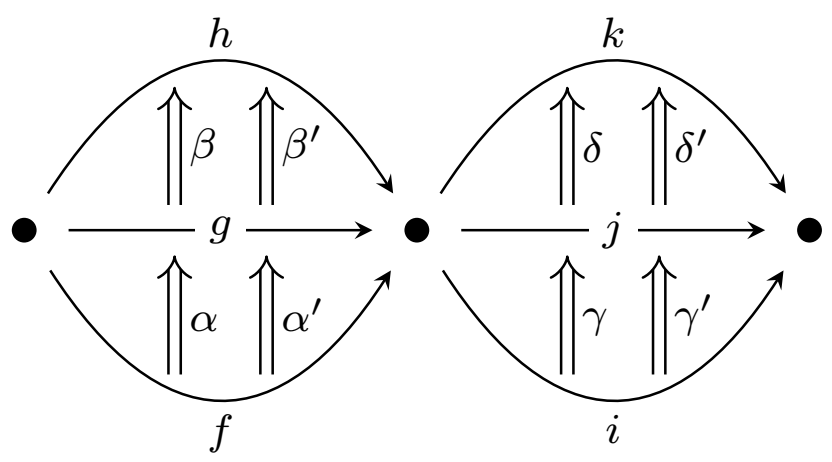

and let $\Delta = \Gamma, (X : \alpha *_0 \delta \to \alpha' *_0 \delta'), (Y : \beta *_0 \gamma \to \beta' *_0 \gamma')$. We then have the following distinct composites:

$$\begin{pmatrix} f *_0 \gamma \\ *_1 \\ X \\ *_1 \\ \beta *_0 k \end{pmatrix} *_2 \begin{pmatrix} \alpha' *_0 i \\ *_1 \\ Y \\ *_1 \\ h *_0 \delta' \end{pmatrix} \not\cong \begin{pmatrix} \alpha *_0 i \\ *_1 \\ Y \\ *_1 \\ h *_0 \delta \end{pmatrix} *_2 \begin{pmatrix} f *_0 \gamma' \\ *_1 \\ X \\ *_1 \\ \beta' *_0 k \end{pmatrix}$$

which are intuitively the composite of $X$ and $Y$ in either order, where $X$ and $Y$ have been whiskered with the appropriate terms. We note that the matrix notation above is only used to aid comprehension, and does not represent the application of any matrix operations. The approach described above of merging together composites would lead to both of the above composites of $X$ and $Y$ being reduced to the same composite over $\Delta$, contradicting the viability of such an approach.

An alternative, non-rewriting based approach could be defined by the following equality rule:

$$\left\{ (\Gamma, s[\![\sigma]\!], t[\![\sigma]\!]) \;\middle|\; \begin{array}{l} s \text{ and } t \text{ are pure composite terms,} \\ s = t \text{ in a strict } \infty\text{-category} \end{array} \right\}$$

where a *pure composite* is a term constructed only using standard composites. Such an approach avoids the counter example above, as the two composites of $X$ and $Y$ are not equal in a strict $\infty$-category, and so would not be equated in the type theory generated by this equality rule set.

We note that due to an algorithm of Makkai [Mak05], which is also described and implemented by Forest [For21], it can be decided whether terms $s$ and $t$ are equal in a strict $\infty$-category. Therefore, to decide equality of the above system, we need a method of finding the correct decomposition of a term into a substitution applied to a purely compositional term. We conjecture that there exists a factorisation system on Catt with the left class of morphisms given by purely compositional substitutions, substitutions whose contained terms are all pure composites, which could be used for this purpose. We leave all details of such a construction for future work.